\documentclass[useAMS,usenatbib]{mn2e}

\usepackage{graphicx}

\usepackage{amsmath}
\usepackage{amssymb}

\newcommand{\apj}{ApJ}
\newcommand{\apjs}{ApJS}
\newcommand{\aj}{AJ}
\newcommand{\aap}{A\&A}
\newcommand{\aaps}{A\&AS}
\newcommand{\apss}{Ap\&SS}
\newcommand{\mnras}{MNRAS}

\newcommand{\im}{i_\mathrm{m}}
\newcommand{\um}{n_1}
\newcommand{\uvm}{n_2}

\title[ETV analysis of eccentric Kepler triples]{Eclipse Timing Variation Analyses of Eccentric Binaries with Close Tertiaries in the {\it Kepler} field}
\author[T. Borkovits et al.]{T. Borkovits$^{1,2}$\thanks{E-mail:
borko@electra.bajaobs.hu (TB)}, S. Rappaport$^3$, T. Hajdu$^4$, J. Sztakovics$^4$ \\
$^1$Baja Astronomical Observatory, H-6500 Baja, Szegedi \'ut, Kt. 766, Hungary\\
$^2$ELTE Gothard-Lend\"ulet Research Group, H-9700 Szombathely, Szent Imre herceg \'ut 112, Hungary \\
$^3$M.I.T. Department of Physics and Kavli Institute for Astrophysics and Space Research, 70 Vassar St.,Cambridge, MA, 02139\\ 
$^4$Astronomical Department of E\"otv\"os University, H-1118 P\'azm\'any P\'eter stny. 1/A, Budapest, Hungary}

\begin{document}

\date{Accepted ??? Received ???; in original form ???}

%\pagerange{\pageref{firstpage}--\pageref{lastpage}} \pubyear{2014}

\maketitle

\label{firstpage}

\begin{abstract}
We report eclipse timing variation analyses of 26 compact hierarchical triple stars comprised of an eccentric eclipsing (`inner') binary and a relatively close tertiary component found in the {\em Kepler} field. We simultaneously fit the primary and secondary $O-C$ curves of each system for the light-travel time effect (LTTE), as well as dynamical perturbations caused by the tertiary on different timescales. For the first time, we include those contributions of three-body interactions which originate from the eccentric nature of the inner binary. These effects manifest themselves both on the period of the triple system, $P_2$, and on the longer ``apse-node" timescale. We demonstrate that consideration of the dynamically forced rapid apsidal motion yields an efficient and independent tool for the determination of the binary orbit's eccentricity and orientation, as well as the 3D configuration of the triple. Modeling the forced apsidal motion also helps to resolve the degeneracy between the shapes of the LTTE and the dynamical delay terms on the $P_2$ timescale, due to the strong dependence of the apsidal motion period on the triple's mass ratio. This can lead to the independent determination of the binary and tertiary masses without the need for independent radial velocity measurements. Through the use of our analytic method for fitting $O-C$ curves we have obtained robust solutions for system parameters for the ten most ideal triples of our sample, and only somewhat less robust, but yet acceptable, fits for the remaining systems.  Finally we study the results of our 26 system parameter fits via a set of distributions of various physically important parameters, including mutual inclination angle, and mass and period ratios.
\end{abstract}

\begin{keywords}
methods: analytical -- stars: multiple -- stars: eclipsing 
\end{keywords}

\section{Introduction}

Amongst the richly populated family of eclipsing binaries (hereafter `EBs') which offer a ``royal road'' to stellar astrophysics \citep{russell48}, eccentric systems represent an especially important subgroup. For example, the rate of apsidal motion (precession of the orbital ellipse) in such systems is a direct observable and, before the dawn of asteroseismology, this phenomenon offered the first observational probe of stellar interiors \citep[see, e.g,][and references therein]{claretgimenez93}. The same effect can also serve as a check on the predictions of general relativity or, more generally, for testing alternative theories of gravity\footnote{Though in recent years such tests are much better done with binary radio pulsars. See, for example, \citet{taylor95} and \citet{weisbergetal10}} \citep[see, e.g.][]{moffat84}. Furthermore, statistical studies of the orbital eccentricity distribution in connection with stellar age and spectral type yield strong constrains on tidal dissipation theories, and thus also for models of stellar interiors \citep[see e.g.,][for a review]{mazeh08}. The mutual interactions of the stellar surfaces, and stellar envelopes as well, with the ever changing gravitational tidal field of the companion stars as they move along their eccentric orbits, lead to additional exotic effects, which have also become observable these days thanks to ultraprecise, space-based photometry.  Such periodically time varying interactions offer a seemingly inexhaustible source of phenomena to be studied with present-day asteroseismology.  ``Heartbeat'' binaries (e.g.,\citealt{thompsonetal12}) represent perhaps the most spectacular class of these effects, but tidally induced stellar oscillations have also been detected in several other systems, even at unexpectedly low orbital eccentricities and large separations. Note that a heartbeat system, of course, does not necessarily exhibit eclipses \citep[e.g., the remarkable KOI-54 itself is seen nearly pole-on;][]{welshetal11}, 
%and tidally induced oscillations can also be observed in non-eclipsing systems, 
but the presence of eclipses strongly constrains several orbital and astrophysical parameters and can therefore provide additional benefits. (On the other hand, however, we note that the eclipses may seriously inhibit the detection and analysis of driven stellar oscillations both in the time and frequency domains because they occur at the same frequencies; see e.g., \citealp{debosscheretal13,dasilvaetal14}, and references therein). Last, but not least, we mention that careful, accurate photometric and spectroscopic analyses of detached EBs may lead to very precise stellar masses and radii, which are important for calibrating and testing stellar structure and evolution theories \citep[see e.g,][and references therein]{torresetal10}.  As eccentric binaries are necessarily detached systems, they may also be good candidates for such investigations. 

There is, however an evident selection effect that has led to the {\em under}representation of eccentric EBs, especially before the era of the long-duration (years), quasi-continuous photometric sky-surveys (such as, e.g., MACHO \citealp{alcocketal93}\footnote{In the context of the classification of LMC EBs in the MACHO database see also \citet{derekasetal07}.}, OGLE, \citealp{udalskietal08} and others), not to mention the space-based photometry over the last few years (e.g., MOST, \citealp{walkeretal03}, CoRoT, \citealp{auvergneetal09}, {\em Kepler}, \citealp{gillilandetal10}). This anti-selection effect arises from the generally longer orbital periods of eccentric EBs, as well as from the fact that, due to the larger orbital separation of the binary members, tidal effects (and other possible star--star interactions) that may induce specific signatures in the out-of-eclipse light curves, are hardly observable in most cases with ground-based photometry; both of these effects reduce the chance of direct or serendipitous discovery. Another fact that strongly works against the discovery of eccentric EBs, especially via ground-based photometric studies of EBs with periods of months and longer, is that for such a system an eclipse event becomes longer than a night, or even a day, which certainly makes it nearly impossible to measure accurate eclipse times for such binaries.

Breaking this disadvantageous trend, the four-year-long, nearly continuous, and high-precision {\it Kepler} observations have led to the discovery of nearly a thousand new eccentric EBs\footnote{The Aug. 22 2014 update of the Kepler Eclipsing Binary Catalog V3 (http://keplerebs.villanova.edu) contained 578 systems with period larger than 15.0 days. A few of them might be either non-eclipsing HB stars, or other false positives, but most of them certainly should be eccentric EB. Furthermore, there are several eccentric systems among the shorter period EBs. For example, the shortest period eccentric EB in our present sample has an orbital period of $P_1=3.99$ days.}. Furthermore, several dozens of these new eccentric EBs have been found to be members of exotic, compact hierarchical triple (hereafter ``CHT'') stellar systems (see, e.g., \citealt{rappaportetal13,conroyetal14}). The compactness, i.e., the small characteristic size of the whole triple -- or, higher multiple -- system (and/or the low outer vs.~inner period ratio) presents new challenges for both star formation and stellar evolution theories and even, in the context of their dynamical evolution and stability, for celestial mechanics. 

One of the new challenges, for example, is whether these recently found CHTs fit in with one of the suggested mechanisms for the formation of close binary systems, or if there is a need for alternate scenarios? The formation of the closest binaries requires one or more effective mechanisms for orbital shrinkage \citep[see][for a short discussion of this question]{fabryckytremaine07}. For young binaries, where neither star is evolved and, therefore, mass-exchange can be ruled out, the most widely accepted model is the Kozai-Lidov-cycle with tidal friction (``KCTF''; \citealt{kozai62,lidov62,kiselevaetal98,eggletonkiseleva01}) mechanism, where a distant third object, in an initially highly inclined orbit forces the orbital shrinkage of the originally wide inner binary.  According to the detailed investigations of \citet{fabryckytremaine07}, and more recently \citet{naozfabrycky14}, this mechanism places statistical constraints on the period and mutual inclination angle distributions of the finally evolved, relaxed hierarchical triples. Unfortunately, for most of the previously known triple systems, the mutual inclination angle of the two orbital planes, which would be a key-parameter for checking model predictions, cannot be determined readily. It can be measured only in an indirect way over a long time interval and with great effort, and only then with the use of high-tech instruments with restricted availability. (The methodology of such measurements, and its obstacles are summarized, e.g., in \citealp{borkovitsetal10}; two examples are given by the pioneering effort of \citealp{lestradeetal93}, and a very recent study by \citealp{laneetal14}.) It is therefore not surprising that there are only about ten triple systems, containing close binaries, where the mutual orbital inclination angle was determined before space missions, which is evidently insufficient for statistical considerations. On the other hand, the compactness of the recently discovered CHTs implies more easily detectable short-term, significant mutual gravitational perturbations.  These allow for a quick and direct determination of the mutual inclination angle as well as the mass ratio, all of which can be extracted from the eclipse timing variations (ETV) of close EBs.

In this context, small outer-vs.~inner-period ratios, and short outer periods combine to make the investigation of such systems considerably more interesting. The lack of ternary components with periods shorter than ($P_2 \lesssim 1000 ~\mathrm{d}$) was noted already by \citet{tokovininetal06}. In his more recent study the same author also notes the complete absence of such third companions for a distance-limited sample of triple systems comprised at least partly of solar-type dwarfs \citep{tokovinin14b}. Even considering triples formed by non-solar type stars (mostly more massive, but excluding non-degenerate stars), only a very limited sample of such short outer period triples was known before the {\em Kepler} era\footnote{Amongst them, $\lambda$~Tauri was considered to be an extreme case both for its very short outer period of $P_2\sim33^\mathrm{d}$, and low period ratio of $P_2/P_1\sim8.3$. After the first four years of {\em Kepler} observations this unique system still guards its first-place status with the shortest outer period; however, KOI-126 has approached very closely with $P_2\sim33.9^\mathrm{d}$ \citealt{carteretal11}.  But, the glory of possessing the smallest period ratio is now held by KIC~07668648 with $P_2/P_1\sim7.3$, a system first identified in our previous work \citep{rappaportetal13}, and which is included also into the present study.}. In the present work, 8 of the 26 systems that we have investigated have outer periods shorter than 1 year, and an additional 8 remain under the ``magic threshold'' of 1000 days. Therefore, these systems can serve as observational probes at the highly underpopulated short-end of the outer period domain in regard to the conclusions of the above mentioned works of \citet{fabryckytremaine07} and \citet{naozfabrycky14}.

Another issue which arises is that the perturbations of such a close ternary may significantly counteract the synchronization and circularization processes of eccentric binary systems. Such effects may not result simply in a delayed orbital circularization, but these perturbations can actually generate highly eccentric orbits even in a previously coplanar and circularized system \citep{lietal14}. Note, a hierarchical triple consisting of a host star and two giant planets, with low mutual inclination angle, but large inner eccentricity (Kepler-419), was found recently by \citealt{dawsonetal14}). Therefore, it is important to obtain some information on the frequency of such CHTs, because in the absence of such information, statistical results related to tidal circularization and synchronization processes should be considered with caution.

For CHTs all the orbital parameters are subject to periodic perturbations on different timescales. Although, these variations naturally affect all kinds of observations (e.g., light and radial velocity curves, etc.), they can be best studied through ETV analyses. With this approach, in theory, we can determine the full spatial configuration of a CHT, which is adequate for modeling the dynamical evolution of individual systems (see discussion in \citealt{borkovitsetal11}). Furthermore, not only the outer mass ratio, but - at least, in some special cases - the individual masses can also be determined from eclipse timing (see \citealt{borkovitsetal13}). 

The analysis of ETVs (or, in the case of exoplanetary systems, TTVs) that are driven by gravitational perturbations can be carried out by following either a numerical or an analytic approach. In the former case, the equations of motion are integrated numerically, and the eclipse timing pattern can thereby be emulated and then compared to the observed ETV curve.  Usually the fitting is done by the use of some bayesian methods (mostly MCMC). (For a very recent example of software operating in this manner see \citealp{borsatoetal14}.) This approach has led to very spectacular results in the identification and confirmation of multiple exoplanetary systems \citep[see, e.g.,][and references thereins]{steffenetal13,mazehetal13}.  Further examples of systems analysed this way would include the above mentioned Kepler-419 \citep{dawsonetal14} and the Solar system analog KIC 11442793 (Kepler-90) \citep{cabreraetal14}. 

Recently, something of a hybrid approach, but still substantially a numerical method, was developed by \citet{decketal14} which integrates an approximate Hamiltonian of the system under investigation instead of the equations of motion, thereby yielding a very fast technique. 

In a purely analytic method, however, there is no need for time-consuming numerical integrations, which must be done for many possible realizations of the system configurations. Rather, the analytic approach provides a theoretical expression for the ETV curves in closed-form analytic (mostly trigonometric) functions of time, where the system parameters occur as additional (time-)dependent variables. In such a way these formulae, as well as  their analytic derivatives with respect to the different system parameters, can be quickly and easily calculated, thereby offering extremely fast parameter inversion methods.  Unfortunately, however, for a typical planetary configuration with comparable separations between the bodies, and also more specifically for the case of mean motion resonances, an analytic description with sufficient accuracy would require an enormous number of higher order (e.g., in eccentricity) trigonometric terms, and therefore the analytic method becomes essentially impractical and unusable. However, for the case of a hierarchical system configuration, the formulae become substantially simpler, as was discussed, e.g., in \citet{borkovitsetal11}. Because triple (and multiple) stellar systems, due to stability criteria, form almost exclusively as hierarchical systems, an analytic approach to the investigation of such systems remains quite effective, fast, and readily-applicable.  In this paper we follow such an analytic method.

Perhaps the most important advantage of the analytic approach to modeling ETV curves is that it allows us to gain a deeper insight into the astrophysics of the problem.  It shows us the functional dependences of the formulae on the different parameters, and may even reveal further qualitative and/or quantitative relationships.

Previous work, in the context of the analytic description of ETVs in hierarchical triple systems, concentrated almost exclusively on the middle of the three classes of timescales for the periodic perturbations occurring in such triples\footnote{According to the original classification of \citet{brown36}, the three categories are the:
\begin{itemize}
\item[--]{{\it short-period perturbations,} for which the typical period is of the order of the binary period, $P_1$, and the amplitude is related to $(P_1/P_2)^2$, where $P_2$ is the period of the outer binary.}
\item[--]{{\it long-period perturbations,} with a characteristic period of $P_2$, and amplitude of $P_1/P_2$ and,}
\item[--]{{\it apse-node terms,} having period about $P_2^2/P_1$, and the amplitude may reach unity.}
\end{itemize}
At this point we emphasize that this classification scheme differs substantially from the more conventional categorization of the perturbations, followed by e.g., Harrington in his pioneering works on the stellar three-body problem \citep{harrington68,harrington69}, and most of his followers (in accordance with the convention of planetary perturbation theory). In this latter theory the $P_2$ time-scale perturbations are also counted within the ``short-period'' category, and our group of ``apse-node'' perturbations are referred as ``long-period'' terms.}. In the present paper we improve the analytical description of ETVs with the inclusion of both the smallest amplitude, shortest period ``short-term'' terms, and the longest period, ``apse-node'' timescale apsidal- and orbital-plane precession terms. This is necessary for a more precise, correct modeling of the continually lengthening data series for dynamically less relaxed, non-coplanar, eccentric CHTs.  For these systems the amplitude of the smallest magnitude and shortest period terms may substantially exceed the detection limit at the short end of the timescales, and the characteristic ``apse-node'' periods may be as short as a few decades at the other end.  As we demonstrate, such improvement in the analytic method results in other benefits as well, since the inclusion of these terms may resolve some degeneracies and ambiguities within parameter space. 

In Sect.~\ref{Sec:analyticmodel} we give a longer summary of our extended analytical model. Then a short description of the numerical code and method are presented in Sect.~\ref{Sec:code}, while the principles of the system selection, and the data preparation are outlined in Sect.~\ref{Sec:dataprep}. Our results and associated discussion are presented in Sects.~\ref{sec:results} and \ref{sec:stat}.  There we present the ETVs and the fitted solutions for 26 CHTs; 10 of these are from \citet{rappaportetal13}, and the remaining 16 are reported here for the first time. Finally, after a short summary (Sect.~\ref{sec:summary}) we give the detailed expressions for the long-term octupole, the short timescale, and the apse-node terms of the analytic model (Appendices~\ref{app:octupole}, \ref{app:shortperiod} and \ref{app:apsidalmotion}, respectively).  We discuss those geometric constrains which are related to the spatial configuration of the system, and also those that follow from the precession of the orbits in Appendix~\ref{app:Sphericaltriangle}. Finally, we describe the extended numerical tests of our fitting process in Appendix \ref{app:numericalanalysis}.

\begin{table}
 \caption{Meaning the symbols used in the paper}
 \label{Tab:symbols}
% \begin{center}
 \begin{tabular}{@{}lll}
  \hline
  Parameter & symbol & explanation \\
\hline
Mass & & \\
~~CB members & $m_\mathrm{A,B}$ &\\
~~total mass of CB & $m_\mathrm{AB}$ & $m_\mathrm{A}+m_\mathrm{B}$ \\
~~ternary's mass & $m_\mathrm{C}$ & \\
~~total mass & $m_\mathrm{ABC}$ & $m_\mathrm{A}+m_\mathrm{B}+m_\mathrm{C}$ \\
~~CB's mass ratio & $q_1$ & $m_\mathrm{B}/m_\mathrm{A}$ \\
~~WB's mass ratio & $q_2$ & $m_\mathrm{C}/m_\mathrm{AB}$ \\ 
\hline
Period &  &\\
~~sidereal    & ${P_\mathrm{s}}_{1,2}$ & \\
~~anomalistic & ${P_\mathrm{a}}_{1,2}$ & \\
\hline
Semi-major axis & & \\
~~relative orbit & $a_{1,2}$ & \\
~~absolute orbit of CB & $a_\mathrm{AB}$ & $m_\mathrm{C}/m_\mathrm{ABC}\cdot a_2$ \\
\hline
eccentricity  & $e_{1,2}$ & \\
\hline
mean anomaly & $l_{1,2}$ & \\
\hline        
true anomaly & $v_{1,2}$ & \\
\hline
true longitude & & see Fig.~\ref{Fig:krsz-ek}, App.~\ref{app:Sphericaltriangle}\\
~~observational & $u_{1,2}$ & $v_{1,2}+\omega_{1,2}$ \\
~~dynamical     & $w_{1,2}$ & $v_{1,2}+g_{1,2}$ \\
&&$u_{1,2}-n_{1,2}+(0,1)\times\pi$ \\
\hline
argument of periastron &  & see Fig.~\ref{Fig:krsz-ek}, App.~\ref{app:Sphericaltriangle}\\
~~observational & $\omega_{1,2}$ &  \\
~~dynamical  & $g_{1,2}$      & $\omega_{1,2}-n_{1,2}+(0,1)\times\pi$ \\
\hline
inclination &  & see Fig.~\ref{Fig:krsz-ek}, App.~\ref{app:Sphericaltriangle}\\
~~observable & $i_{1,2}$ & \\
~~dynamical  & $j_{1,2}$ & \\
~~mutual (relative) & $\im$ & $j_1+j_2$ \\
& $I$ & $\cos\im$ \\
~~invariable plane to the sky & $i_0$ & \\
\hline
ascending node & & see Fig.~\ref{Fig:krsz-ek}, App.~\ref{app:Sphericaltriangle}\\
~~observational & $\Omega_{1,2}$ & \\
& $\Delta\Omega$ & $\Omega_2-\Omega_1$ \\
~~dynamical & $h$ & \\
~~sky -- dyn. nodes angle & $n_{1,2}$ & \\
& $\alpha$ & $\uvm-\um$ \\
& $\beta$ & $\uvm+\um$ \\
\hline
time of periastron passage & $\tau_{1,2}$ & \\
\hline
speed of light & $c$ & \\ 
Gravity constant & $G$ & \\
\hline
\end{tabular}
%\end{center}

Note, CB and WB are abbreviations for close (i.e., inner) and wide (outer) binaries, respectively.
\end{table}

\section{Outlines of the Analysis}
\label{Sec:analyticmodel}

\subsection{General remarks}

The present paper is a natural continuation and extension of the previous work of \citet{borkovitsetal03,borkovitsetal07,borkovitsetal11} from the theoretical side, and of \citet{rappaportetal13} in terms of the application of the analytic perturbation theory for analyzing close hierarchical triple star systems discovered by the {\it Kepler} spacecraft in recent years. In this series of previous papers we gave detailed descriptions both of the fundamentals of the applied physical model, i.e., the hierarchical stellar three-body problem (including historical references), and the method of calculation of the different contributions to the ETV. Therefore, we give only a brief summary here. 

Multiple stellar systems almost exclusively exhibit hierarchical configurations. Restricting ourselves to triple stars, `hierarchical' means that one of the three distances which can be formed mutually among the three constituent stars remains substantially smaller (by at least an order of magnitude) than the other two distances during the whole life-time of the system. In such cases the motion of the three stars can be more or less well approximated by two 2-body (or Keplerian) systems. Therefore, this problem can be discussed in the framework of the (perturbed) motion of two binaries: an `inner', or close binary formed by the two closer members, and an `outer', or wide binary consisting of the more distant third star, and the center of mass of the inner binary. Then, the usual sets of orbital elements can be defined for both orbits, and the time-dependent variations of these elements describe the orbital behavior. Here, as before, we study the variations of the orbital elements of the inner, eclipsing binary, in the context of their effect on the occurrence and variations of the mid-eclipse times.  These ETVs can be accurately determined from the unprecedentedly precise, and nearly continuous four year-long observations of {\it Kepler}.

Before enumerating the different effects affecting the ETVs, we comment on the notations that we have followed. In the formulae below, different sets of orbital elements will appear. Subscript `1' refers to the orbital elements and related quantities of the inner orbit (the eclipsing binary) or, more precisely, the relative orbit of the secondary component of the eclipsing binary around the primary star of the binary. Similarly, subscript `2' denotes the orbital elements of the ternary's relative orbit around the center-of-mass of the close binary. Furthermore, since the occurrences of the eclipses depend mainly on the relative positions of the bodies with respect to the observer, while the gravitational perturbations depend on their relative positions with respect to each other, two different sets of the angular orbital elements appear in the equations. For example, $\omega_i$ will denote the argument of periastron in the observational frame (i.e., measured from the ascending node of the $i$-th orbital plane and the plane of the sky), while $g_i$ will refer to the corresponding quantity in the dynamical frame (i.e., measured from the ascending node of the orbital plane, and the system's invariable plane). We summarize the quantities that are used in Table~\ref{Tab:symbols}. Furthermore, the meaning and relation among the different elements can be seen in Fig.~\ref{Fig:krsz-ek}, and are also given in Appendix~\ref{app:Sphericaltriangle}.

\begin{figure}
\includegraphics[width=84mm]{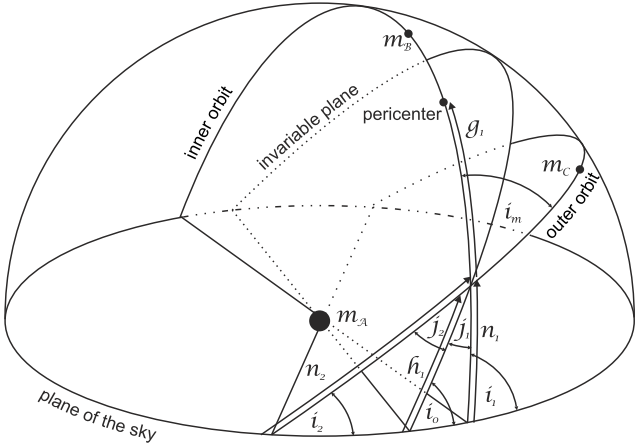}
 \caption{The meaning of the different kinds of angular elements}
 \label{Fig:krsz-ek}
\end{figure}

\subsection{The contributions of Eclipse Timing Variations}

We define the general form of the ETV as follows:
\begin{eqnarray}
\Delta&=&T(E)-T(0)-P_\mathrm{s}E \nonumber \\
&=&\sum_{i=0}^{2}c_iE^i+\left[\Delta_\mathrm{LTTE}+\Delta_\mathrm{dyn}+\Delta_\mathrm{apse}\right]_0^E,
\end{eqnarray}
where, on the first row, $T(E)$ denotes the observed time of the $E$-th eclipse, $T(0)=T_0$ indicates the reference epoch, i.e., the observed time of the ``zeroth'' eclipse, while the constant $P_\mathrm{s}$ stands for the sidereal (or eclipsing) period. Furthermore, the $c_0$, $c_1$ coefficients give corrections in $T_0$ and $P_\mathrm{s}$, respectively, while $c_2$ is equal to half of the constant period-variation rate per cycle ($\Delta P/2$), independent of its origin. Finally, $\Delta_\mathrm{LTTE}$, $\Delta_\mathrm{dyn}$ and $\Delta_\mathrm{apse}$ refer to the contributions of light-travel time effect (LTTE), short period dynamical perturbations, and apsidal motion effect (AME, including longer time-scale dynamical perturbations) to the ETVs, respectively. Note, the integer values of cycle number $E$ refer to the primary, and half-integers to the secondary eclipses. 
In the following we briefly discuss each of the above mentioned components.

%\begin{enumerate}
{\it (i) Light Travel Time effect (LTTE):}
This is the classical Roemer delay that arises from the changing distance of the eclipsing binary from the observer during its revolution around the center of mass (CM) of the triple system. This effect is well-observed in hundreds of eclipsing systems. LTTE is a close analog of the Doppler shift in the radial velocities in binaries.  It produces exactly the same information which can be obtained from an SB1 radial velocity curve. It can be written as
\begin{equation}
\Delta_\mathrm{LTTE}=-\frac{a_\mathrm{AB}\sin i_2}{c}\frac{\left(1-e_2^2\right)\sin(v_2+\omega_2)}{1+e_2\cos v_2},
\label{Eq:LITE} 
\end{equation}
Note, the negative sign on the r.h.s.~comes from the fact that in the LTTE-term the motion of the binary is reflected, and we applied the relation $\omega_\mathrm{AB}=\omega_2+180\degr$.
By the use of Kepler's third law, the mass function can be defined as
\begin{equation}
f(m_\mathrm{C})=\frac{m_\mathrm{C}^3\sin^3i_2}{m_\mathrm{ABC}^2}=\frac{4\pi^2a_\mathrm{AB}^3\sin^3i_2}{GP_2^2}
\label{Eq:mass-function-def}
\end{equation}
and thus, the amplitude of LTTE can be written as
\begin{eqnarray}
{\cal{A}}_\mathrm{LTTE}&=&\frac{G^{1/3}}{c}\left(\frac{P_2}{2\pi}\right)^{2/3}f(m_\mathrm{C})^{1/3}\sqrt{1-e_2^2\cos^2\omega_2} \nonumber \\
&\approx&1.1\times10^{-4}\frac{m_\mathrm{C}\sin i_2}{m_\mathrm{ABC}^{2/3}}P_2^{2/3}\sqrt{1-e_2^2\cos^2\omega_2},
\label{A_LTTE}
\end{eqnarray}
where, in the last row, masses are given in units of $M_\odot$, the period in days, and the amplitude is also expressed in days.

The LTTE term carries information about the following parameters: $P_2$, $e_2$, $\omega_2$, $\tau_2$ (or its equivalents), and the projected semi-major axis $a_\mathrm{AB}\sin i_2$ or, the mass function $f(m_\mathrm{C})$. 

{\it (ii) $P_2$ time-scale dynamical effects:} 
This is the medium of the three classes of periodic perturbations (both in amplitude, and period) in hierarchical triple configurations. Such perturbations in the context of eclipse timing variations were first analyzed by \citet{soderhjelm75}, and \citet{mayer90}. Later, a corrected and easily applicable form was given in \citet{borkovitsetal03} which was adequate insofar as the inner binary had a circular orbit and, furthermore, the third companion was sufficiently distant that terms of higher order than quadruple of the perturbing potential (or force) would be negligible. \citet{rappaportetal13} successfully applied the combination of LTTE and their dynamical model for the identification and preliminary orbital parameter determination of 39 close hierarchical triple systems amongst {\it Kepler} eclipsing binary stars. A natural extension of the previous model for eccentric inner binaries was carried out by \citet{borkovitsetal11}. For eccentric inner binaries the equations (up to the first order in the $a_1/a_2$ ratio) take the following form:
\begin{eqnarray}
\Delta_\mathrm{1}&=&\frac{P_1}{2\pi}A_\mathrm{L1}\left(1-e_1^2\right)^{1/2}\left\{\left[\frac{8}{15}f_1+\frac{4}{5}K_1\right]\right.{\cal{M}} \nonumber \\
&&+\left(1+I\right)\left[K_{11}{\cal{S}}(2u_2-2\alpha)-K_{12}{\cal{C}}(2u_2-2\alpha)\right] \nonumber \\
&&+\left(1-I\right)\left[K_{11}{\cal{S}}(2u_2-2\beta)+K_{12}{\cal{C}}(2u_2-2\beta)\right] \nonumber \\
&&+\sin^2\im\left(K_{11}\cos2\um+K_{12}\sin2\um\right. \nonumber \\ &&\left.\left.-\frac{2}{5}f_1-\frac{3}{5}K_1\right)\left[2{\cal{M}}-{\cal{S}}(2u_2-2\uvm)\right]\right\} \nonumber \\
&&+\Delta_1^*(\sin\im\cot{i_1}),
\label{Eq:dyn1}
\end{eqnarray}
where the dimensionless, $P_2$ (i.e., long-) timescale dynamical amplitude is
\begin{equation}
A_\mathrm{L1}=\frac{15}{8}\frac{m_\mathrm{C}}{m_\mathrm{ABC}}\frac{P_1}{P_2}\left(1-e_2^2\right)^{-3/2},
\end{equation}
while
\begin{eqnarray}
{\cal{M}}&=&v_2-l_2+e_2\sin v_2, \nonumber \\
{\cal{S}}(2u_2)&=&\sin{2u_2}+e_2\left[\sin(u_2+\omega_2)+\frac{1}{3}\sin(3u_2-\omega_2)\right], \nonumber \\
{\cal{C}}(2u_2)&=&\cos{2u_2}+e_2\left[\cos(u_2+\omega_2)+\frac{1}{3}\cos(3u_2-\omega_2)\right]. \nonumber \\
\label{Eq:dyntrigdef1}
\end{eqnarray}
Furthermore,
\begin{eqnarray}
f_1&=&1+\frac{25}{8}e_1^2+\frac{15}{8}e_1^4+\frac{95}{64}e_1^6+{\cal{O}}(e_1^8),\\
K_1&=&\mp e_1\sin\omega_1+\frac{3}{4}e_1^2\cos2\omega_1+{\cal{O}}(e_1^3),\\
K_{11}&=&\frac{3}{4}e_1^2\pm e_1\sin\omega_1+\frac{51}{40}e_1^2\cos2\omega_1+{\cal{O}}(e_1^3), \\
K_{12}&=&\mp e_1\cos\omega_1+\frac{51}{40}e_1^2\sin2\omega_1+{\cal{O}}(e_1^3).
\label{Eq:K12def}
\end{eqnarray}
The functions $K_n(e_1,\omega_1)$ are also given up to higher orders in $e_1$ in Eqs.~(\ref{Eq:K1ndef}) of Appendix \ref{app:octupole}. It is important to bear in mind that, in these formulae and throughout this paper, $\omega_i$'s are used in the sense of the argument of periastron of the secondary's orbit relative to its primary. With such a choice, the upper signs represent primary minima. Finally, $\Delta_1^*(\sin\im\cot{i_1})$ stands for the terms which arise directly from the nodal precession for non-coplanar systems. As these terms are multiplied by $\cot{i_1}$ they give a substantially smaller contribution for eclipsing binary systems seen nearly edge on. For the sake of completeness, however, we retained and included these terms in the analysis. They are given in Eq.~(\ref{Eq:Delta1*}) of Appendix \ref{app:octupole}.

Equation (\ref{Eq:dyn1}) is equivalent to Eq.~(B.15) of \citet{borkovitsetal11}. However, the use of true longitude ($u_2$), and node-like azimuthal angles ($\um$, $\uvm$) instead of true anomaly ($v_2$) and dynamical argument of periastrons ($g_1$, $g_2$), as was done previously, has the advantage that, in such a way, Eq.~(\ref{Eq:dyn1}) remains valid even for the special cases of circular orbits, and for coplanar configurations as well. (For this latter case we note that, although $\um$ and $\uvm$ have no meaning for coplanar orbits, it can be easily seen that for $\im\rightarrow0\degr$ $\alpha\rightarrow0\degr$, and for $\im\rightarrow180\degr$ $\beta\rightarrow180\degr$ and thus, the non-vanishing components of Eq.~[\ref{Eq:dyn1}] really retain their well defined meanings.)

These terms yield information on the mass-ratio $m_\mathrm{C}/m_\mathrm{ABC}$, and (either directly, or indirectly) on the orbital elements of both the inner and the outer orbits.  This includes angles referring to both the celestial (or observational) and the relative (or dynamical) frames of reference, namely: $P_1$, $e_1$, $\omega_1$, $P_2$, $e_2$, $\omega_2$, $\tau_2$, $\im$ and (via the node-like angles $n_1$ and $n_2$), $\Delta\Omega$, $g_1$, $g_2$, $h_1$, $h_2$, and furthermore, theoretically, even on all the observable ($i_1$, $i_2$) and dynamical inclinations (i.e., orbital inclinations relative to the fundamental plane of the system, $j_1$, $j_2$). Finally, the inclination of the invariable plane relative to the plane of the sky ($i_0$) can also be determined and therefore, the complete spatial configuration can be inferred as well. For all these calculations, and therefore for the entire orbital parameter fitting process, the spherical triangle(s) formed by the two orbital planes and the plane of the sky on the celestial sphere, and also its two constituents, i.e., the analogous triangles formed by one or the other of the orbital planes with the invariable plane and the sky (see Fig.~\ref{Fig:krsz-ek}), have fundamental importance. A thorough discussion of these spherical triangles including all the possible information to extract in all possible configurations is given in Appendix~\ref{app:Sphericaltriangle}. 

Some of the recently discovered {\it Kepler}-triples that we have investigated were found to be such compact systems that we have decided to include additional terms which are second order in the $a_1/a_2$ ratio. These contributions occur when the octupole term of the perturbing potential function (or its equivalent perturbing force components) are included. In order to improve our model with these contributions, we used a similar method to that for the case of the quadruple approximation in our previous work. Therefore, we do not include the calculations here. The complete formula is so lengthy that it is given only in Eq.~(\ref{Eq:dyn2}) of Appendix \ref{app:octupole}. At this point we discuss only some general properties, and describe a few special cases, where the expressions become substantially simpler.

The amplitude of the second order term is
\begin{equation}
A_\mathrm{L2}=\frac{1-q_1}{1+q_1}\left(1-\frac{m_\mathrm{C}}{m_\mathrm{ABC}}\right)^{1/3}\left(\frac{P_1}{P_2}\right)^{2/3}\frac{A_\mathrm{L1}}{1-e_2^2},
\end{equation}
which reveals, that this term introduces one additional parameter, which is the mass ratio of the inner binary ($q_1$). Furthermore, when $q_1$ tends to be unity (i.e., the two components of the inner binary tend to have equal masses), the octupole terms tend to be zero. As will be discussed later, despite the relatively small contribution of the octupole terms to the whole ETV, the resultant mass ratios were found, in most cases, to be in qualitative accord with the expected values. By this we mean that for systems exhibiting more or less similarly deep primary and secondary eclipses, the resultant mass ratios were found to be near unity, while in cases of highly unequal eclipse depths, smaller mass ratios were obtained.

For a coplanar configuration, the equations reduce significantly. For such a scenario Eq.~(\ref{Eq:dyn1}) simplifies to
\begin{eqnarray}
\Delta_1^\mathrm{pro}&=&\frac{1}{2\pi}\frac{m_\mathrm{C}}{m_\mathrm{ABC}}\frac{P_1^2}{P_2}\frac{\left(1-e_1^2\right)^{1/2}}{\left(1-e_2^2\right)^{3/2}}\left[\left(1\mp\frac{3}{2}e_1\sin\omega_1\right){\cal{M}}\right. \nonumber \\
&&\left.\pm2e_1{\cal{C}}(2u_2-\omega_1)\right]+{\cal{O}}(e_1^2), \nonumber \\
\Delta_1^\mathrm{ret}&=&\frac{1}{2\pi}\frac{m_\mathrm{C}}{m_\mathrm{ABC}}\frac{P_1^2}{P_2}\frac{\left(1-e_1^2\right)^{1/2}}{\left(1-e_2^2\right)^{3/2}}\left[\left(1\mp\frac{3}{2}e_1\sin\omega_1\right){\cal{M}}\right. \nonumber \\
&&\left.\mp2e_1{\cal{C}}(2u_2+\omega_1)\right]+{\cal{O}}(e_1^2)
\end{eqnarray}
for prograde and retrograde configurations, respectively. In the same case the octupole term, i. e. Eq.~(\ref{Eq:dyn2}) takes the following form:
\begin{eqnarray}
\Delta_2^\mathrm{pro}&=&\frac{P_1}{2\pi}A_\mathrm{L2}\left(1-e_1^2\right)^{1/2}\left[\mp{\cal{C}}_{21}(u_2)\right. \nonumber \\
&&-e_1\cos\omega_1{\cal{S}}_{21}(u_2)-\frac{57}{20}e_1{\cal{S}}_{21}(u_2-\omega_1) \nonumber \\
&&\left.-\frac{1}{4}e_1{\cal{S}}_{21}(u_2-3\omega_1)\right]+{\cal{O}}(e_1^2), \nonumber \\
\Delta_2^\mathrm{ret}&=&\frac{P_1}{2\pi}A_\mathrm{L2}\left(1-e_1^2\right)^{1/2}\left[\mp{\cal{C}}_{21}(u_2)\right. \nonumber \\
&&+e_1\cos\omega_1{\cal{S}}_{21}(u_2)+\frac{57}{20}e_1{\cal{S}}_{21}(u_2+\omega_1) \nonumber \\
&&\left.+\frac{1}{4}e_1{\cal{S}}_{21}(u_2+3\omega_1)\right]+{\cal{O}}(e_1^2). 
\end{eqnarray}
A further reduction of this situation for a circular inner orbit leads to the approximation used by \citet{agoletal05}. (Note, these authors considered the first order, or quadrupole terms only.)

Second, if both orbits are circular, but the orbital planes are inclined, Eqs.~(\ref{Eq:dyn1}) and (\ref{Eq:dyn2}) reduce to
\begin{eqnarray}
\Delta_1&=&\frac{3}{8\pi}\frac{m_\mathrm{C}}{m_\mathrm{ABC}}\frac{P_1^2}{P_2}\sin^2\im\sin(2u_2-2\uvm), \label{Eq:O-Cdyn1circ} \\
\Delta_2&=&\pm\frac{P_1}{4\pi}A_\mathrm{L2}\left\{\left[(1-I)\cos(u_2-\beta)\right.\right.\nonumber \\
&&\left.-(1+I)\cos(u_2-\alpha)\right] \nonumber \\
&&+\frac{5}{2}\sin^2\im\left\{(1+I)\left[\cos(u_2-\alpha)\right.\right. \nonumber \\
&&\left.+\frac{1}{2}\cos(u_2-\beta)-\frac{1}{2}\cos(3u_2-2\alpha-\beta)\right] \nonumber \\
&&-(1-I)\left[\cos(u_2-\beta)+\frac{1}{2}\cos(u_2-\alpha)\right. \nonumber \\
&&\left.\left.\left.-\frac{1}{2}\cos(3u_2-2\beta-\alpha)\right]\right\}\right\}. 
\label{Eq:O-Cdyn2circ}
\end{eqnarray}
Insofar as we consider only the first order approximation (Eq.~\ref{Eq:O-Cdyn1circ}), one can see that in this case the dynamical term has a unique period of $1/2P_2$; therefore, this is the only scenario where the LTTE and dynamical perturbation are clearly separable in Fourier-space. In the coplanar and circular binary cases, however, the first order dynamical contribution vanishes. This situation occurs in the triply eclipsing system HD~181068 \citep{borkovitsetal13}. By contrast, in the octuple approximation, even in this latter scenario, there remains a small amplitude, non-zero contribution with period equal to $P_2$, as long as the inner binary members have unequal masses. Moreover, the octupole terms result in a small phase displacement, and might also break the degeneracy between the primary and secondary eclipses even in (originally) circular cases.

Finally, we note that the dimensionless dynamical amplitudes in the ETV functions considered above scale with the inner binary's orbital period ($P_1$), and take on their time dimensions in the following form: 
\begin{equation}
{\cal{A}}_\mathrm{L1,L2}=\frac{P_1}{2\pi}A_\mathrm{L1,L2}.
\end{equation}

{\it (iii) $P_2$ time-scale residuals of the $P_1$ time-scale dynamical effects:}
For most of the systems, investigated in this paper, the $P_1/P_2$ ratio is between $10^{-1}$ and $10^{-2}$ and, therefore the amplitude of the shortest period perturbations may reach nearly 10\% of the longer period ones and, therefore, their effects should also be considered. These perturbations, however, due to our natural sampling process in the eclipse minima, will also produce additional, smaller amplitude, $P_2$ time-scale terms. The highly simplified, approximate nature of the way they are calculated is discussed in Appendix~\ref{app:shortperiod}. The somewhat lengthy complete form of this perturbative term is also given there.

The dimensionless amplitude of these terms becomes
\begin{equation}
A_\mathrm{S}=\frac{P_1}{P_2}\frac{A_\mathrm{L1}}{\left(1-e_2^2\right)^{3/2}},
\end{equation}
which, in the ETV curves, scales as
\begin{equation}
{\cal{A}}_\mathrm{S}=\frac{P_1}{2\pi}A_\mathrm{S}.
\end{equation}
For coplanar configuration we obtain that
\begin{eqnarray}
\Delta_\mathrm{S}^\mathrm{pro}&=&\frac{P_1}{\pi}A_\mathrm{S}\left(1-e_1^2\right)^{1/2}(1+e_2\cos{v_2})^3\left\{-\frac{11}{30}\sin(2u_2) \right. \nonumber \\
&&\pm e_1\left[\cos\omega_1+\frac{4}{5}\cos(2u_2-\omega_1)\right. \nonumber \\
&&\left.\left.+\frac{8}{15}\cos(2u_2+\omega_1)\right]\right\}+{\cal{O}}[e_1^2,(P_1/P_2)^3], \nonumber \\
\Delta_\mathrm{S}^\mathrm{ret}&=&\frac{P_1}{\pi}A_\mathrm{S}\left(1-e_1^2\right)^{1/2}(1+e_2\cos{v_2})^3\left\{\frac{11}{30}\sin(2u_2)\right. \nonumber \\
&&\pm e_1\left[\cos\omega_1+\frac{4}{5}\cos(2u_2+\omega_1)\right. \nonumber \\
&&\left.\left.+\frac{8}{15}\cos(2u_2-\omega_1)\right]\right\}+{\cal{O}}[e_1^2,(P_1/P_2)^3],
\end{eqnarray}
while the non-coplanar, doubly circular scenario results in
\begin{eqnarray}
\Delta_\mathrm{S}&=&\frac{11}{32\pi}\frac{m_\mathrm{C}}{m_\mathrm{ABC}}\frac{P_1^3}{P_2^2}\left\{-(1+I)\sin(2u_2-2\alpha) \right. \nonumber \\
&&+(1-I)\sin(2u_2-2\beta) \nonumber \\
&&\left.-\sin^2\im\sin2\um[1+\cos(2u_2-2\uvm)]\right\} \nonumber \\
&&+{\cal{O}}[(P_1/P_2)^3]
\end{eqnarray}
The above equations reveal that if the third star is close enough, even the quadrupole approximation can result in a significant non-vanishing component in the doubly circular, coplanar case, with period of $P_2/2$. Furthermore, as will be discussed, the inclusion of these terms substantially improved our fits in the low-amplitude mutual inclination regime, even in the case where both the inner and outer orbits had significant eccentricity.

{\it (iv) Apse-node time-scale effects:} In the case of an eccentric eclipsing binary, the orientation of the orbit with respect to the observer strongly affects the orbital phase and, therefore, the time when eclipses occur. The apsidal motion contribution to the ETV can be calculated from Kepler's equation in a straightforward manner and, although it is usually given in a trigonometric series of $\omega_1$ \citep[see e.g.][]{gimenezgarcia83}, it has an exact, analytical form, as follows:
\begin{eqnarray}
\Delta_\mathrm{apse}&=&\frac{P_1}{2\pi}\left[2\arctan\left(\frac{\pm e_1\cos\omega_1}{1+\sqrt{1-e_1^2}\mp e_1\sin\omega_1}\right)\right. \nonumber \\
&&\left.\pm\sqrt{1-e_1^2}\frac{e_1\cos\omega_1}{1\mp e_1\sin\omega_1}\right],
\label{Eq:apsidalmotion}
\end{eqnarray}

Apsidal motion studies of eccentric eclipsing binaries have been carried out for more than 75 years \citep[see][]{cowling38,sterne39}. In all the previously observed systems the apsidal motion has arisen from the tidally deformed (i.e., oblate) stellar shapes, and/or from relativistic effects. There were no systems involving non-degenerate stars, however, where forced apsidal motion due to dynamical perturbations was previously detected. (On the other hand, the perturbing effects of an unseen third body have been suggested for explaining the anomalously slow apsidal motion of a few binaries, \citealp[see e.g.,][for DI Her]{khaliullinetal91}, and \citealp[for AS Cam]{khodykinvedeneyev97}. Note, that even for these two systems, recent investigations have shown that the origin of the unexpectedly slow apsidal motion may be explained by the misalignment of the spin axes, instead of the effects of a third body (\citealp[see][for the two systems, respectively]{albrechtetal09,pavlovskietal11}).  

In the case of relativistic apsidal motion the apsidal line rotates with a constant angular velocity in the direction of the orbital motion. ``Pseudo-synchronously'' rotating oblate stars with negligible, or weak tidally induced oscillations also produce similar kinds of apsidal motion. Therefore, for most of the previously known binaries we can simply write that
\begin{equation}
\omega_1(E)=\omega_1(0)+\Delta\omega_1E,
\label{Eq:omega1def_tidal}
\end{equation}
where $\Delta\omega_1$ denotes the apsidal advance rate for one orbital period, and therefore, the apsidal motion contribution to the ETV can be modeled in a simple way by substituting Eq.~(\ref{Eq:omega1def_tidal}) into (\ref{Eq:apsidalmotion}).

For third-body forced apsidal motion, the situation is substantially more complicated. In this case, in general, none of the orbital parameters (except the semi-major axes) remains constant. It is especially true for high mutual inclination systems with negligible tidal oblateness, where an initially very low, or even zero eccentricity may grow up to near unity. (This is the so-called Kozai-Lidov mechanism, which have been investigated and discussed in several recent papers, \citealp[see e.g.][and further references therein]{naozetal13}.) 

In this paper, however, we leave out a rigorous consideration of such a scenario, and restrict ourselves to the case where the variation in the inner eccentricity is small, or negligible. (We will verify this choice in the discussion.) This situation was elaborately investigated by \citet{borkovitsetal07}, where references to previous works were also given. For the sake of completeness, the basic steps, and additional discussion are included in Appendix~\ref{app:apsidalmotion}.

In this simplified approach the apsidal motion was modeled in three different ways: (i) the apsidal advance rate $\Delta\omega_1$ (or $\dot{\omega}_1$) is taken to be an unconstrained constant which is an adjustable additional parameter of the solution; or (ii) $\dot{\omega}_1$ is a constrained constant whose numerical value is calculated from other system parameters according to Eq.~(\ref{Eq:om1pont}); or (iii) $\dot{\omega}_1$ is no longer considered to be constant, but a time-dependent quantity. In this third case we have no need to calculate $\dot{\omega}_1$ (or $\Delta\omega_1$), because the instantaneous value of $\omega_1$ can be directly calculated from the time-dependent value of the dynamical apse and node $g_1$, $h_1$ via the approximate quadrupole analytic model that is described in Appendix~\ref{app:apsidalmotion}. 

Because some of our systems exhibit rapid eclipse depth variations, which are a clear indicator of the varying  inclination angle of the inner binary due to precession of its orbital plane, we also modeled this effect. This phenomenon does not influence the ETVs in a similarly expressive way as for apsidal motion. Its direct contribution to the ETVs is multiplied by $\cot i_1$ and, therefore, becomes negligible for our nearly edge-on systems. On the other hand, apsidal motion substantially affects the dynamical apsidal motion rates, and also the $n_{1,2}$ node-like quantities.  Another significant contribution of its effect to the ETVs comes from the variations of the $n_1$ and $n_2$ angles which appear explicitly in the $P_2$ period dynamical terms ($\Delta_1$, $\Delta_2$). In modeling the orbital precession, we applied the same approximations as in the case of apsidal motion.  Specifically, in the case of a constant apsidal motion rate, the nodal regression (or progression) rate ($\Delta h$) was also considered to be constant, being either unconstrained, or constrained with a value obtained from Eq.~(\ref{Eq:hpont}), and was substituted into the equation
\begin{equation}
h(E)=h(0)-\Delta hE,
\label{Eq:hdef_prec}
\end{equation}
or, when $\omega_1$ was calculated according to the first order analytic solution, the same was done for $h_1$. (A detailed description is given in Appendix~\ref{app:apsidalmotion}.) Then, when the actual value of the dynamical node ($h$) for cycle number $E$ was calculated, the corresponding $n_i$'s were computed using the theorems of spherical triangles. The straightforward calculation, and its not-so-straightforward discussion, are given in Appendix~\ref{app:Sphericaltriangle}.

Equation (\ref{Eq:apsidalmotion}) provides very strong constraints on $e_1\cos\omega_1$ and, especially for the shorter-period apsidal motion systems (i.e., where a relatively larger portion of a complete cycle is covered), on the apsidal motion period as well. Furthermore, this period provides further constraints on the following parameters: $e_1$, $g_1$, $e_2$, $\im$, $P_1/P_2$ as well as, via the apsidal motion period, the mass ratio $m_\mathrm{C}/m_\mathrm{ABC}$, thereby establishing a connection between the amplitude of the $P_2$-period dynamical perturbations and the apsidal motion (and orbital precession) periods. These provide important additional constraints for a physically reliable solution.

The orbital precession terms have lesser direct influence due to their moderate contribution to the ETVs; however, there is a strong connection among the rate and amplitude of the eclipse depth variations and the precession rate, the mutual inclination, and the relative orientation of the orbits with respect to the observer (as is discussed in Appendices~\ref{app:apsidalmotion} and \ref{app:Sphericaltriangle}). Later, in Sect.~\ref{sec:results} we use these connections to verify or reject certain ETV solutions, or to choose between alternative, ambiguous solutions of the same triples.

{\it (iv) Other, small effects:}
There are several additional, usually negligibly small amplitude effects, two of which we nonetheless mention.

First is the intrinsic light-travel-time effect for the two components of the inner binary. For short (few-day) period eclipsing binaries, which form the large majority of EBs discovered incidentally by astronomers in previous centuries, its effect, due to the small orbital separation, remains below the accuracy of ground-based timing measurements.  However, for longer period eclipsing binaries, observed especially with the accuracy of {\it Kepler} photometry this effect becomes detectable. Therefore, as far as we know, this may be the reason why this effect was not considered before the {\it Kepler} era. A simplified form for circular orbits was first used by \citet{kaplan10}, while the general, eccentric form is given in \citet{fabrycky10}. These papers, however give only the differential form of the effect, i.e., the displacement of the secondary eclipses with respect to the primary eclipses. Here we list the formula separately for the two types of eclipses:
\begin{equation}
\Delta_\mathrm{LTTEin}=\pm\frac{1}{c}\frac{q_1-1}{q_1+1}\frac{a_1\sin i_1\left(1-e_1^2\right)}{1\mp e_1\sin\omega_1}.
\label{Eq:LTTEindef}
\end{equation}

The other small effect is due to the slight inclination ($i_1$) dependence of the occurrences of the eclipse events for eccentric orbits. This effect was discussed in detail by \citet{gimenezgarcia83}, for example. These authors also gave the mathematical form of the ETVs due to apsidal motion with the extension of this inclination dependence. These formulae are in use up to the present time. For a recent paper on this topic see \citet{wolfetal13}. However, this effect can be included into our equations by simply redefining the observable argument of periastron formally as follows:
\begin{equation}
\omega_1^*=\omega_1\mp\frac{e_1\cos\omega_1\cos^2i_1}{\sin^2i_1\mp e_1\sin\omega_1},
\end{equation}
which is the first order approximation of Eq.~(10) of \citet{gimenezgarcia83}. 

{\it (v) Other effects, not taken into account:} We have left out of the present considerations the changes or perturbations in the outer orbital elements. This was done for two reasons. First, the amplitudes of the variations in the outer orbital elements for hierarchical systems usually remain much lower than those of the inner orbital elements \citep[see, e.g.][]{harrington68,harrington69}. Furthermore, the perturbations in the outer elements affect the binary motion and, therefore the ETV curves, in an indirect way which would appear only in higher-order approximations. Despite this, as a forthcoming step, we plan the inclusion of these terms for a better modeling of the most compact triples.

We also omitted those apsidal motion and orbital precession effects which would arise from tidal or relativistic interactions. In Sect.~\ref{sec:results}, in light of our results, we justify this decision. The tidal effects, however, are discussed briefly in Appendix~\ref{app:apsidalmotion}.

Finally, we also neglected the octupole ``apse-node'' timescale perturbation terms.  Our experience from the present study is that these latter terms certainly have to be included for better future modeling of the most compact systems.

\section{Analysis code}
\label{Sec:code}

In the parameter search for the sample of 26 hierarchical triple systems studied in this work we departed from the method followed in the previous study of \citet{rappaportetal13}. In that paper it was stated that due to the strong and highly nonlinear correlations between several parameters, the conventional Levenberg-Marquardt (LM) fitting procedure was not adequate for the exploration of the parameter phase space. The main source of this difficulty originated from the non-orthogonality of the two functions describing the LTTE and the dynamical terms, which each contributed ETV curves that could be more or less comparable in magnitude and shape, for their sample of CHTs. In contrast, in our current collection of eccentric triples, the $P_2$ time-scale quadrupole dynamical terms highly dominate over all the other contributors for all systems. Furthermore, the inclusion of additional relations and information allows for strict constraints to be set on some of the combinations of parameters and, therefore, reduces the degeneracies. Here we refer, e.g., to the connections between the different inclination and node-like parameters (which provide additional constraints on even the dynamical angular elements and the masses), and the apsidal motion terms, the latter of which very strictly constrain not only the $e_1\cos\omega_1$-term, but in the cases of several systems, both $e_1$ and $\omega_1$ individually, and finally the outer mass ratio. Therefore, we decided to apply a combination of the LM fitting procedure and a grid-search method in our parameter adjustment process. 

In its present state, the code contains 20 adjustable parameters; however, because of the different kinds of interconnections we do not allow for the adjustment of all these parameters in the same run. (For example, from the six angles and node-like arcs of the spherical triangle discussed in Appendix~\ref{app:Sphericaltriangle}, only three are allowed to be included in the fitting process.) There are some additional flags which choose the actual working mode of the code (i.e., which terms to be included, or not, and which additional constraints to be applied, or not)\footnote{In that sense the code follows a similar philosophy to that of the renowned Wilson-Devinney eclipsing binary lightcurve program \citep[e.g.,][]{wilsondevinney71,wilson79}}. 
The adjustable parameters, collecting them into five groups, are as follows:
\begin{enumerate}
\item[(i)]{$V_\gamma$ -- systemic radial velocity (not used in this work)}
\item[(ii)]{$c_0$, $c_1$, $c_2$ -- coefficients of the polynomial contribution which are used in part for determining the refined value of the epoch $T_0$ and sidereal period $P_\mathrm{s1}$.  The fitting of the quadratic coefficient $c_2$ can be, and was, disabled for all but one of the runs presented here.}
\item[(iii)]{$e_1$, $\omega_1$, $\Delta\omega_1$ (the last of which, i.e., the apsidal advance rate, may be either unconstrained, or calculated according to one of the two methods as discussed)}
\item[(iv)]{$a_\mathrm{AB}\sin{i_2}$, $a_\mathrm{C}\sin{i_2}$, $P_2$, $e_2$, $\omega_2$, $\tau_2$, $q_1$ -- or, optionally, some physical (but not numerical) equivalents, e.g., the mass function, $f(m_\mathrm{C})$, and mass ratio, $m_\mathrm{C}/m_\mathrm{ABC}$.}
\item[(v)]{$\im$, $i_1$, $i_2$, $n_1$, $n_2$, $\Delta{h}$ (where two from the first five are computed from the other three, while the orbital precession rate may be either unconstrained, or calculated in a similar manner, as was discussed for the apsidal advance)}
\end{enumerate}
During a fitting run session one of the following six possibilities was applied for each of the parameters: it was (i) kept fixed at its initial value; (ii) kept fixed at different, equally spaced initial values (grid search); (iii) adjusted by the LM process, starting from a single initial value; (iv) LM-adjusted, starting from several equally spaced initial values; (v) calculated (constrained) from other parameters; or (vi) not considered, according to the respective model. A detailed description of the code will be presented elsewhere in a technical paper; here we discuss only that part which is relevant for the present work.

In order to check the analytical formulae on one hand, and the numerical behavior of our parameter adjustment process, as well as the uniqueness of the solutions, on the other hand, we have carried out various tests. Basically, these investigations have two separate parts. First, we obtained solutions for actual {\em Kepler} ETV curves by utilizing different model approximations. Then the solutions that were obtained were used as initial parameters for a 3-body numerical integration from which we generated the associated artificial ETV curves.  We then compared these to the actual, observed ETVs.  The consistency of this loop is a direct measure of how good the solution is.  Second, we obtained fitted solutions for these numerically generated ETVs, and compared the solution parameters with the known initial values. Furthermore, we have also varied some of the input parameters to check the solutions' behavior and dependence upon some of the model parameters. Moreover, the same test runs were used to check the reliability of the formal errors calculated from the covariance matrices of the LM-solutions with the empirical rms scatter in the different solutions we investigated.

It is clear, however, that despite the speed and effectiveness of this method for exploring how well the analytic fits work, it has some inevitable disadvantages.  In particular, the LM portion of the fits does not explore non-ellipsoidal correlations in multi-dimensional $\chi^2$ space, while the grid portion of the search excludes certain physically unrealistic regions of parameter space.  Therefore, instead of automatically accepting the formal errors obtained from our fits, we use the solutions recovered from the numerically generated ETVs with known parameters to demonstrate the overall reliability of our methods, and for the estimation of more conservative uncertainties for some of the parameters. In Appendix~\ref{app:numericalanalysis} we discuss the steps of the complete investigation for a few systems in detail, which also provides us some insight into the methodology of the analysis.

\begin{table*}
\begin{center}
\caption{Properties of the investigated systems} 
\label{Tab:Systemproperties}  
\begin{tabular}{lcccccccc} 
\hline
KIC No. & $P_\mathrm{s1}$&$T_0$ & $K_\mathrm{p}$&$T_\mathrm{eff}$ &$\log{g}$&$P_2/P_1$&Ecl.depths&Tertiary\\
        & (days)         &(MBJD)&     (mag)     &     (K)         &         &         & variation&eclipses\\
\hline
04940201$^a$& 8.816578 & 54967.276926 & 14.98 & 5284 & 4.61 & 41.4 & ... & ... \\
05255552&32.448635 & 54970.636491 & 15.21 & 4775 & 4.59 & 26.5 & (j+)& yes \\
05653126&38.493382 & 54985.913152 & 13.17 & 5766 & 3.81 & 25.1 & +a2 & ... \\
06545018$^a$& 3.991460 & 54965.835642 & 13.75 & 5594 & 4.46 & 22.7 & ... & ... \\
07289157$^a$& 5.266425 & 54969.966600 & 12.95 & 6013 & 4.19 & 46.2 &  -  & yes \\
07812175&17.793925 & 55002.612666 & 16.33 &  NA  &  NA  & 32.7 &  NA & ... \\ 
08023317$^a$&16.579002 & 54979.733478 & 12.89 & 5625 & 4.05 & 36.8 &  +  & ... \\  
08210721&22.672816 & 54971.157082 & 14.27 & 5412 & 4.28 & 34.8 & ... & ... \\
08938628$^a$& 6.862216 & 54966.603088 & 13.68 & 5602 & 4.29 & 56.6 &  c- & ... \\
09714358$^a$& 6.474177 & 54967.395501 & 15.00 & 4825 & 4.55 & 16.02& ... & ... \\
\hline
05771589$^a$& 10.739142 & 54962.130765 & 11.81 & 5927 & 4.23 & 10.5 & -+ & ... \\
06964043    &  5.362659 & 55291.992805 & 15.61 & 5374 & 4.44 & 22.3 & +- & yes \\
07668648$^a$& 27.818590 & 54963.315401 & 15.32 & 5875 & 4.52 &  7.3 & +,x & yes\\
07955301$^a$& 15.326340 & 54967.950750 & 12.67 & 4821 & 3.12 & 13.6 & +c,x & ...\\
\hline	  
04769799& 21.929314 & 54968.505532 & 10.95 & 4911 & 3.57 & 56.1 & j-;d2 & ...\\
05003117& 37.613001 & 54986.092638 & 14.03 & 5387 & 4.49 & 57.2 & c(-) & ...\\
05731312& 7.946382  & 54968.093163 & 13.81 & 4658 & 4.49 & 114.1& (j-) & ... \\
07670617& 24.703160 & 54969.139216 & 15.52 & 4876 & 4.73 & 130.9&  j-  & ...\\ 
08143170& 28.785943 & 54970.113064 & 12.85 & 4957 & 3.68 & 59.4 & (c+) & ... \\ 
09715925& 6.308199  & 54998.939653 & 16.52 & 4891 & 4.46 & 116.7& c+   & ... \\
09963009& 40.069657 & 54986.018248 & 14.46 & 5653 & 4.33 & 94.1 & c-2  & ...\\ 
10268809& 24.708999 & 54971.999951 & 13.74 & 5787 & 4.42 & 283.3& j-;x & ...\\ 
10319590$^a$& 21.320459 & 54965.716743 & 13.73 & 5518 & 4.37 & 21.1 & -d & ...\\ 
10979716& 10.684056 & 54967.082259 & 15.77 & 3932 & 4.61 & 97.8 & ... & ... \\ 
11519226& 22.160715 & 54973.018008 & 13.03 & 5646 & 4.54 & 64.6 & ... & ... \\ 
12356914& 27.307455 & 54976.508322 & 15.53 & 5368 & 4.58 & 66.5 & ... & ... \\ 
\hline	 
\end{tabular}
\end{center}

{\bf Notes.} (1) Sidereal period ($P_\mathrm{s1}$) and epoch ($T_0$) were used for plotting $O-C$ curves. $T_0$ was also used as reference epochs for most of the parameters listed in Tables~\ref{Tab:Orbelem}--\ref{Tab:Massetal}. (The exceptions are noted in each table.) (2) {\it Kepler} magnitude, effective temperature and $\log g$ were taken from the Kepler Input Catalog. (3) Further notes for column `KIC numbers': $a$: listed in \citet{rappaportetal13} -- for column `eclipse depth variations': +/-: continuous increase/decrease; j: sudden jump; c: constant (marked only if the eclipse depth remains constant during a portion, but not the whole time-span, of the observations); a2/d2: appearance/disappearance of secondary eclipses; d: eclipses disappeared; x: exchange of the amplitudes in primary/secondary eclipses; (marks in parenthesis): slight/uncertain variation; ... : no eclipse depth variation.
\end{table*}

\section{System selection and data preparation}
\label{Sec:dataprep}

The present version of the {\em Kepler} Eclipsing Binary Catalog\footnote{http://keplerebs.villanova.edu/} \citep{conroyetal14} contains 2645 EBs. We selected our systems from that sample. We started our search for the appropriate CHTs with the construction of $O-C$ (`observed minus calculated' eclipse times) curves for the primary and, when possible, the secondary, eclipses for all 2645 binaries.  At the same time we also produced folded light curves for each binary. In all, we found some 400 binary systems that have interesting (i.e., non-linear) $O-C$ curves (see also \citealt{rappaportetal13,conroyetal14}).  However, most of these tend to be either parabolically shaped or have sinusoidal shapes with a period comparable to, or longer than, the {\em Kepler} mission.  The majority of these are probably triple systems, as indicated by the presence of perturbations that are likely due to a third body in the system, but are otherwise not particularly interesting for the present study. We then restricted our attention only to the subset of these systems which satisfied the following three criteria:
\begin{itemize}
\item[(i)] The inner eclipsing binary should have an eccentric orbit. 

A good indicator of an EB's eccentricity in the light curve is the displacement of the secondary eclipse from the mid-time of two consecutive primary minima. For systems, however, with small eccentricity, and/or semi-major axes lying almost along the line of sight (i.e., $\omega_1\approx\pm90\degr$), the eccentricity might go unnoticed. An equivalent sign of eccentricity in the $O-C$ curves occurs when the (averaged) primary and secondary $O-C$ curves do not overlap, or they even converge toward, or diverge from, each other (due to apsidal motion).   

\item[(ii)] The ETV curve should show clear signatures of third-body perturbations.

These signatures can range from quasi-periodic modulations to relatively abrupt jumps in the ETV curves. Examples of both kinds will be shown later. Note, that light curves can also exhibit features which most likely come from a third component. These signs include {\it (a)} extra eclipses, especially when these extra events show definite variability in their shapes, and even in occurrences; {\it (b)} variations in the eclipse depths, which can be associated with the disappearance (or appearance) of one or the other or both eclipses, the change in the shape and duration of the eclipses, and even an exchange of the eclipse depths between the primary and secondary\footnote{Note, in principle, a non-aligned spin-axis of one or both binary members may also cause orbital plane precession and, therefore, eclipse depth variations with all of the above listed properties; such an effect was observed, e.g., in the case of the hot Jupiter Kepler-13b \citep[see, e.g][]{szaboetal12}. The efficiency of this effect, however, is strongly related to the tidal timescale which, for the systems we investigated (as will be shown below), exceeds the dynamical timescales by orders of magnitudes.  Therefore, in the present study, this possibility does not play a significant role.}; or {\it (c)} a rapid variation of the time-lag between the primary and consecutive secondary eclipses.

\item[(iii)] Both primary and secondary minima should have been observed, at least during a part of the {\em Kepler} mission. 

This last point is a technical requirement because, in the absence of secondary eclipses (and of course coupled with the lack of radial velocity measurements), we cannot constrain $e_1$, $\omega_1$ and $\Delta\omega_1$ from the apse-node timescale terms, and therefore, our solution would be strongly under-determined. There was only one supposed triple system, which was omitted according to this criterion.  It was KIC~07837302\footnote{Omission of this triple was a hard decision, especially since it shows a very nice and large amplitude ETV, which during the second half of {\it Kepler}'s mission significantly departed from the solution which was found in the above cited paper.} which was included in the paper of \citet{rappaportetal13}. Note, that due to this requirement we did not check additional systems in the catalog which exhibit only one eclipse, therefore, we cannot exclude the presence of additional interesting triples amongst the EBs showing only one eclipse per orbit.
\end{itemize}

According to these criteria we first selected 10 EBs from the 39 systems investigated in \citet{rappaportetal13}. We then made an extended search for other additional systems that fulfill our criteria.  This included inspecting each of the 2645 ETV curves and associated folded light curves to find good candidates.
%An additional, independent search was made also by checking manually all phased light curves in the Kepler Eclipsing Binary Catalog in the $P>3^\mathrm{d}$ regime for systems with displaced secondary minima. Downloading the light curves of all the positive candidates, we determined their ETV curves again, using the more accurate, interactive method described in \citet{borkovitsetal14}. 
Combining the results of these searches, we made a final selection of 26 systems for further analysis -- of which the remaining 16 are newly reported here. The important parameters for these 26 hierarchical triples are listed in Table~\ref{Tab:Systemproperties}. 

During the course of our analyses we realized that our sample should be divided into three subgroups, which are separated with two horizontal lines in the tables. The first group contains ten systems which were found to be the most ideal for our purposes, and therefore yielded the most reliable solutions. The four triples in the second group were found to be too close (i.e., quite compact) and therefore, our analytic fitting model was somewhat less satisfactory.  For the remaining dozen systems, the largest uncertainties in the system parameters should arise from the insufficient {\em Kepler} coverage of the outer period. A detailed discussion is presented in Sect.~\ref{sec:results}.)

We used the entire Q0--17 long cadence (LC) datasets of {\em Kepler} eclipsing binaries. In order to have a unified treatment of the data, for the final runs we downloaded the complete, detrended LC light curves of all the selected systems from the Villanova site\footnote{http://keplerebs.villanova.edu/}, and determined the times of minima and their uncertainties by the use of the first (BJD), seventh (detrended relative flux) and eighth (flux uncertainty) columns of these files. 

{\em Determination of Eclipse Times:} The procedure for determining accurate eclipse times was done as follows. First, we calculated a phase-folded, binned and averaged light curve for each system. We found that the use of 1\,000 orbital phase bins (of equal duration) was appropriate.  For a few systems where substantially better-sampled short-cadence (SC) data were also available, we made similar folded and binned light curves for the SC data with 2\,000 phase bins. (Note, in the case where tertiary eclipses were also present in the light curves, we naturally eliminated those intervals.)  We then used these folded, averaged light curves to calculate templates for both for the primary and secondary eclipses in the form of polynomials. In principle, our code allows for a maximum 20$^{\rm th}$ order polynomial, but in most cases a $6^{\rm th}$- or $8^{\rm th}$-degree polynomial template was used. 

Our code scanned the entire dataset for a given target, and identified possible eclipse events according to one of the following two preliminary criteria.  We (i) searched around the expected mid-eclipse phases, determined from the template, and (ii) used a simple preset flux limit.  After the identification of possible eclipse events, we made a three-parameter Levenberg-Marquardt fit with the appropriate (primary or secondary) template, optimizing the flux vs.~phase function (over a preset phase-interval) in the form of $f=A\sum_i c_i(\phi-\phi_0)^i+B$, where the $c_i$'s describe the coefficients of the template polynomials. Of the three adjusted parameters ($\phi_0$, $A$, $B$) the one of greatest interest is $\phi_0$, which gives the phase-lag between the template and the current eclipse. (Considering the multiplicative $A$ and additive $B$ parameters, if the eclipse depths and the out-of-eclipse light levels would remain unchanged, the values of $A=1$ and $B=0$ should be constant. By allowing for their adjustment, however, we found that this kind of template-fitting results in fast and accurate eclipse time determinations even for light curves with the most significantly varying eclipse depths and durations.) Finally, the entire process was reiterated (usually 5-times) in order to further improve the accuracy of the mid-eclipse times. 

As an alternative method, instead of full, higher-order template fittings, the times of eclipses were determined by fitting a simple quadratic function to all individual eclipses. We found that for systems with small or even moderate variations in eclipse shape and duration  the accuracy of the full template-fitting method was superior to the simple quadratic function. The great advantage of the simple parabolic fit is that it can be used even in cases where the light curve properties vary considerably so that a fixed `template' makes little sense. 

As an additional by-product of the above eclipse timing analysis we were able measure the eclipse depth variations that are present in some of our systems in an easy, approximate, and rather accurate manner. We utilized the fact that the detrended LC curves which were used for our analysis had already been corrected previously for instrumental\footnote{The one exception was the special case of KIC~07812175. See the text below for details.} and other longer time-scale effects, and, furthermore, neither stellar pulsations nor significant modulations due to starspots were observable. Therefore, due to the nearly constant out-of-eclipse light levels, the minimum value (in relative flux) of the fitted template curve for each eclipse event resulted in a good measure of the eclipse depth. We were thus able to follow these eclipse depth variations with time. Note that, although in principle the inclusion of the modeling of these eclipse depth variations into our analytic and numerical fitting would have improved the orbital solutions, for reasons which will be discussed shortly in the Conclusions section, we did not make use of this additional information.

In Appendix~\ref{app:ToMtables} we list all the calculated eclipse times for all 26 systems in this study. There, the estimated uncertainty of each individual eclipse time is also given. The uncertainties were calculated in two independent ways. In one, a `boot-strapping' approach, we repeated the fitting process a hundred times for {\em each eclipse event} by adding random scatter to the individual flux data points, assuming a Gaussian distribution with the value of $\sigma$ taken to match that of the data and, furthermore, omitting from each eclipse between 0 and 3 data points randomly. Then, supposing a Gaussian distribution for the resultant eclipse times, we calculated their standard deviation. In a second, simpler method, the formal error on the $\phi_0$ parameter from the LM-fitting, or the corresponding error of the linear least-squares quadratic fit, was calculated.  We found that the error bars in the quadratic fit overestimated the uncertainties by an order of magnitude, on average, while the bootstrapping and the LM-fit errors yielded uncertainties consistent with each other. These uncertainties were then also used to exclude a few data points whose error bars were larger than 3 rms standard deviations from the others. Note, that the power of {\em Kepler} photometry is well illustrated by the fact that, despite our relatively simple eclipse-time determination process, in most cases we were able to reach a typical accuracy of $\sigma \sim 10-50$ sec\footnote{At first glance this accuracy does not seem to be unusually good, especially if we consider that equally good (or better) eclipse times are often achieved with small-aperture ground-based telescopes.  However, a quick comparison of any ground-based $O-C$ curves with those obtained from {\em Kepler} observations clearly demonstrates that the error estimations of the ground-based eclipse times are usually too optimistic. On the other hand, the ground-based observations are not limited by the typical 29.4-min {\em Kepler} long-cadence integration time.}. Naturally, the accuracy of the eclipse-time determination depends on several factors, especially the depth (in particular, relative to the amplitude of the other light curve variations, being either real or instrumental, on a comparable time-scale), the shape, and the duration of the eclipses. According to our experience, neither a shallow depth nor a short (but still significantly longer than the cadence time) duration of the eclipses, or even a combination of the two, significantly reduced the accuracy of our process. In contrast, for shallow, total, (i.e., flat bottomed) eclipses (such as the secondary eclipses of KIC~08023317) we obtained only much more limited accuracy. 

We found that both kinds of error estimation for the eclipse times (LM formal errors and bootstrapping) showed an over-sensitivity to the eclipse depths. For one thing, the calculated uncertainties in some cases are clearly too large in comparison with a visual inspections of the scatter of some of the ETV curves, especially for the shallower eclipses.  In addition, there were some technical issues for the longer outer period triples, resulting in noticable underweighting of either some secondary ETV curves in systems with highly unequal eclipse depths, or even over different sections of the same ETV curve in systems exhibiting significant eclipse depth variations. Therefore, in the final analysis we used two different kinds of uncertainties for the ETV times. Besides using individual ETV point uncertainties, we carried out additional system parameter fits by using a system-specific global uncertainty for the eclipse times for an entire ETV curve.  Finally, independent of which kind of uncertainties were used, in order to obtain physically meaningful error estimations for the parameters, in the final stage, when necessary, the uncertainties were rescaled in order to normalize $\chi^2$ to $\approx1$.

\begin{figure*}
\includegraphics[width=84mm]{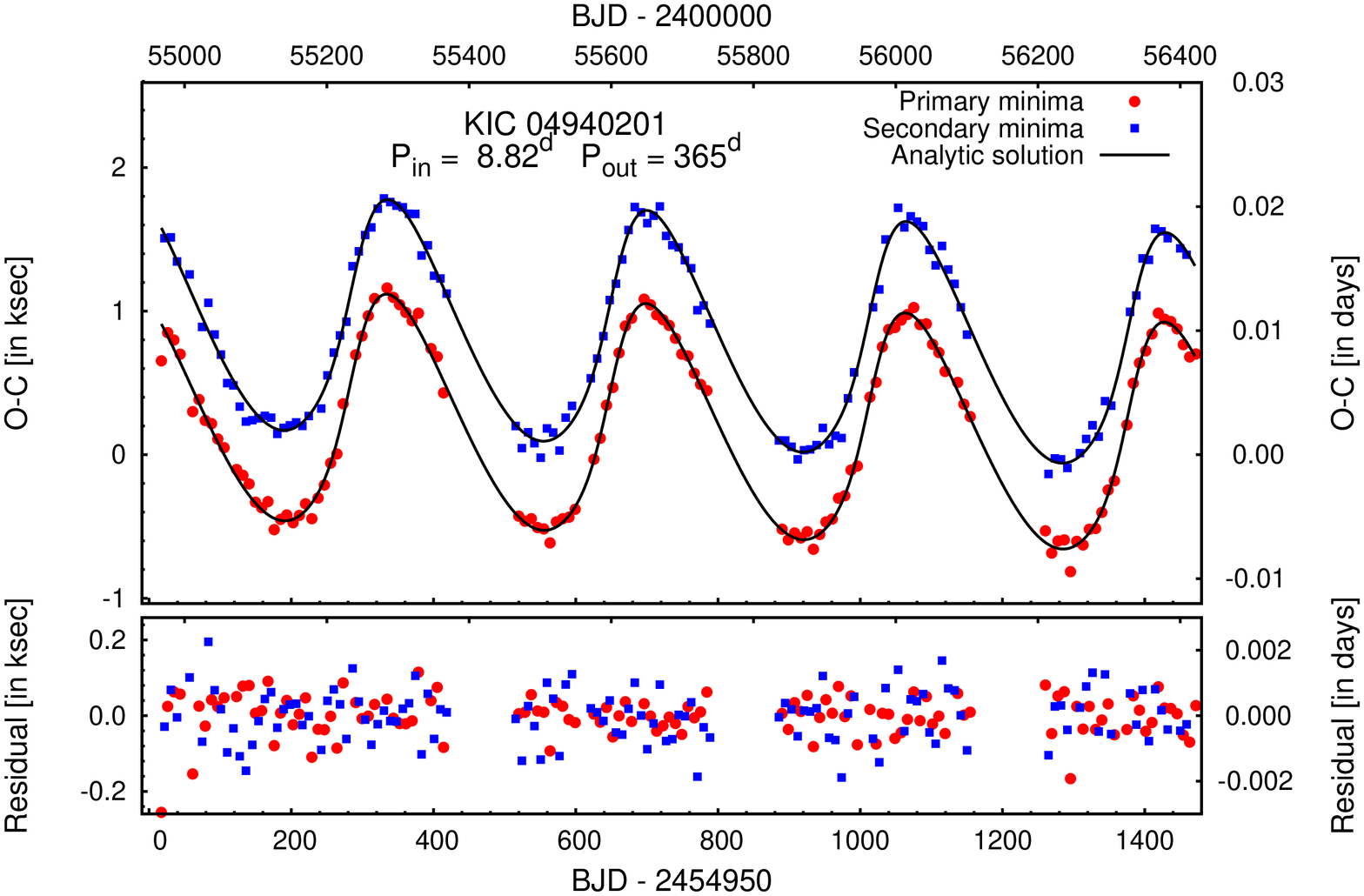}\includegraphics[width=84mm]{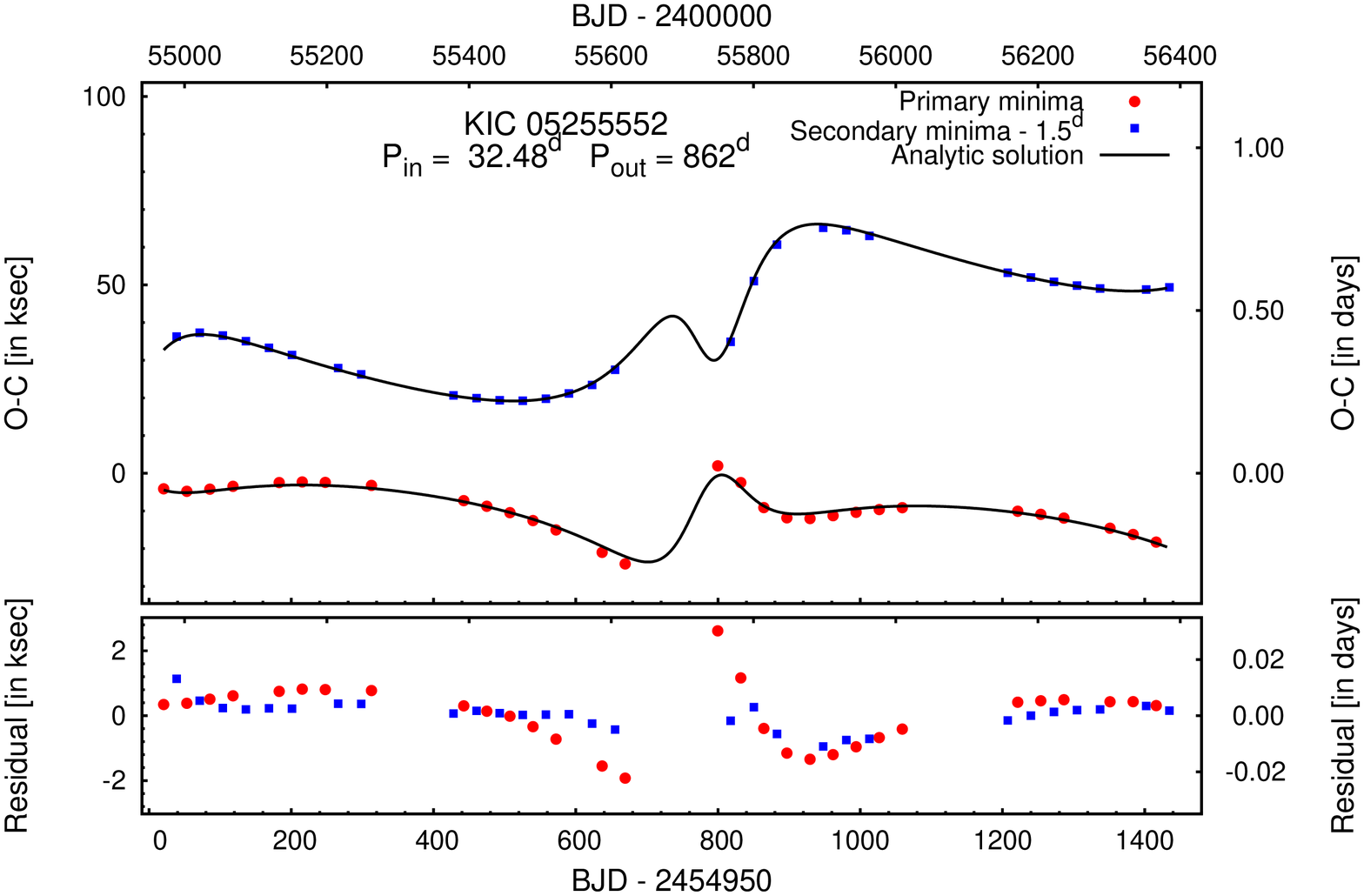}
\includegraphics[width=84mm]{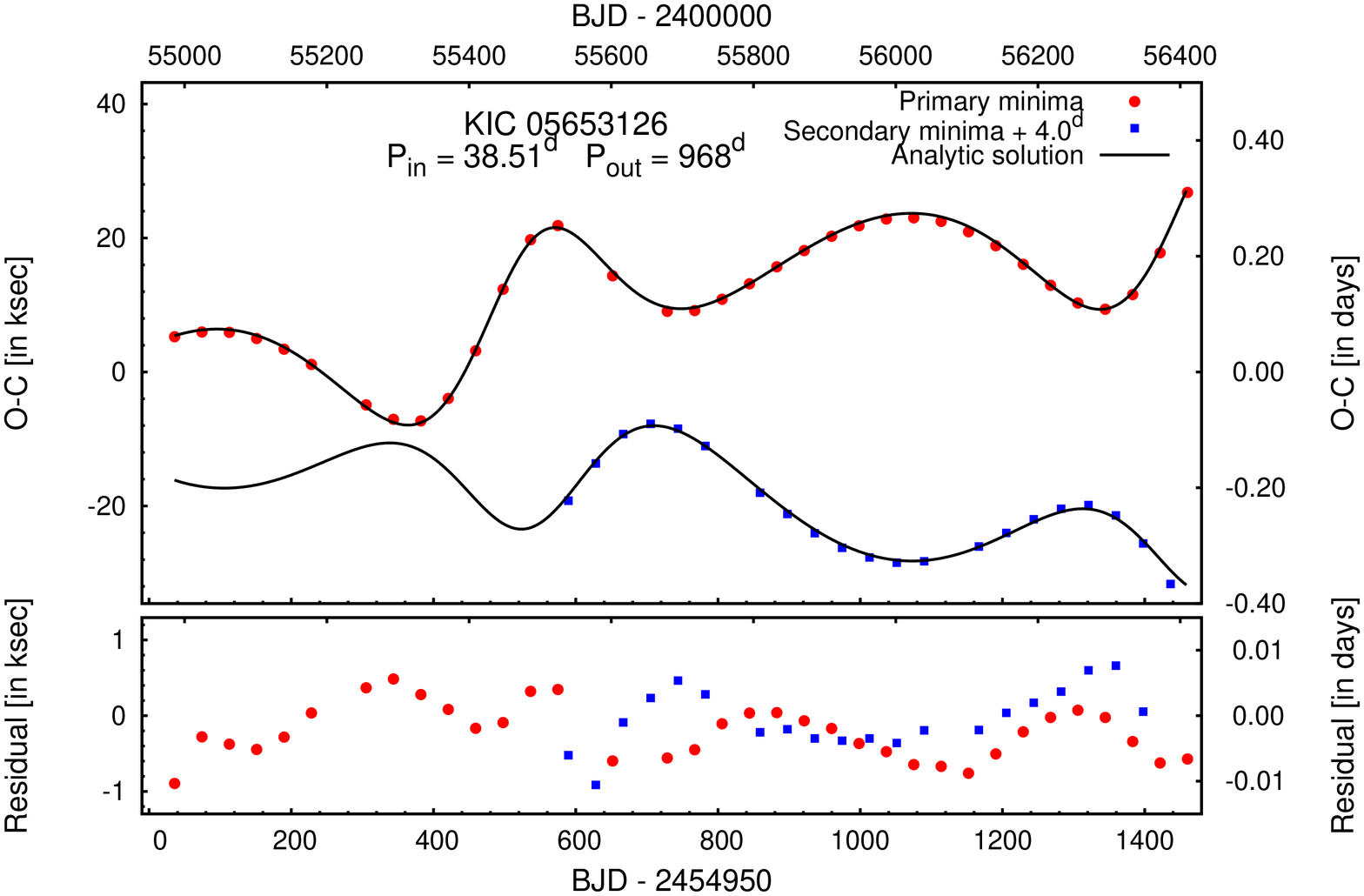}\includegraphics[width=84mm]{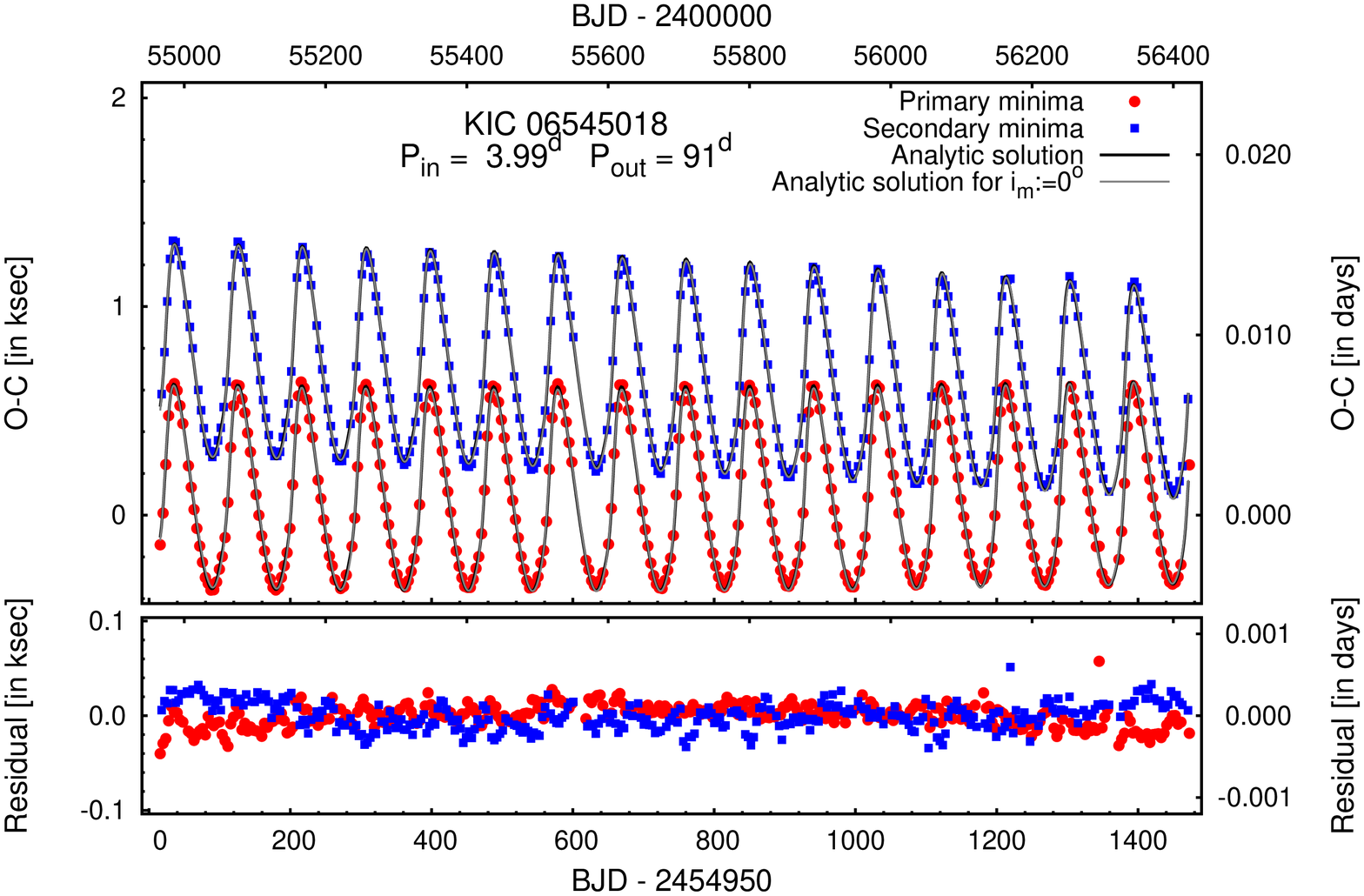}
\includegraphics[width=84mm]{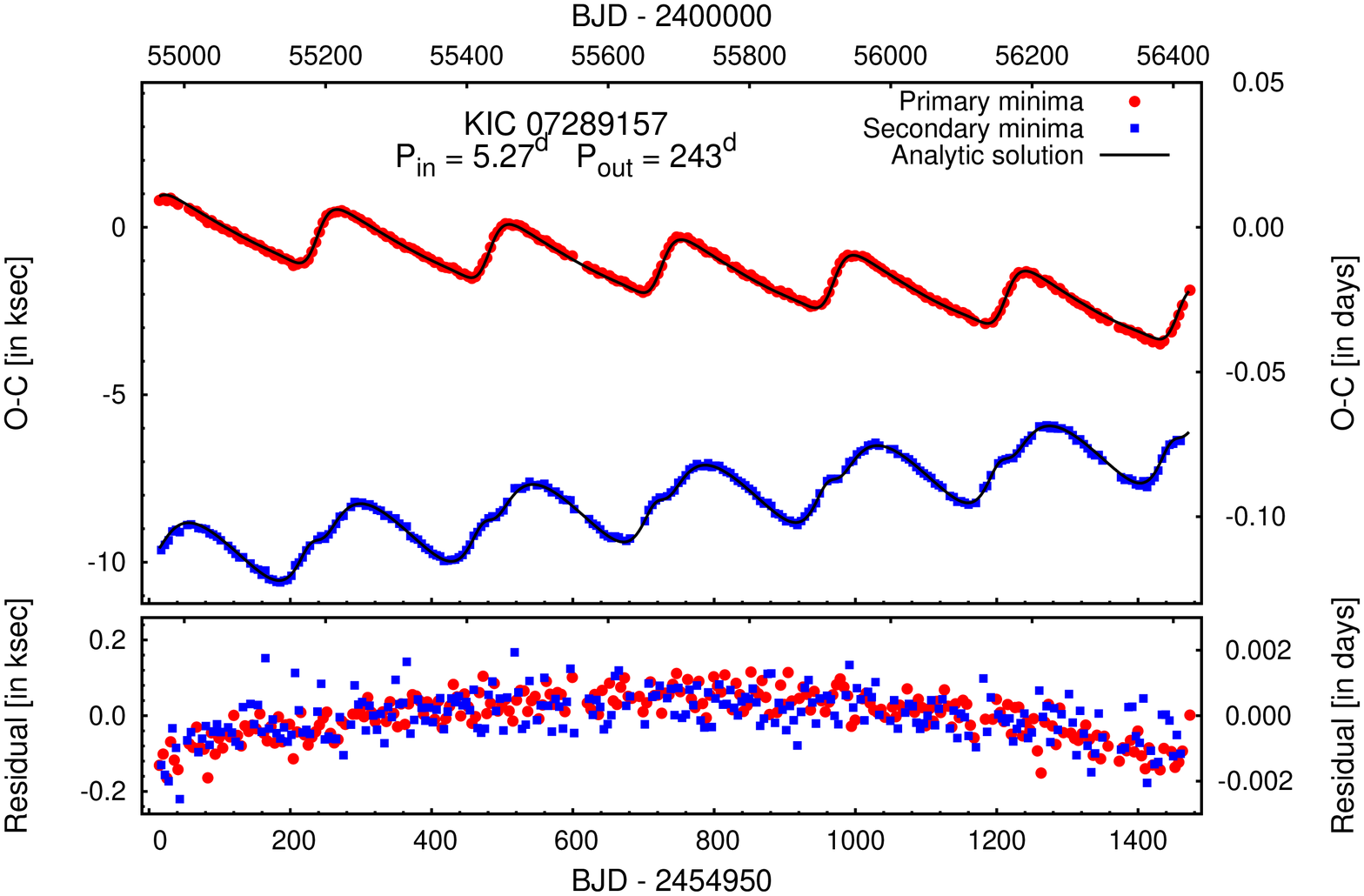}\includegraphics[width=84mm]{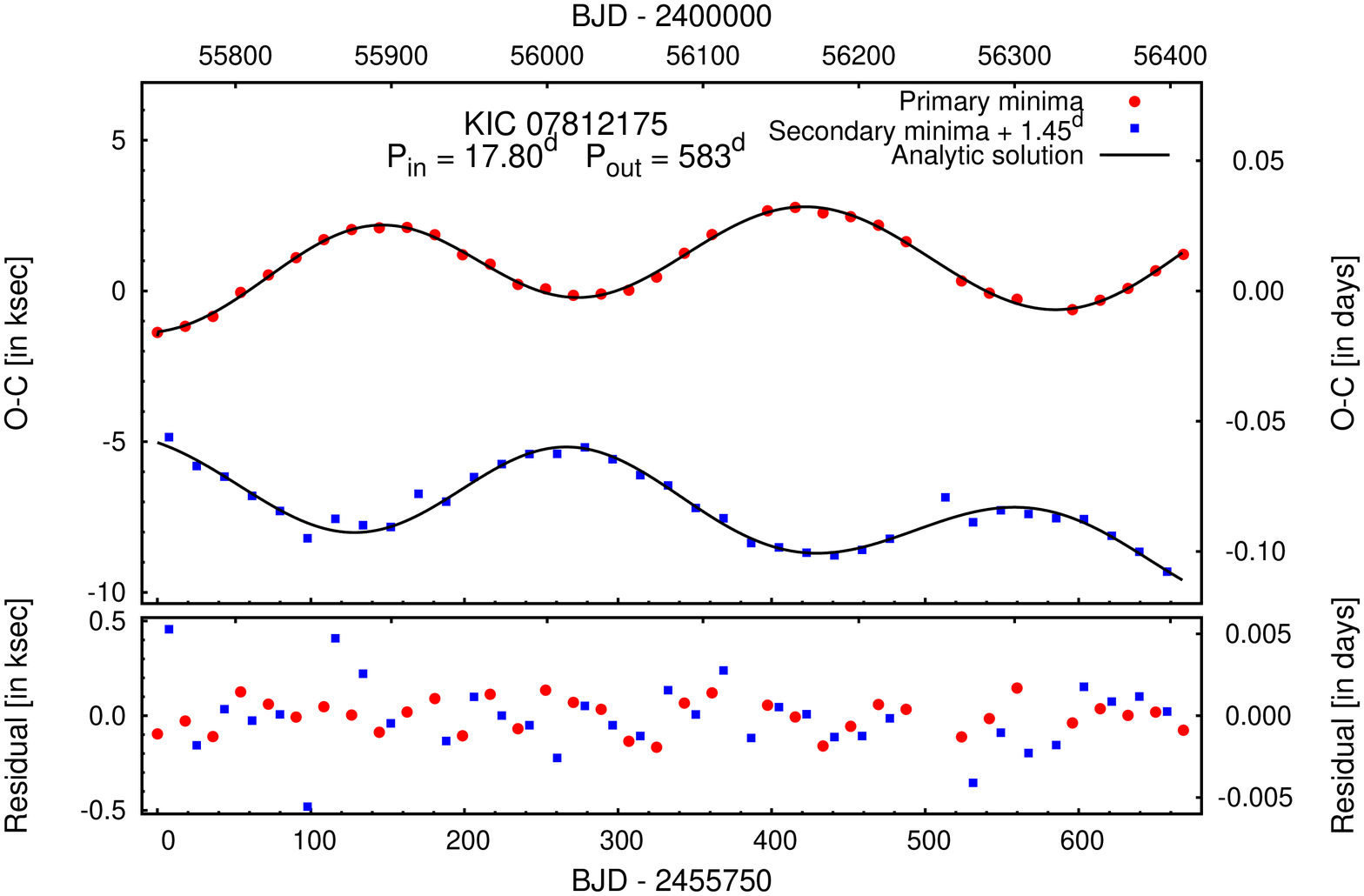}
 \caption{Eclipse Timing Variations (also known as $O-C$ curves) with fitted solutions for the first 6 of the 10 best-modeled systems {\it (top)} and the residual curves {\it (bottom)}. Each panel contains the KIC number of the system, as well as the inner binary period and outer triple period.  The points for the primary eclipses are shown in red, while the blue points are for the secondary eclipses.  The fitted analytic solution is shown with the continuous black curves. (Note, we shifted the secondary curves toward the primaries, where it was necessary for a better visualisation; however, we strictly preserved the order (i.e., the sign of the displacement) of the two curves and, consequently, rigorously avoided introducing false intersections between the curves (an error made in some recent papers). Where there has been a shift, it is listed in the panel's legend.)}
 \label{Fig:ETVall1}
\end{figure*}

\begin{figure*}
\includegraphics[width=84mm]{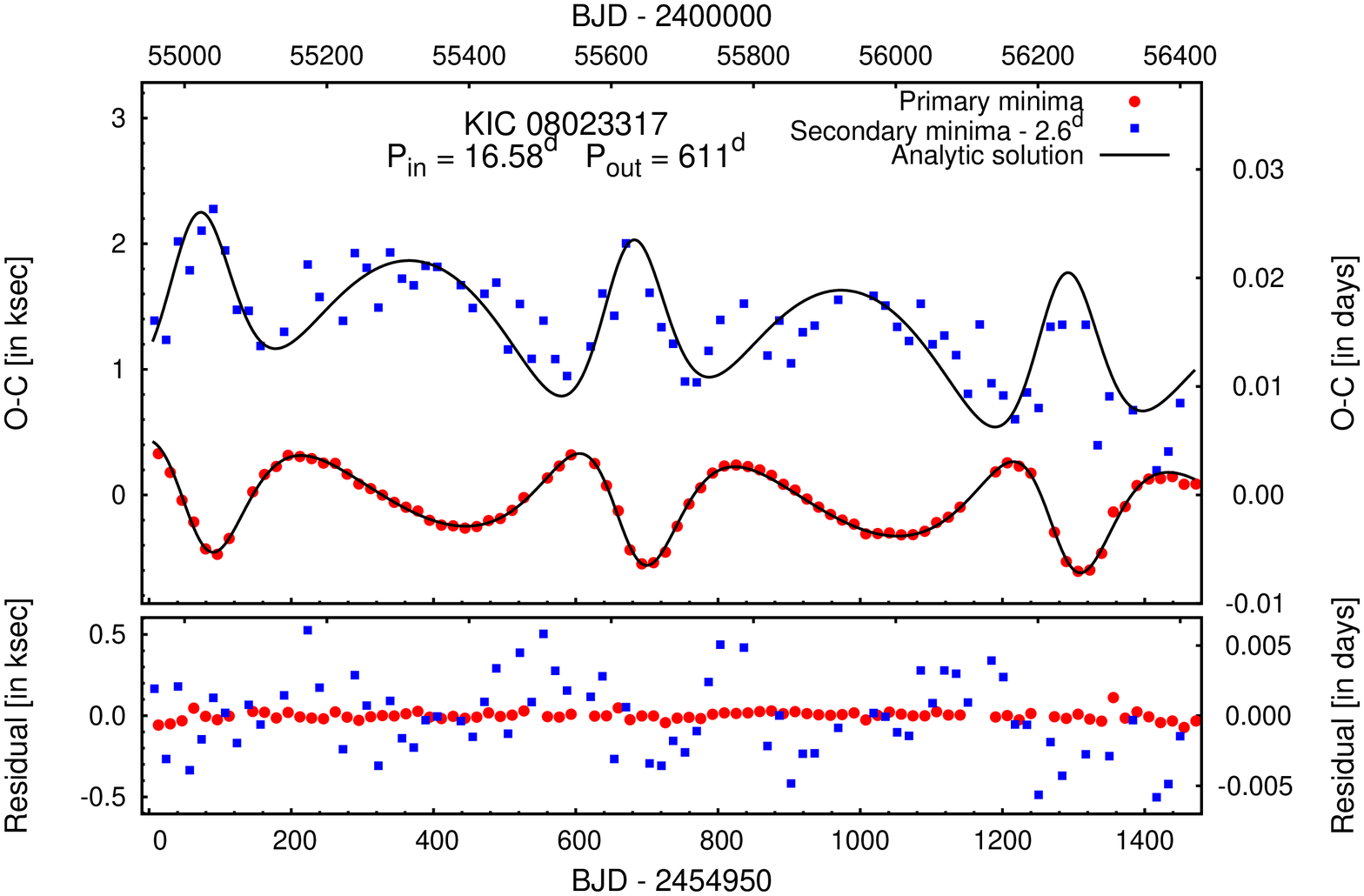}\includegraphics[width=84mm]{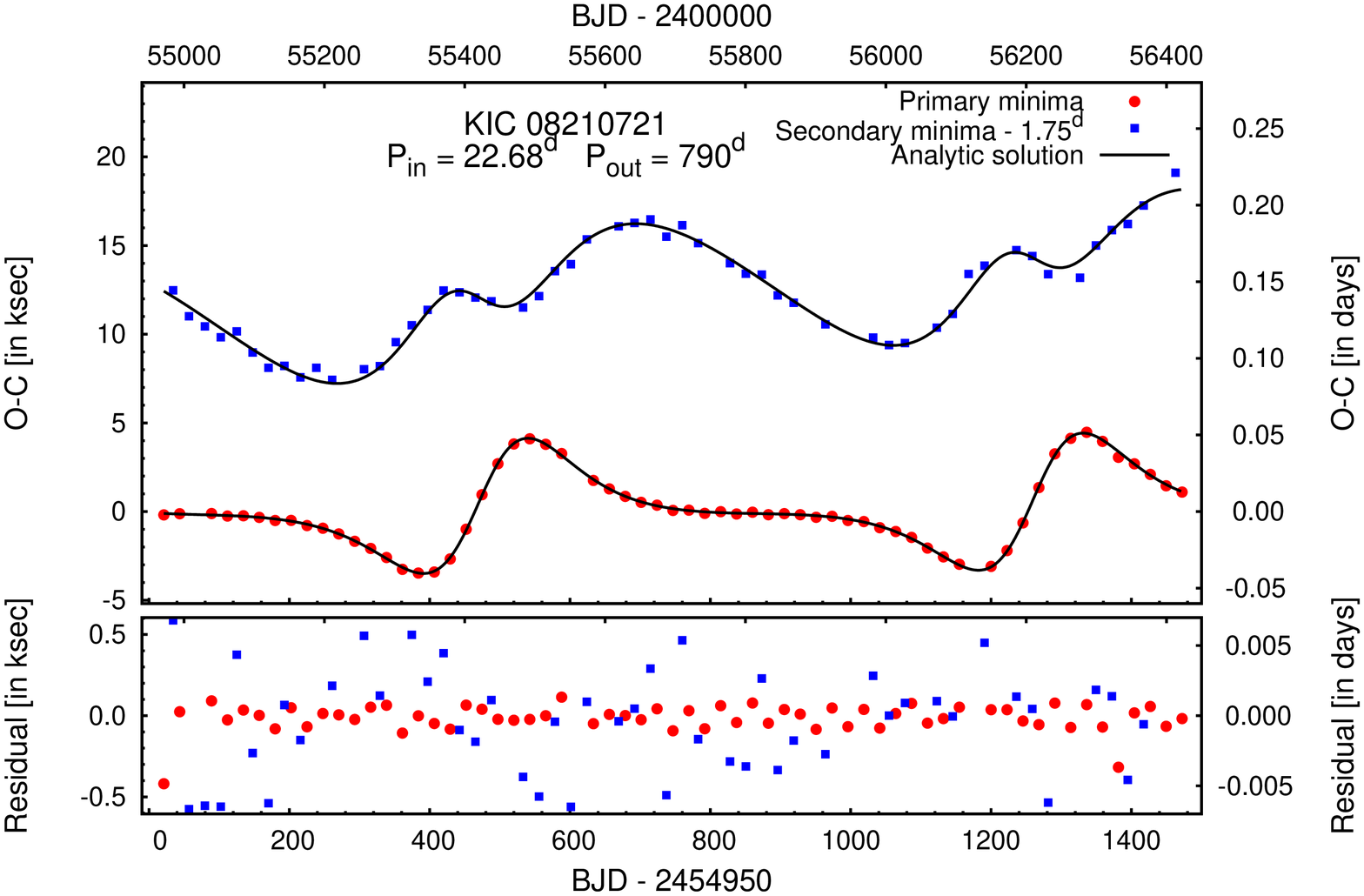}
\includegraphics[width=84mm]{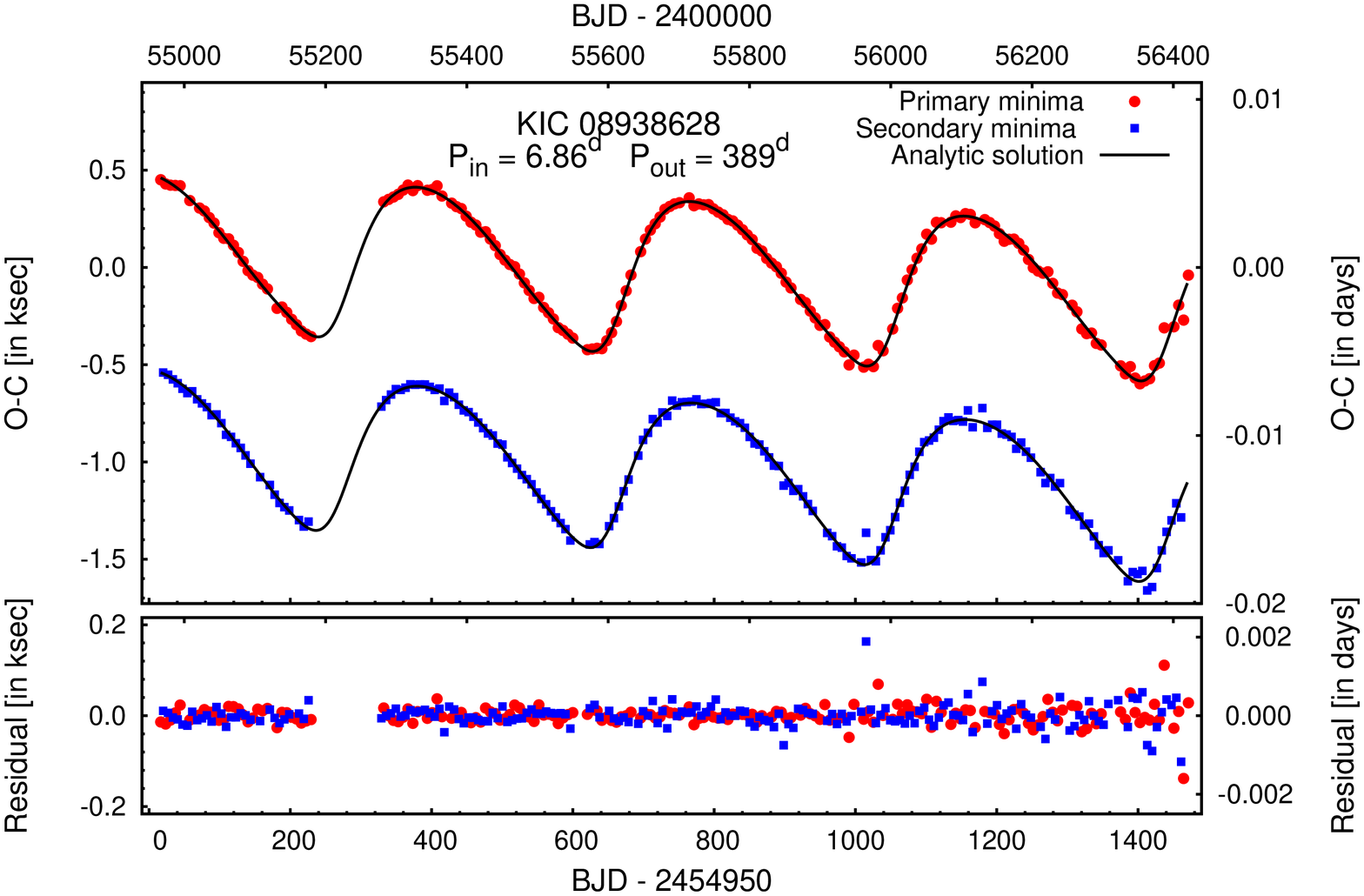}\includegraphics[width=84mm]{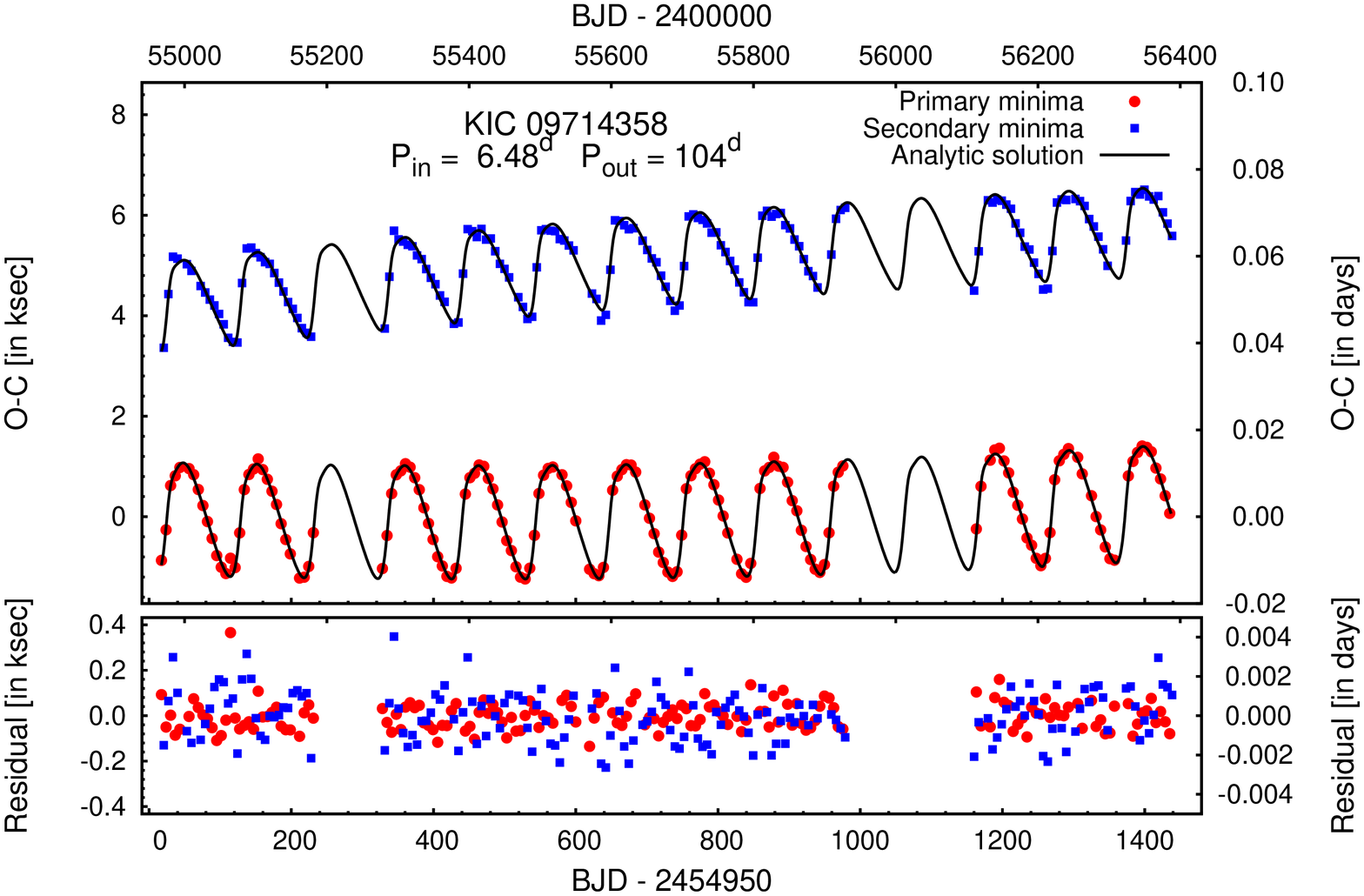}
 \caption{Eclipse Timing Variations with fitted solutions and residuals for the remaining 4 best-modeled systems. The specifications are otherwise the same as in Fig.~\ref{Fig:ETVall1}.}
 \label{Fig:ETVall1b}
\end{figure*}

\begin{figure*}
\includegraphics[width=84mm]{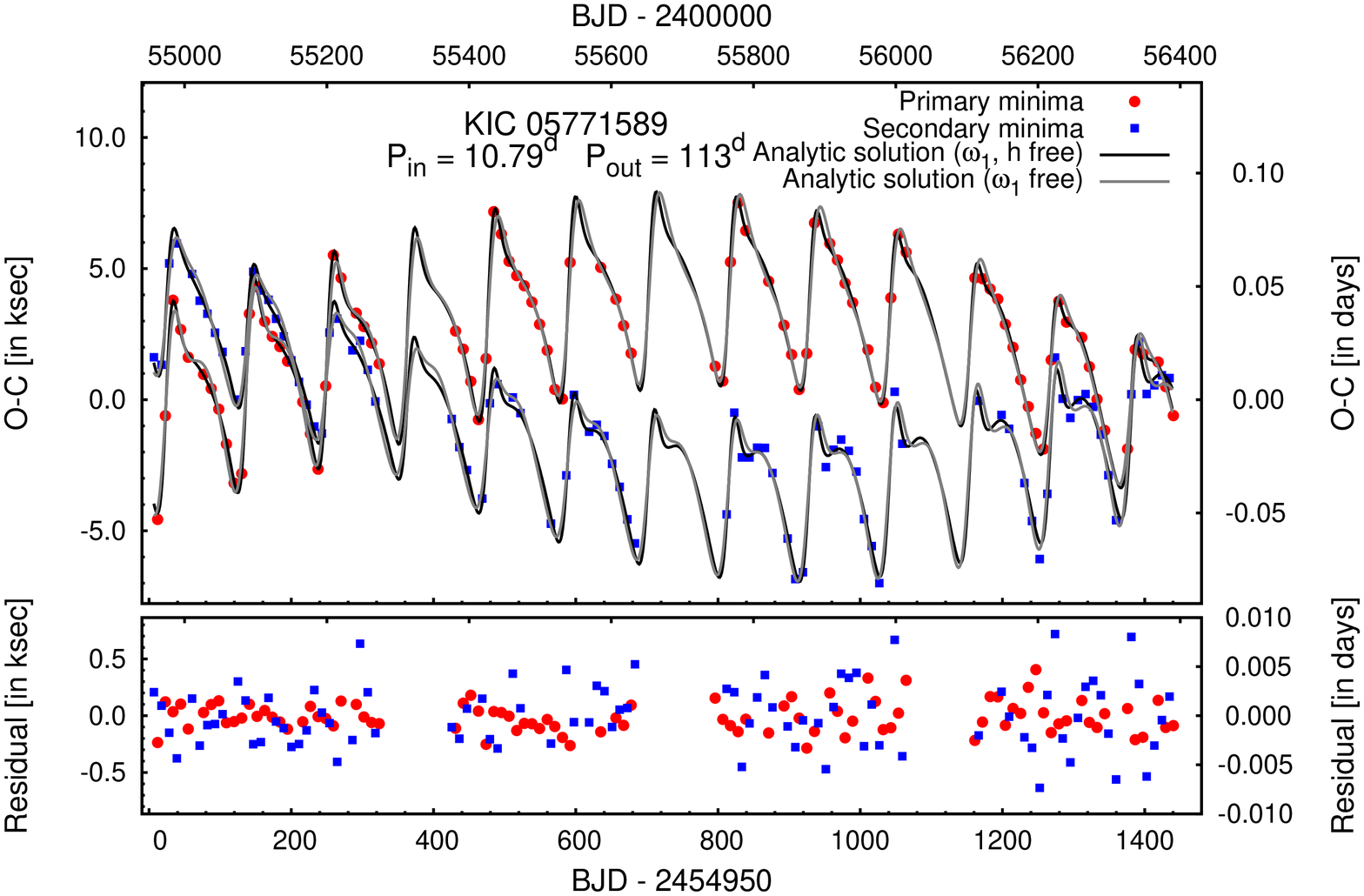}\includegraphics[width=84mm]{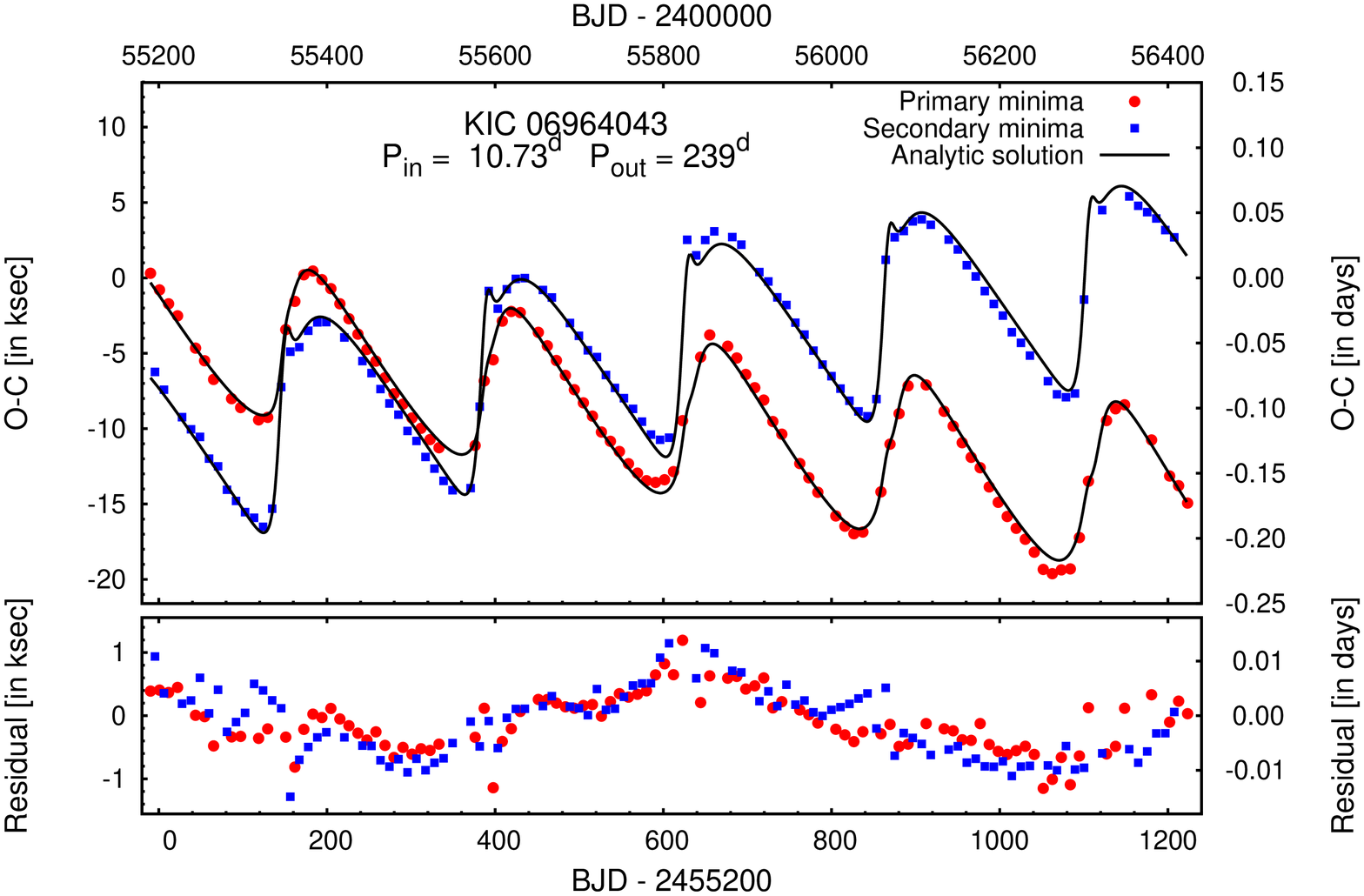}
\includegraphics[width=84mm]{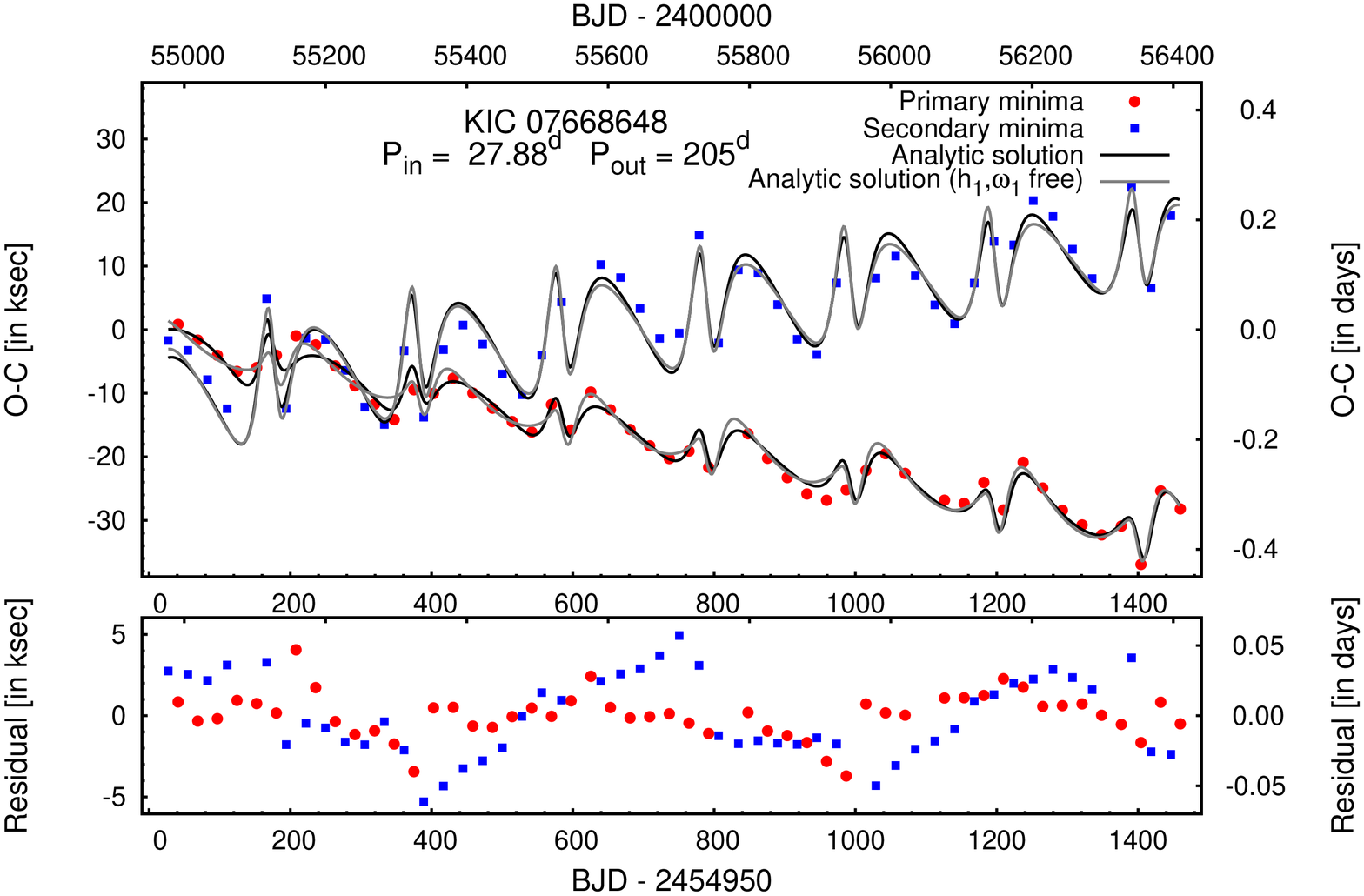}\includegraphics[width=84mm]{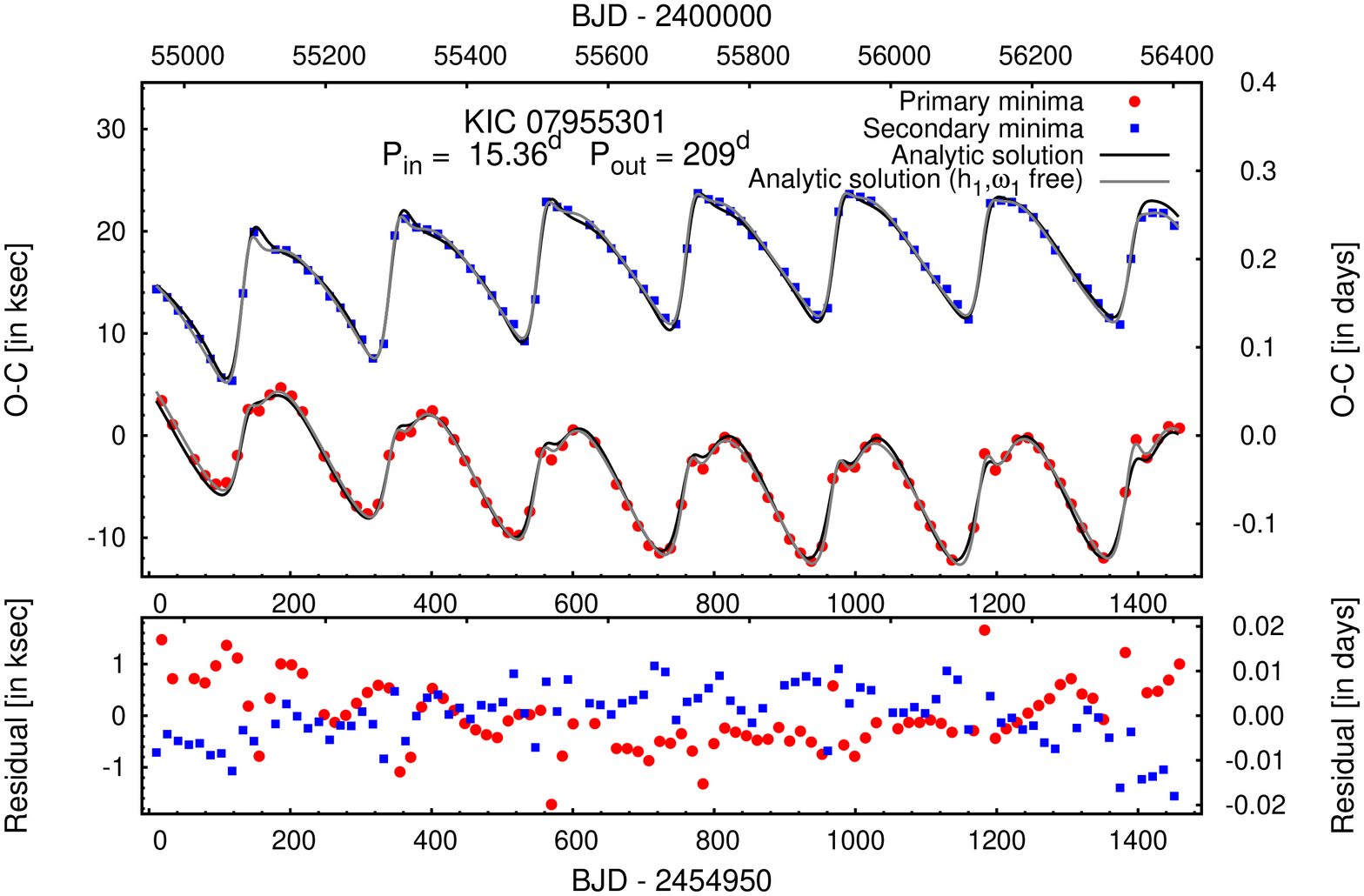}
 \caption{Eclipse Timing Variations with fitted solutions for the closest systems, i.e., those with the smallest ratios of $P_2/P_1$. The specifications are otherwise the same as in Fig.~\ref{Fig:ETVall1}. (Note, where two solutions are given, the residual curve is shown only for the first one listed in the Tables.)}
 \label{Fig:ETVall2}
\end{figure*}

\begin{figure*}
\includegraphics[width=84mm]{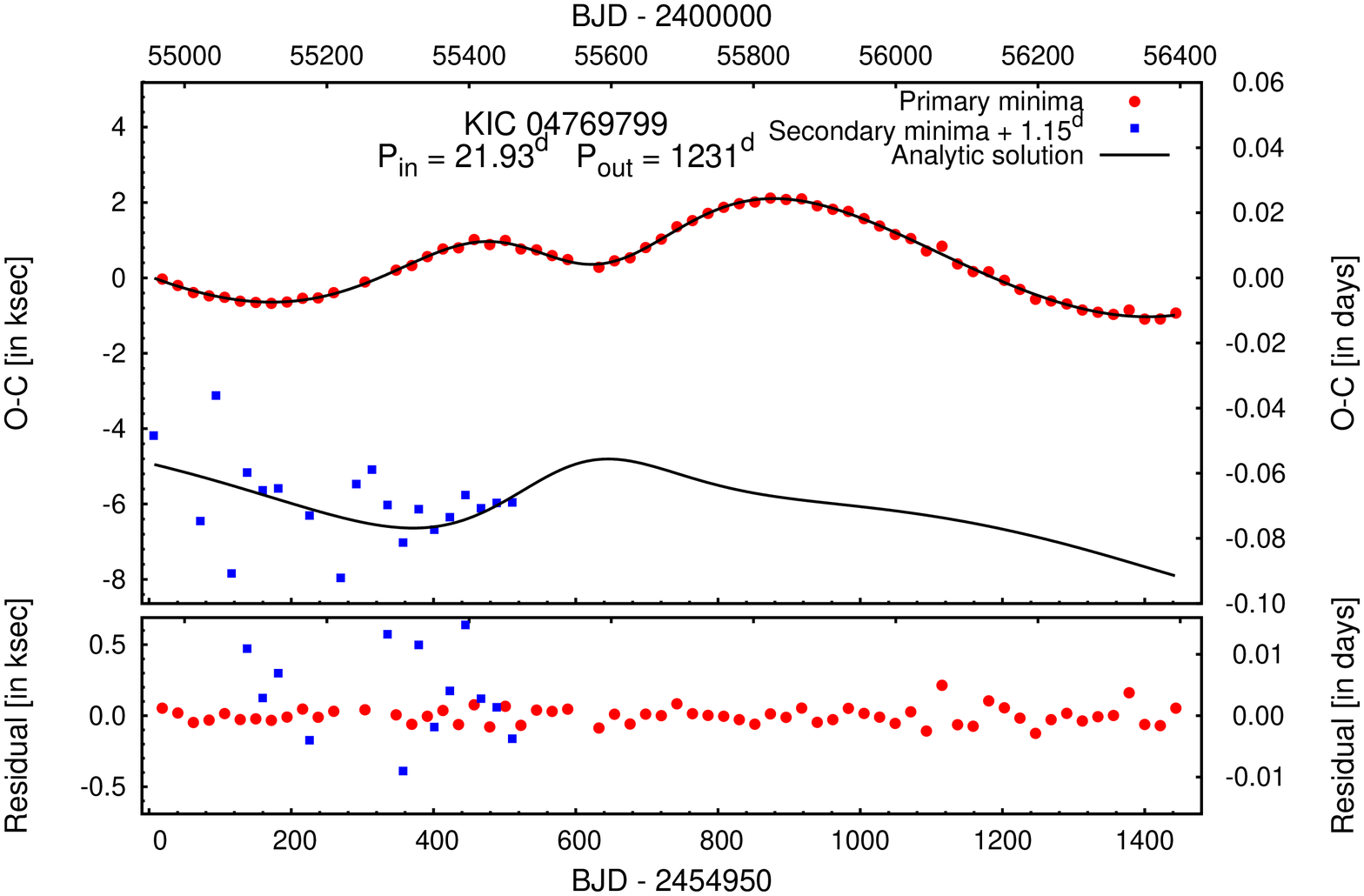}\includegraphics[width=84mm]{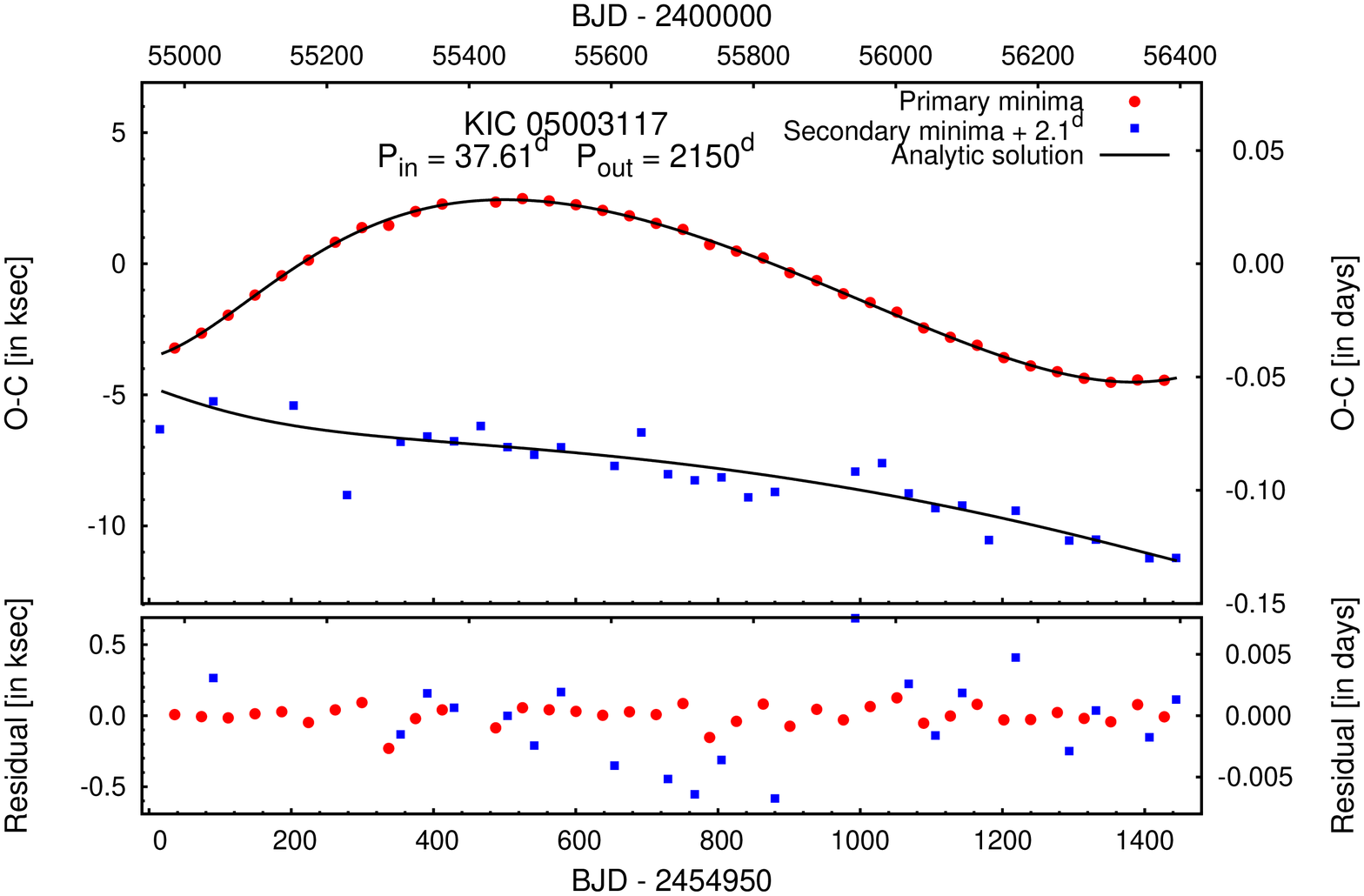}
\includegraphics[width=84mm]{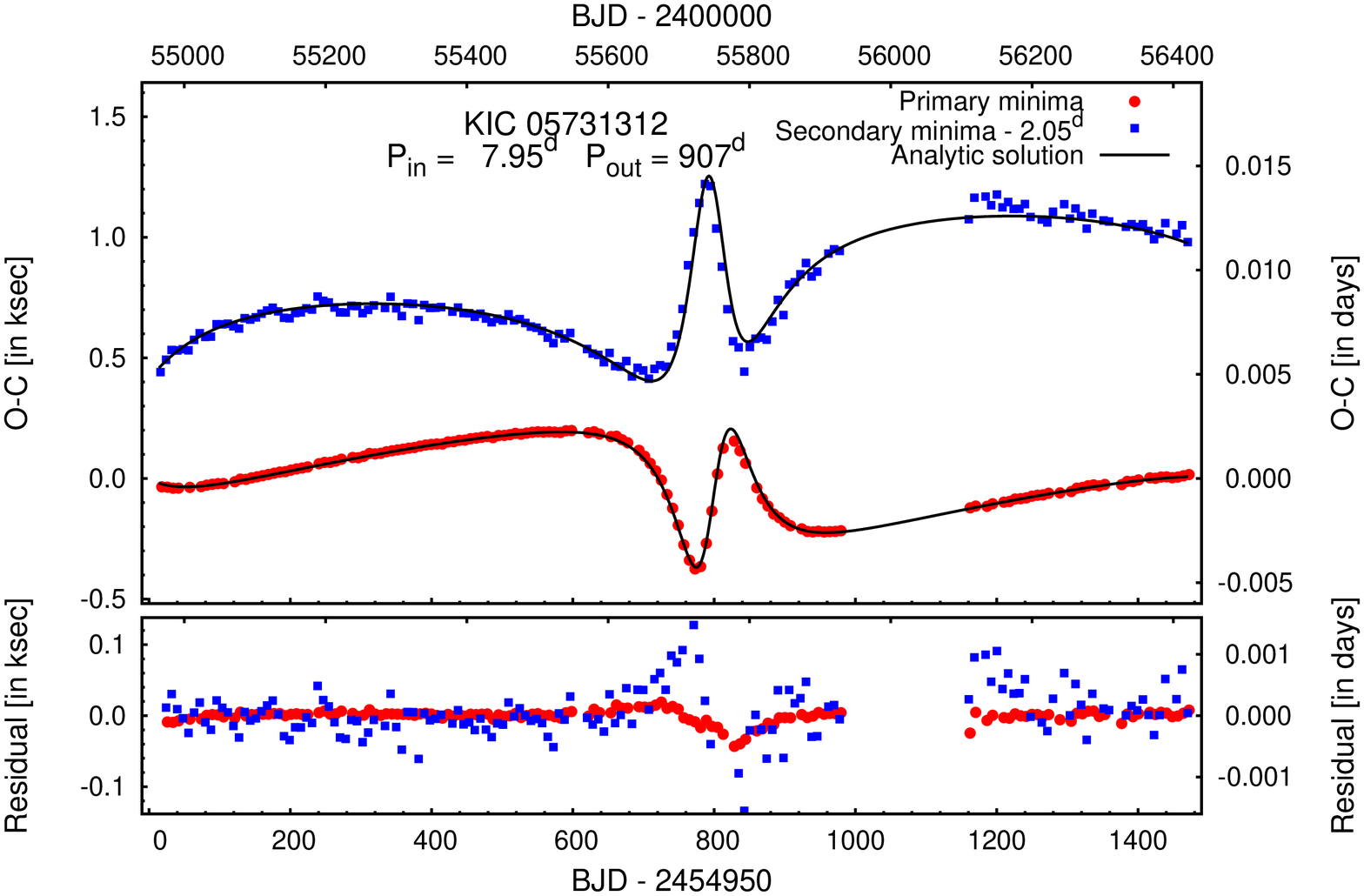}\includegraphics[width=84mm]{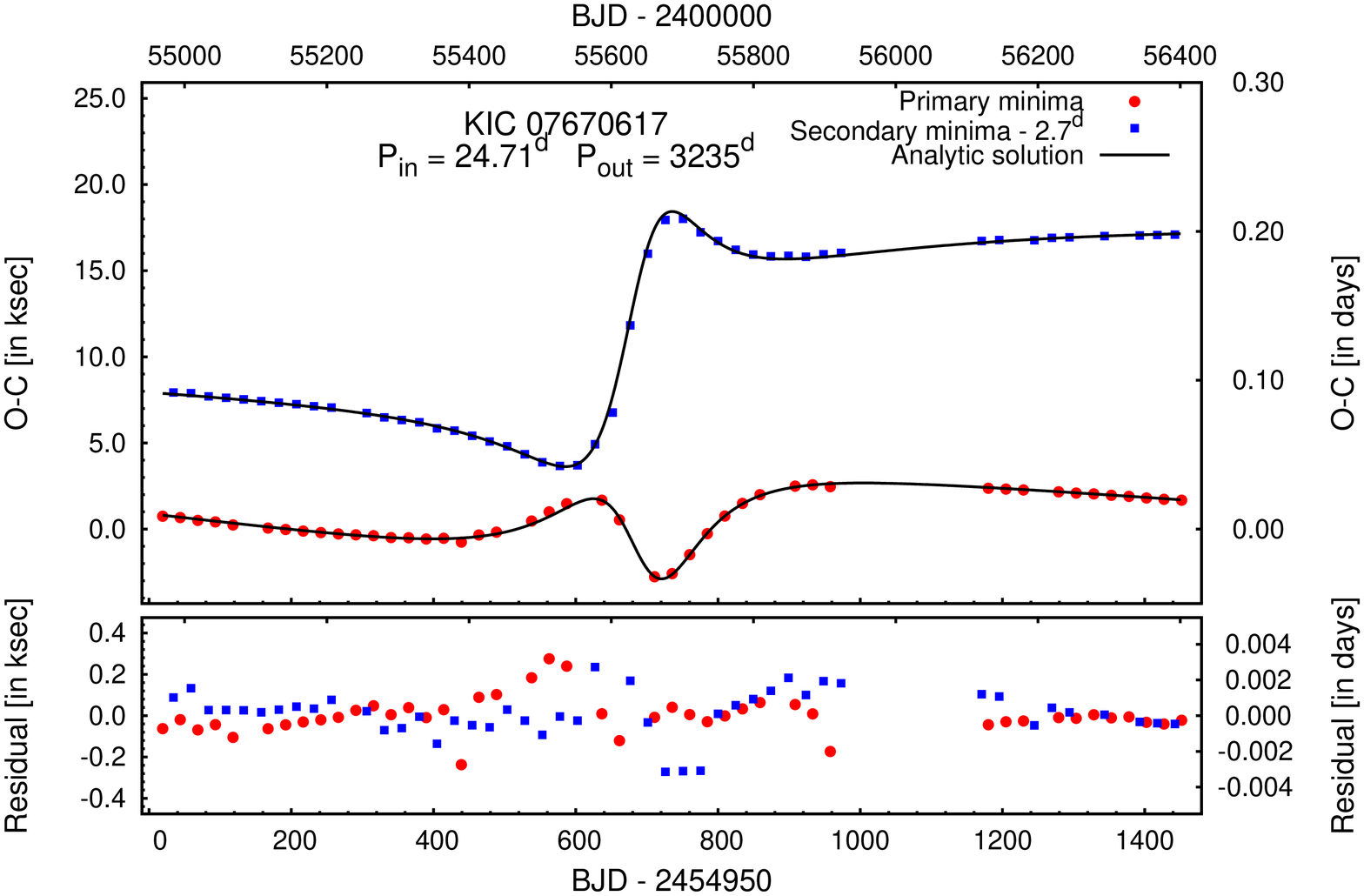}
\includegraphics[width=84mm]{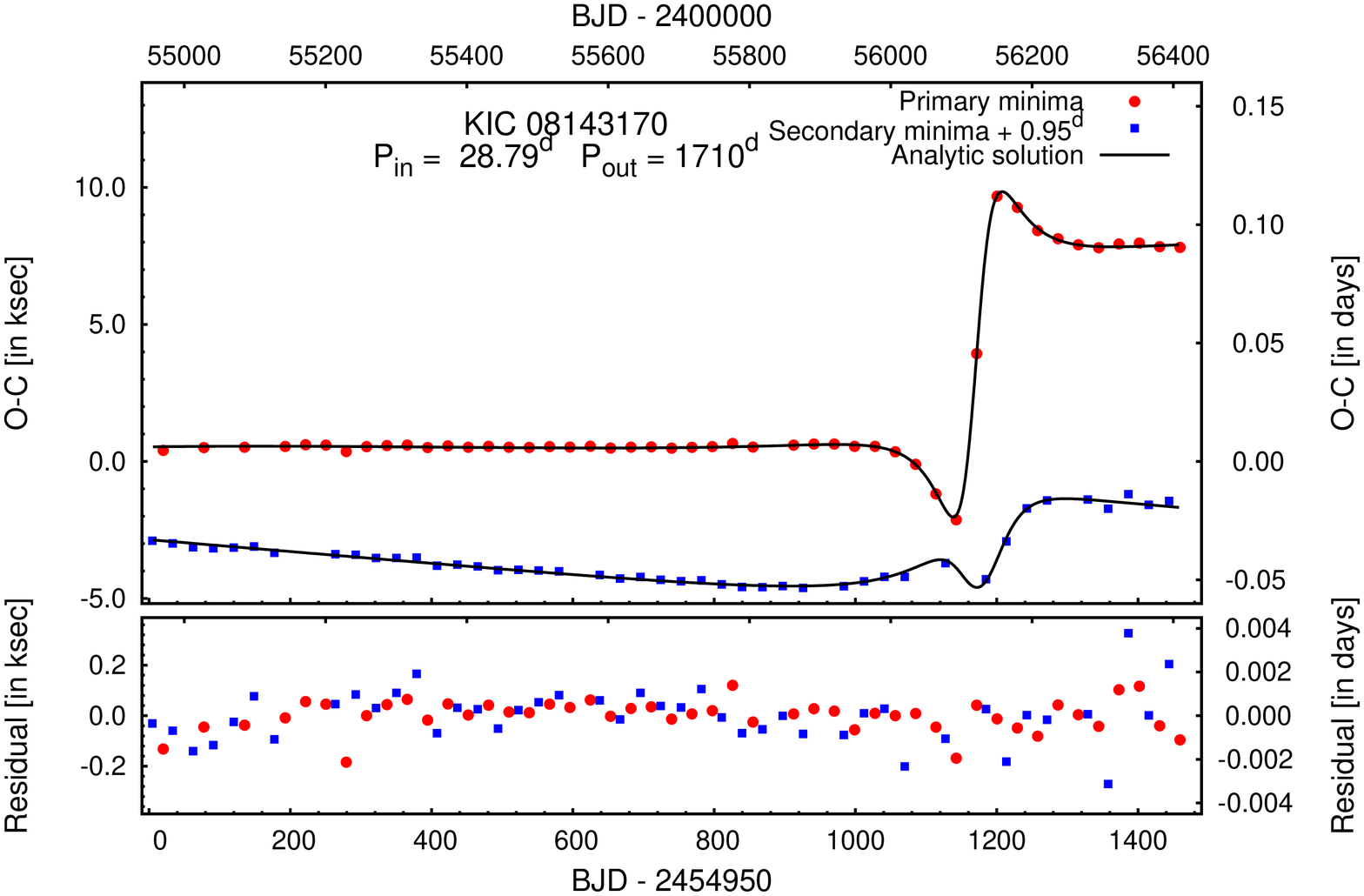}\includegraphics[width=84mm]{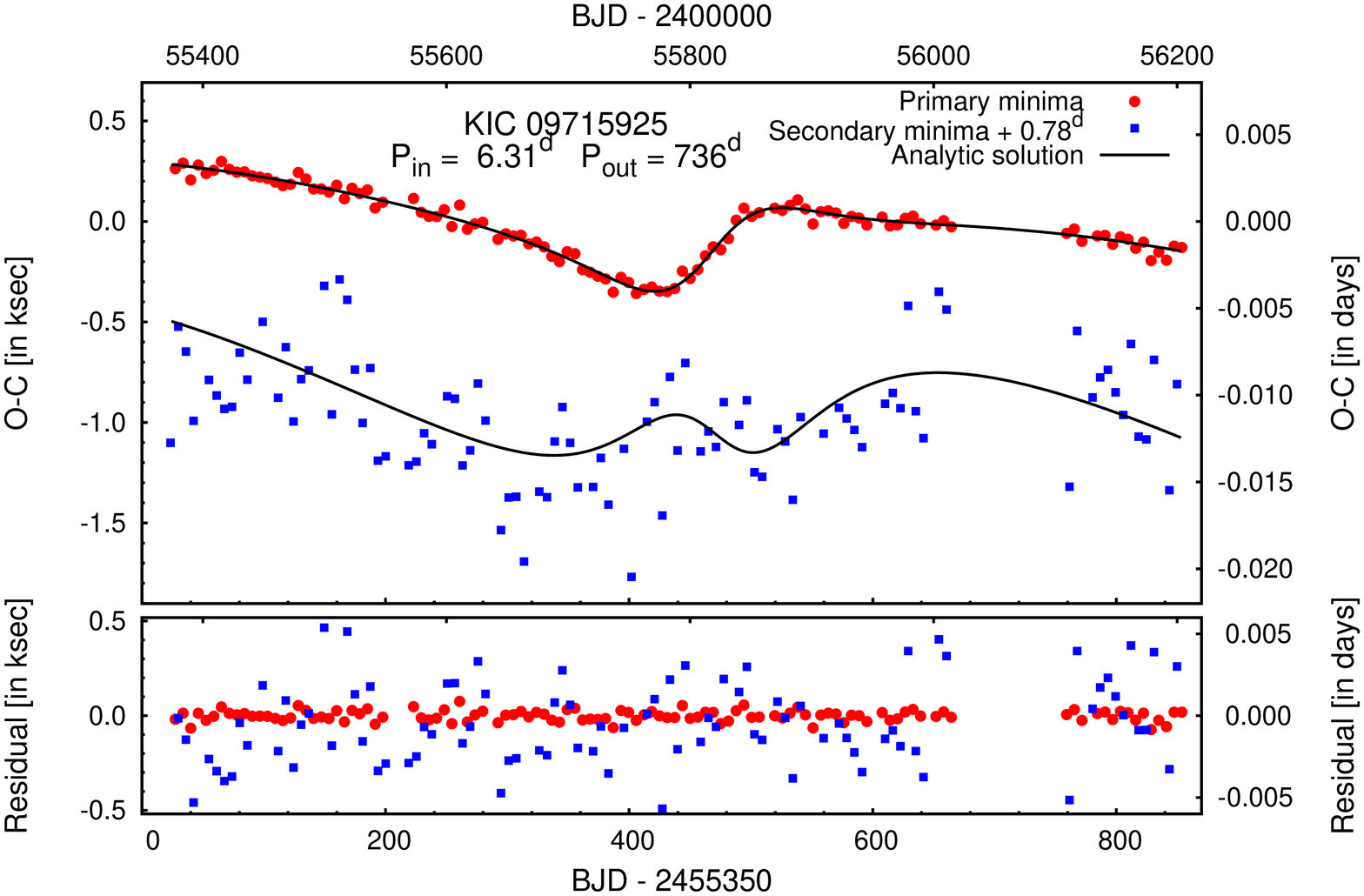}
 \caption{Eclipse Timing Variations with fitted solutions for the first 6 of our 12 systems with {\em Kepler} coverage that spans close to only one outer period or even less in some cases. The specifications are otherwise the same as in Fig.~\ref{Fig:ETVall1}.}
 \label{Fig:ETVall3a}
\end{figure*}

\begin{figure*}
\includegraphics[width=84mm]{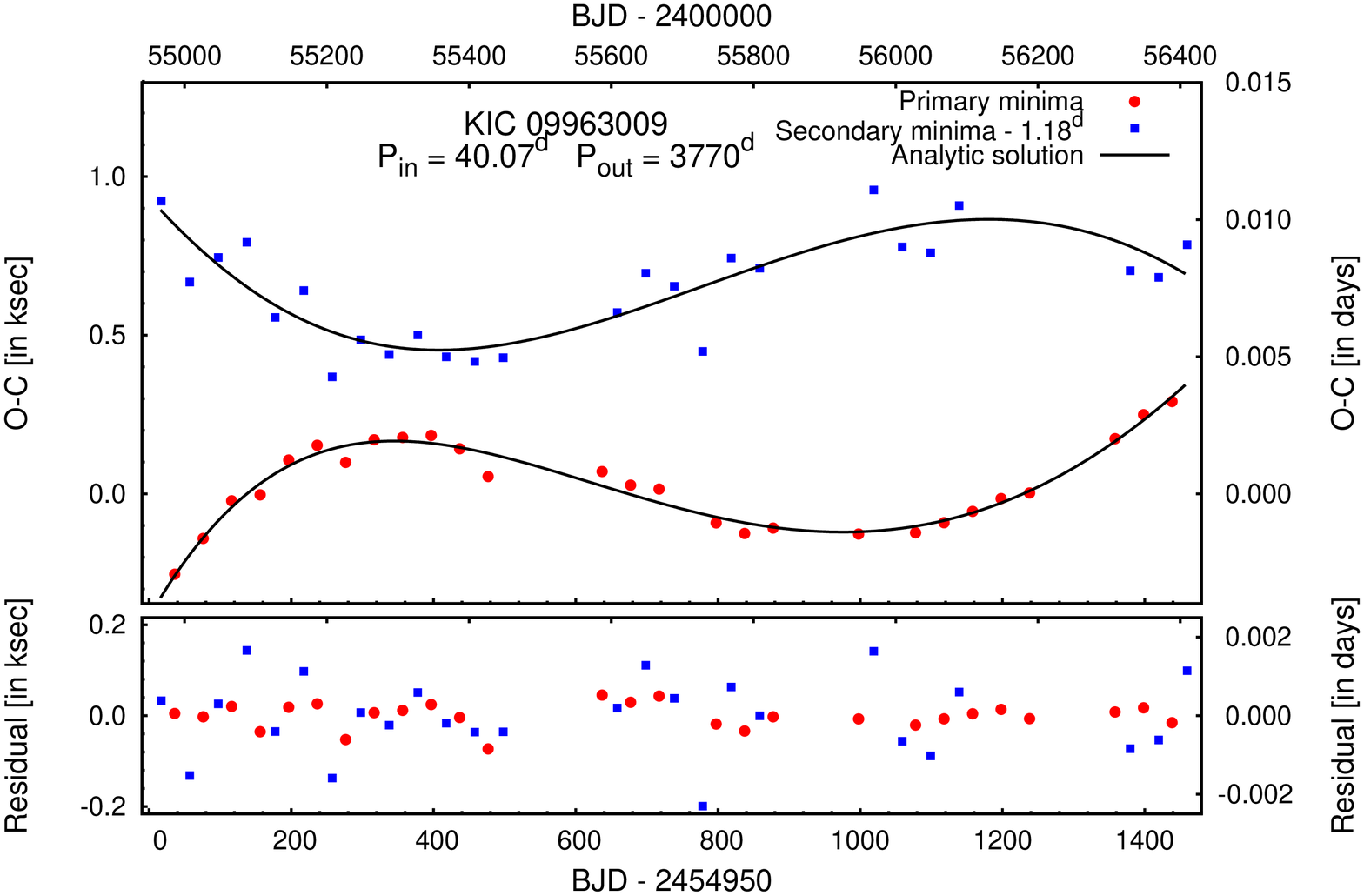}\includegraphics[width=84mm]{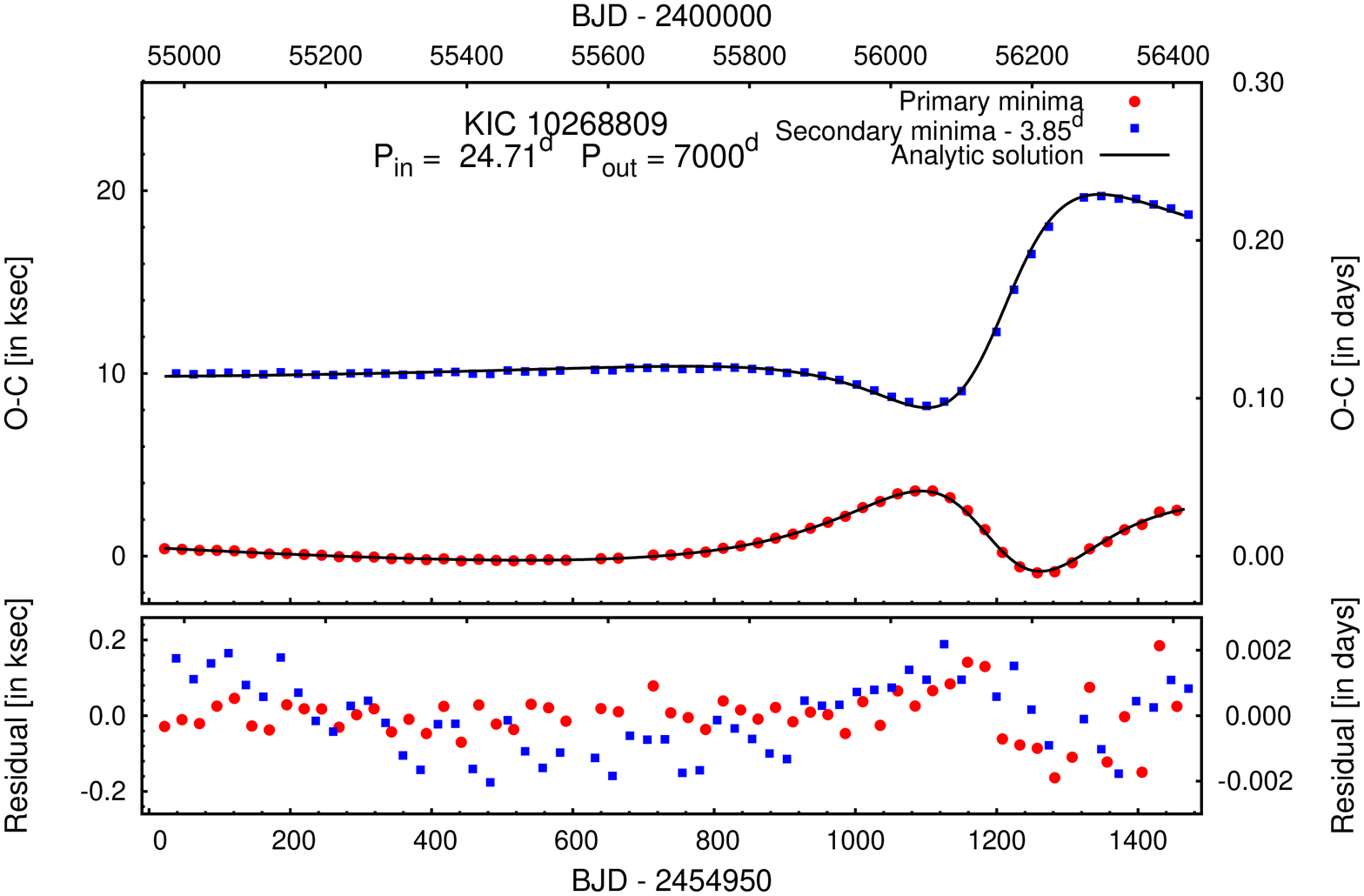}
\includegraphics[width=84mm]{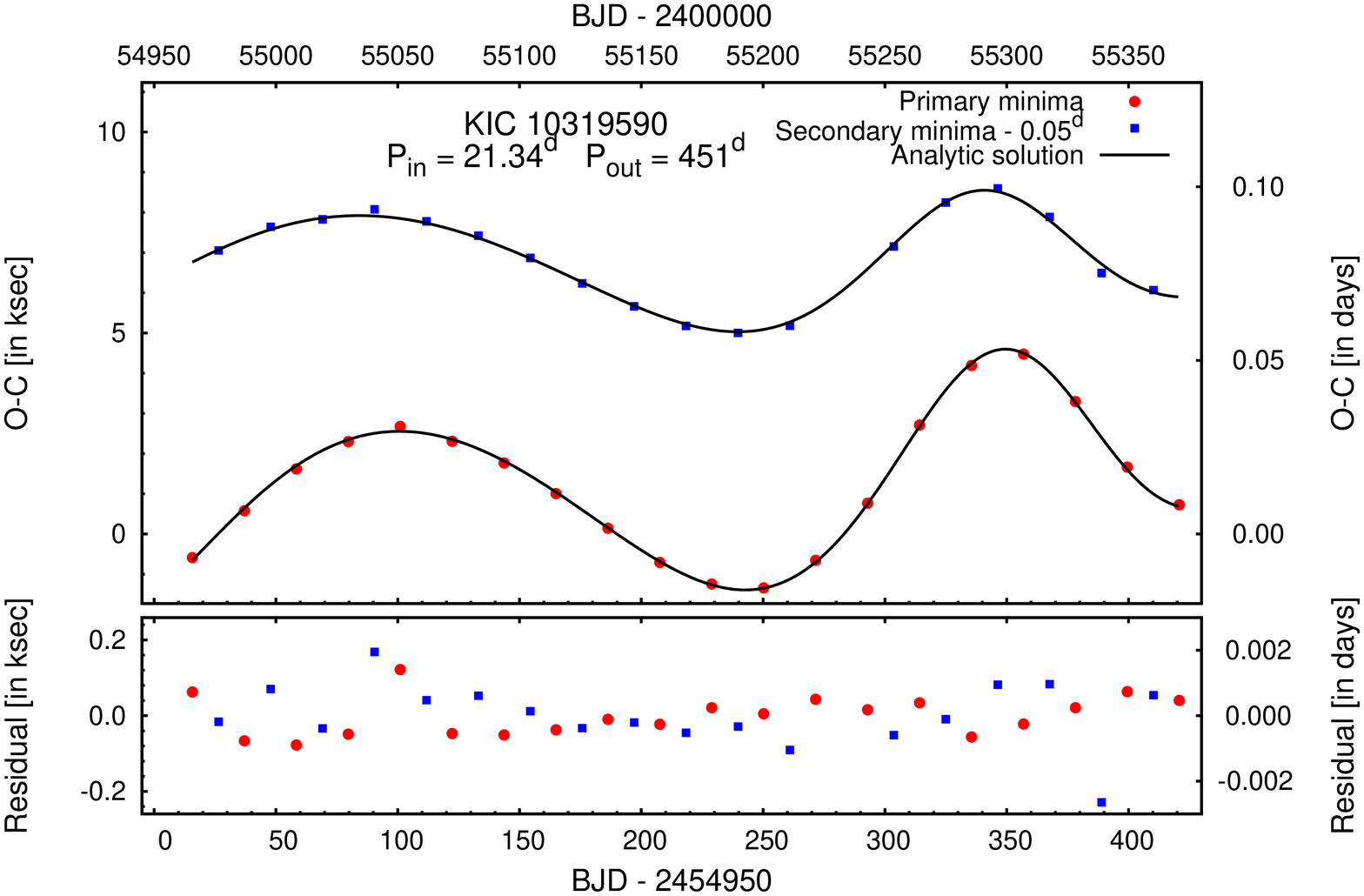}\includegraphics[width=84mm]{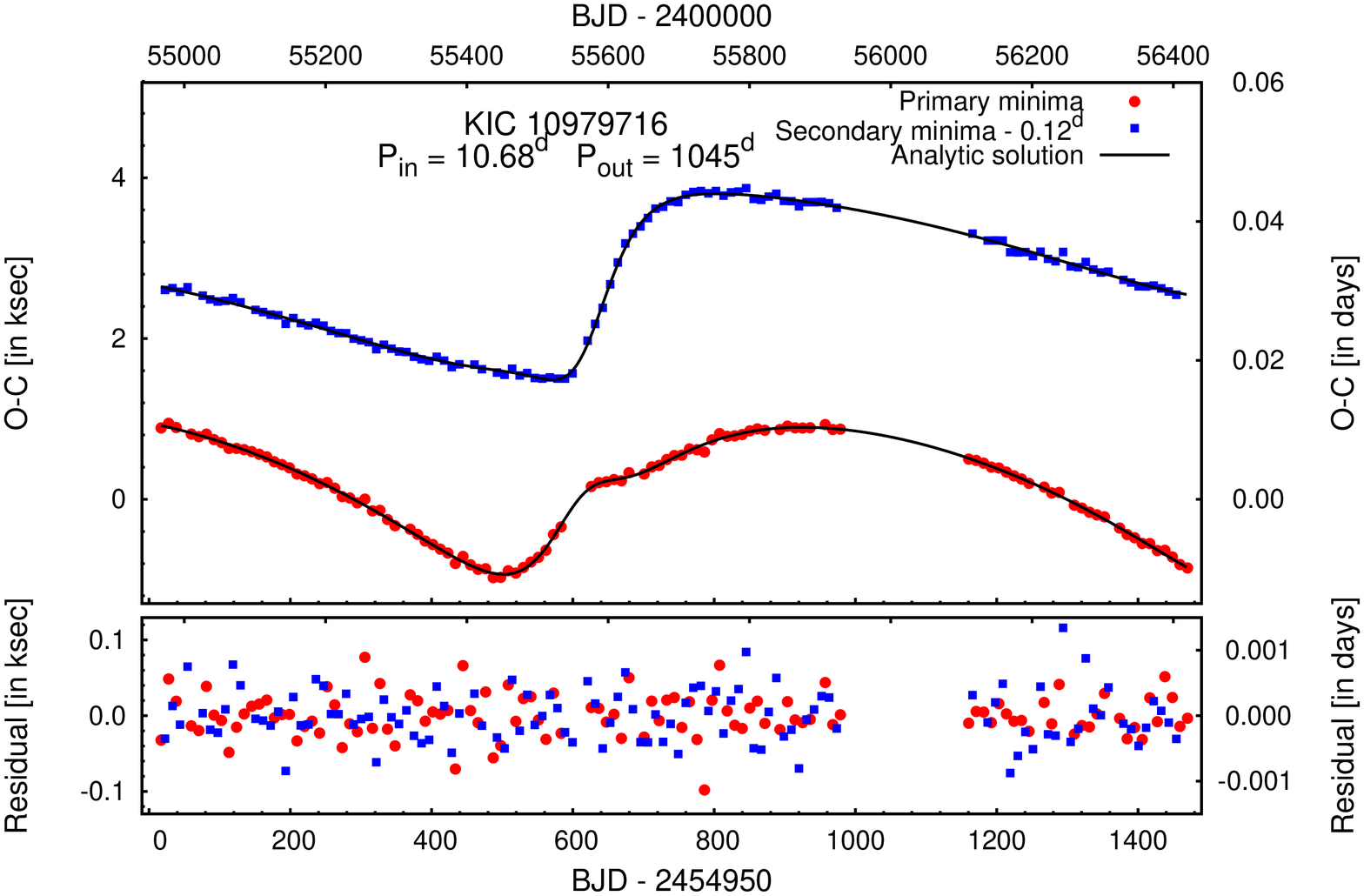}
\includegraphics[width=84mm]{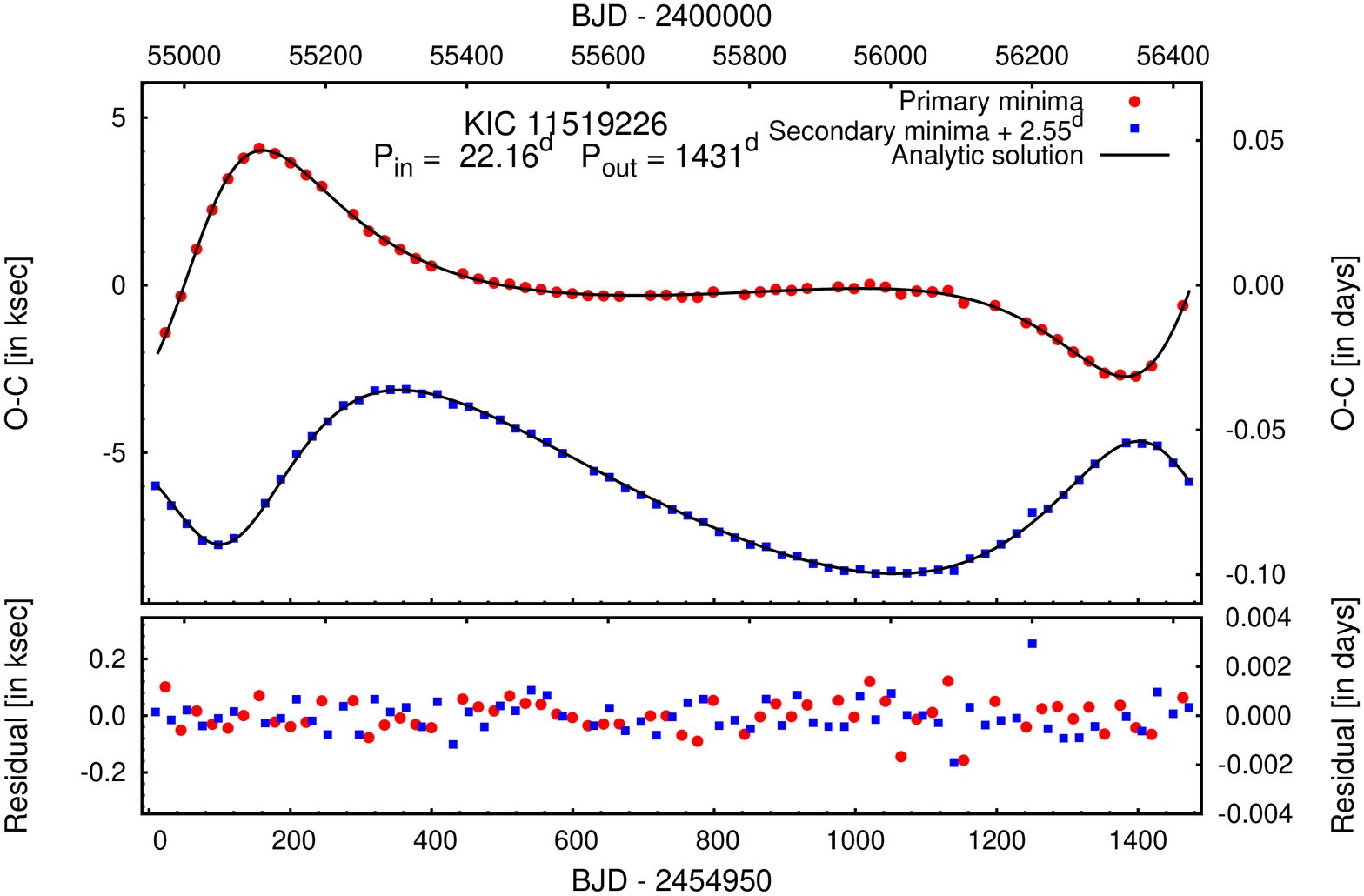}\includegraphics[width=84mm]{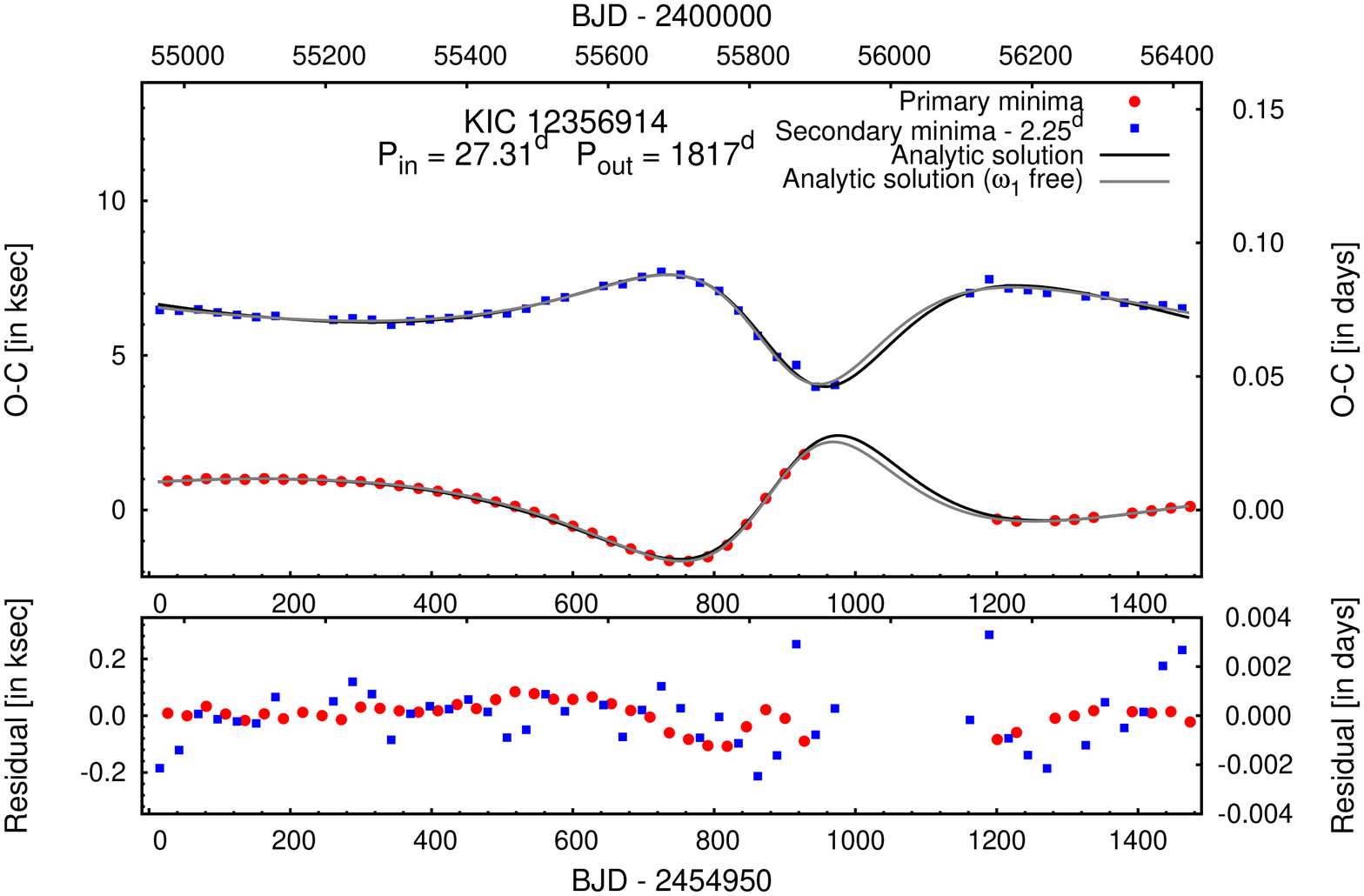}
 \caption{Eclipse Timing Variations with fitted solutions for the remaining 6 systems with {\em Kepler} coverage that spans close to only one outer period, or even less in some cases. The specifications are otherwise the same as in Fig.~\ref{Fig:ETVall1}. (Note, where two solutions are given, the residual curve is shown only for the first labeled one.)}
 \label{Fig:ETVall3b}
\end{figure*}

\section{Results for the 26 {\em Kepler} Compact Hierarchical Triples}
\label{sec:results}

\subsection{Overview of the Results}
\label{sec:overview}

The ETV ($O-C$) curves for the collection of 10 CHTs with the most robustly fitting solutions are shown in Figs.~\ref{Fig:ETVall1},~\ref{Fig:ETVall1b}. As is the case for the other 16 systems, the fitted parameters as well as some additional interesting derived quantities are given in four tables. Table \ref{Tab:Orbelem} contains the orbital elements for both orbits\footnote{Note, the second column gives the derived anomalistic period $P_1$ of the inner binary and, therefore, it is not a redundant parameter with the preliminary sidereal (or eclipsing) period $P_\mathrm{s1}$ listed in Table~\ref{Tab:Systemproperties}.  The latter period was used for the computation of the ETV (or $O-C$) curves.}. Table~\ref{Tab:3Delems}, lists the 3D orbital orientations with respect to both the observational and the dynamical frames of references. Moreover, the ratio of the orbital angular momenta for the inner and outer binaries, $C_1/C_2$, is also given. In the most conservative hierarchical three-body approximation this value is considered to be small (or, asymptotically zero). Here, for retrograde systems, we somewhat arbitrarily use negative signs. Table~\ref{Tab:Massetal} lists the constituent masses, mass ratios, LTTE-derived mass function and amplitudes of the separate constituents of the ETVs. It should be kept in mind that, in contrast to the ${\cal{A}}_\mathrm{LTTE}$ term, which gives the unique amplitude of the LTTE, the dynamical amplitudes may be, and are often strongly altered by the other orbital elements, as well as by the spatial configurations.  Therefore, the latter are given only for a rough comparison. Finally, Table~\ref{Tab:apsenodeetal} lists different periods or timescales of the observable and dynamical apsidal motions and orbital precession, as well as other quantities which characterize the secular evolution of the systems. Most of the listed orbital elements are doubly averaged osculating orbital elements, which are calculated for the epoch $T_0$ which is given in the third column of Table~\ref{Tab:Systemproperties}. Exceptions are the angular variables ($\omega_1$, $g_1$, $h_1$, $i_1$, $i_2$) for which their values are given both for the moment of the first and the last observed eclipse. Realistic estimated parameter uncertainties are also listed in the tables. The uncertainties are given only for those variables that were included in either the LM fitting process or the grid search. The majority of the cited uncertainties come from either the covariance matrix of the LM-solution, or the step-size of the grid-search process; however, in a few cases we adopted more conservative errors, as will be discussed in the next section. For quantities fixed for all the runs, this is also noted on the same lines of the tables, and denoted by the letter `f' after the numerical value.  The 10 systems shown in Figs.~\ref{Fig:ETVall1} and~\ref{Fig:ETVall1b} are listed above the first horizontal break line in these tables.

The next four systems listed in Tables \ref{Tab:Orbelem}-\ref{Tab:Massetal} between the two horizontal break lines have their $O-C$ curves shown in Fig.~\ref{Fig:ETVall2}.  These are the systems for which the ratio of $P_2$ to $P_1$ is smallest among the systems we considered, and lies in the range of $7 \lesssim P_2/P_1 \lesssim 22$. This is to be compared with the range of $P_2/P_1$ of $\sim16-57$ for the ten systems in Fig.~\ref{Fig:ETVall1}.  Our fits for these systems are significantly weaker, in our opinion, mainly due to insufficient modeling of the apsidal motion. In the case of three of these four systems we also give alternative solutions. A comparison of these solutions, however, reveals that despite the large uncertainties, our fits might yet be acceptable in the sense that the derived parameters can at least be used for statistical purposes (see Sect.~\ref{sec:stat}).

Finally, in Figs.~\ref{Fig:ETVall3a} and~\ref{Fig:ETVall3b} we present the fitted ETVs for the dozen remaining systems. The fitted parameters for these are given in Tables \ref{Tab:Orbelem}-\ref{Tab:apsenodeetal} below the final horizontal break line. These systems also yield remarkably good fits to the analytic models, but they have outer periods, $P_2$, with only three exceptions, that are longer than 1000 days; half of the systems have $P_2$ longer than the {\em Kepler} mission.  It is these longer periods that make the fits somewhat less reliable than for the systems shown in Figs.~\ref{Fig:ETVall1},~\ref{Fig:ETVall1b}.  Nonetheless, the incomplete orbital coverage is somewhat compensated for by the fact that 3/4 of these systems have a clear periastron passage of the outer orbit during the course of the {\em Kepler} mission. Note, for one of these latter systems, an alternative solution is also given. 

\setlength{\tabcolsep}{3.5pt}
\begin{table*}
\begin{center}
\caption{Orbital Elements} 
\label{Tab:Orbelem}  
\begin{tabular}{lccccccccccc} 
\hline
KIC No. & $P_1$   &$e_1$& $\omega_1$ &   $g_1$    & $\tau_1$&$P_2$ &   $a_2$   &$e_2$&$\omega_2$&$g_2$&$\tau_2$\\
        & (day)   &     & (deg)      &   (deg)    & (MBJD)  &(day) &(R$_\odot$)&     &   (deg)  & (deg) & (MBJD) \\
\hline
04940201&  8.8183 &0.001(1)&194--202(16)& 93--111&4965.42(38) &364.9(3)&278(24) &0.24(2)&247(5)&326&4864(7)\\
05255552& 32.4787 &0.307(1)&105--109(1) &311--322&4956.79(20) &862.2(2.9)&510(35) &0.43(1)&37(1) & 62&4875(4)\\
05653126$^a$&38.5071&0.272(010)&307--312(1)&220--231&4988.14(43)&968.4(1.4)&571(11)&0.19(1)&321(1)& 53&5467(3)\\
06545018&  3.9928 &0.003(1)&180--225(1) &247--355&4964.84(1) &90.60(1) &122(2)  &0.25(1)&226(1)&114&4970.2(3)\\
        &  3.9915 &0.003(1)&179--228(1) &  $-$   &4964.83(1) &90.58(1) &119(1)  &0.24(1)&229(1)&$-$&4970.4(1)\\
07289157&  5.2674 &0.083(1)& 65--81(1)  &215--249&4972.19(5) &243.4(1)&215(2)  &0.31(1)&157(1)&127&4941.6(6)\\
07812175& 17.7967 &0.160(4)&326--328(2) &257--263&5004.63(8) &582.5(1.8)&367(50) &0.031(4)&213(6)&318&4790(11)\\
08023317& 16.5778 &0.251(1)&178--175(1) &82.3--82.0&4976.81(4)&610.6(5)&242(11) &0.25(1)&164(1)&249&5014(3)\\
08210721& 22.6771 &0.142(1)&156--159(1) &103--112&4964.93(10)  &789.8(4)&498(11) &0.26(1)&211(1)&333&4628(4)\\
08938628&  6.8630 &0.003(1)&345--351(3) & 51--67 &4968.03(5) &388.5(3)&298(20) &0.20(1)& 56(2)&304&4822(4)\\ 
09714358&  6.4783 &0.015(1)&142--379(1) &   $-$  &4965.11(1) &103.78(2)&116(5)  &0.30(1)&120(2)&$-$&4977.4(6)\\
\hline
05771589& 10.7866 &0.013(1)&236--457(1) &134--398&4961.14(1)&113.14(1)&127(6)&0.23(1)&287(1)&  7&4976.1(5)\\
        & 10.7866 &0.013(1)&234--458(1) &174--583&4961.14(3)&112.97(3)&152(8)&0.13(1)&291(1)& 50&4977.9(5)\\
06964043& 10.7372 &0.055(1)& 77--115(1) &162--245&5195.10(1)&239.1(2)  &248(14)&0.52(1)&311(2)&216&5110(2)\\
07668648& 27.8764 &0.065(5)& 85--117(1) &$-32$--53&4976.75(8)&204.8(4) &192(11)&0.33(2)&352(4)& 57&4918(3)\\
        & 27.8592 &0.091(9)& 88--108(1) & 24--51 &4976.97(6)&203.77(37)&204(11)&0.37(1)&341(4)& 87&4922(3) \\
07955301& 15.3633 &0.031(1)&117--201(1) &182--341&4961.57(2)&209.43(14)&228(16)&0.28(1)&300(1)&189&4877.8(1.1)\\
        & 15.3714 &0.029(1)&114--215(1) & 37--152&4961.45(3)&209.06(10)&232(5)&0.32(1)&307(1)& 47&4879.0(1.2)\\
\hline
04769799& 21.9302 &0.101(21)&330--331(21)& 9--12 &4971.59(1.14)&1231(8)  & 653(74)&0.19(1)&233(9)& 95&5542(40)\\
05003117& 37.6137 &0.145(33)&308--309(9)&  4--6  &4989.04(80)&2150(100) &823(141)&0.26(1)&191(5)& 75&4743(46)\\
05731312$^b$&7.9461&0.420(1)&   184(1)  &255--256&4967.20(1)&906.7(3.4)&423(42)    &0.58(1)& 26(1)&282&4842(3)\\
07670617$^a$&24.7050&0.246(5)&136--138(1)&53--57 &4961.53(10)&3235(108) &1041(29)&0.70(1)& 86(1)&167&5641(36)\\
08143170$^a$&28.7868&0.146(3)&291--293(1)&160--163&4971.38(3)&1710(36)  &864(19) &0.70(1)&109(1)&163&6121(27)\\
09715925&  6.3082 &0.201(8)&    354(18) &278--279&5000.01(28)&736(36)   &325(56) &0.38(2)&136(7)&232&5082(42)\\
09963009$^a$&40.0714&0.224(102)&258--257(5)&258--257&4985.19(39)&3770(10) &1425(170)&0.24(6)&189(6)& 9&4073(79)\\
10268809& 24.7093 &0.314(2)&143--145(1) &258--260&4965.57(3)&7000(1000)&2208(60)&0.74(1)&293(1)&227&6147(169)\\
10319590$^a$&21.3370&0.026(1)&249--254(1)&155--162&4964.52(3)&451(3)    &287(11) &0.17(1)&336(2)& 67&4858(3)\\
10979716& 10.6835 &0.074(1)&106--108(1) & 97--101&4962.31(6)&1045(4)   &521(7)   &0.44(1)&61(1)&231&4520(3)\\
11519226& 22.1631 &0.187(1)&359--360(1) & 93--97 &4977.11(4)&1431(1)  &745(8)   &0.33(1)&322(1)&236&5010(2)\\
12356914$^a$&27.3080&0.403(30)& 108(7)  &251--252&4965.71(7)&1817(26)  &948(36)  &0.37(1) & 10(1)&234&5876(18)\\
        &27.3081  &0.325(3)&    113(1)  &251--252&4966.04(4)&1811(26)  &629(76)  &0.39(1) & 35(1)&346&5862(18)\\
\hline
\end{tabular}
\end{center}
{\bf Notes.} {(1) Single-valued columns represent doubly averaged osculating orbital elements for epoch $T_0$ (which is given in Table~\ref{Tab:Systemproperties}). (2)~$\mathrm{MBJD}=\mathrm{BJD}-2\,450\,000$. (3) Double-valued columns give the corresponding orbital elements at the times of the first and the last eclipse observations. (4) Uncertainties in the last digits of the fitted parameters are given in parenthesis. (5) Blank spaces in the KIC-number column indicate an alternative solution for the same system denoted in the previous row. \\ 
$^a:$ uniformly, and equally weighted primary and secondary eclipses; $^b:$ corrected secondary uncertainties (see text for details)}
\end{table*}

\setlength{\tabcolsep}{5pt}
\begin{table*}
\begin{center}
\caption{3D Orbital Orientations} 
\label{Tab:3Delems}  
\begin{tabular}{lccccccccccc} 
\hline 
KIC No. &$\im$& $i_1$    & $i_2$    &$n_1$&$n_2$&$\Delta\Omega$&$i_0$&$j_1$&$j_2$&   $h_1$   & $C_1/C_2$\\
        &(deg)& (deg)    & (deg)    &(deg)&(deg)&   (deg)      &(deg)&(deg)&(deg)&   (deg)   &  \\
\hline
04940201& 5.9(1.9) &85.0--85.9f&86.1--86.0&100.7&101(52)&    $-5.9$    & 85.9&  5.0& 0.9 &101.1--90.7 & 0.190 \\
05255552& 6.4(2.2) &83.7--84.1&89.5--89.4f&154.7&15(20)&    $-2.7$    & 88.5&  5.3& 1.1 &154.9--147.3& 0.212 \\
05653126&11.0(1.0) &87.0--88.1f&86.6--86.4&87.3&88(1) &   $-11.0$    & 86.6&  9.6& 1.4 & 87.9--81.2 & 0.151 \\
06545018&11.2(3) &86.0--77.2f&81.7--84.6&113.0&112(2) &     10.4     & 82.7&  8.5& 2.8 &292.3--228.6& 0.326 \\
        & 0.0f        &   88.0    &   88.0   & $-$ & $-$        &      0.0     & 88.0&  0.0& 0.0 &    $-$     & 0.321 \\
07289157& 4.3(1.3) &85.8--85.3&89.5--89.6f& 30.1& 30(10)&      2.2     & 89.1&  3.9& 0.4 &210.0--192.0& 0.116 \\
07812175&15.4(2.5) &85.9--86.7f&80.4--80.3& 72.6&7(10) &   $-14.9$    & 81.2& 13.1& 2.3 & 70.0--66.2 & 0.180 \\
08023317&49.5(6) &88.0--89.4f&92.9--92.1& 95.5& 95.1(7)&  $-49.3$    & 91.1& 31.0&18.5 & 95.8--93.1 & 0.617 \\
08210721&13.7(1.0) &89.5--90.5f&81.3--81.2& 52.7& 54(14)&   $-11.0$    & 81.9& 12.6& 1.1 & 53.5--47.3 & 0.085 \\
08938628&13.3(1.0) &87.0--85.0f&81.8--82.0&113.4&112(5) &     12.3     & 82.2& 12.2& 1.1 &344.8--350.8& 0.091 \\
09714358&0.0f         &  83.0f    &   83.0   & $-$ & $-$        &      0.0     & 83.0&  0.0& 0.0 &    $-$     & 0.455 \\
\hline
05771589&21.9(4) &85.8--98.3f&90.0--86.5&100.0&101(4) &   $-21.6$    & 89.1& 17.2& 4.7 &101.4--57.6 & 0.280 \\
        & 7.9(1.4) &85.9--78.7f&82.1--83.2& 60.1& 61(12)&    $-6.9$    & 82.6&  6.9& 1.0 &421.0--234.2& 0.145 \\
06964043&19.2(1.6) &91.2--79.2&89.5--91.6f& 94.9& 95(5) &     19.1     & 89.7& 16.4& 2.7 &275.0--229.2& 0.169 \\
07668648&40.9(1.5) &84.1--105.2f&102.5--85.2&117.3&115(2)&  $-36.6$    & 94.5& 22.5&18.4 &117.6--60.7 & 0.825 \\
        &42.3(1.5) &83.9--86.3f&68.0--65.7& 63.7& 74(9) &   $-40.6$    & 74.6& 22.6&19.8 & 67.8--60.6 & 0.881 \\
07955301&19.1(7) &83.0--63.6f&75.4--78.5&114.9&112(3) &     17.9     & 76.4& 16.5& 2.6 &292.1--216.2& 0.161 \\
        &19.3(8) &83.1--87.1f&79.2--78.4& 77.3& 80(6) &   $-19.2$    & 79.8& 16.1& 3.2 & 79.5--64.3 & 0.200 \\
\hline
04769799&21.7(2.1) &86.0--85.7f&69.4--69.5&141.0&138(35)&    14.4    & 72.1& 18.1& 3.6 &318.8--317.1& 0.204 \\
05003117&44.0(1.0) &89.0--88.7f&66.3--66.3&124.1&115(9) &    38.9    & 68.4& 39.1& 4.9 &297.2--296.6& 0.136 \\
05731312&37.8(4) &88.5--88.0f&77.4--77.6&108.9&104(2) &    36.4    & 79.8& 28.3& 9.5 &286.1--285.0& 0.347 \\
07670617&147.1(5)&86.0--84.8f&   89.3   & 82.4&98.6(9)&  $-147.5$  & 88.6&142.9& 4.3 & 98.5--100.4&$-0.124$\\
08143170&38.5(3) &89.0--89.6f&113.6--113.3&131.7&125.5(5)&$-30.5$  &105.0& 24.4&14.1 &129.4--127.6& 0.591 \\
09715925&36.9(2.3) &83.2--83.6f&   76.1   & 76.2&83(10) &  $-36.9$   & 76.6& 33.0& 3.9 & 81.8--81.2 & 0.125 \\
09963009&33.7(2.8) &   89.0f   &   55.3   &  0.0& 0(3) &   $-0.0$   & 57.6& 31.4& 2.3 & 0.0--$-0.1$& 0.077 \\
10268809&23.7(4) &84.0--83.3f&   93.8   & 66.1&65.7(1.3)&    21.6    & 93.2& 22.2& 1.5 &245.6--243.7& 0.071 \\
10319590&135.4(3)$^a$&88.0--85.5f&94.0--94.4&93.7&88.8(8)& $-135.4$  & 94.1&128.5& 6.9 & 89.3--92.5 &$-0.154$\\
10979716&9.0(1.3)     &86.0--86.0f&77.2--77.2&  9.5&  9.7(9.3) & $-1.5$& 78.0&  8.1& 0.9 &  9.7--7.9  & 0.110 \\
11519226&17.0(3)    &88.0--87.2f&89.3--89.4& 85.7& 85(1) & 17.0  & 89.2& 15.6& 1.4 &265.4--262.5& 0.091 \\
12356914&143.1(1.0)   &88.0--88.5f&120.4--120.3&37.2&135.6(6)& 155.1 &126.8&133.9& 9.2 &311.1--312.2&$-0.223$\\
        & 40.2(3)   &88.0--87.5f&116.7--116.8&42.5&49.0(5) &  29.2 &111.0& 31.7& 8.5 &226.3--225.0& 0.281 \\
\hline
\end{tabular}
\end{center}
$a$: numerical checks have also resulted in prograde solutions with $\im=43\pm2\degr$ 
\end{table*}

\begin{table*}
\begin{center}
\caption{Mass related and other quantities} 
\label{Tab:Massetal}  
\begin{tabular}{lcccccccccc} 
\hline 
KIC No. & $f(m_\mathrm{C})$&$m_\mathrm{C}/m_\mathrm{ABC}$&$q_1$&$m_\mathrm{AB}$&$m_\mathrm{C}$&${\cal{A}}_\mathrm{LTTE}$&${\cal{A}}_\mathrm{L1}$&${\cal{A}}_\mathrm{L2}$&${\cal{A}}_\mathrm{S}$\\
        &(M$_\odot$)&     &     &(M$_\odot$)&(M$_\odot$)&(ksec)&(ksec)&(ksec)&(ksec)\\
\hline
04940201&   0.062 &0.307(90)&0.90f        &1.50(62) & 0.66(26)    & 0.196& 1.845& 0.008& 0.049 \\
05255552&   0.061 &0.294(10)&0.50(5)&1.69(72) & 0.71(36)    & 0.327&12.679& 0.521& 0.653 \\
05653126&   0.098 &0.334(20)&0.37(10) &1.77(17) & 0.89(7)    & 0.436&13.935& 0.677& 0.587 \\
06545018&   0.036 &0.232(10)&0.80(1)&2.29(13) & 0.69(5)    & 0.064& 1.161& 0.016& 0.057 \\
        &   0.036 &0.235(10)&0.84(1)&2.11(11) & 0.65(4)    & 0.064& 1.167& 0.012& 0.056 \\
07289157&   0.139 &0.395(10)&0.48(1)&1.37(9) & 0.89(4)    & 0.189& 1.348& 0.034& 0.034 \\
07812175&   0.050 &0.327(15)&0.85(1)&1.02(74) & 0.49(31)    & 0.252& 4.585& 0.032& 0.140 \\
08023317&   0.002 &0.103(30)&0.53(1)&1.29(15) & 0.15(5)    & 0.079& 1.311& 0.037& 0.039 \\
08210721&   0.097 &0.335(20)&0.16(1)&1.77(20) & 0.89(8)    & 0.373& 6.252& 0.397& 0.199 \\
08938628&   0.244 &0.474(70)&0.91(1)&1.25(54) & 1.12(28)    & 0.323& 1.571& 0.004& 0.029 \\
09714358&   0.010 &0.172(50)&0.45(1)&1.65(29) & 0.34(10)    & 0.046& 2.068& 0.127& 0.149 \\ 
\hline
05771589&   0.067 &0.314(50)&1.45(5)&1.48(33) & 0.68(14)    & 0.093& 8.962&$-0.319$&0.923 \\ 
        &   0.443 &0.497(60)&1.20(10)&1.88(71) & 1.85(37)    & 0.173&13.410&$-0.205$&1.307 \\
06964043&   0.271 &0.424(10)&0.85(5)&2.06(6) & 1.51(6)    & 0.229& 8.488& 0.099  & 0.614 \\
07668648&   0.006 &0.144(60)&0.69(1)&1.96(42) & 0.33(14)    & 0.059&16.730& 0.867& 2.714 \\
        &   0.005 &0.132(50)&0.58(1)&2.38(46) & 0.36(16)    & 0.054&16.082& 1.249& 2.726 \\
07955301&   0.306 &0.453(70)&0.97(1)&1.99(85) & 1.65(42)    & 0.230&14.869& 0.014& 1.235 \\
        &   0.224 &0.393(20)&0.85(5)&1.27(29) & 1.53(13)    & 0.205&13.374& 0.179& 1.151 \\
\hline
04769799&   0.036 &0.262(110)&0.80f        &1.83(92)&0.65(36)& 0.369& 2.786& 0.020& 0.052 \\ 
05003117&   0.045 &0.332(170)&0.50(10) &1.08(62)&0.53(40)& 0.562& 6.233& 0.131& 0.121 \\
05731312&   0.001 &0.109(100)&0.30(1)&1.10(39)&0.13(13)& 0.089& 0.365& 0.012& 0.006 \\
07670617&   0.085 &0.389(30)&0.90f        &0.89(17)&0.56(8)& 0.939& 5.238& 0.018& 0.111 \\
08143170&   0.005 &0.126(20)&0.55(5)&2.59(24)&0.37(7)& 0.226& 4.401& 0.159& 0.207 \\
09715925&   0.007 &0.206(170)&0.20f        &0.68(48)&0.18(17)& 0.145& 0.361& 0.011& 0.004 \\
09963009&   0.104 &0.410(10)&0.40(10)&1.61(32)&1.12(28)& 1.083& 4.918& 0.091& 0.057 \\
10268809&   0.317 &0.477(30)&0.70f        &1.55(97)&1.41(43)& 2.238& 3.470& 0.025& 0.040 \\
10319590&   0.098 &0.398(40)&0.60(5)&0.94(20)&0.62(9)& 0.262&10.790& 0.306& 0.532 \\
10979716&   0.098 &0.394(10)&0.96(1)&1.05(8)&0.69(3)  & 0.453& 1.533& 0.002& 0.022 \\
11519226&   0.267 &0.463(10)&1.23(1)&1.44(9)&1.25(5)  & 0.773& 4.867&$-0.028$&0.089\\
12356914&   0.020 &0.208(40)&0.50(10)&2.75(44)&0.72(16)  & 0.368& 2.741& 0.060& 0.051 \\ 
        &   0.005 &0.190(120)&0.70(1)&0.83(40)&0.19(14)  & 0.235& 2.575& 0.030& 0.050\\
\hline
\end{tabular}
\end{center}
\end{table*}

\begin{table*}
\begin{center}
\caption{Secular evolution related quantities} 
\label{Tab:apsenodeetal}  
\begin{tabular}{lcccccc} 
\hline 
KIC No. &$P_{\omega_1}$&$P_h$&$P_{g_1}^\mathrm{inst}$&$P_{g_1}^\mathrm{mod}$&$\varepsilon$&$P_{GR+tide}/P_{3b}$\\
        & (years)  & (years)&(years) & (years)\\
\hline
04940201& 171.7      & 138.8   &  76.9  &  75.9  & 0.01 & 608  \\
05255552& 238.7      & 123.7   &  81.2  &  90.0  &$-0.12$&4472 \\
05653126& 272.1      & 199.2   & 115.9  & 115.0  &$-0.06$&4295 \\
06545018&  33.2      & 23.0    &  13.6  &  13.2  & 0.04 & 234  \\
        &  29.2      & 22.1    &  12.6  &  12.6  & 0.00 & 300  \\
07289157&  90.6      &  79.8   &  42.4  &  42.3  & 0.00 & 387  \\
07812175& 284.2      & 170.2   & 108.6  & 104.3  & 0.05 & 1925 \\
08023317& $-595.3$   & 588.2   &$-4558.5$&1051.2$^a$&1.20& 33  \\
08210721& 344.3      & 235.8   &  142.1 & 136.7  & 0.04 & 1523 \\
08938628& 178.2      & 150.7   &   82.9 &  81.8  & 0.07 & 408  \\
09714358&  30.6      &  21.0   &   12.5 &  12.5  & 0.00 & 1731 \\ 
\hline
05771589& 6.5$^b$    &32.4$^b$ &   6.6  &   6.8   & 0.18 &10456 \\ 
        & 6.5$^c$    &  7.5    &   4.0  &   4.1   & 0.02 &14243 \\
06964043&  27.0      & 26.0    &  13.5  &  15.2   & 0.14 & 3810 \\
07668648&  41.6      & 24.7    &  16.1  &  27.9   & 0.68 & 18146\\
        &  62.9$^d$  &193.2$^d$&  14.4  &  31.0   & 0.72 & 17554\\
07955301&  18.1      & 18.7    &   9.5  &  10.9   & 0.14 & 10942\\
        &  14.7$^e$  &96.7$^e$ &  11.4  &  11.9   & 0.14 & 7979 \\
\hline
04769799& 814.5      &  818.8  &  423.0 & 494.0   & 0.18 & 548 \\ 
05003117& 835.6      & 1556.1  &  591.2 & 1707.5  & 1.27 & 1196\\
05731312&$-5613.1$   & 1005.3  & 1441.6 & 1165.3  & 0.25 & 25  \\
07670617& 926.8$^f$  &$-1623.7$& 1382.2 & 1367.1  & 0.51 & 278 \\
08143170& 929.0      & 890.3   &  475.1 & 847.8   & 0.59 & 524 \\
09715925&$-3182.2$   & 1163.4  & 2545.3 & 1270.0  & 0.63 & 18  \\
09963009&$-19194.3$  & 2672.4  & 3789.5 & 2426.2  & 0.48 & 154 \\
10268809&1830.4$^g$  & 3333.5  & 2799.2 & 2600.1  & 0.08 &  90 \\
10319590&  79.4      &$-131.2$ &  57.7  &  80.8   & 1.86 & 5322\\
10979716& 750.6      & 611.1   & 340.5  & 333.3   & 0.02 & 256 \\
11519226& 954.8      & 510.3   & 340.6  & 319.9   & 0.06 & 693 \\
12356914&$-8378.2$   &$-1499.5$& 3036.1 & 2019.1  & 0.59 &  60 \\ 
        &$-10311.8^h$& 1335.8  & 2455.9 & 1653.6  & 0.57 & 184 \\
\hline
\end{tabular}
\end{center}
{{\bf Notes.} (1) For the definition and a detailed discussion of the quantities listed in the Table see Appendix ~\ref{app:apsidalmotion}. (2) A negative sign in the apsidal motion periods $P_{g_1}$ and $P_{\omega_1}$ indicates retrograde apsidal motion. (3) A negative orbital precession period $P_h$ denotes {\it prograde} orbital precession, i.e., nodal progression.  \\
$a$: dynamical apse librates; $b$: unconstrained apsidal motion and nodal regression (constrained values are $P_{\omega_1}=15.6$\,y, and $P_h=10.9$\,y, respectively); $c$: unconstrained apsidal motion (theoretical $P_{\omega_1}=8.3$\,y); $d$: unconstrained apsidal motion and nodal regression (constrained values are $P_{\omega_1}=29.6$\,y, and $P_h=25.5$\,y, respectively); $e$: unconstrained apsidal motion and nodal regression (constrained values are $P_{\omega_1}=24.8$\,y, and $P_h=20.1$\,y, respectively); $f$: unconstrained apsidal motion (constrained $P_{\omega_1}=4366.1$\,y); $g$: unconstrained apsidal motion (constrained $P_{\omega_1}=13392.6$\,y); $h$: unconstrained apsidal motion (constrained $P_{\omega_1}=-4168.7$\,y)}
\end{table*}

\subsection{Robustness of System Parameter Determinations}
\label{sec:parameterconstraints}

The amplitudes or, more strictly, the magnitude of the amplitudes, of the different contributors to the ETV curves (tabulated in columns 7--10 of Table~\ref{Tab:Massetal}) reveal that in our entire sample, the quadrupole-level dynamical perturbations clearly dominate the ETV on the time-scale of the outer orbital period. 

A review of the parameter uncertainties, listed in Tables~\ref{Tab:Orbelem}--\ref{Tab:Massetal}, reveals that the most robustly determined parameters are the eccentricities, $e_{1,2}$; arguments of periastron in the observational frame, $\omega_{1,2}$; the times of periastron passage of the outer orbit, $\tau_2$; and naturally, for fully covered outer orbits, the outer orbital period, $P_2$\footnote{At this point we note that if an ETV curve represents the combination of AME and pure LTTE (as is the case, e.g., for AO Mon; \citealt{wolfetal10}) this would allow us to determine the same orbital elements. The determination of further system parameters, however, is only possible because of the dynamical contribution.}. In the case of $e_1$ and $\omega_1$ the high accuracy and robustness arises mainly from the precisely measured displacement of the secondary eclipses from a phase of $\phi=0.5$, which leads directly to a very high accuracy determination of the quantity $e_1\cos\omega_1$. Then, when at least one or two full outer orbital periods are observed, the shape of the ETV curve breaks the $e_1$--$\omega_1$ degeneracy, which results in a typical-LM uncertainty of $\delta\omega_1\sim0.1\degr$ which implies an uncertainty in $\delta e_1$ of $\sim10^{-4}$. A good example from our sample is KIC~04940201 (see left upper panel of Fig.~\ref{Fig:ETVall1}) which demonstrates clearly that an inner binary eccentricity of $e_1\simeq 0.001$ results in a clearly distinguishable and measurable displacement of the secondary eclipse curve. Despite this, however, we decided to adopt the somewhat more conservative uncertainty limits of $\delta e_1\ge0.001$ and $\delta\omega_1\ge1\degr$ for these two orbital parameters. One reason is, of course, the possibly of systematic effects which will certainly arise from the approximate nature of our analytic model. The other reason has to do with the perturbed and, therefore, continuously varying nature of these elements. As will be illustrated in Appendix~\ref{app:numericalanalysis} the amplitudes of these perturbations may exceed even these more conservative error limits over a timescale of days. Furthermore, in accord with our numerical tests in Appendix~\ref{app:numericalanalysis} we also introduce another conservative error limit for the outer eccentricity by fixing $\delta e_2\ge0.01$.

An additional important parameter is the mutual inclination, $\im$. As will be discussed in the Statistical Results section (Sect.~\ref{sec:stat}), this parameter exhibits a bimodal distribution. Here we concentrate only on the characteristics of the uncertainties in different regimes of $\im$. The LM-uncertainties in this quantity are generally around $1-2\degr$. On the other hand, from the additional numerical runs that we made, we find significantly larger uncertainties for low mutual inclination systems than for systems with higher mutual inclinations (i.e. $\im>30\degr$). This should result from the fact that, for nearly coplanar configurations, the $\im$-dependence is much shallower than for higher mutual inclinations. Moreover, note that in this regime the short-term and octupole perturbations have a relatively higher weight, in contrast to the quadrupole terms. This was our main reason for introducing these latter terms, and in this way we were actually able to reduce the uncertainties in $\im$ in this regime. (They had been even larger before.) In conclusion, we can say that, even in spite of the somewhat larger uncertainties in $\im$ for the near-coplanar situations, the accuracy that was obtained is evidently sufficient to be able to make statistically meaningful statements concerning the mutual inclination distribution of our sample. 

Considering the other parameters that describe the complete spatial configuration of the systems, the node-like angle, $n_2$, was adjusted either with the Levenberg-Marquardt fitting procedure, or with a grid-search, while its inner orbit counterpart, $n_1$, was constrained in a geometrical way from the corresponding spherical triangle, as is discussed in Appendix~\ref{app:Sphericaltriangle}. The estimated uncertainties for this parameter cover a substantially wider range than that for the other parameters mentioned above.  Note also, that for some triples, this parameter was controlled subjectively to exclude solutions which would have resulted in inclination angle variations that were too rapid, or opposite in direction to those which were observed. The larger uncertainties may arise primarily from the fact that these parameters, in most cases, appear in linear combinations: $\alpha=n_2-n_1$, $\beta=n_1+n_2$, where $\alpha$ and $\beta$ are typically not determined with the same accuracy. (For the prograde case, $\alpha$ is much more constrained than $\beta$, while for a retrograde configuration, the reverse is true.) Therefore, their individual values remain somewhat undetermined.

Consider, finally, the mass-related parameters, which are the most important---at least from an astrophysical point of view.  Unfortunately, here the picture is not so sanguine. As was expected, the mass ratio of the outer orbit, being relatively well determined, was found to have an uncertainty of a few to ten percent in most systems. The mass ratio of the inner binary appears only in the octupole terms.  Due to the smaller-amplitude contribution of these terms, we did not expect highly robust results for this parameter. Qualitatively, however, in most cases our solutions have led to physically reliable values, seemingly in accord with the crude preliminary estimates deduced from the flux ratios of the primary and secondary eclipses. Nevertheless, on the other hand, considerable caution is required, since in some cases we arrived at implausibly small or large numerical values; this could be a consequence of the physical model which might be insufficient in some extreme situations, as will be discussed in the context of KIC~07670617 in Appendix~\ref{app:numericalanalysis}. Although, a combination of the dynamical effects and the LTTE, in principle, should lead to physical mass determinations for the system components, our results, unfortunately, show some of the largest uncertainties in these quantities. This is, however, not so surprising. In the systems we investigated, the LTTE provides only a minor contribution to the ETV pattern. The one exception is KIC~10268809 where the amplitude ratio was found to be ${\cal{A}}_\mathrm{L1}/{\cal{A}}_\mathrm{LTTE}\sim0.64$. We carried out some tests with that solution to check whether we were able to recover the individual masses in such a case (see Appendix~\ref{app:numericalanalysis}). We found that, unfortunately, even if the relevant parameters, $a_\mathrm{AB}\sin{i_2}$ and $m_\mathrm{C}/m_\mathrm{ABC}$, are obtained with a few percent accuracy, the uncertainty obtained for the individual masses becomes as high as a 10\%-20\%. While this level of accuracy may be insufficient for some astrophysical considerations; it should be acceptable for statistical purposes as well as for initial estimates used in future follow-up studies.

\subsection{General Properties of the CHTs}
\label{sec:properties}

The ETV curves of our 26 systems show significant diversity. In the simplest cases, both the primary and secondary ETV curves look similar to one another and are purely `monoperiodic' (i.e., have only one maximum per outer period; KICs 04940201, 0654508, 08938628).  However, there is a nearly continuous transition from this type of system to much more complex situations.  These include cases where the two curves are still predominantly monoperiodic, but an extra hump begins to appear in one or both ETV curves and, in a parallel way, the primary and secondary ETV curves become more and more distinct from each other (e.g., KICs 07289157, 10979716, 09714358, 07955301, 05771589, 06964043).  Then there are the most peculiar ones, having a `double periodicity' (two distinct maxima per outer period), and/or anti-phased primary and secondary ETV minima.  Finally, there are ETV curves that are almost flat, but with a simple jump (which typically have remarkably different amplitudes in the primary and secondary ETV curves). 

As discussed in \citet{borkovitsetal11} these features are mostly governed in a complex way by the eccentricities on the one hand, and the spatial orientations of the orbits with respect to each other and to the observer, on the other hand, while the mass and period ratios only scale the ETV curves in amplitude and time, respectively.  We emphasize again that here we primarily refer to the $P_2$ time-scale quadrupole perturbation term, which, however dominates the ETVs in all of the 26 systems. Due to this complexity, we need considerable caution in making any qualitative assessment of a system's orbital properties from only the morphology of its ETV curve. For example, it can be said that a triple with small mutual inclination and negligible inner eccentricity always produces a monoperiodic ETV but the reverse statement would already be false, as was illustrated in Fig.~5l of \citet{borkovitsetal11}, which also shows monoperiodic ETVs for a perpendicular configuration. 

Some statements about the systems' general characteristics, however, can be made from a perusal of the overall qualitative structure of the ETV curves. Thus, if during the most rapidly varying portion of the ETVs, i.e., around periastron passage of the wide orbit, the net variation is positive, i.e., the system clock becomes more delayed (as is the case in all of the systems that we investigated), we can exclude a near-perpendicular configuration in any of our investigated systems. (The analytical verification of this is given in Appendix~\ref{app:octupole}.) In such a case, a double periodicity suggests significant mutual orbital inclinations, but, simultaneously, only moderate or small eccentricities. (As for a high-eccentricity case, the period ``doubling'' would reduce to a ``spike'' in the ETV curve.)  Another, natural conclusion, is that the steeper a short section of the ETV curve is, the higher one or both of the eccentricities are.  

Considering the apsidal motion contribution to the ETVs, the situation is simpler. If the secondary (blue) ETV curve is located below the primary (red) ETV curve, it indicates that the secondary eclipses are ``in a hurry'' with respect to the primary's (i.e., the secondary eclipses occur before orbital phase 0.5). Consequently, in these cases the periastron passage of the inner binary orbit occurs between the primary and secondary eclipses (i.e., in the range of $-90\degr<\omega_1<90\degr$). Since the maximum separation between the primary and secondary ETV curves occur when $\omega_1=0\degr$ (secondary ETV curve is below), or $\omega_1=180\degr$ (secondary ETV curve is above), the divergence or convergence of the two curves provides some initial information on the location of $\omega_1$ in the various quadrants, as well. These preliminary assessments not only provide us with general insights into the relations among, and values of, the system properties and ETV curve morphologies, but they may also help in finding reliable initial parameters for the LM-fitting processes. A detailed discussion is given in Appendix~\ref{app:numericalanalysis}.

\begin{figure}
\includegraphics[width=84mm]{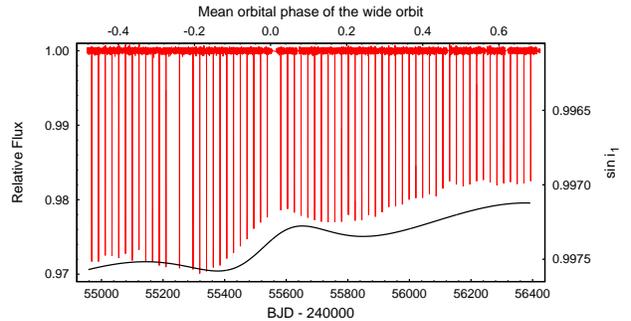}
 \caption{The {\it Kepler} light curve of KIC 04769799 (red) exhibits characteristic, uneven eclipse depth variations. The black line represents the rescaled and mirrored value of $\sin{i_1}$, calculated from our ETV solution. The similiraty between the two nicely illustrates that the eclipse depth variations can be attributed to the net effect of secular nodal regression and the $P_2$-period perturbations of the dynamical node. Note again, that none of the lightcurve characteristics (including eclipse depth variations) are directly built into our analytic model and, therefore, do not constrain  our ETV solution.}
 \label{Fig:K4769799ecldepth}
\end{figure}

\subsection{Notes on individual systems}
\label{sec:individual_notes}

\subsubsection{Group I systems}

{\bf KIC 04940201.} This system has the lowest inner-binary eccentricity of our sample. The absence of eclipse depth variations suggests that the inferred mutual inclination angle of $\im=5.9\degr\pm1.9\degr$ may be a bit overestimated. (The possible reasons were discussed earlier in Sect.~\ref{sec:parameterconstraints}.) This system was included in the early {\it Kepler}-field low-mass eclipsing binary study of \citet{coughlinetal11}.

\noindent
{\bf KIC 05255552.} A rather unique, triply-eclipsing system, with one of the largest amplitude ETV curves. Amongst the recently discovered triply eclipsing systems it has the longest outer orbital period ($P_2=862\pm3$~days). Moreover, both orbits are remarkably eccentric, i.e., $e_1=0.307\pm0.001$ which is among the largest in our sample, and $e_2=0.43\pm0.01$. This CHT is also quite compact $P_2/P_1\sim27$. Unfortunately, {\em Kepler} observations ended right during the third set of outer eclipsing events, and the only outer periastron passage in the {\em Kepler}-era occurred during a gap in the data.

\noindent
{\bf KIC 05653126.} This system exhibits remarkably high amplitude, rapid eclipse depth variations which are in good correspondence with the inferred system geometry parameters. Secondary eclipses occur only a few months after the outer periastron passage. For this system we used constant and equal uncertainties for both the primary and secondary ETV points, instead of individual uncertainties, because the use of the latter clearly overweights the primary curve with respect to the secondary one.

\noindent
{\bf KIC 06545018.} This system has the shortest inner period ($P_1=3.99$ days) as well as outer period ($P_2=90.6$ days) of all the systems in our sample. We give two solutions. In the first, $\im$, as usual, was allowed to vary freely. The result is $\im=11.2\degr\pm0.3\degr$ which clearly contradicts the lack of eclipse depth variations. Therefore, we give an alternative, coplanar (i.e, $\im\equiv0\degr$) solution. Note, despite the fact that the out-of-eclipse light curve clearly reveals tidally-induced ellipsoidal effects, the apsidal motion even here is also clearly dominated by dynamical effects, similar to the other systems we have investigated.

\noindent
{\bf KIC 07289157.} Another exemplary triply eclipsing system. A detailed description of its analysis is given in Appendix~\ref{app:numericalanalysis}. The residuals of our solution show a clearly parabolic trend. Therefore, we carried out an additional fitting run, including a second-order polynomial together with all the other parameters adjusted previously. This second fit resulted in a substantially lower $\chi^2$ value (half of the previous one), but all of the previously adjusted parameter values were preserved to within the standard errors of the first solution. From this combined parabolic and third-body solution we obtained a (constant) rate of period change of $\Delta P_1=-9.3\pm0.3\times10^{-8}$~day/cycle. This inferred period decrease might be a slight indicator of either an additional, more distant component in the system, or some orbital shrinking of the binary; however, we cannot exclude the possibility that it is a pure consequence of the imperfect modeling of the secular third-body perturbations. 

\noindent
{\bf KIC 07812175.} This system is located nearly midway between two substantially brighter KIC objects (07812167 and 07812179) that are separated by about 15$''$. This is likely the cause of some spurious eclipse depth variations in this system during alternating {\em Kepler} quarters. However, this does not materially affect our ETV analysis. It has the lowest outer eccentricity among our collection of triples ($e_2 = 0.031\pm0.004$).

\noindent
{\bf KIC 08023317.} The highest mutual inclination triple ($\im=49.5\pm0.6\degr$) in the sample. It also holds claim to the lowest outer mass ratio of $m_\mathrm{C}/m_\mathrm{ABC}=0.103\pm0.03$ or, $q_2\sim0.11$. As a consequence, in contrast to the majority of hierarchical systems, the orbital angular momentum is predominantly stored in the inner orbit. The $C_1/C_2=0.617$ ratio is also one of the highest amongst our 26 systems. We have chosen this unusual system to illustrate the operation of our analysis in the medium mutual inclination regime, and further details can be found in Appendix~\ref{app:numericalanalysis}.

\noindent
{\bf KIC 08210721.} Another totally eclipsing binary system with primary transits and very shallow secondary occultations. Our solution indicates a rapid variation in inclination ($\Delta i_1\sim1\degr$); however, the totality of the eclipses might explain the absence of the eclipse depth variations. On the other hand, a more detailed inspection reveals that the eclipse durations also remain unchanged; therefore, the mutual inclination angle that we obtained is probably overestimated a bit.

\noindent
{\bf KIC 08938628.} The inner binary shows rapidly decreasing eclipse depths, which is in accord with our solution.

\noindent
{\bf KIC 09714358.} One of the four CHTs in the sample with a period ratio $P_2/P_1<20$. Because of the lack of eclipse depth variations, we searched only for co-planar solutions. Despite the small period ratio, we found the solution we obtained to be reliable enough to rank this CHT in the first group of our systems. Note that the out-of-eclipse light curve reveals other stellar variability, as was reported also by \citet{debosscheretal11}.

\subsubsection{Group II systems}

{\bf KIC 05771589.} Shallow eclipses with first decreasing, and then increasing, depths. The out-of-eclipse light curve sections are also distorted. We give two alternative solutions for this system.  In the first, both the nodal regression and apsidal motion rates are unconstrained, while in the second case, nodal regression was constrained. The fit is quite poor in both cases, and neither of them is in accord with the characteristics of the eclipse depth variations. The orbital elements, however, with the exception of the outer eccentricity, are similar in the two solutions. In this triple, as in the other three systems, which were categorized into group II, a more sophisticated modeling is necessary.

\noindent
{\bf KIC 06964043.} Triply eclipsing system with shallow inner eclipses, and substantially deeper outer ones with variegated structures. The (inner) eclipse depth variation is opposite to that of the previously discussed system, i.e., an interval of increasing eclipse depths is followed by one with decreasing depths. For this system we were able to find a constrained solution (both in the apsidal advance and nodal regression); however, its reliability is similarly questionable, as in the previous case.

\noindent
{\bf KIC 07668648.} The light curve reveals continuously increasing eclipse depths, and even the primary and secondary eclipses are interchanged after a few cycles. A few additional shallow eclipses, not due to the binary, are also observed.  Since these occurrences match the outer companion's period and phase, we can conclude that this is not a blending of two systems in the {\em Kepler} photometric aperture for this object, but rather KIC 07668648 is also a marginally triply eclipsing system. It is also has the lowest period ratio ($P_2/P_1<7.3$), not only in our sample, but amongst all the known triple systems which contain an eclipsing inner binary.  Accordingly, both of our fits (with constrained and unconstrained secular effects) are quite poor and, therefore, this triple system also requires an improved analysis. 

\noindent
{\bf KIC 07955301.} This system exhibited rapid and remarkable eclipse depth growth during the first two years of the {\em Kepler} observations. Furthermore, similarly to the previous triple, the depths of the two kinds of eclipses also interchanged after the first few eclipses. (For this reason, in this most recent analysis we also interchanged the two kinds of eclipses. Therefore, the eclipse timings which were analyzed in \citealt{rappaportetal13} as primary eclipses, are now considered to be secondary eclipses.) This system was also analysed by \citet{gaulmeetal13} who found that the light curve shows clear red giant pulsations, from which they concluded that the tertiary should be a red giant with $M_3=1.2\pm0.1\,M_\odot$, and $R_3=5.9\pm0.2\,R_\odot$. The presence of such a luminous tertiary explains the shallow binary eclipses, as well. We present two solutions, first with constrained apsidal motion and orbital precession, and the second, without. 

\subsubsection{Group III systems}

{\bf KIC 04769799.} The very shallow secondary eclipses disappear completely before the periastron passage of the outer system. The amplitude of the primary eclipses also nicely exhibits the inclination jump on the $P_2$-timescale, with a net secular decrease.  Our fit seems to be quite reliable, even in the sense of reproducing the light curve features (see Fig.~\ref{Fig:K4769799ecldepth}); however, because of the absence of the information afforded by the secondary eclipses during a substantial portion of the {\em Kepler} observations, we classified this doubly low-eccentricity triple with the group of uncertain cases.

\noindent
{\bf KIC 05003117.} Slightly decreasing eclipse depths, and shallow secondary eclipses. The outer period is found to be substantially longer than the observed time-span of the {\em Kepler} observations. 

\noindent
{\bf KIC 05731312.} Despite its relatively short inner period ($P_1\simeq 7.95$ days), this binary has the largest inner eccentricity ($e_1=0.420\pm0.001$). The outer eccentricity is also amongst the highest ($e_2=0.58\pm0.01$). As a consequence, the ETV shows a marked spike during the outer periastron passage. The amplitude of the ETV, however, is low (less than 0.01 days even for the higher amplitude secondary ETV curve).  That is a consequence of the relatively wide separation of the triple ($P_2/P_1=114.1$, which is one of the highest in the sample) on the one hand, and the low mass of the ternary component ($m_\mathrm{C}/m_\mathrm{ABC}=0.11\pm0.01$ is the second smallest ratio) on the other hand. In theory, more than one and a half outer cycles are covered but, unfortunately, only one periastron passage is located very near the middle of the {\em Kepler} measurements, and with the exception of this $\sim400$-day-long interval, the ETV is almost featureless before and after, during the remaining 1000 days. Despite this, our solution seems to be reasonable (including the reconstruction of the jump-like eclipse depth decrease near the periastron passage). Note, this is another system where the use of the individual ETV point uncertainties resulted in an overweighted primary ETV curve and, therefore, the secondary ETV curve was not well fit. This is especially the case for the final {\em Kepler} quarters when, after periastron passage, the eclipse depths decreased significantly. As a compromise, we somewhat arbitrarily reduced the uncertainties in the secondary points by a factor of three.

\noindent
{\bf KIC 07670617.} This system shows a marked periastron passage with a sharp step-like feature in the ETV of the secondary, which is clearly also reflected in the light curve (see Fig.~\ref{Fig:K7670617lc}, in Appendix~\ref{app:numericalanalysis}). The outer binary is one of our three highest eccentricity systems ($e_2=0.70\pm0.01$). The reliability of our retrograde solution is discussed in Appendix~\ref{app:numericalanalysis}. For this system we also preferred the use of constant and equal uncertainties in the ETV points.

\noindent
{\bf KIC 08143170.} The ETV curves, at first sight, show remarkable similarities to the previously discussed system. The inner binary light curve, however, in the present case exhibits only a minor, almost unnoticeable eclipse depth variation (increase) around periastron passage. The former property (i.e., the similarity of the ETV curve segments) can be explained by the similarly high outer eccentricities ($e_2=0.70\pm0.01$), and also with their similar mutual inclinations (this is in spite of the fact that this system has a prograde outer orbital configuration, in contrast to the retrograde outer orbit of KIC 07670617). On the other hand the minor eclipse depth variations here might be explained by the totally eclipsing nature of KIC 08143170, in which case the eclipse depths are less sensitive to small variations in the inclination. This is especially true for occultations where the inclination affects only the eclipse duration. This is another triple for which the equal, global ETV uncertainties mode was used. 

\noindent
{\bf KIC 09715925.} The light curve exhibits a step-like, moderate eclipse depth growth during the periastron passage. The very shallow and short secondary eclipses (which in most cases contain only 3--4 {\em Kepler} long cadence points) makes our solution less reliable. However, despite the large uncertainty in the outer period, most of the orbital elements, and the mutual inclination also seem to be reliable, at least qualitatively. 

\noindent
{\bf KIC 09963009.} This system has the longest period inner binary ($P_1\simeq 40.1$ days) in our sample. Our solution points to one of the longest outer periods ($P_2\sim3770$ days) which suggests a wide, weakly interacting system ($P_2/P_1\sim94$). The relatively short coverage of the outer orbit, the low ETV amplitude, and the lack of marked features in the ETV curves make our solution somewhat untrustworthy. On the other hand, this triple allows us to illustrate the worthiness of including additional available information into the solution process. For this system we found two strongly different solutions with similar $\chi^2$ values; however, taking into consideration some of the lightcurve properties, we were able to eliminate one of these alternative solutions. This was because the {\em Kepler} light curve exhibited constant primary eclipse depths, but decreasing secondary depths, from which we concluded that the inner orbit's periastron passage should be located closer to the primary than the secondary eclipse. Therefore we rejected the solution which resulted in $\omega_1=106\pm10\degr$, and retained the one with $\omega_1=258\pm5\degr$. (Here, again, we utilized the global and equal ETV uncertainty mode.)  

\noindent
{\bf KIC 10268809.} Opposite to the case of the previous triple, these primary eclipses exhibit marked depth variations (similar to those shown in Fig.~\ref{Fig:K4769799ecldepth}), while the secondary eclipses exhibit only minor decreases in depth. The amplitudes of the two kinds of eclipses also interchange around BJD 2\,455\,800. Our finding of $\omega_1=143\pm1\degr$ is in accord with this fact but, for the very long inferred outer period ($P_2=7\,000$ days), some caution is needed concerning the reliability of these results. Note, this is the only triple where the LTTE amplitude is comparable to the quadrupole dynamical term. Therefore, we use our solution in Appendix~\ref{app:numericalanalysis} to test whether the individual stellar masses can be recovered in such circumstances.

\noindent
{\bf KIC 10319590.} As was already reported in \citet{rappaportetal13}, the eclipses completely disappeared after the first 400 days of the {\it Kepler} observations. Previously SS Lac \citep{zakarovazimov90} and V907 Sco \citep{lacyetal99} were the only two EBs where the same phenomenon was documented\footnote{There is a third such system, namely HS Hya, where the cessation of eclipses is predicted to occur in the near future \citep{zaschepaschke12}.}. Our solution resulted in an outer orbit with a period, $P_2=451\pm3$ days, that is very close to the observational window where the eclipses were present. The double periodicity in the ETV curves, however, is clearly seen and therefore we believe that this finding for $P_2$ is reliable. Our original solution revealed a retrograde orbit with $\im=135.4\pm0.3\degr$, but our numerical tests have shown that it cannot be distinguished significantly from the corresponding prograde solutions. An indirect verification of our solution (independent of its prograde or retrograde nature) comes from the fact that the node-like parameter $n_1$ was found to be around $90\degr$. This provides the fastest instantaneous precession rate during the entire nodal period (see Eq.~[\ref{Eq:dotcosi1}]) and, therefore, produces $\Delta i_1=-2.5\degr$ inclination variation during only 1.1 year of a $\sim131$-year-long precession-cycle, in good correspondence with the relatively rapid disappearance of the eclipses. (Because of the rapid eclipse depth variations, we utilized equal global uncertainties for the ETV points in order to avoid underweighting those points derived from the last, just-disappearing, very shallow, grazing eclipses.) 

\noindent
{\bf KIC 10979716.} The ETV curves for this system, and also the resultant fitted parameters, resemble in most aspects those of KIC~07289157. The main exception is in the lack of eclipse depth variation, which is well explained by the fact that the two orbits intersect each other close to the celestial plane $h_1=9.7\degr-7.9\degr$, in which case the observable precession rate becomes practically zero. This system was placed in the third group only because of the absence a second outer periastron passage; however, our solution should have almost the same robustness as for the group I systems.

\noindent
{\bf KIC 11519226.} Besides KIC 05771589, this is the only system, where a mass ratio of $q_1>1$ was found. Since there is only a minor difference in depths between the two eclipses in the binary, the $q_1=1.23\pm0.01$ value might even be correct. Our solution has also reveals that the binary line of the apsides lies almost perfectly in the celestial plane ($\omega_1\sim359\degr\pm1\degr$). The durations of the two eclipses are similar which is also in accord with this result. The weakness of our fit is, however, that our solution would predict an inclination angle variation of nearly $\Delta i_1\sim-0.8\degr$ during the {\em Kepler} prime mission, but the light curve, despite the partial eclipses, does not show eclipse depth variations.

\noindent
{\bf KIC 12356914.} The light curve exhibits deep primary eclipses and significantly shallower, flat-bottomed secondary eclipses that are longer in duration. There are no eclipse depth variations. We give two alternative solutions. The constrained apsidal motion fit resulted in a retrograde solution, while the unconstrained fit yielded a prograde solution. The main weakness of both solutions is that they put the inner periastron passage close to the secondary eclipse which seems to contradict the fact that the secondary event is substantially longer, (i.e., occurs when the inner binary components are moving the slowest). Therefore, the reliability of these results could be questionable. Equal ETV uncertainties were used in this case as well.)

\section{Statistical Results for the 26 Triples}
\label{sec:stat}

\begin{figure}
\includegraphics[width=85mm]{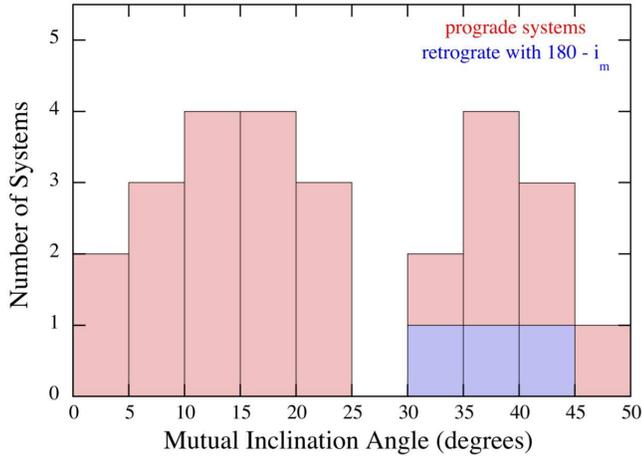}
 \caption{The distribution of mutual orbital inclination angle, $\im$.  In cases where the orbit is retrograde, i.e., $\im > 90\degr$ we have plotted the value $180\degr - \im$, but indicated those systems with blue shading.  There are two fairly clear peaks, one centered near $13\degr$ and the other near $38\degr$. See text for a discussion.}
 \label{Fig:im}
\end{figure}

\begin{figure}
\includegraphics[width=85mm]{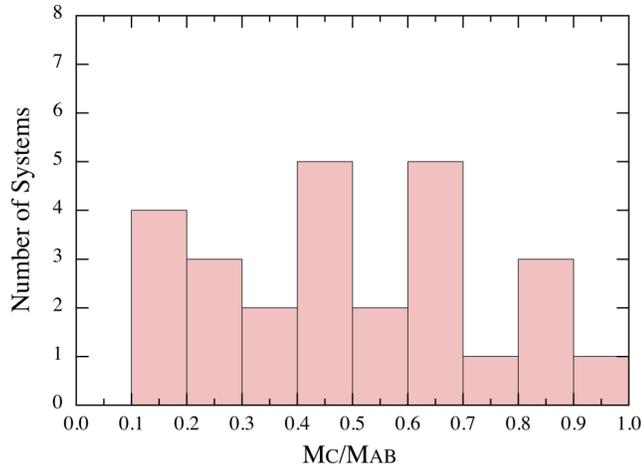}
 \caption{Distribution of the ratio the mass of the third body, $m_\mathrm{C}$, to the mass of the inner binary, $m_{\rm AB}$. A ratio of 0.5 corresponds to the case where all three masses in the system could be similar.  Only a relative handful of systems have distinctly low-mass tertiary companions; likewise, none has a third body which dominates the system.}
 \label{Fig:masses}
\end{figure}

\begin{figure}
\includegraphics[width=85mm]{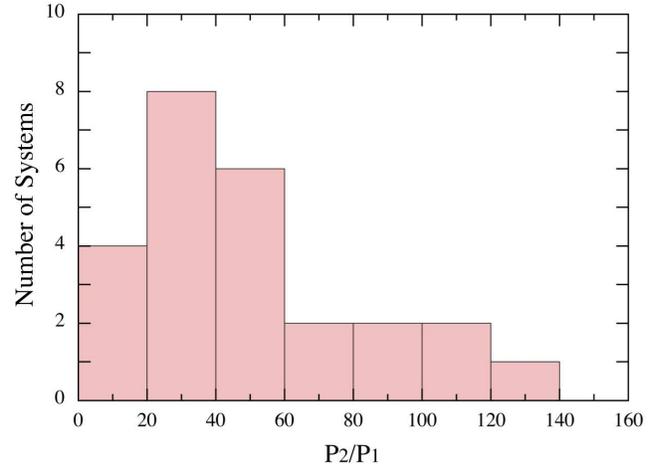}
 \caption{Distribution of the ratio of the outer (triple) period to the inner (binary period, $P_2$ vs.~$P_1$.  We have removed one of the 26 systems from this distribution, KIC 10268809, because its outer period greatly exceeds the span of {\em Kepler} observations, and is therefore highly uncertain.   The bulk of the remaining systems lie in the range $10 \lesssim P_2/P_1 \lesssim 100$.  The upper limit is a selection effect due to the overall 4-year duration of the {\em Kepler} mission.}
 \label{Fig:periods}
\end{figure}

\begin{figure}
\includegraphics[width=85mm]{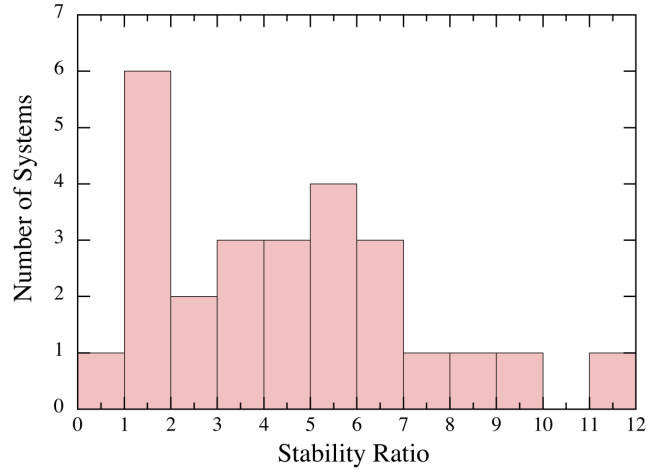}
 \caption{Distribution of the orbital stability ratio, defined as $(P_2/P_1)/(P_2/P_1)_{\rm stab}$, where $P_2$ and $P_1$ are the triple and binary periods, respectively, and $(P_2/P_1)_{\rm stab}$ is the minimum ratio of these two periods required for stability.  We utilize Eqn.~(90) from \citet{mardlingaarseth01} for the stability criterion (see text for details).}
 \label{Fig:stability}
\end{figure}

\begin{figure}
\includegraphics[width=85mm]{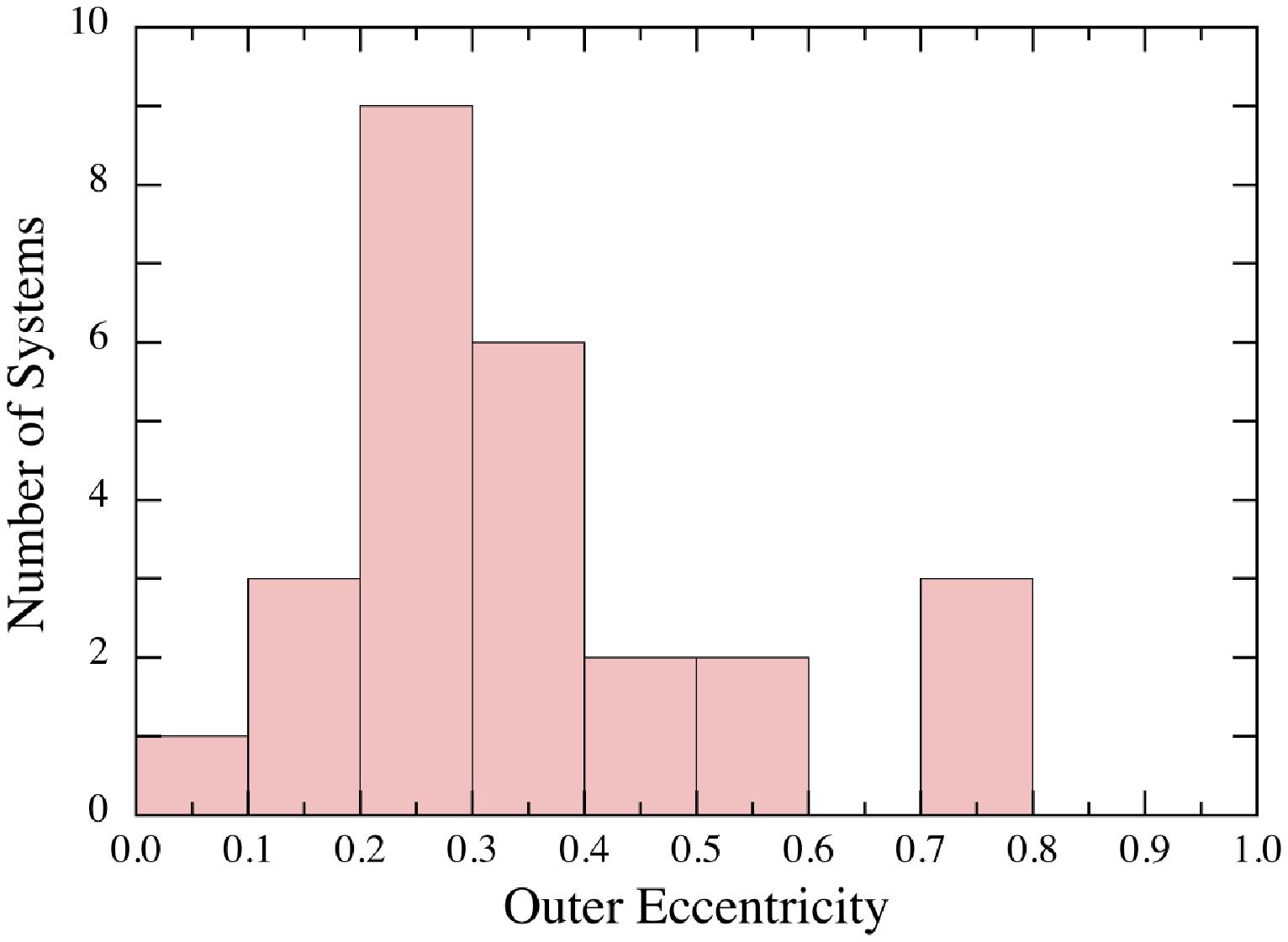}
 \caption{Distribution of the eccentricity of the outer orbit, i.e., that of the triple ($e_2$). The triples mostly have modest eccentricities in the range of 0.2--0.5, though 3 of the 26 systems have $e_2 > 0.7$.}
 \label{Fig:ecc}
\end{figure}

	Our sample of 26 compact hierarchical triples (CHTs) with well measured system parameters is not large, but is considerably more than has been heretofore available for statistical analysis. It is especially true, if we take into account also the small characteristic sizes, i.e., the short outer period of many of the investigated systems. (The rarity of systems with outer period less than 1000 days was discussed in the Introduction.) Some important relationships among the system parameters that we can examine are the ratios between the outer and inner periods ($P_2$ and $P_1$),  the ratios of masses ($m_\mathrm{C}$ and $m_\mathrm{AB}$), the mutual inclination angles, $i_m$, and eccentricities, in particular that of the outer orbit. We also compare the period ratios, $P_2/P_1$, with an approximate expression for mass ratios leading to dynamical stability.
	
We start with the mutual inclination angle (i.e., the angle between the plane of the binary and that of the outer third body). Figure \ref{Fig:im} shows the distribution of the mutual inclination angle for our 26 CHTs. The three systems marked in blue are those with retrograde orbits ($\im > 90\degr$), and here we have displayed them as $180\degr - \im$  for convenience.  In spite of the limited statistics, there are two clear groups of systems, one centered at $\sim$$13\degr$ and the other at $\sim$$38\degr$. This seems to be consistent with two populations, one that was born with more coplanar orbits, and which have remained that way over time; the other which was born with potentially quite large mutual inclination angles $$\sin^2 \im > 2/5~~~{\rm or}~~~ 39.2\degr \lesssim \im \lesssim 140.8\degr$$ and which were subject to the Kozai-Lidov oscillation cycles with tidal damping \citep{kozai62,lidov62,kiselevaetal98,eggletonkiseleva01}. In this latter group, the Kozai-Lidov cycles might drive the inner binary to be tighter, and tidal dissipation in that binary can terminate the Kozai-Lidov cycles leading to a preferential set of mutual inclination angles in the range of $35\degr$ to $50\degr$ \citep[see, e.g.,][]{fabryckytremaine07}, just what is seen in Fig.~\ref{Fig:im}. 
	
Among the many other measured system parameters, we also consider the ratio of the third body mass to the mass of the inner binary. Figure \ref{Fig:masses} shows the distribution of $m_\mathrm{C}/m_\mathrm{AB}$. All the systems have $m_\mathrm{C}/m_\mathrm{AB} < 1$.  Otherwise the distribution of $m_\mathrm{C}/m_\mathrm{AB}$ is flat to within the small-number statistics.  Roughly half the systems have $m_\mathrm{C}/m_\mathrm{AB} = 0.55 \pm 0.15$ which is consistent, within uncertainties, with all three constituent masses being comparable.   
	
The inner binary periods of the systems we have investigated are between $\sim$4 and 40 days. Note that, according to the convention followed, e.g., by \citet{naozfabrycky14} only those systems with $P_\mathrm{orb}<16 \,\mathrm{d}$ are grouped into the category of ``close binaries''.  By that strict definition, 11 inner binaries of our sample fall into this category\footnote{A natural problem of course with all of such kinds of simple delimiters is how to categorize borderline systems. For example, in our present sample KIC~07955301 has a period of $P_1=15.36$ days and therefore would be classified as a close binary, while KIC~08023317 with its period of $P_1=16.58$ days is left out of this category. On the other hand, however, this latter binary has an eccentricity of $e_1=0.25$ in contrast to the much smaller eccentricity ($e_1=0.029$) of the former system, and therefore, in the sense of periastron distance, KIC~08023317 is the closer.}. The outer period range extends from $\sim$90 to $\sim$7000 days, although this latter limit, being substantially longer than the observational interval, is quite approximate. In Figure \ref{Fig:periods} we plot the distribution of the ratio of the outer (i.e., triple) period to the period of the inner binary.  The vast majority of the systems lie in the range of $10 \lesssim P_2/P_1 \lesssim 100$.  The systems with large ratios of $P_2/P_1$ are clearly in the dynamically stable regime.  However, we next investigate this issue of stability somewhat more quantitatively. 

A number of groups have investigated the long-term dynamical stability of hierarchical three-body systems, and have developed approximate `empirical' expressions for stability.  \citet{mikkola08} conveniently summarizes four of these expressions for stability.  Somewhat arbitrarily, we have chosen the expression of Mardling \& Aarseth (2001) to use for comparison with our observed period ratios.  Their expression can be rewritten as:
\begin{equation}
\left(\frac{P_2}{P_1}\right)_{\rm stab} \gtrsim 4.68\left(\frac{m_\mathrm{C}}{m_{\rm AB}} \right)^{1/10} \frac{(1+e_2)^{3/5}}{(1-e_2)^{9/5}}
\label{eqn:stability}
\end{equation}
which holds for a wide range of mass ratios and eccentricities (but note that it depends only on $e_2$). Its main limitation is that it is valid specifically only for coplanar orbits, and we apply it to our systems with this caveat in mind.  In Figure \ref{Fig:stability} we show the distribution of the observed ratio ($P_2/P_1$) in units of the ratio for stability, $(P_2/P_1)_{\rm stab}$ given by Eqn.~(\ref{eqn:stability}), for our sample of 26 CHTs.  One of the systems, KIC 06964043, has a stability ratio only marginally more than unity (1.02), while another, KIC 07668648, has a stability ratio of only 0.77 and, as noted above, it has the smallest ratio of $P_2/P_1$ at 7.3. The fact, that this system is far from a coplanar configuration, ($\im\sim40\degr$) might play an important role in its stability. In any case, the system is manifestly dynamically stable, and so this may just point to some minor limitations of the \citet{mardlingaarseth01} expression.  The remaining 24 systems have stability ratios (defined above) as greater than unity.
	
In this regard, it is also interesting to note that 17 of our 26 CHT systems have outer periods of less than 1000 days, and for 8 of the triples $P_2$ is even shorter than one year. Taking into consideration the works of \citet{carteretal11,derekasetal11,steffenetal11,giesetal12,leeetal13,rappaportetal13}; and \citet{conroyetal14}, which have reported an additional 51 secure short outer period triple star systems with $P_2<1000$ days in the {\it Kepler} field, this brings the total number of such systems to at least 68. Therefore, we can made a crude estimate of the frequency of such short outer-period triples. Accepting the stability criteria of \citet{mardlingaarseth01}, and neglecting its eccentricity and mass ratio dependence, the upper limit on the inner period for a 1000 day-long outer orbit becomes $P_1\sim215$ days. The Kepler EB catalog contains 2582 entries of binaries with shorter periods. This inner period limit of 215 days, however, could be further restricted considering the fact that the typical detection method for these short outer period {\it Kepler} triples is via an ETV analysis. Based on our experience, in order to have a reliable ETV analysis, at least 15-20 eclipse timing data points from the two eclipses combined are required, though this is being somewhat optimistic. (In our sample, the 38 eclipse times for KIC~10319590 could be considered minimal.) This latter requirement considerably reduces the useable range of $P_1$ to $\sim 1/10$th of the Q0--Q17 time-span (as one orbit of the inner binary produces two eclipses), i.e., $P_1\lesssim145$ days.  If we consider, however, that a period ratio $P_2/P_1>20$ is observationally much more typical, we actually need to take into account only eclipsing systems with $P_1<50$ days.  There are 2458 such entries in the {\em Kepler} binary catalog. 

From the above discussion we can conclude that, very roughly, 2.7\% of {\em Kepler} binaries with $P_1 \lesssim 50$ days have third-body stellar mass companions with relatively short outer periods (i.e., $P_2 \lesssim 1000$ days). As noted in the Introduction, in a recent paper \citet{tokovinin14b} reports the complete lack of ternaries with $P_2<1000$~d orbital period in his distance-limited solar-like (or less massive) triples. Since a substantial portion of our triples sample, and similarly of the other reported {\em Kepler} triples (referenced above), are supposed to be comprised nearly exclusively of solar- and lower-mass stars \citep{coughlinetal11}, our results seem highly appropriate for comparison with the \citet{tokovinin14b} findings. The complete sample of \citet{tokovinin14a,tokovinin14b} contains only $\sim$200 binaries with $P<50$~days and, therefore, statistically we would expect only $\sim$6 short-period triples amongst them. Furthermore, there is an evident selection effect (not mentioned in the discussion of \citealp{tokovinin14b}, although, in a different context, it was discussed in \citealp{tokovinin14a}), which further reduces the expected number of such short-period triples in his sample. This is so because the discovery probability of a close binary sub-system within a previously known few-hundred day wider binary via radial velocity measurements, is evidently biased toward those systems where the more luminous (and therefore usually the more massive) component forms the close binary. In turn, this is true because, if the {\em secondary} component of a wider, single-lined spectroscopic binary (SB1) were a close binary, it could easily remain unnoticed. A good example is HD~181068 which, before the {\em Kepler} discovery of its unique triply eclipsing nature \citep{derekasetal11} was categorized as a simple, ``boring'' SB1 binary \citep{guilloutetal09}. An effect which can further strengthen this bias is that for such a bright ternary as can be found, e.g. both in HD~181068 and KOI-126 \citep{carteretal11} the photometric signatures of a faint close binary subsystem also would remain hidden from ground-based observations. (Note, these selection effects may also provide a natural explanation for the fact that, amongst more massive binaries, short period ternaries were already known.) Therefore, the absence of $P_2<1000$ days system in the \citet{tokovinin14b} sample is not in serious disagreement with the frequency of such systems in the substantially larger {\em Kepler} sample; but, the question of whether it is a selection effect, or not, still requires further investigation.
	
Finally, we show in Figure \ref{Fig:ecc} the distribution of the eccentricity of the outer binary, i.e., the triple.  The outer eccentricities show significant diversity, from the almost circular outer orbit of KIC~07812175 to those three CHTs: KICs~07670617, 08143170, and 10268809, where $e_2>0.7$.  The eccentricities of the inner binaries lie in the range of $0.001 \lesssim e_1 \lesssim 0.42$. 
	
In the case of eccentricities (both inner and outer), however, it should keep in mind that our sample has substantial selection and observational biases.  First, considering the inner eccentricities, in the extreme limit it is evident that we excluded systems with very nearly circular inner orbits.  The negative bias toward the high inner-eccentricity end manifests itself in the fact that the higher the inner eccentricity the larger the possibility of the occurrence of only a single binary eclipse event instead of two, and such systems were also excluded from our analysis.  Regarding the outer eccentricities, there are clearly counteracting biases. Since the full amplitudes of the dynamical ETV contributions are proportional to $(1-e_2^2)^{-3/2}$, or even higher (negative) powers, it is evident that a higher outer eccentricity results in a larger amplitude and, therefore, more readily detectable ETV. On the other hand, if the outer period significantly exceeds the duration of the data set, another, counteracting selection effect becomes increasingly important. Namely, for highly eccentric outer orbits, the ETV curves reduce to a spike or jump around periastron passage (the larger $e_2$ is, the more narrow in orbital phase and larger in amplitude is this feature). In this case, during most of the outer orbit -- in the absence of significant dynamical perturbations -- the inner period remains almost constant and, therefore, the ETV curves also become plain and featureless. As a consequence, for systems with an outer period significantly longer than the observing window, the detectability of medium or small outer eccentricity systems mainly depends on the ETV amplitude, while for the high outer eccentricity systems it is more limited by the orbital-phase coverage. (This was nicely illustrated in Figs.~6 and 10 of \citealt{borkovitsetal11}.) Therefore, apart from the inner eccentricity related selection effects, we suspect that there are not many additional {\em Kepler} systems with $P_2 \lesssim 1000$ and medium outer eccentricity, such as KICs~05003117, 09963009, and 1235694 that would have remained undetected in our search, with the possible exception of those having a substantially lower-mass tertiary star.  On the other hand, we cannot exclude the presence of undetected high outer eccentricity systems, similar to e.g. KICs~07670617, 08143170, 10268809. Note, a more comprehensive and detailed discussion of the detectability limits and some of the selection effects connected to specific parameters of the ETV curves are given in \citet{borkovitsetal11}.

We also looked at correlations between $e_2$ vs.~$e_1$, $e_2$ vs.~$P_2$, $e_2$ vs.~$\im$, and $e_1$ vs.~$\im$, but none of them was particularly statistically significant.

\section{Summary and Conclusions}
\label{sec:summary}

We have analyzed the eclipse timing variations of a sample of 26 eccentric eclipsing binaries in the original {\it Kepler} field which were suspected of being members of highly gravitationally interacting, compact, hierarchical triple stellar systems. The investigation has followed a distinctly analytical approach. We have improved and extended the analytical description of the effects of the $P_2$-time-scale third-body perturbations on the ETV curve(s) of an eclipsing binary orbiting in a hierarchical configuration \citep{borkovitsetal03,borkovitsetal11}.  We have also included, for the first time, the long-timescale octupole and short-timescale quadrupole perturbation terms, and connected them to the longest period apsidal motion and orbital precession effects. For these latter two effects, the quadrupole-level third-body perturbations were also taken into account. Our approach made it possible to simultaneously determine most of the orbital parameters of the inner and the outer binaries, both in the observational and the dynamical frames of references, as well as the complex 3D orientation of the orbits, both with respect to each other and to an Earth-based observer. The model was implemented in a computer code which has also been described in this work.

We used our analytic approach to fitting ETV curves simultaneously for both the primary and secondary eclipses for all 26 of the compact hierarchical triple systems we selected from the {\em Kepler} sample.  We broke this up into three sub-groups: those with complete information needed for robust determination of the system parameters; those systems that were sufficiently `close' (in the sense of having a small $P_2/P_1$ ratio) that the analytic model is imperfect; and finally those where the model is quite adequate but either the {\em Kepler} data train did not cover a sufficient portion of the outer triple orbit, or some other technical issue limited the determination of some of the system parameters.

{\em Group I Systems:}  By the use of this analytic approach for representing the ETV curves we were able to determine reliable and robust system parameters for 10 of the 26 systems we analyzed. For those 10 systems (comprising group I), at least one and a half outer orbits were covered by the observations (the outer period range for these systems was $104~\mathrm{d}\geq P_2\geq 968~\mathrm{d}$), and the ratio of the inner and outer periods (with one exception) was $P_2/P_1>20$. Especially noteworthy from this group is KIC~05255552, which has the longest known outer period among any triply eclipsing hierarchical triple system. With such a long period, it is remarkable that the system parameters can be fairly securely determined with eclipse timing, and without the need for a multiseasonal ground-based RV study.

{\em Group II Systems:} Our most ``insufficiently modeled'' sub-group contains the three most compact systems in our sample.  For these systems, our model needs some improvement, especially in regard to describing the secular apsidal motion, in order to obtain a more acceptable solution. The fourth triple in group II, KIC~06964043, has a somewhat larger period ratio ($P_2/P_1\simeq22.3$) and hence, semi-major axis ($a_2/a_1\simeq9.8$) ratio, but due to its large outer eccentricity ($e_2\simeq0.51$) and therefore, the rather small separation near periastron passage of the outer binary, our model description was also somewhat inadequate. Among this group, KIC~07668648 is unique in having the most `compact' configuration of all such CHTs known, with $P_2/P_1 \simeq 7.3$ and $a_2/a_1 \simeq 4.0$!

{\em Group III Systems:}  For the remaining 12 triples (our group III systems) the somewhat less robust solutions result mainly from observational, or other technical, reasons. In these systems the observations, with one exception, do not span at least one and a half orbital periods. Despite the less robust solutions, as we have discussed in detail, for most of these systems the inferred parameters are in accord with additional system properties which can be deduced independently from the light curves.  This supplemental information includes the directions and the rates of eclipse depth variations, and also the different properties of the primary and secondary eclipses in each system. Therefore, we can conclude, that our results for the group III triples, though less accurate than the group I systems, should still be reliable for most of these systems. Among the Group III systems, KIC~07670617 and KIC~10319590 stand out for having clearly retrograde outer orbits.

Our results confirm the bimodal mutual inclination angle distribution of hierarchical triples with relatively short inner periods, in good agreement with the predictions of \citet{fabryckytremaine07}, and they therefore, indirectly, support the KCTF theory of the formation of close binary systems. However, we hasten to add that most of the 26 systems we studied (16/26) have relatively low mutual inclination angles, likely formed that way, and were not subject to KCTF.  Since \citet{fabryckytremaine07} started with an isotropic distribution of mutual inclination angles, it is reasonable that their final distribution would have a minimum at $\im = 0\degr$ since that is the least likely a priori angle to start with.  However, to the extent that the initial distribution of $\im$ is itself bimodal, with a good fraction of the systems (at least those with $P_{\rm 2, init} \lesssim 2000$ days) having $\im \lesssim 39\degr$, then the final distribution will be a blend of their Fig.~7 with an additional peak at low $\im$ representing those systems that were born that way (as we see in our Fig.~\ref{Fig:im}).

From among our other statistical results we find, in nearly all cases, that $10 \lesssim P_2/P_1 \lesssim 100$.  Also, the mass ratios of 19 of the 26 systems in our sample have $1/3 \lesssim m_\mathrm{C}/m_\mathrm{AB} \lesssim 1$, indicating that the third (outer) body is neither particularly massive nor light.

Furthermore, our findings substantially increase the number of the shortest outer period triple systems known and, therefore, can serve as observational probes of the highly underpopulated short-end of the outer period regime of hierarchical triples. These may be essential for understanding the different theories of close binary formation.

\begin{figure}
\includegraphics[width=84mm]{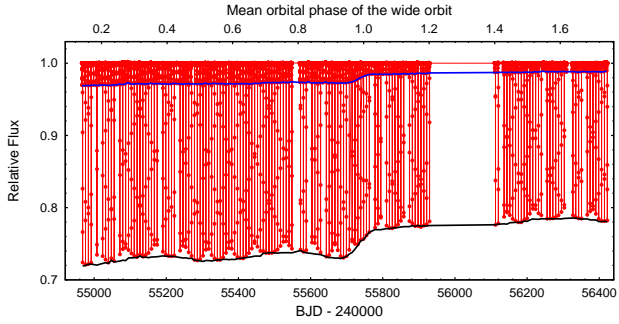}
 \caption{The {\em Kepler} light curve of KIC 05731312 (red), which exhibits characteristic, uneven eclipse depth variations. The black and blue lines connect the approximate brightness (relative flux) levels at the mid-eclipse points for primary and secondary eclipses, respectively. Their values were determined as a by-product of the ETV calculations (i.e., eclipse template  fitting), as was described in detail in the text. As the out-of-eclipse flux level clearly remains constant during the entire observing interval, these black and blue curves could be directly used for quantifying the eclipse depth variations. Note also, that the red points show the individual LC flux points, and their Moir{\'e} pattern helps us to visualize the beating effect between the data sampling frequency and the orbital period. It is apparent that the template-fitting procedure has immunity from this beating effect.}
 \label{Fig:K5731312ecldepthcount}
\end{figure}

Finally, we discuss two issues related to our approach to fitting for orbital solutions in CHTs. First is the omission of such additional information as the eclipse depth and duration variations which can also be deduced from {\em Kepler} observations, and which could (in some cases) have dramatically improved our solutions, and/or made them more unique. In one sense this is admittedly a weakness of our approach.  On the other hand, the inclusion of such effects into an otherwise fast and analytic method, has both theoretical and practical obstacles.  Consider first the eclipse depth variations. In this case, despite the relatively low frequency of the LC measurements, which for most systems results in a very weak sampling of each individual eclipse, the varying depths of the eclipses can be easily measured with considerable accuracy, as was mentioned previously in Sect.~\ref{Sec:dataprep}. We illustrate this for the case of KIC~05731312 in Fig.~\ref{Fig:K5731312ecldepthcount}. A substantial difficulty arises, however, from the theoretical side.  Namely, the eclipse depth variations are highly sensitive not only to the geometric properties of the inner orbit, but also to stellar parameters such as their radii relative to each other and to the orbital separation, tidally and rotationally distorted shapes, $T_{\rm eff}$,  and limb-darkening (see \citealt{csizmadiaetal13} for a recent study of the influence of limb-darkening models on transiting exoplanet light curve solutions). Furthermore, in the case of partial eclipses, the functional dependences are quite complex and far from trivial\footnote{By contrast, for total eclipses the formulae become remarkably simpler \citep[see, e.g.][]{seagermallen-ornelas03}, but in such cases the eclipse depth variations do not yield any additional valuable information, as they either reflect only the variation of the limb-darkening in different regions of the transited star, or remain constant (during occultation).} (for details, see, e.g., Chapter~IV in \citealt{kopal79}).  

Turning to the eclipse durations, the problem here is more of a practical one than theoretical.  Since the theoretical aspect of eclipse durations was comprehensively studied by \citet{kipping10}, we comment here only the practical obstacles, i.e. observational aspects, which prevent us from utilizing varying eclipse durations in our fits.  The {\em Kepler} sampling time of nearly half an hour, in our opinion, makes it very difficult, if not impossible, to find the subtle changes in times of the first and last (or fourth) contacts of each individual eclipse with sufficient accuracy, at least not without the use of an a priori (or, a posteriori) physical-geometrical light curve model which we have intentionally avoided.

Furthermore, regarding our approach to finding orbital solutions for CHTs, we note that the sole use of easily obtainable and accurate eclipse timing data makes our algorithm more generally applicable, not only for {\em Kepler} data, but even for non-homogenous observational data sets. For example, ground-based (follow up) observations at different locations and times, and with different instruments\footnote{Including even moderate and small aperture telescopes, operated by not only professional, but even non-professional, backyard astronomers.}, can produce easily comparable and sufficiently accurate mid-eclipse times, with a little effort.  
%, and, for shorter period binaries, during a few hour-long observation (which can be definitely shorter than the whole eclipse event). SR - I don't get the point here!
By contrast, for measuring the depth of an individual eclipse, and more specifically its duration, a substantially longer observing window is necessary, which may dramatically limit the available events from a given geographical observing site. Furthermore, in the case of eclipse depth measurements, the strong wavelength-dependence also makes them somewhat more difficult to compare with other measurements. In conclusion, the fact that our algorithm is exclusively based on eclipse timing data, makes it readily applicable to any other ETV data.

Second in regard to our fitting approach, we note that the orbital solutions, could have been found by direct 3-body numerical integrations coupled to Levenberg-Marquardt and/or MCMC parameter estimation schemes.  The advantage of using the admittedly complicated analytic expressions to model the ETV curves is that they provide us with key insight into what kinds of information are required for complete and for partial orbital solutions. The analytical ETV expressions also show us the functional dependences on the various physical parameters of the system. In particular, we now understand how the use of both the primary and secondary ETV curves for an eccentric inner binary can break the near degeneracy between the LTTE and physical delay effects contributing to the ETV curves. In addition, as discussed in the Introduction, fitting ETV curves via an analytic approach is considerably faster than running a numerical integration for many possible system configurations.  In the {\em Kepler} era and beyond, where there promise to be hundreds of such hierarchical triples, speed in deducing system parameters may be of some importance.  A numerical approach to fitting TTVs is likely to remain the appropriate technique in the exoplanet realm, but the analytic approach promises to remain effective for hierarchical triples. 

\section*{Acknowledgements}

We thank Alan Levine for sharing his $O-C$ curves of the {\em Kepler} binaries with us; this enabled us to identify several of the CHTs discussed in this work. The authors are greatful to Emese Forg\'acs-Dajka for drawing Figure~1.
This project has been partially supported by the Hungarian OTKA Grant K113117.
This research has made use of data collected by the {\em Kepler} mission, which is funded by the NASA Science Mission directorate.  Some of the data presented in this paper were obtained from the Mikulski Archive for Space Telescopes (MAST).  STScI is operated by the Association of Universities for Research in Astronomy, Inc., under NASA contract NAS5-26555.  Support for MAST for non-HST data is provided by the NASA Office of Space Science via grant NNX13AC07G and by other grants and contracts. T.\,B. would like to thank the City of Szombathely for support under Agreement No. S-11-1027.

\onecolumn
\appendix

\section[Long-term octupole terms]{Long-term octupole terms\footnotemark}
\label{app:octupole}

\footnotetext{These terms are not to be confused with the ``apse-node'' time-scale octupole terms, which were calculated in their full complexity by \citet{krymolowskimazeh99} for the first time, but, because of the use of another naming convention, these were also referred to as ``long-term'' in their paper.}

The most common procedure for calculating analytic perturbation formulae in orbital dynamics is the use of canonical equations on perturbations to the Hamiltonian function. For the hierarchical stellar three-body problem, the pioneering work was carried out in a series of papers by \citet{harrington68,harrington69} who used the renowned von Zeipel averaging technique. For some practical reasons which were enumerated by \citet{borkovitsetal03} we followed a different, but equivalent method. Instead of the perturbing {\em potential} and, therefore, the Hamiltonian, we calculated the perturbing {\em force}.  From that, we obtained the direct, analytic perturbation equation for the ETVs with the use of the perturbation equations expressed via the force components \citep[see e.g.,][]{milanietal87}. In this Appendix we do not repeat the individual steps (which were described in e.g.~\citealt{borkovitsetal11}), and only list the perturbing force up to second order in $a_1/a_2$, and also give the general final-form of the long-term octupole contribution of the ETV.  We felt that these expressions would be too lengthy for inclusion in the main body of the text.

Thus restricting ourselves, the perturbing force components are as follows:
\begin{equation}
\vec{f}=\frac{Gm_\mathrm{C}}{\rho_2^3}\left\{\left[\sum_{n=0}^\infty\left(\frac{m_\mathrm{A}}{m_\mathrm{AB}}\right)^n\left(\frac{\rho_1}{\rho_2}\right)^nP_n(\lambda)\right]^3\vec{r}_\mathrm{BC}-\left[\sum_{n=0}^\infty(-1)^n\left(\frac{m_\mathrm{B}}{m_\mathrm{AB}}\right)^n\left(\frac{\rho_1}{\rho_2}\right)^nP_n(\lambda)\right]^3\vec{r}_\mathrm{AC}\right\}\;,
\end{equation}
where $\rho_{1,2}$, are the lengths of the first two Jacobian position vectors (i.e., essentially the separations of the inner and outer binary), $P_n$ indicates the $n$-th Lagrangian polynomial, while $\lambda$ is the direction cosine between the (Jacobian) radius vectors ($\vec{\rho}_1$ and $\vec{\rho}_2$) of the two binaries. The radial, tangential, and normal components of this perturbing force vector up to the octupole order\footnote{Note, that the ``extra'' $\rho_2^{-2}$ multiplicator will disappear from the long-period perturbation equations at the step when the independent variable will be changed from time to the true anomaly $v_2$ of the outer orbit.} are
\begin{eqnarray}
f_\mathrm{r}&=&\frac{Gm_\mathrm{C}}{\rho_2^2}\left\{2\left(\frac{\rho_1}{\rho_2}\right)P_2(\lambda)+3\frac{m_\mathrm{A}-m_\mathrm{B}}{m_\mathrm{A}+m_\mathrm{B}}\left(\frac{\rho_1}{\rho_2}\right)^2P_3(\lambda)\right\}~, \\
f_\mathrm{t}&=&\frac{Gm_\mathrm{C}}{\rho_2^2}\left\{3\left(\frac{\rho_1}{\rho_2}\right)\lambda+3\frac{m_\mathrm{A}-m_\mathrm{B}}{m_\mathrm{A}+m_\mathrm{B}}\left(\frac{\rho_1}{\rho_2}\right)^2\left[\frac{5}{2}\lambda^2-\frac{1}{2}\right]\right\}\mu~, \\
f_\mathrm{n}&=&\frac{Gm_\mathrm{C}}{\rho_2^2}\left\{3\left(\frac{\rho_1}{\rho_2}\right)\lambda+3\frac{m_\mathrm{A}-m_\mathrm{B}}{m_\mathrm{A}+m_\mathrm{B}}\left(\frac{\rho_1}{\rho_2}\right)^2\left[\frac{5}{2}\lambda^2-\frac{1}{2}\right]\right\}\nu~,
\end{eqnarray}
respectively, where the direction cosines are
\begin{eqnarray}
\lambda&=&\cos{w_1}\cos{w_2}+\sin{w_1}\sin{w_2}\cos{i}_\mathrm{m}, \label{lambdadef} \\
\mu&=&-\sin{w_1}\cos{w_2}+\cos{w_1}\sin{w_2}\cos{i}_\mathrm{m}, \label{mudef} \\
\nu&=&\sin{w_2}\sin{i}_\mathrm{m}.
\label{nudef}
\end{eqnarray}
A straightforward calculation leads to the following results:
\begin{eqnarray}
f_\mathrm{r}&=&\frac{3}{8}\frac{Gm_\mathrm{C}}{\rho_2^2}\frac{\rho_1}{\rho_2}\left[(1+I)^2\cos(2w_2-2w_1)+(1-I)^2\cos(2w_2+2w_1)+2(1-I^2)(\cos2w_1+\cos2w_2)+2\left(I^2-\frac{1}{3}\right)\right] \nonumber \\
&&+\frac{15}{64}\frac{Gm_\mathrm{C}}{\rho_2^2}\left(\frac{\rho_1}{\rho_2}\right)^2\frac{m_\mathrm{A}-m_\mathrm{B}}{m_\mathrm{A}+m_\mathrm{B}}\left\{(1+I)^3\cos(3w_2-3w_1)+(1-I)^3\cos(3w_2+3w_1)\right. \nonumber \\
&&+3\left(1-I^2\right)\left\{(1+I)\left[\cos(3w_2-w_1)+\cos(w_2-3w_1)\right]+(1-I)\left[\cos(3w_2+w_1)+\cos(w_2+3w_1)\right]\right\}\nonumber \\
&&\left.-\frac{3}{5}\left(1+11I-5I^2-15I^3\right)\cos(w_2-w_1)-\frac{3}{5}\left(1-11I-5I^2+15I^3\right)\cos(w_2+w_1)\right\}, \label{fr} \\
f_\mathrm{t}&=&\frac{3}{8}\frac{Gm_\mathrm{C}}{\rho_2^2}\frac{\rho_1}{\rho_2}[(1+I)^2\sin(2w_2-2w_1)-(1-I)^2\sin(2w_2+2w_1)-2(1-I^2)\sin2w_1] \nonumber \\
&&+\frac{15}{64}\frac{Gm_\mathrm{C}}{\rho_2^2}\left(\frac{\rho_1}{\rho_2}\right)^2\frac{m_\mathrm{A}-m_\mathrm{B}}{m_\mathrm{A}+m_\mathrm{B}}\left\{(1+I)^3\sin(3w_2-3w_1)-(1-I)^3\sin(3w_2+3w_1)\right. \nonumber \\
&&+\left(1-I^2\right)\left\{(1+I)\left[\sin(3w_2-w_1)+3\sin(w_2-3w_1)\right]-(1-I)\left[\sin(3w_2+w_1)+3\sin(w_2+3w_1)\right]\right\}\nonumber \\
&&\left.-\frac{1}{5}\left(1+11I-5I^2-15I^3\right)\sin(w_2-w_1)+\frac{1}{5}\left(1-11I-5I^2+15I^3\right)\sin(w_2+w_1)\right\}, \label{ft} \\
f_\mathrm{n}&=&\frac{3}{4}\frac{Gm_\mathrm{C}}{\rho_2^2}\frac{\rho_1}{\rho_2}[2\cos{w_1}\sin2w_2\sin{i}_\mathrm{m}+(1-\cos2w_2)\sin{w_1}\sin2i_\mathrm{m}] \nonumber \\
&&+\frac{15}{32}\frac{Gm_\mathrm{C}}{\rho_2^2}\left(\frac{\rho_1}{\rho_2}\right)^2\frac{m_\mathrm{A}-m_\mathrm{B}}{m_\mathrm{A}+m_\mathrm{B}}\sin\im\left[(1+I)^2\sin(3w_2-2w_1)+(1-I)^2\sin(3w_2+2w_1)+2\left(1-I^2\right)\sin3w_2\right. \nonumber \\
&&\left.+\left(1-2I-3I^2\right)\sin(w_2-2w_1)+\left(1+2I-3I^2\right)\sin(w_2+2w_1)-\frac{6}{5}\left(1-5I^2\right)\sin{w_2}\right].
\label{fn}
\end{eqnarray}
In each of the three above equations, the first line represents the quadrupole forces, while the others indicate the octupole contributions. We call attention to the last term of the quadrupole contribution to the radial component of perturbation force. This is the only term which does not depend upon any of the $w$ angles, i.e., the relative positions of the bodies with respect to the intersections of the orbital planes, but depends only upon their actual distance ratio. Therefore, it acts to effectively modify the mass of the inner binary in a time-dependent way. However, as far as $I^2<1/3$ this extra radial force is always directed outward and results in a reduced mass, and therefore a longer inner binary period, on average, while for highly inclined configurations (i.e., for $\sim54.736\degr<\im<\sim125.264\degr$) the net effect is the opposite. As a natural consequence, around the periastron advance of an eccentric outer binary, (i.e., when the ratio of $\rho_1/\rho_2$ is the smallest and, therefore this effect is the largest) the lower mutual inclination systems produce a rapid change in time delay, which in the ETV curve morphology manifests itself by a steep and short ascending branch of the sinusoid; several examples of this can be seen in Figs.~\ref{Fig:ETVall1}--\ref{Fig:ETVall3b}. In the opposite sense, eccentric systems with near perpendicular orbits would produce short, steep descending features of the ETV curves, as are illustrated in some figures of \citet{borkovitsetal11}.   

Including these octupole terms into the process described in the above cited papers, we arrive at the following result:

\begin{eqnarray}
\Delta_\mathrm{2}&=&\frac{P_1}{2\pi}A_\mathrm{L2}\left(1-e_1^2\right)^{1/2}\left\{2(1+I)\left\{K_{21}{\cal{C}}_{21}(u_2-\alpha)+K_{22}{\cal{S}}_{21}(u_2-\alpha)-\frac{1}{3}\left[K_{23}{\cal{C}}_{23}(3u_2-3\alpha)+K_{24}{\cal{S}}_{23}(3u_2-3\alpha)\right]\right\}\right. \nonumber \\
&&+2(1-I)\left\{-K_{21}{\cal{C}}_{21}(u_2-\beta)+K_{22}{\cal{S}}_{21}(u_2-\beta)-\frac{1}{3}\left[-K_{23}{\cal{C}}_{23}(3u_2-3\beta)+K_{24}{\cal{S}}_{23}(3u_2-3\beta)\right]\right\} \nonumber \\
&&+\sin^2\im\left\{5(1+I)\left\{-K_{21}{\cal{C}}_{21}(u_2-\alpha)-K_{22}{\cal{S}}_{21}(u_2-\alpha)+\frac{1}{2}\left[-K_{21}{\cal{C}}_{21}(u_2-\beta)+K_{22}{\cal{S}}_{21}(u_2-\beta)\right]\right.\right. \nonumber \\
&&-\frac{1}{10}\left[K_{23}{\cal{C}}_{21}(u_2-2\alpha+\beta)+K_{24}{\cal{S}}_{21}(u_2-2\alpha+\beta)\right]+\frac{1}{2}\left[K_{21}{\cal{C}}_{23}(3u_2-2\alpha-\beta)+K_{22}{\cal{S}}_{23}(3u_2-2\alpha-\beta)\right] \nonumber \\
&&\left.+\frac{1}{15}\left[K_{23}{\cal{C}}_{23}(3u_2-3\alpha)+K_{24}{\cal{S}}_{23}(3u_2-3\alpha)\right]+\frac{1}{30}\left[-K_{23}{\cal{C}}_{23}(3u_2-3\beta)+K_{24}{\cal{S}}_{23}(3u_2-3\beta)\right]\right\} \nonumber \\
&&+5(1-I)\left\{K_{21}{\cal{C}}_{21}(u_2-\beta)-K_{22}{\cal{S}}_{21}(u_2-\beta)+\frac{1}{2}\left[K_{21}{\cal{C}}_{21}(u_2-\alpha)+K_{22}{\cal{S}}_{21}(u_2-\alpha)\right]\right. \nonumber \\
&&+\frac{1}{10}\left[K_{23}{\cal{C}}_{21}(u_2-2\beta+\alpha)-K_{24}{\cal{S}}_{21}(u_2-2\beta+\alpha)\right]-\frac{1}{2}\left[K_{21}{\cal{C}}_{23}(3u_2-2\beta-\alpha)-K_{22}{\cal{S}}_{23}(3u_2-2\beta-\alpha)\right] \nonumber \\
&&\left.\left.\left.+\frac{1}{15}\left[-K_{23}{\cal{C}}_{23}(3u_2-3\beta)+K_{24}{\cal{S}}_{23}(3u_2-3\beta)\right]+\frac{1}{30}\left[K_{23}{\cal{C}}_{23}(3u_2-3\alpha)+K_{24}{\cal{S}}_{23}(3u_2-3\alpha)\right]\right\}\right\}\right\} \nonumber \\
&&+\Delta_2^*(\sin\im\cot{i_1}).
\label{Eq:dyn2}
\end{eqnarray}
The amplitude of the second order term is
\begin{equation}
A_\mathrm{L2}=\frac{m_\mathrm{A}-m_\mathrm{B}}{m_\mathrm{AB}}\left(\frac{m_\mathrm{AB}}{m_\mathrm{ABC}}\right)^{1/3}\left(\frac{P_1}{P_2}\right)^{2/3}\frac{A_\mathrm{L1}}{1-e_2^2},
\end{equation}
while the integrals of the trigonometric functions of revolution of the wide binary lead to
\begin{eqnarray}
{\cal{S}}_{21}(u_2)&=&\left(1+\frac{1}{2}e_2^2\right)\sin u_2+\frac{1}{2}e_2\sin(2u_2-\omega_2)+\frac{1}{4}e_2^2\sin(u_2-2\omega_2)+\frac{1}{12}e_2^2\sin(3u_2-2\omega_2)+e_2\cos\omega_2(v_2-l_2), \nonumber \\
{\cal{C}}_{21}(u_2)&=&\left(1+\frac{1}{2}e_2^2\right)\cos u_2+\frac{1}{2}e_2\cos(2u_2-\omega_2)-\frac{1}{4}e_2^2\cos(u_2-2\omega_2)+\frac{1}{12}e_2^2\cos(3u_2-2\omega_2)-e_2\sin\omega_2(v_2-l_2), \nonumber \\
{\cal{S}}_{23}(3u_2)&=&\frac{1}{3}\left(1+\frac{1}{2}e_2^2\right)\sin3u_2+\frac{1}{2}e_2\sin(2u_2+\omega_2)+\frac{1}{4}e_2\sin(4u_2-\omega_2)+\frac{1}{4}e_2^2\sin(u_2+2\omega_2)\nonumber \\
&&+\frac{1}{20}e_2^2\sin(5u_2-2\omega_2) \nonumber \\
{\cal{C}}_{23}(3u_2)&=&\frac{1}{3}\left(1+\frac{1}{2}e_2^2\right)\cos3u_2+\frac{1}{2}e_2\cos(2u_2+\omega_2)+\frac{1}{4}e_2\cos(4u_2-\omega_2)+\frac{1}{4}e_2^2\cos(u_2+2\omega_2)\nonumber \\
&&+\frac{1}{20}e_2^2\cos(5u_2-2\omega_2) \label{Eq:dyntrigdef2}
\end{eqnarray}
Furthermore,
\begin{eqnarray}
K_{21}&=&\mp\left(\frac{1}{4}+\frac{5}{16}e_1^2\right)+\frac{57}{80}e_1\sin\omega_1\pm\frac{5}{16}e_1^2\cos2\omega_1+\frac{1}{16}e_1\sin3\omega_1\mp\frac{1}{16}e_1^2\cos4\omega_1+{\cal{O}}(e_1^3), \nonumber \\
K_{22}&=&-\frac{77}{80}e_1\cos\omega_1\pm\frac{3}{16}e_1^2\sin2\omega_1-\frac{1}{16}e_1\cos3\omega_1\mp\frac{1}{16}e_1^2\sin4\omega_1+{\cal{O}}(e_1^3), \nonumber \\
K_{23}&=&\pm\frac{105}{16}e_1^2\cos2\omega_1+{\cal{O}}(e_1^3), \nonumber \\
K_{24}&=&\pm\frac{105}{16}e_1^2\sin2\omega_1+{\cal{O}}(e_1^3).
\label{Eq:K2ndef}
\end{eqnarray}

Finally, in order to save space in the main body of the text, we list here some additional quantities connected with the quadrupole term, given by Eq.~(\ref{Eq:dyn1}).  First, there are the quadrupole-level auxiliary functions, $K(e_1,\omega_1)$, which we list here up to the seventh order in the inner eccentricity.  Second, we give the complete expression for the last term of Eq.~(\ref{Eq:dyn1}), i.e., $\Delta_1^*(\sin\im\cot{i_1})$, which describes those parts of the dynamical perturbations of the ETV which arise directly from the precession of the orbital plane of the inner binary due to an inclined ternary component. Note, since this expression evidently vanishes for coplanar configurations (i.e., when $\sin\im=0$), it was not worth converting to a form where all the variables would retain their meaning even for $\sin\im=0$ (as was done for the other components of Eq.~[\ref{Eq:dyn1}]). Therefore, the expression below (apart from a corrected sign error) is practically identical to the corresponding part of Eq.~(B.15) in Appendix~B of \citet{borkovitsetal11}.

\begin{eqnarray}
\Delta_1^*&=&\frac{P_1}{2\pi}A_\mathrm{L1}\left(1-e_1^2\right)^{-1/2}\sin\im\cot{i_1}(1-2K_1)\left\{\left[\frac{2}{5}\left(1+\frac{3}{2}e_1^2\right)\cos\um-e_1^2\cos(2\omega_1-\um)\right]\cos\im\left[{\cal{M}}-\frac{1}{2}{\cal{S}}(2u_2-2\uvm)\right]\right. \nonumber \\
&&\left.-\frac{1}{2}\left[\frac{2}{5}\left(1+\frac{3}{2}e_1^2\right)\sin\um+e_1^2\sin(2\omega_1-\um)\right]{\cal{C}}(2u_2-2\uvm)\right\}.
\label{Eq:Delta1*}
\end{eqnarray}

\begin{eqnarray}
K_1(e_1,\omega_1)&=&\mp e_1\sin\omega_1+\left(\frac{3}{4}e_1^2+\frac{1}{8}e_1^4+\frac{3}{64}e_1^6\right)\cos2\omega_1\pm\left(\frac{1}{2}e_1^3+\frac{3}{16}e_1^5\right)\sin3\omega_1-\left(\frac{5}{16}e_1^4+\frac{3}{16}e_1^6\right)\cos4\omega_1\nonumber \\
&&\mp\frac{3}{16}e_1^5\sin5\omega_1+\frac{7}{64}e_1^6\cos6\omega_1+{\cal{O}}(e_1^7), \nonumber \\
K_{11}(e_1,\omega_1)&=&\frac{3}{4}e_1^2+\frac{3}{16}e_1^4+\frac{3}{32}e_1^6\pm\left(e_1+\frac{1}{2}e_1^3+\frac{1}{4}e_1^5\right)\sin\omega_1+\left(\frac{51}{40}e_1^2+\frac{37}{80}e_1^4+\frac{241}{640}e_1^6\right)\cos2\omega_1\mp\frac{3}{16}e_1^3\sin3\omega_1\nonumber \\
&&-\left(\frac{1}{16}e_1^4-\frac{1}{16}e_1^6\right)\cos4\omega_1\mp\frac{1}{16}e_1^5\sin5\omega_1+\frac{3}{64}e_1^6\cos6\omega_1+{\cal{O}}(e_1^7), \nonumber \\
K_{12}(e_1,\omega_1)&=&\mp\left(e_1-\frac{1}{2}e_1^3-\frac{1}{4}e_1^5\right)\cos\omega_1+\left(\frac{51}{40}e_1^2+\frac{87}{80}e_1^4+\frac{541}{640}e_1^6\right)\sin2\omega_1\mp\frac{3}{16}e_1^3\cos3\omega_1\nonumber \\
&&-\left(\frac{1}{16}e_1^4+\frac{5}{32}e_1^6\right)\sin4\omega_1\pm\frac{1}{16}e_1^5\cos5\omega_1+\frac{3}{64}e_1^6\sin6\omega_1+{\cal{O}}(e_1^7).
\label{Eq:K1ndef}
\end{eqnarray}

\section{Short period terms}
\label{app:shortperiod}

For an approximative calculation of the contribution of the short-period terms (i.e., those that contain the inner true anomaly, $v_1$ in their arguments) of the perturbation equations, we integrated these terms formally with respect to the true longitude-like quantity ($u_1$). To the extent that all the parameters on the right-hand sides of the perturbation equations, with the exception of $v_1$ (or $u_1$) and $v_2$, can be considered to be constant, then the interesting terms take the following forms: $\sin(ku_1+nv_2+const)(1+e_2\cos{v_2})^3$, or $\cos(ku_1+nv_2+const)(1+e_2\cos{v_2})^3$, where $k$ is a non-zero integer, while $n=0$, or $n=2$. In order to integrate these equations, we have to express $v_2$ as a function of $u_1$ (or $v_1$). This can be done with two consecutive applications of the Kepler equation, as
\begin{equation}
l_2=\frac{P_1}{P_2}l_1+(l_2)_0
\end{equation}
and then,
\begin{eqnarray}
v_2&=&l_2+2e_2\left(1-\frac{e_2^2}{8}\right)\sin{l_2}+\frac{1}{2}e_2^2\sin2l_2+\frac{3}{8}e_2^3\sin3l_2+{\cal{O}}\left(e_2^4\right), \nonumber \\
l_1&=&v_1-2e_1\sin{v_1}+\frac{3}{4}e_1^2\sin2v_1-\frac{1}{3}e_1^3\sin3v_1+{\cal{O}}\left(e_1^4\right).
\end{eqnarray}
Then, substituting the corresponding trigonometric functions of $v_2$ into the equations, we integrate them formally, and take the lower and upper limits to be $u=u_0$ and $u=u_0+2\pi N$, where $N$ is an integer (essentially the cycle number), and $u_0=\mp\pi$ for the primary and secondary eclipses, respectively.

In such a manner we arrive at the following result:
\begin{eqnarray}
\delta({O-C})_\mathrm{dir2S}&=&\frac{P_1}{2\pi}A_\mathrm{S}\left(1-e_1^2\right)^{1/2}\left\{\pm2e_1{\cal{C}}_0^1(-\omega_1)-\frac{5}{3}e_1^2{\cal{S}}_0^2(-2\omega_1)\right. \nonumber \\
&&+(1+I)\left[\frac{11}{15}\left(1-\frac{7}{22}e_1^2\right){\cal{S}}_2^{-2}(2u_2-2\alpha)\mp\frac{4}{5}e_1{\cal{C}}_2^{-1}(2u_2-2\alpha-\omega_1)\right. \nonumber \\
&&\left.\mp\frac{8}{5}e_1{\cal{C}}_2^{-3}(2u_2-2\alpha+\omega_1)-\frac{13}{6}e_1^2{\cal{S}}_2^{-4}(2u_2-2\alpha+2\omega_1)\right] \nonumber \\
&&+(1-I)\left[\frac{11}{15}\left(1-\frac{7}{22}e_1^2\right){\cal{S}}_2^2(2u_2-2\beta)\pm\frac{4}{5}e_1{\cal{C}}_2^1(2u_2-2\beta+\omega_1)\right.\nonumber \\
&&\left.\pm\frac{8}{5}e_1{\cal{C}}_2^3(2u_2-2\beta-\omega_1)-\frac{13}{6}e_1^2{\cal{S}}_2^4(2u_2-2\beta-2\omega_1)\right] \nonumber \\
&&+\sin^2\im\left[\mp3e_1{\cal{C}}_0^1(-\omega_1)+\frac{5}{2}e_1^2{\cal{S}}_0^2(-2\omega_1)\right. \nonumber \\
&&+\frac{11}{15}\left(1-\frac{7}{22}e_1^2\right){\cal{S}}_0^2(-2\um)\pm\frac{4}{5}e_1{\cal{C}}_0^1(\omega_1-2\um)\pm\frac{8}{5}e_1{\cal{C}}_0^3(-\omega_1-2\um)-\frac{13}{6}e_1^2{\cal{S}}_0^4(-2\omega_1-2\um)\nonumber \\
&&\pm\frac{3}{2}e_1{\cal{C}}_2^1(2u_2-2\uvm-\omega_1)\mp\frac{3}{2}e_1{\cal{C}}_2^{-1}(2u_2-2\uvm+\omega_1)-\frac{5}{4}e_1^2{\cal{S}}_2^2(2u_2-2\uvm-2\omega_1)\nonumber \\
&&-\frac{5}{4}e_1^2{\cal{S}}_2^{-2}(2u_2-2\uvm+2\omega_1)-\frac{11}{30}\left(1-\frac{7}{22}e_1^2\right){\cal{S}}_2^{-2}(2u_2-2\alpha)\pm\frac{2}{5}e_1{\cal{C}}_2^{-1}(2u_2-2\alpha-\omega_1)\nonumber \\
&&\pm\frac{4}{5}e_1{\cal{C}}_2^{-3}(2u_2-2\alpha+\omega_1)+\frac{13}{12}e_1^2{\cal{S}}_2^{-4}(2u_2-2\alpha+2\omega_1)-\frac{11}{30}\left(1-\frac{7}{22}e_1^2\right){\cal{S}}_2^2(2u_2-2\beta) \nonumber \\
&&\left.\left.\mp\frac{2}{5}e_1{\cal{C}}_2^1(2u_2-2\beta+\omega_1)\mp\frac{4}{5}e_1{\cal{C}}_2^3(2u_2-2\beta-\omega_1)+\frac{13}{12}e_1^2{\cal{S}}_2^4(2u_2-2\beta-2\omega_1)\right]\right\}, \nonumber \\
\end{eqnarray}
where
\begin{eqnarray}
{\cal{S}}_0^n&=&-\frac{\nu}{n^2-\nu^2}\left[3e_2\left(1+3e_2^2\frac{n^2+2\nu^2}{n^2-4\nu^2}\right)\sin v_2+3e_2^2\frac{2n^2+\nu^2}{n^2-4\nu^2}\sin2v_2+\frac{3}{2}e_2^3\frac{3n^4+31n^2\nu^2+8\nu^4}{(n^2-4\nu^2)(n^2-9\nu^2)}\sin3v_2\right], \nonumber \\
{\cal{C}}_0^n&=&\frac{1}{n}\left[1+\frac{3}{2}e_2^2\frac{n^2+\nu^2}{n^2-\nu^2}+3e_2\frac{n^2}{n^2-\nu^2}\left(1+\frac{1}{4}e_2^2\frac{n^2+32\nu^2}{n^2-4\nu^2}\right)\cos{v_2}+\frac{3}{2}e_2^2\frac{n^2}{n^2-\nu^2}\frac{n^2+5\nu^2}{n^2-4\nu^2}\cos2v_2\right. \nonumber \\
&&\left.+\frac{1}{4}e_2^3\frac{n^2}{n^2-\nu^2}\frac{n^4+91n^2\nu^2+160\nu^4}{(n^2-4\nu^2)(n^2-9\nu^2)}\cos3v_2\right]+{\cal{O}}(e_2^4), \\
{\cal{S}}_2^n&=&\frac{1}{n+2\nu}\left\{\left[1+\frac{1}{2}e_2^2\frac{3n^2-8n\nu-9\nu^2}{(n+\nu)(n+3\nu)}\right]\sin2u_2+\frac{1}{2}e_2\left[\frac{3n+2\nu}{n+\nu}+\frac{3}{4}e_2^2\frac{n^2-22n\nu-12\nu^2}{(n+\nu)(n+3\nu)}\right]\sin(u_2+\omega_2)\right.\nonumber \\
&&+\frac{1}{2}e_2\left[\frac{3n+2\nu}{n+3\nu}+\frac{3}{4}e_2^2\frac{n^3-42n^2\nu-52n\nu^2-16\nu^3}{(n+\nu)(n+3\nu)(n+4\nu)}\right]\sin(3u_2-\omega_2)\nonumber \\
&&+\frac{1}{4}e_2^2\left[\frac{3n+\nu}{n+\nu}\sin2\omega_2+\frac{3n^2-5n\nu}{(n+3\nu)(n+4\nu)}\sin(4u_2-2\omega_2)\right] \nonumber \\
&&\left.-\frac{1}{8}e_2^3\frac{n^2}{(n+\nu)(n-\nu)}\sin(u_2-3\omega_2)+\frac{1}{8}e_2^3\frac{n^3-24n^2\nu+24n\nu^2}{(n+3\nu)(n+4\nu)(n+5\nu)}\sin(5u_2-3\omega_2)\right\} +{\cal{O}}(e_2^4), \\
\end{eqnarray}
where
\begin{equation}
\nu=\frac{P_1}{P_2}.
\end{equation}
Note that, e.g. 
\begin{eqnarray}
{\cal{C}}_0^1(-\omega_1)&=&{\cal{C}}_0^1\cos\omega_1+{\cal{S}}_0^1\sin\omega_1, \nonumber \\
{\cal{S}}_2^2(2u_2-2\beta)&=&{\cal{S}}_2^2\cos2\beta-{\cal{C}}_2^2\sin2\beta.
\end{eqnarray}

\section{Apsidal motion}
\label{app:apsidalmotion}

While calculating both the LTTE and the $P_2$ time-scale perturbations we assumed that the orbital elements remain constant in time. This is not strictly the case; however, it is a plausible approximation for certain restricted intervals, e.g., not substantially longer than the outer period. As is known from the theory of the dynamics of hierarchical triple systems, the highest amplitude periodic perturbations have the longest timescale. The characteristic timescale of these, so-called ``apse-node'' terms is on the order of $U\sim P_2^2/P_1$ \citep[see e.g.,][]{brown36}. For most of the triple stellar systems known before the {\it Kepler}--era this timescale exceeds centuries. In sharp contrast to this, the same timescale for some recently discovered systems, which are investigated in this paper, does not exceed 20-30 years. The consequence of this situation is that during the four-year-long observations of {\it Kepler}, the argument of periastron ($\omega_1$) of such a binary should have changed by $50\degr-70\degr$ (see Table~\ref{Tab:Orbelem}). It should also be borne in mind, that these ``apse-node'' time-scale perturbations are not restricted to apsidal motion and nodal regression, but occur in nearly all the orbital elements (with the exception of the semi-major axes). Here we mainly concentrate on apsidal motion and orbital plane precession\footnote{This latter effect is typically referred to as `nodal regression'. There are, however, a few systems in our sample where our solutions resulted in nodal {\em progression}, instead of regression; therefore, we simply use the term `precession'.}, however, a short discussion of other effects will also be presented later.

The angular velocity of the observable apsidal motion (averaged over one binary period) can be written as
\begin{eqnarray}
\dot{\omega}_1&=&\dot{g}_1+\dot{n}_1 \nonumber \\
&=&\dot g_1+\dot{h}_1\cos{j_1}-\dot{\Omega}_1\cos{i_1}.
\label{Eq:dotomega_general}
\end{eqnarray}
Here, the expression on the first row can be seen directly in Fig.~\ref{Fig:krsz-ek}, while the expression on the second line comes from the theorem of spherical triangles (see Appendix~\ref{app:Sphericaltriangle}). Eq.~(\ref{Eq:dotomega_general}) illustrates that the observed apsidal motion will be a combination of the apsidal motion in the dynamical frame, and the nodal regression (or, in other words, precession of the orbital planes). In the following we omit the very last term, because for eclipsing binary orbits viewed nearly edge-on, its contribution is negligible. Restricting ourselves to the quadrupole approximation, the perturbation equations which are the most interesting for us are as follows:
%the dynamical apsidal advance and nodal regression rates, in general, can be written, as:%
\begin{eqnarray}
\frac{P_1}{2\pi}\dot{g}_1&=&A+B\cos2g_1, \label{Eq:gpont}\\
\frac{P_1}{2\pi}\dot{h}_1&=&A_\mathrm{h}+B_\mathrm{h}\cos2g_1, 
\label{Eq:hpont}
\end{eqnarray}
and similarly,
\begin{eqnarray}
\frac{P_1}{2\pi}\dot\omega_1&=&A_\mathrm{o}+B_\mathrm{o}\cos2g_1, 
\label{Eq:om1pont}
\end{eqnarray}
where
\begin{eqnarray}
A&=&A_\mathrm{GR}+A_\mathrm{tidal}+A_\mathrm{3b}, \nonumber \\
B&=&B_\mathrm{3b} \nonumber \\
&=&B_\mathrm{o}-B_\mathrm{n}.
\end{eqnarray}
The individual contributions of the relativistic, (equilibrium) tide, and third-body effects are as follows:
\begin{eqnarray}
A_\mathrm{GR}&=&3\frac{Gm_\mathrm{AB}}{c^2a_1\left(1-e_1^2\right)}, \label{Eq:Arel}\\
A_\mathrm{tide}&=&\frac{5{\cal{T}}}{2a_1^5}\frac{1+\frac{3}{2}e_1^2+\frac{1}{8}e_1^4}{\left(1-e_1^2\right)^5}+\frac{{\cal{R}}}{a_1^2\left(1-e_1^2\right)^2}, \\
A_\mathrm{3b}&=&A_\mathrm{sec1}\left(1-e_1^2\right)^{-1/2}\left[I^2-\frac{1}{5}\left(1-e_1^2\right)+\frac{2}{5}\left(1+\frac{3}{2}e_1^2\right)\frac{C_1}{C_2}I\right], \\
B_\mathrm{3b}&=&A_\mathrm{sec1}\left(1-e_1^2\right)^{-1/2}\left[1-e_1^2-I^2-e_1^2\frac{C_1}{C_2}I\right], \label{Eq:B3b} \nonumber \\
A_\mathrm{h}&=&-\frac{2}{5}\frac{A_\mathrm{sec1}}{\cos j_1}\frac{1+\frac{3}{2}e_1^2}{\left(1-e_1^2\right)^{1/2}}\left(I^2+\frac{C_1}{C_2}I\right), \\
B_\mathrm{h}&=&\frac{A_\mathrm{sec1}}{\cos j_1}\frac{e_1^2}{\left(1-e_1^2\right)^{1/2}}\left(I^2+\frac{C_1}{C_2}I\right), \\
A_\mathrm{o}&=&A+A_\mathrm{h}\cos{j_1} \nonumber \\
&=&A_\mathrm{rel}+A_\mathrm{tidal}+A_\mathrm{sec1}\left(1-e_1^2\right)^{1/2}\frac{3}{5}\left(I^2-\frac{1}{3}\right),\nonumber \\
 \\
B_\mathrm{o}&=&B+B_\mathrm{h}\cos{j_1} \nonumber \\
&=&A_\mathrm{sec1}\left(1-e_1^2\right)^{1/2}\sin^2\im, \\
B_\mathrm{n}&=&\cos j_1B_\mathrm{h}.
\end{eqnarray}
Here we have introduced
\begin{equation}
A_\mathrm{sec1}=A_\mathrm{L1}\frac{P_1}{P_2}
\end{equation}
as the characteristic dimensionless amplitude of the ``apse-node'' timescale (sometimes called ``secular'') quadrupole perturbations.  Furthermore, the orbital angular momenta of the two orbits are
\begin{eqnarray}
C_1&=&\frac{m_\mathrm{A}m_\mathrm{B}}{m_\mathrm{AB}}\sqrt{Gm_\mathrm{AB}a_1\left(1-e_1^2\right)}, \\
C_2&=&\frac{m_\mathrm{AB}m_\mathrm{C}}{m_\mathrm{ABC}}\sqrt{Gm_\mathrm{ABC}a_2\left(1-e_2^2\right)},
\end{eqnarray}
and, moreover, the coefficients of the (lowest order equilibrium) tidal and (aligned) rotational oblateness are
\begin{eqnarray}
{\cal{T}}&=&6\left(\frac{m_\mathrm{B}}{m_\mathrm{A}}k_2^\mathrm{A}R^5_\mathrm{A}+\frac{m_\mathrm{A}}{m_\mathrm{B}}k_2^\mathrm{B}R^5_\mathrm{B}\right) \\
{\cal{R}}&=&\frac{k_2^\mathrm{A}R^5_\mathrm{A}s^2_\mathrm{A}}{Gm_\mathrm{A}}+\frac{k_2^\mathrm{B}R^5_\mathrm{B}s^2_\mathrm{B}}{Gm_\mathrm{B}}.
\end{eqnarray}
In the above equations $k_2$, $R$ and $s$ refer to the first apsidal motion constants, radii, and rotational angular velocities of stars $A$ and $B$. 

By the use of the quantities defined in Eqs.~(\ref{Eq:Arel}--\ref{Eq:B3b}) we can readily determine the apsidal motion period or, more strictly speaking, approximate timescales for the different contributions to the apsidal motion phenomena. Therefore, the ratio of the dynamical timescale to the sum of relativistic and (simplified, quasi-synchronuously rotating, equilibrium) tidal timescales can be defined as:
\begin{equation}
\frac{P_\mathrm{GR+tide}}{P_\mathrm{3b}}=\frac{\sqrt{|A_\mathrm{3b}^2-B_\mathrm{3b}^2|}}{A_\mathrm{GR}+A_\mathrm{tide}}.
\label{Eq:apstimescales}
\end{equation}
where the subscript ``$\mathrm{3b}$'' refers to the dynamically driven apsidal motion. This ratio was calculated for all our system solutions, and is listed in the last column of Table~\ref{Tab:apsenodeetal}. As one can see, the smallest ratio is $\sim18$, but in most cases it exceeds one hundred, which means that typically the strength of the dynamical effect is two orders of magnitude greater than for the GR and tidal effects. We feel that this nicely justifies our the omission of the non-dynamical contributions to the apsidal motion during the entire analysis.

In contrast to the relativistic and tidal effects, the dynamical apsidal motion comprises a `circulating' (or secular) and a librating (i.e., sine-like) component. We define the amplitude ratio of the two components, as
\begin{equation}
\varepsilon=\frac{B}{A} 
\label{Eq:epsilondef}
\end{equation}
(libration over circulation) which can be used for the qualitative description of the main characteristics of the apsidal motion effect. As one can see, for small values of $\varepsilon$, the characteristics of the dynamical apsidal motion remain similar to the tidal, and/or relativistically dominated scenarios, i.e., a (nearly) pure circulation occurs with a well-defined, constant period. The larger the (absolute value of the) ratio, the larger the libration contribution, and for $\varepsilon^2=1$ the characteristics of the apsidal motion vary substantially. If we omit the non-third-body contributions, this happens when
\begin{equation}
I^2+\frac{1}{5}\left(1+4e_1^2\right)\frac{C_1}{C_2}I=\frac{3}{5}\left(1-e_1^2\right),
\end{equation}
which, for the asymptotic approximation of $C_1/C_2\equiv0$ recovers the ``switching on'' condition of the Kozai-Lidov cycles. Note that adding the tidal and relativistic contributions to the dynamical effect usually\footnote{We say ``usually'' and not ``always'' since, for large mutual inclinations, $A_\mathrm{3b}$ may be negative and, therefore, in such a case it can happen that the net denominator becomes smaller.} increases the denominator in $\varepsilon$ and, therefore, reduces its value. Importantly, this may lead to the cancellation of the Kozai-Lidov cycles, as was first pointed out by \citet{soderhjelm84}.
    
The $\varepsilon$ parameter however, as well as the individual $A$, $B$ coefficients are primarily dependent on $e_1$ and $\im$, or $j_1$ and, therefore, they do not necessarily remain constant in time. The perturbation equations of these latter variables are as follows:
\begin{eqnarray}
\frac{P_1}{2\pi}\dot{e}_1&=&e_1B_\mathrm{o}\sin2g_1, \label{Eq:epont}\\
\cot j_1\frac{P_1}{2\pi}\dot{j}_1&=&B_\mathrm{n}\sin2g_1.
\label{Eq:j1pont}
\end{eqnarray}
The form of these equations reveals that, for small inner eccentricities and/or small mutual inclinations, there are only minor variations in these quantities and, therefore also for the right-hand-side coefficients of Eqs.~(\ref{Eq:gpont}--\ref{Eq:om1pont}). Insofar as the right-hand-side coefficients of Eqs.~(\ref{Eq:gpont}--\ref{Eq:om1pont}) and Eqs.~(\ref{Eq:epont}--\ref{Eq:j1pont}) are considered to be constant, this system of differential equations has simple, closed-form, analytic solutions. This solution was first given in papers by \citet{mazehshaham79,soderhjelm82}. For $\varepsilon^2<1$ the solution takes the following form:
\begin{eqnarray}
g_1&=&\arctan\left(\sqrt{\frac{1+\varepsilon}{1-\varepsilon}}\tan{\cal{G}}\right), \label{Eq:g1(t)}\\
h_1&=&{h_1}_0+\frac{B_\mathrm{h}}{B}(g_1-{g_1}_0)+\left(A_\mathrm{h}-\frac{B_\mathrm{h}}{\varepsilon}\right)(u_1-{u_1}_0), \\
e_1&=&{e_1}_0\sqrt{\left(\frac{1-\varepsilon\cos2{\cal{G}}}{1-\varepsilon\cos2{\cal{G}}_0}\right)^{B_\mathrm{o}/B}},  \label{Eq:e1(t)} \\
\sin j_1&=&{\sin j_1}_0\sqrt{\left(\frac{1-\varepsilon\cos2{\cal{G}}}{1-\varepsilon\cos2{\cal{G}}_0}\right)^{B_\mathrm{n}/B}},
\end{eqnarray}
and in a similar manner, for the observable argument of periastron
\begin{eqnarray}
\omega_1&=&{\omega_1}_0+\frac{B_\mathrm{o}}{B}(g_1-{g_1}_0)+\left(A_\mathrm{o}-\frac{B_\mathrm{o}}{\varepsilon}\right)(u_1-{u_1}_0).
\end{eqnarray}
Here we have introduced the quantity ${\cal{G}}$ which, by formal analogy to the Kepler-problem, can be called the `mean dynamical argument of periastron', and is defined as
\begin{eqnarray}
{\cal{G}}&=&\arctan\left(\sqrt{\frac{1-\varepsilon}{1+\varepsilon}}\tan{g_1}\right) \nonumber \\
&=&{\cal{G}}_0+\Pi(u_1-{u_1}_0),
\end{eqnarray}
while the apsidal advance rate, averaged over one binary orbit, is
\begin{equation}
\Pi=\sqrt{A^2-B^2}.
\label{Eq:Pidef}
\end{equation}
In such a way, the dynamical apsidal motion period becomes
\begin{equation}
P_\mathrm{g_1}=\frac{P_1}{\Pi}.
\label{Eq:Pg1model}
\end{equation}

Eq.~(\ref{Eq:e1(t)}) reveals that as long as $\varepsilon$ is small, the variation in the inner eccentricity and, similarly for the dynamical inclination, remains small and, therefore in such cases, there are also only minor variations in $\varepsilon$. 

There is also a similar analytic solution for the ``hyperbolic'' case, i.e., when $\varepsilon^2>1$; however, we do not list it since, in that case, due to the significant eccentricity variation, the constancy of the right-hand-sides does not hold.

We also note that the equations above are valid, of course, not only for the purely dynamical case, but retain the same form when all the apsidal motion contributions are considered \citep[see][]{borkovitsetal07}.

In the next to last column of Table~\ref{Tab:apsenodeetal} we list the values of $\varepsilon$. For 10 (or 11) systems (naturally, for the lowest mutual inclination triples), this value remains under $0.1$, while for an additional five (or four) systems it is smaller than $0.2$ (or its negative counterpart). On the other hand, most of the remaining systems, for which $0.25<\varepsilon<1.86$, the apsidal advance periods usually\footnote{Evident exceptions are the ``too-close'' systems of the second group, especially KIC 07668648, and from the third group, KIC 10319590.} are substantially longer than the observational window and, therefore, despite the expected larger variations in the orbital elements and other apsidal motion parameters, a linear approximation over such a relatively short timescale can also be sufficient.

Taking into account the above considerations, our software, in the present case operates in four different apsidal motion calculation modes, as follows:

\begin{itemize}
\item[AP1:]{Unconstrained, constant apsidal motion rate.}
In this mode the apsidal advance rate, $\Delta\omega$, is considered to be an independent constant, and can be fitted accordingly.
\item[AP2:]{Constrained, constant apsidal motion rate.}
Here the program calculates the instantaneous observed apsidal motion period as
\begin{equation}
P_{\omega_1}=\frac{P_1}{A_\mathrm{o}+B_\mathrm{o}\cos2(g_1)_0},
\label{Eq:Pomega1inst}
\end{equation}
and the variable $\Delta\omega$ is set accordingly. If the apsidal motion rate is also included in the LM process, the software takes into account the functional dependences of this parameter on the other fitted elements, and builds them into the analytical derivatives of the other elements being fitted.
\item[AP3:]{Constrained, according to the first-order analytical model.}
In this mode the former Eqs.~(\ref{Eq:g1(t)}--\ref{Eq:Pg1model}), or their large mutual inclination counterparts are used to compute both the observed apsidal motion and the precession rates. Furthermore, the secular variation of $e_1$ is also computed, optionally.
\end{itemize}
 
Similarly the dynamical precession rate can also be (i) unconstrained, (ii) a constrained constant according to the instantaneous period of
\begin{equation}
P_\mathrm{h}=\frac{P_1}{A_\mathrm{h}+B_\mathrm{h}\cos2(g_1)_0},
\label{Eq:Phinst}
\end{equation}
or (iii) calculated via the process described above. Then, after obtaining the dynamical precession rate, its realization in the observable quantities is calculated according to the description in the forthcoming Appendix.

\section{Constraints of geometry: the spherical triangle(s) formed by inclination angles and nodal arcs}
\label{app:Sphericaltriangle}

As was mentioned in the main text, the spherical triangle that is formed by the intersections of the two orbital planes and the plane of the sky on the abstract celestial sphere, carries an extraordinary importance in describing and constraining not only the complete three-dimensional configuration of an actual triple system with respect to both its invariable plane, and the observer, but even the orbital dynamics of the triple system. Stirctly speaking we can identify four such triangles in each hemisphere, as is nicely illustrated in Fig.~\ref{Fig:krsz-ek}. From these eight triangles, in what follows, we choose and discuss the one whose three vertices are the celestial ascending nodes of the two orbits, and consecutive projected intersection of the two orbits. Then, two of the inner vertices are the mutual inclination angle ($\im$), and one of the observable inclinations, while the other observable inclination becomes the external angle of the third vertex.  Moreover, the three arcs are the nodal-like quantities $n_1$, $n_2$, and also $\Delta\Omega=\Omega_2-\Omega_1$. Furthermore, with such a choice, the invariable plane of the triple also cuts across our spherical triangle, and divides it into two smaller triangles.  In these smaller triangles one arc and vertex are common with the parent triangle (i.e., $n_1$ and $i_1$, or $n_2$ and $i_2$), while the corresponding dynamical inclination ($j_{1,2}$) substitutes for $\im$ and also, the corresponding $\Omega_{1,2}$ (measured arbitrarily from the node of the invariable plane) replaces $\Delta\Omega$, while the third vertices and arcs (i.e., the ``other'' observable inclination and node are replaced with the ``observable'' inclination of the invariable plane ($i_0$), and dynamical node $h$. (We intentionally omit the subscript for this quantity because it will be the subject of the forthcoming discussion of whether it is $h_1$ or $h_2$.) 

First consider the ``large'' triangles. From their arcs, both $n_1$ and $n_2$ occur directly in the perturbation equations. Furthermore, these angles establish a connection between the observable arguments of periastrons $\omega_1$ and $\omega_2$, from which the first is really an observable (via not only the apsidal part of the ETV analysis, but even with radial velocity measurements or, photometrically, with light-curve analyses), and their dynamical counterparts ($g_1$, $g_2$), which play a substantial role in the secular dynamics of triple systems \citep[see e.g,][]{fordetal00}. Moreover, as $g_1$ is a necessary, important parameter for determining the apsidal advance rate, the spherical triangles constrain even the calculation of the apsidal motion as well (see Appendix~\ref{app:apsidalmotion}). 

Among the angles, $\im$ is another key parameter not only for the $P_2$-timescale perturbations, but also for the secular, or ``apse-node''-type perturbation equations. From the two observable inclinations, $i_1$ is present directly only in the nodal terms, via its cotangent and, therefore, it has only a weak direct effect on the ETV curves. By contrast, however, the eclipsing nature of our systems yields very strong constraints for its numerical value, which is especially true for the longer period binaries. The outer inclination $i_2$ appears directly in the amplitude of the LTTE contribution (via its sine) but, on one hand, this contribution is small, and therefore, has only minor importance for the systems we are investigating, and on the other hand, it also has a complicated and somewhat degenerate connection with the stellar masses and their ratios. In the case of outer-orbit eclipses, however, $i_2$ is more strongly constrained than $i_1$, due to the substantially larger outer-orbit separations.

As is known from the theory of spherical harmonics, any combinations of the three parameters out of the six constituents determine the given spherical triangle, although, in most configurations, the solution is ambiguous. As a consequence, only three of the five above listed parameters can be set, or adjusted freely, the other two (and also $\Delta\Omega$, not mentioned above) is then already determined (with ambiguity) via the theorem of spherical triangles, and this fact provides strict constraints for our solutions.

The question naturally arises as to which combination(s) are to be used. A complete discussion would be too lengthy to undertake here. Here we show two examples, as follows:

\begin{itemize}
\item[a.)]{\it Systems exhibiting outer-orbit eclipses (free parameters: $i_2$, $\im$, $n_2$).}
In this situation $i_2$, $\im$ and $n_2$ were chosen. The presence of outer eclipses very strongly constrains the outer inclination, therefore it can remain reasonably fixed around a value close to $90\degr$, and in such a way controls the physical reliability of the solution. Furthermore, these parameters form one arc and the two vertices lying on that arc of the triangle, which offer one of the simplest computations of the other vertex and arcs. The calculation, however, requires a careful discussion which we present here.

As was mentioned above, one observable inclination is the inner, and the other is outer angle of the triangle. For example, in the scenario plotted in Fig.~\ref{Fig:krsz-ek}, $i_2$ is the inner, while $i_1$ is the outer angle. First we have to resolve this ambiguity. It can also be seen in that figure, that for the illustrated situation $\sin\Delta\Omega<0$. (It also can be shown more generally, that in such a case $-90\degr\leq\Delta\Omega<0\degr$ for prograde, and $180\degr<\Delta\Omega\leq270\degr$ for retrograde configurations.) Therefore, we can use the sign of $\sin\Delta\Omega$ symbolically, for the separation of the two cases. Thus, our equations will be step-by-step as follows:
\begin{eqnarray}
\cos{i_1}_{a,b}&=&\cos i_2\cos\im+\mathrm{sgn}(\sin\Delta\Omega)\sin i_2\sin\im\cos n_2, \\
\sin{i_1}_{a,b}&=&\sqrt{1-\cos^2{i_1}_{a,b}}, \\
\sin{n_1}_{a,b}&=&\frac{\sin i_2}{\sin{i_1}_{a,b}}\sin n_2,\\
\sin\Delta\Omega_{a,b}&=&\mp\frac{\sin\im}{\sin{i_1}_{a,b}}\sin n_2, \nonumber \\
&=&\mp\frac{\sin\im}{\sin{i_2}}\sin{n_1}_{a,b}, \\
\cos\Delta\Omega_{a,b}&=&\frac{\cos\im-\cos i_2\cos{i_1}_{a,b}}{\sin i_2\sin{i_1}_{a,b}}, \nonumber \\
&=&\frac{\cos\im\sin i_2-\mathrm{sgn}(\sin\Delta\Omega)\sin\im\cos i_2\cos n_2}{\sin{i_1}_{a,b}}, \\
\cos{n_1}_{a,b}&=&\frac{\cos\Delta\Omega_{a,b}-\sin n_2\sin{n_1}_{a,b}\cos\im}{\cos n_2}, \nonumber \\
&=&\cos n_2\cos\Delta\Omega_{a,b}-\sin n_2\sin\Delta\Omega_{a,b}\cos i_2~. 
\end{eqnarray}
As one can see, in a few cases we have given alternative forms, from which we can choose the one that is more appropriate (e.g., in its numerical behavior). In such a way, we can find two solutions. The software can be set to compute either both solutions, or only one of them; in the program terminology --  the ``first'' (i.e., $\sin\Delta\Omega<0$), or the ``second''.  (In the case where both calculated solutions are allowed, the program compares the $\chi^2$ values, and automatically chooses the better one.)  As we shall discuss a bit later, in most cases, one of the solutions results in decreasing eclipse depths with time, while the other in increasing depths. Finally, we also note that similar, but re-labeled equations can be used in the situation where $i_1$, $\im$, $n_1$ are chosen. 
\item[b.)]{\it Systems with low inner eccentricity and without outer eclipses ($i_1$, $\im$, $n_2$).}
If the outer binary does not produce eclipses, $i_2$ is no longer constrained in such a way. Therefore, we can choose $i_1$ instead\footnote{Because of the smaller orbital separations, the presence of eclipses does not yield such strong constraints for $i_1$ as for $i_2$ in the previous case.  Although, for example, total eclipses in a binary with $P_1\sim16$ days, and possibly consisting of dwarf components (the case of KIC~08023317) may provide a sufficiently certain estimation.}. In such a case, one can choose the set of $i_1$, $\im$ and $n_1$ and, therefore similar equations can be applied as before; however, as one can see from Eq.~\ref{Eq:dyn1}, the role of $n_1$ and $n_2$ is a bit different in the most prominent quadrupole term. Namely, all terms which are connected to $n_1$ are multiplied by $e_1$ and, therefore, for small inner eccentricities, $n_1$ gives only a small contribution. In contrast, the leading $n_2$-dependent term remains present even in the doubly circular, non-coplanar case as well. Therefore, it could be better to choose the trio of $i_1$, $\im$ and $n_2$. Unfortunately, from the sense of the spherical triangle theorem, this is not among the most fortuitous of cases, and the process becomes somewhat more complex. The steps are as follows: 
\begin{eqnarray}
\sin\Delta\Omega_{a,b}&=&\mp\frac{\sin\im}{\sin i_1}\sin n_2, \\
\cos\Delta\Omega&=&\sqrt{1-\sin^2\Delta\Omega}, \\
\cos{i_2}_{a,b}&=&\frac{\cos i_1\cos\im-\mathrm{sgn}(\sin\Delta\Omega_{a,b})\sin i_1\sin\im\cos\Delta\Omega\cos n_2}{1-\sin^2n_1\sin^2\im}, \label{Eq:cosi2ab}\\
\sin{i_2}_{a,b}&=&\sqrt{1-\cos^2{i_2}_{a,b}}, \\
\cos\Delta\Omega&=&\frac{\cos\im-\cos i_1\cos{i_2}_a}{\sin i_1\sin{i_2}_a} \nonumber \\
&=&\frac{\cos\im-\cos i_1\cos{i_2}_a}{\sin i_1\sin{i_2}_a}~.
\end{eqnarray}
At this step, if $\cos\Delta\Omega<0$ then labels of $i_2$ should be interchanged. Finally,
\begin{eqnarray}
\cos{n_1}_{a,b}&=&\cos\Delta\Omega\cos n_2-\sin\Delta\Omega_{a,b}\cos{i_2}_{a,b}, \\
\sin{n_1}_{a,b}&=&\frac{\sin{i_2}_{a,b}}{\sin i_1}\sin n2 \nonumber \\
&=&\sqrt{1-\cos^2{n_1}_{a,b}}~.
\end{eqnarray}
Note, that for $\im=n_2=90\degr$ the right hand side of Eq.~(\ref{Eq:cosi2ab}) takes the form of zero divided by zero. This situation can happen only when $i_1=i_2=n_1=90\degr$, and $\cos\Delta\Omega=0$, i.e., when the two orbits and the plane of the sky are orthogonal to each other. Such a situation is very far from any of the systems we investigated. Interestingly, on the other hand, the archetypical triple system Algol itself has a configuration not so far from this extreme \citep[see e.g.,][~and further references therein]{zavalaetal10}.

\end{itemize}

Now, we can turn to the other two smaller triangles, which define the position of the invariable plane with respect to the sky, and also make it possible to transform the orbital precession into the observational frame. The unknown vertices and arcs can be calculated, e.g., in the following simple manner.
\begin{eqnarray}
\cos i_0&=&\frac{C_1}{C}\cos i_1+\frac{C_2}{C}\cos i_2, \\
\sin i_0&=&\sqrt{1-\cos^2i_0}, \\
\cos h&=&\frac{C_1}{C}\frac{\sin i_1}{\sin i_0}\cos n_1+\frac{C_2}{C}\frac{\sin i_2}{\sin i_0}, \\
\sin h&=&\frac{\sin i_1}{\sin i_0}\sin n_1=\frac{\sin i_2}{\sin i_0}\sin n_2, \\
\cos j_1&=&\frac{C_1}{C_2}+\frac{C_2}{C}\cos\im, \\
\sin j_1&=&\frac{C_2}{C}\sin\im, \\
\cos\Omega_1&=&\frac{C_1}{C}\frac{\sin i_1}{\sin i_0}+\frac{C_2}{C}\frac{\sin i_2}{\sin i_0}\cos\Delta\Omega, \\
j_2&=&\im-j_1~,
\end{eqnarray}
where $C_1$, $C_2$, as before, denote the absolute value of the inner and outer angular momenta, and the net orbital angular momentum is calculated as:
\begin{eqnarray}
C&=&C_2\sqrt{1+2\frac{C_1}{C_2}\cos\im+\left(\frac{C_1}{C_2}\right)^2}~.
\end{eqnarray}

If once the dynamical inclinations and node are determined, the variations of the observable inclinations ($i_1$, $i_2$), which are forced by the orbital precession (and which are the direct sources of the eclipse depth variations observed in several systems), can be easily calculated by application of the identities obtained from the theorems of spherical triangles. Namely,
\begin{eqnarray}
\cos i_1&=&\cos i_0\cos j_1+\mathrm{sgn}(\sin\Delta\Omega)\sin i_0\sin j_1\cos h, \label{Eq:cosi1}\\
\cos i_2&=&\cos i_0\cos j_2-\mathrm{sgn}(\sin\Delta\Omega)\sin i_0\sin j_2\cos h, \\
\sin i_1&=&\sqrt{1-\cos^2i_1}, \\
\sin i_2&=&\sqrt{1-\cos^2i_2}, \\
\sin n_1&=&\frac{\sin i_0}{\sin i_1}\sin h, \\
\sin n_2&=&\frac{\sin i_0}{\sin i_2}\sin h, \\
\sin\Omega_1&=&-\mathrm{sgn}(\sin\Delta\Omega)\frac{\sin j_1}{\sin i_1}\sin h, \\
\sin\Omega_2&=&+\mathrm{sgn}(\sin\Delta\Omega)\frac{\sin j_2}{\sin i_2}\sin h, \\
\cos\Omega_1&=&\frac{\cos j_1-\cos i_0\cos i_1}{\sin i_0\sin i_1}=\frac{\cos j_1\sin i_0-\mathrm{sgn}(\sin\Delta\Omega)\sin j_1\cos i_0\cos h}{\sin i_1}, \\
\cos\Omega_2&=&\frac{\cos j_2-\cos i_0\cos i_2}{\sin i_0\sin i_2}=\frac{\cos j_2\sin i_0-\mathrm{sgn}(\sin\Delta\Omega)\sin j_2\cos i_0\cos h}{\sin i_2}, \\
\cos n_1&=&\frac{\cos\Omega_1-\sin h\sin n_1\cos j_1}{\cos h}=\cos h\cos\Omega_1+\sin h\sin\Omega_1\cos i_0, \\
\cos n_2&=&\frac{\cos\Omega_2-\sin h\sin n_2\cos j_2}{\cos h}=\cos h\cos\Omega_2-\sin h\sin\Omega_2\cos i_0~.
\end{eqnarray}

In the above expression, the meaning of the dynamical node ($h$) requires some additional discussion. First, we had left open the question of whether is it $h_1$, or $h_2$. The answer, naturally, is that it depends on the sign of $\Delta\Omega$. As far as $\sin\Delta\Omega<0$, as is the case in Fig.~\ref{Fig:krsz-ek}, the third vertex of the triangle represents the dynamical ascending node of the inner orbit and, therefore $h=h_1$, while in the opposite case $h=h_2=h_1+180\degr$. Another consequence is, that similarly, for the $\sin\Delta\Omega<0$ case $g_1=\omega_1-n_1$, $g_2=\omega_2-n_2-180\degr$, while if $\sin\Delta\Omega>0$ then $g_1=\omega_1-n_1-180\degr$, $g_2=\omega_2-n_2$.

Finally, we use Eq.~\ref{Eq:cosi1} to enable us to make a few simple, qualitative statements about the eclipse depth variations. Deriving this equation, taking into account that $\dot{\iota_0}\equiv0$, and using additional identities based on the theorems of spherical triangles, we obtain the expressions:
\begin{equation}
\dot{\iota_1}=\dot{j_1}\cos n_1+\mathrm{sgn}(\sin\Delta\Omega)\dot{h}\sin n_1\sin j_1=\dot{j_1}\cos n_1-\dot{h}\sin\Omega_1\sin i_0~.
\label{Eq:dotcosi1}
\end{equation}
In our simple model we omit the usually small, cyclic variation in the dynamical inclination ($j_1$), as well. In such a way, the last term remains. This shows clearly that an alternation between the two ambiguous solutions (discussed above) changes the sign of the variation of $i_1$ and, therefore, the observed direction of the eclipse depth variation (if detected) allows us to resolve the ambiguity.

\section{Details of the numerical analysis}
\label{app:numericalanalysis}

\begin{figure*}
\begin{center}
\includegraphics[width=125mm]{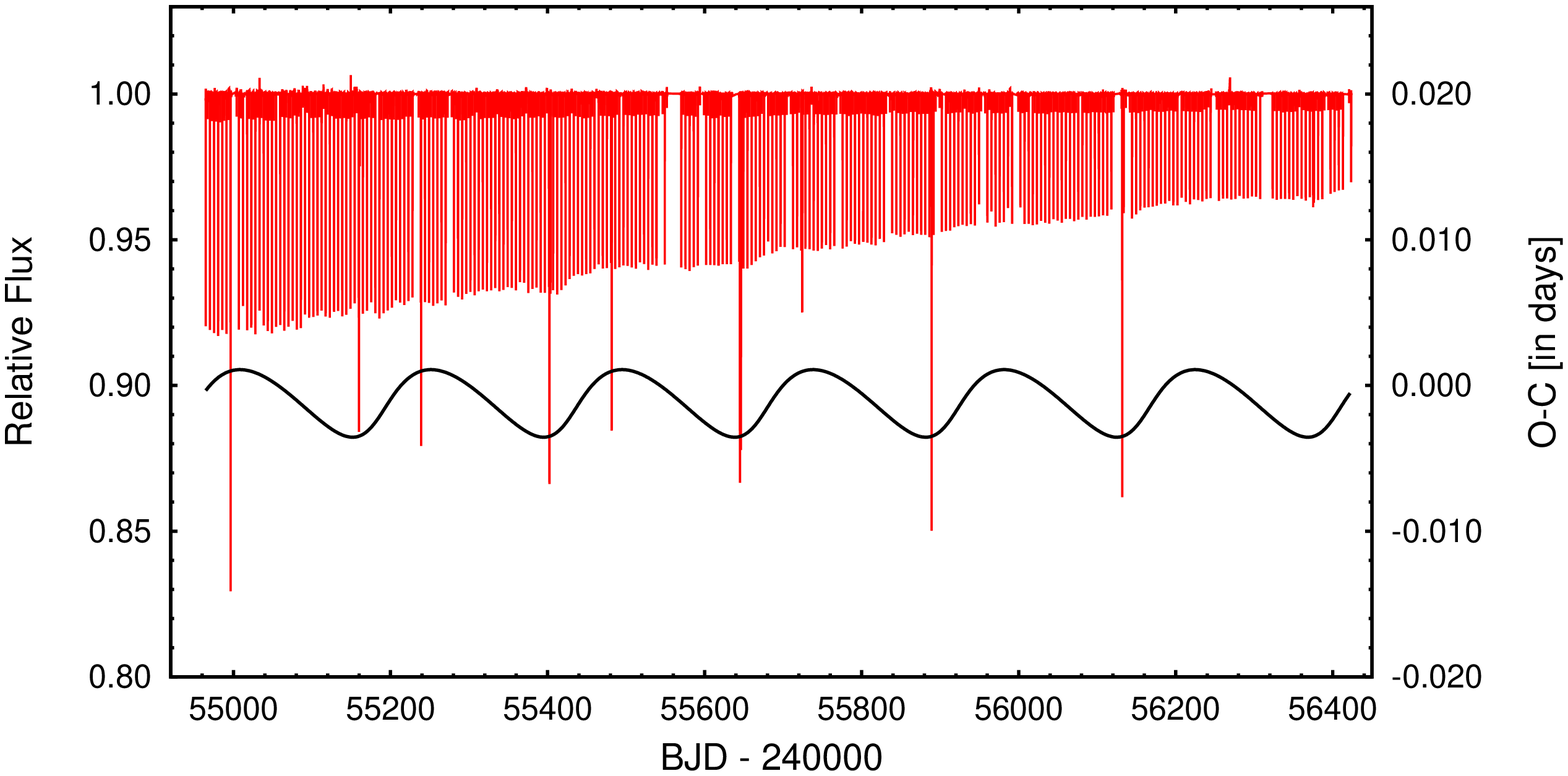}
 \caption{{\it Red curve:} The detrended long cadence light curve of KIC 07289157. The varying eclipse depths as well as the `anomalous' outer eclipses are readily discerned. {\it Black curve:} The light-time effect (or Roemer delay) contribution to the complete ETV solution. It nicely illustrates that the extra eclipses occur near the extrema of the ETV curve. At the upper extrema events, the third star eclipses one (or both) binary members, while in the lower extrema cases the third star is eclipsed.}
 \label{Fig:K7289157lc}
 \end{center}
\end{figure*}

In this Appendix we describe in detail how the fits of our analytic ETV functions, with their many system parameters, were fit to the observed ETV curves.  We utilize 4 of the 26 triple systems in this study to illustrate how the fitting was done in practice.  One of these systems has a low mutual inclination angle, one has a medium inclination angle, while the third is undergoing retrograde motion. The fourth system also has a low mutual inclination angle, but since it is the only triple in our sample where the LTTE term was found to be comparable with the $P_2$ timescale quadrupole term, it was chosen for testing the reliability of obtaining individual stellar masses in such a scenario.

\subsection{The low mutual inclination regime: the case of KIC~07289157}

\begin{figure*}
\begin{center}
\includegraphics[width=125mm]{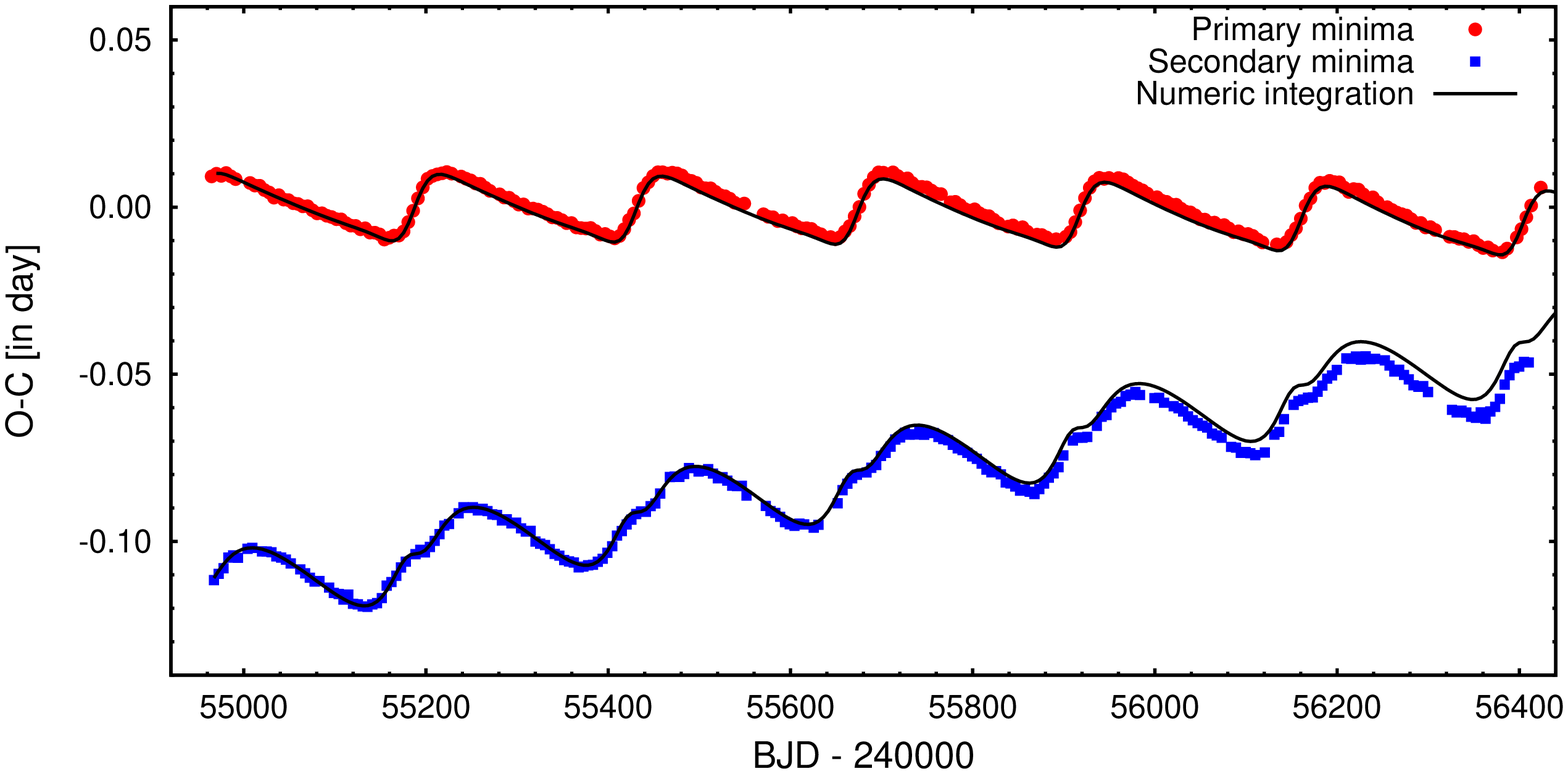}
 \caption{The ETV curves for KIC 07289157 (red and blue points for primary and secondary minima, respectively) together with the numerically generated ETV curve with the initial parameters of the OS3 solution of Table\,\ref{Tab:KIC7289157} (black lines).} 
 \label{Fig:K7289157numanalfits}
\end{center}
\end{figure*}

The triply eclipsing nature of KIC~07289157 makes it extraordinarily useful for testing our analytic fitting model. In Figs.~\ref{Fig:K7289157lc} and \ref{Fig:K7289157numanalfits} its long cadence {\it Kepler} light curve and ETV curves (determined from the same lightcurves) are shown. The two remarkable features of its light curve are the occurrence of extra, or outer, eclipses and the continuously varying eclipse depths. The period of the cyclic variations of the ETV curve is in accord with the occurrence of the extra eclipses, and clearly reflects the orbital period of the outer binary. The effect of rapid apsidal motion is also readily visible (mostly in the converging primary and secondary ETV curves). 

In what follows, we first illustrate the method by which we estimate reliable input parameters for the initial fitting runs. KIC~07289157 has one of the shortest inner periods ($P_1\sim5.27$~days) in our sample. The observable argument of periastron ($\omega_1$) and eccentricity ($e_1$) of the inner binary can be well approximated from the ETV. The fact that the secondary $O-C$ curve (blue points in Fig.~\ref{Fig:K7289157numanalfits}) is located below the primary $O-C$ curve indicates that the interval between a primary eclipse and the consecutive secondary eclipse is shorter than half of the binary period, i.e., the periastron passage occurs between the primary and the secondary eclipses. Therefore, $\omega_1$ should be located in the fourth or the first quarter, i.e., $-90\degr<\omega_1<90\degr$, where the maximum separation occurs at $\omega_1=0\degr$ (i.e., when the line of the apsides lies in the plane of the sky). Because the two $O-C$ curves are converging in phase, it follows that in the present situation, as long as we assume prograde apsidal motion, $0\degr<\omega_1<90\degr$ is the appropriate quadrant. Therefore, setting $\omega_1=45\degr$, we find an appropriate initial guess\footnote{In a purely mathematical vein this simply comes from the fact that the difference between the primary and secondary eclipses is well-approximated by the $\cos\omega_1$ function} for the Levenberg-Marquardt (`LM') fitting procedure. This latter set of arguments also provides a crude, but satisfactory, initial guess for $e_1$, as the separation of the primary and secondary ETVs is related to $\sim 2P_1/\pi e_1\cos\omega_1$.  But, as we found, there was no need for this latter estimation for $e_1$ since the use of any arbitrary, not-too-extreme initial guess was satisfactory. 

Considering the initial parameters for the outer orbit, we have no similar crutches to rely on for its parameter estimations. However, if the $P_2$-period contribution to the ETV has a substantially asymmetric shape (which suggests a significant outer eccentricity), the periastron passage time $\tau_2$ can be well approximated from the location of the fastest varying portion of the ETV curve. Fortunately, we have found that it was sufficient in most cases to simply guess the correct quadrant for $\omega_2$. In other words, we prescribe four initial guesses for the LM fitting of $\omega_2$ as $45\degr$, $135\degr$, $225\degr$ and $315\degr$, respectively. Considering the spatial configuration of the system, the triply eclipsing nature gives a very strict constraint on the various inclination angles, especially for $i_2$. Namely, for such a long period eclipsing system we expect the outer inclination angle to be very close to $90\degr$. Therefore, although $i_2$ has no direct effect on the shape of the ETV, we have chosen that version of the program where $i_2$, $\im$ and $n_2$ determine the spatial orientation of the triple system, and therefore, $i_1$, $n_1$ and $\Delta\Omega$ can no longer be adjusted. Therefore, we set $i_2=89.5\degr$. To find an appropriate initial guess for the mutual inclination ($\im$) we realized that neither the shape of the ETV nor the rate of the eclipse depth variations suggests a substantial mutual inclination angle; therefore, we can set a low initial value, say, between $5\degr$ and $15\degr$, as well as on its retrograde counterparts. On the other hand, we had only a very rough idea for the initial constraints on $n_2$; however, we found that six evenly spaced initial values in the range, e.g., $30\degr$ -- $180\degr$ could be appropriate. 

Finally, some initial estimation of the amplitudes, i.e., masses and mass ratios were also necessary. As is expected, the dominant contribution should be the quadrupole dynamical terms; therefore, its amplitude is the most important. In our experience, setting the $m_\mathrm{C}/m_\mathrm{ABC}$ parameter to be initially around $1/3$ (which would be the accurate value for three equal masses) was an appropriate choice. Considering the other quantities, related to the LTTE and octupole terms, respectively, we can make some other initial estimates, but they have no crucial significance for the final fit. The Kepler Input Catalog contains the following data for KIC~07289157: $T_\mathrm{eff}=6013$\,K, $\log g=4.188$, which suggest that the brightest member of the system may be of approximately a solar mass, and a somewhat evolved star. The close binary exhibits very shallow secondary minima, despite the fact that the secondary eclipses are closer to the periastron passage; therefore, it follows that the surface brightness of the secondary star should be low relative to the primary one. If both components are main sequence stars then we may conclude that the secondary might be a faint, M or K star, which results in a small inner mass ratio. On the other hand, the extra anomalous eclipses suggest that the primary and the third, distant star might have similar surface brightnesses. Therefore, as an initial estimate, we assumed that these latter components might have nearly equal masses of $\sim1$\,M$_\odot$. Thus, we set the appropriate mass parameters accordingly.

In conclusion, in our first trial run the following parameters were adjusted in the LM process: $c_0$, $c_1$, $e_1$, $\omega_1$, $P_2$, $e_2$, $\omega_2$ (with 4 initial values), $\tau_2$, $m_\mathrm{C}/m_\mathrm{ABC}$ (with 3 initial values, e.g., in the range of $0.3-0.5$), $\im$ (1 or 2 initial values), $n_2$ (6 initial values). In most cases, such a first run was sufficient to locate the appropriate subspace within the larger parameter space. Then, in the following stages of the fitting, most of the parameters above were also LM-adjusted, but now with better initial values, while additional parameters of $a_\mathrm{AB}\sin{i_2}$, $q_1$, and even $i_1$ or $i_2$ were also included via the grid-search method. In some systems we obtained a better end-result by disabling the LM-adjustments of $\im$ and $n_2$ during the final fitting, and simply refining them via the grid-search method. In that phase, the grid-searched parameters (i.e., $a_\mathrm{AB}\sin{i_2}$, $q_1$, $\im$, $n_2$ and sometimes $i_2$, or $i_1$) were used also for controlling the physical reliability of the solutions in an interactive, subjective way. We accepted solutions only with plausible masses and inclinations, and it was also expected that the parameters would be at least marginally consistent with the observed eclipse-depth variations. In the first column of Table~\ref{Tab:KIC7289157} we list the main parameters of our solution for KIC~07289157. A more detailed list of the parameters (both adjusted, and derived) obtained from a similar fit\footnote{Here we list the parameters obtained from a fit utilizing equal, global uncertainties for the ETV-points, while in the main text, we tabulated the results found by using individual uncertainties for the ETV-points. Therefore, a comparison of the two parameter sets carries additional information on the weighting-scheme dependence and the robustness of our solutions.}, is listed in Tables~\ref{Tab:Orbelem}--\ref{Tab:Massetal}. We repeat these results here for an easier comparison with the numerical test results that are described below.

In order to check the reliability of our solutions, we carried out a number of numerical tests. For this (and the forthcoming) numerical integrations we used the code described in \citet{borkovitsetal04}. Although, our code is able to take into account tidal effects as well, for our present purpose we used simply the three-body point-mass approximation. We generated ETV curves directly from 3-body numerical integrations using the fitted solutions as initial conditions; in the next stage, these were used as ``observed'' input ETV curves for additional tests of the analytical formulae and fitting processes.

Before any detailed discussion, however, we make note of the difficulty in comparing orbital elements obtained from numerical and analytical solutions. The principal problem is that the analytical formulae (in general) use doubly averaged orbital elements (first for the inner, and then, second, for the outer orbital period), while the numerical integrator, which works directly with the rectangular Jacobian coordinates and velocity vectors, requires instantaneous osculating orbital elements for its initialization. The conversion between the two sets of elements is far from trivial, and would require detailed theoretical considerations. Some short discussion of this issue can be found, e.g., in \citet{kiselevaetal98} and \citet{borkovitsetal02}. However, this is only of minor importance for our study, and we did not deal with it in full detail, choosing instead to use only a simplified procedure. By the use of the $P_2$ time-scale perturbation equations for each orbital element (see, e.g., Eqs.~[8]--[15] in \citealp{borkovitsetal11}) we corrected the doubly averaged outputs of our code for $P_2$-period perturbations. Then the numerical integrator was initialized with these values. In such an approach, one of the largest amplitude effects was neglected. The remaining other systematic discrepancies which were caused by the incomplete initialization are mainly realized in somewhat different and initial-epoch-dependent orbital and eclipsing periods and phases. However, in conclusion, these discrepancies do not substantially affect any of our main results.  

\begin{table*}  
%\begin{center}
\caption{Different model solutions for KIC~07289157.} 
\label{Tab:KIC7289157}  
\begin{tabular}{lll|lll} 
\hline 
\multicolumn{2}{l}{Parameters} &OS3&    N0F1   &   N0S3    & NsimS3    \\
\hline
$P_1$      & (d)       & 5.267371  & 5.268198  & 5.268189  & 5.268188  \\
$e_1$      &           & 0.0849    & 0.0917    & 0.0951    & 0.0957    \\
$\omega_1$ & ($\degr$) & 66.018    & 67.588    & 68.498    & 68.638    \\
$\tau_1$   & (MBJD)    & 54972.199 & 54972.233 & 54972.247 & 54972.249 \\
\hline
$P_2$      & (d)       & 243.328   & 242.595   & 242.642   & 242.666   \\
$a_2$      &(R$_\odot$)& 232.047   & 224.069   & 231.847   & 232.438   \\
$e_2$      &           & 0.309     & 0.309     & 0.315     & 0.317     \\
$\omega_2$ & ($\degr$) & 156.330   & 159.754   & 157.834   & 157.669   \\
$\tau_2$   & (MBJD)    & 54941.615 & 54942.569 & 54941.769 & 54941.711 \\
\hline
$\im$ & ($\degr$)          &  4.630 &  9.596 &  5.900 &  6.191 \\
$i_1$ & ($\degr$)          & 85.123 & 99.014 & 83.859 & 83.550 \\
$i_2$ & ($\degr$)          & 89.500 & 89.500 & 89.500 & 89.500 \\
$n_1$ & ($\degr$)          & 19.071 &172.468 & 17.100 & 16.104 \\
$n_2$ & ($\degr$)          & 19.000 &172.561 & 17.000 & 16.000 \\
$\Delta\Omega$ & ($\degr$) &  1.511 &  1.252 &  1.732 &  1.714 \\
\hline
$m_\mathrm{C}/m_\mathrm{ABC}$&& 0.388 & 0.400 & 0.380 & 0.379 \\
$m_\mathrm{B}/m_\mathrm{A}$ & & 0.480 &  $-$  & 0.450 & 0.440 \\
$m_\mathrm{A}$ & (M$_\odot$)  & 1.173 & 1.453 & 1.217 & 1.236 \\
$m_\mathrm{B}$ & (M$_\odot$)  & 0.563 &       & 0.548 & 0.544 \\
$m_\mathrm{C}$ & (M$_\odot$)  & 1.100 & 0.970 & 1.079 & 1.085 \\
\hline
$\chi^2$ &                   & 1.9695 & 0.3183& 0.0631& 0.3671\\
\hline
\end{tabular} 
%\end{center}  
\end{table*}

\begin{figure*}
\includegraphics[width=86mm]{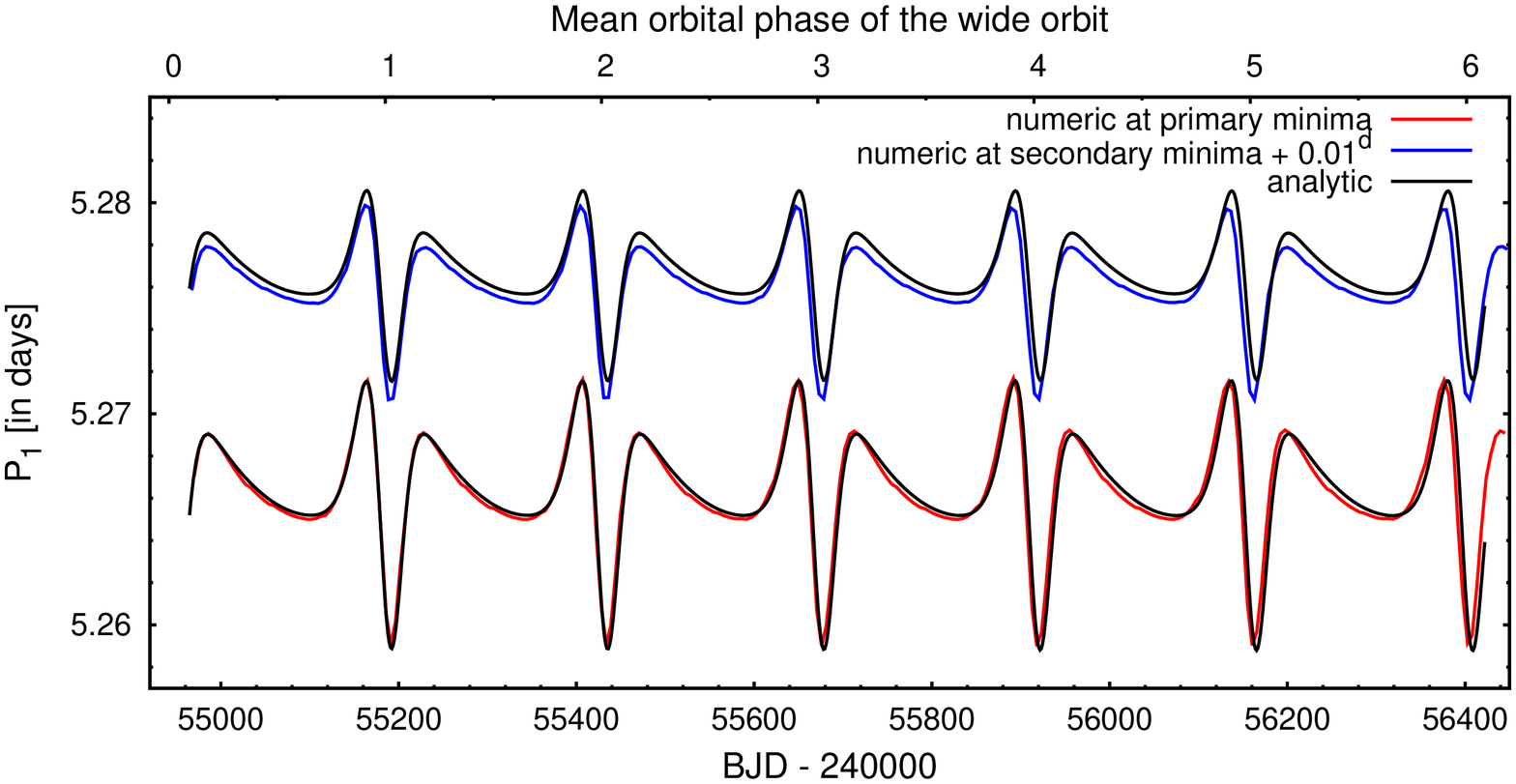}\includegraphics[width=86mm]{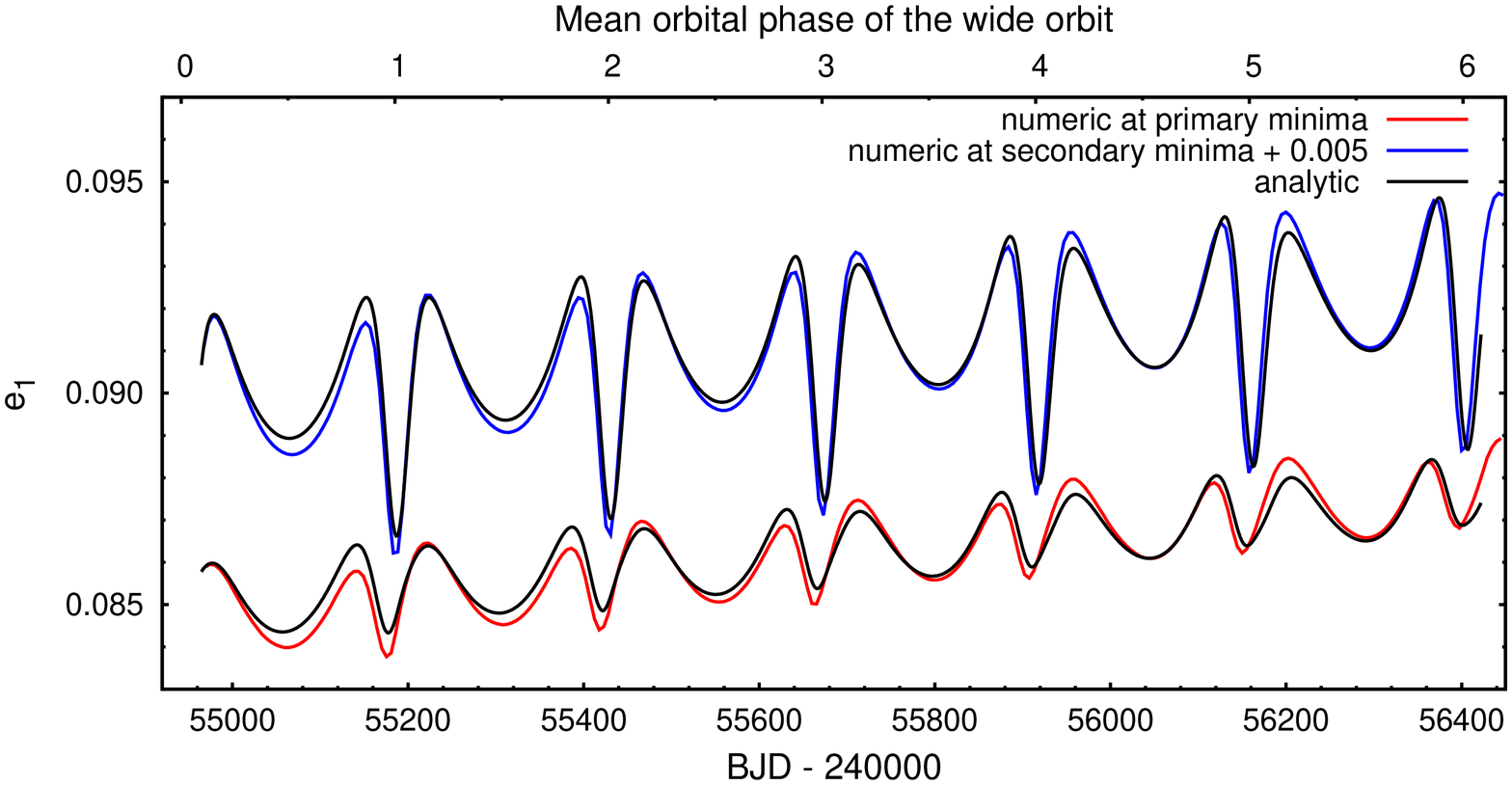}
\includegraphics[width=86mm]{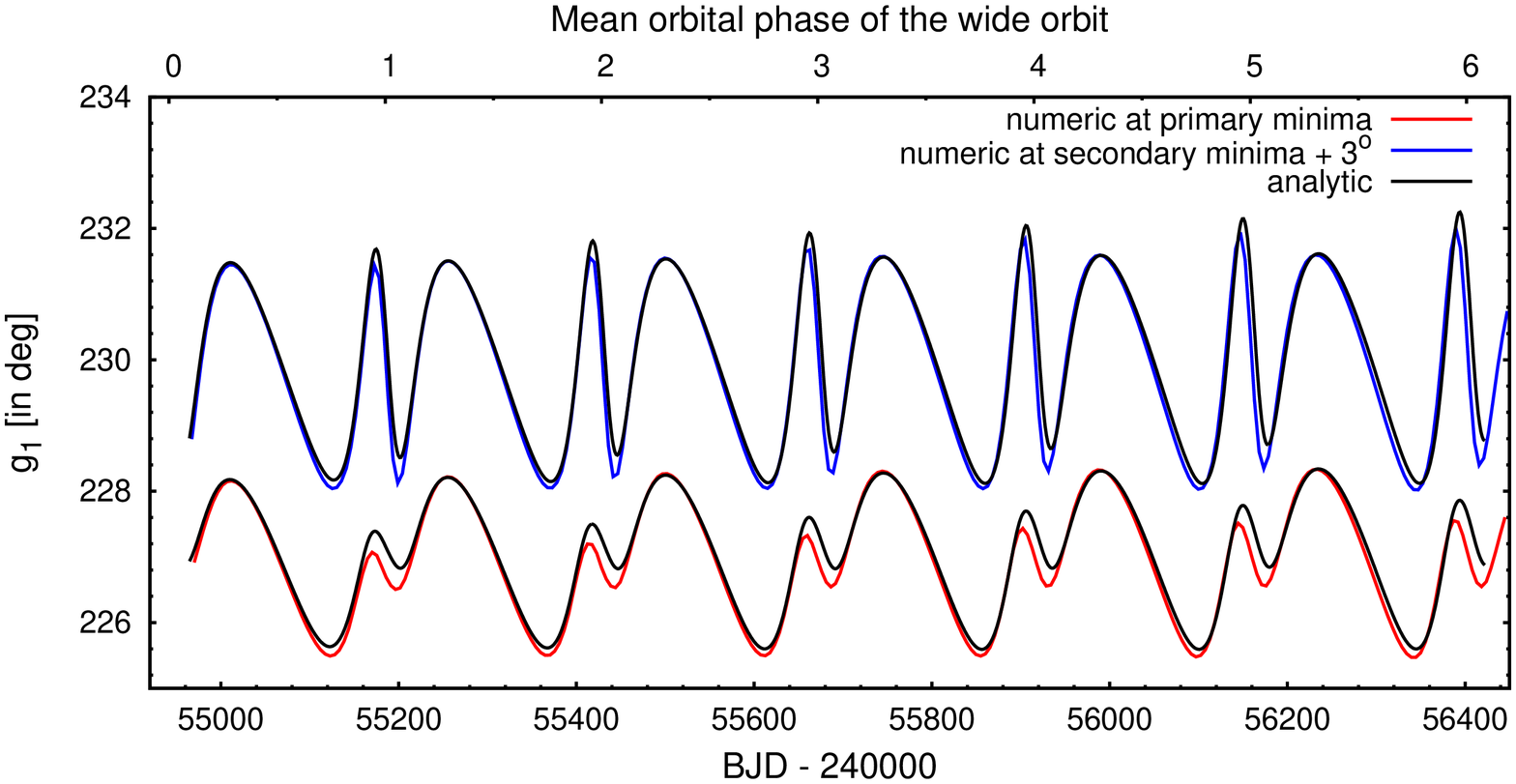}\includegraphics[width=86mm]{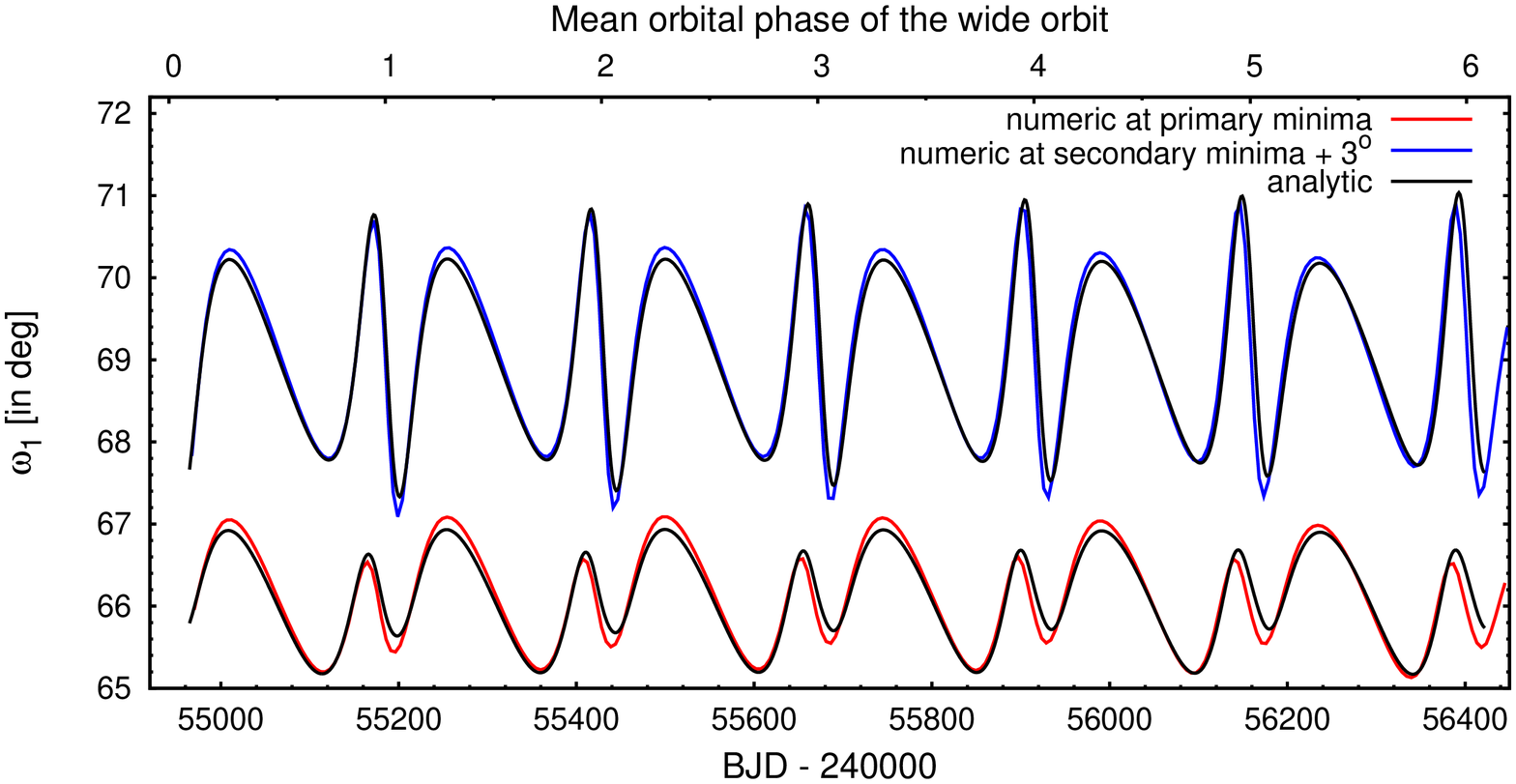}
 \caption{Analytically computed (black), and numerically integrated (red, blue) orbital element variations as a function of time during the OS3 solution, and the equivalent numerical integration for KIC~07289157. Red and blue curves show the osculating instantaneous orbital elements calculated from the integration steps closest to the primary (red) and secondary (blue) eclipses respectively, and demonstrates nicely the phase dependence of such sampling, which can be well modeled by the inclusion of the shortest timescale perturbations (see references in text). The individual panels from left to right, and top to bottom are as follows: {\it first row:} instantaneous period (which have short-term contribution exclusively, $P_1$); eccentricity ($e_1$); {\it second row:} the long- and short-term contribution of the dynamical argument of periastron ($g_1$); the same for the observable argument of periastron ($\omega_1$). }
%{\it third row:} the observable ($\omega_1$ -- red) and dynamical ($g_1$ -- magenta) arguments of periastron (where, in order to improve the appearance, the former were sampled at primary eclipses, and the latter at secondary eclipses, only); the observable inclinations ($i_1$ -- red and $i_2$ -- magenta); {\it fourth row:} the dynamical node ($h_1$); and the same, after subtracting its secular variation; {\it fifth row:} the observable node ($\Delta\Omega_1$); and the same, after subtracting its secular variation.}
 \label{Fig:K7289157orbelems1}
\end{figure*}

\begin{figure*}
\includegraphics[width=86mm]{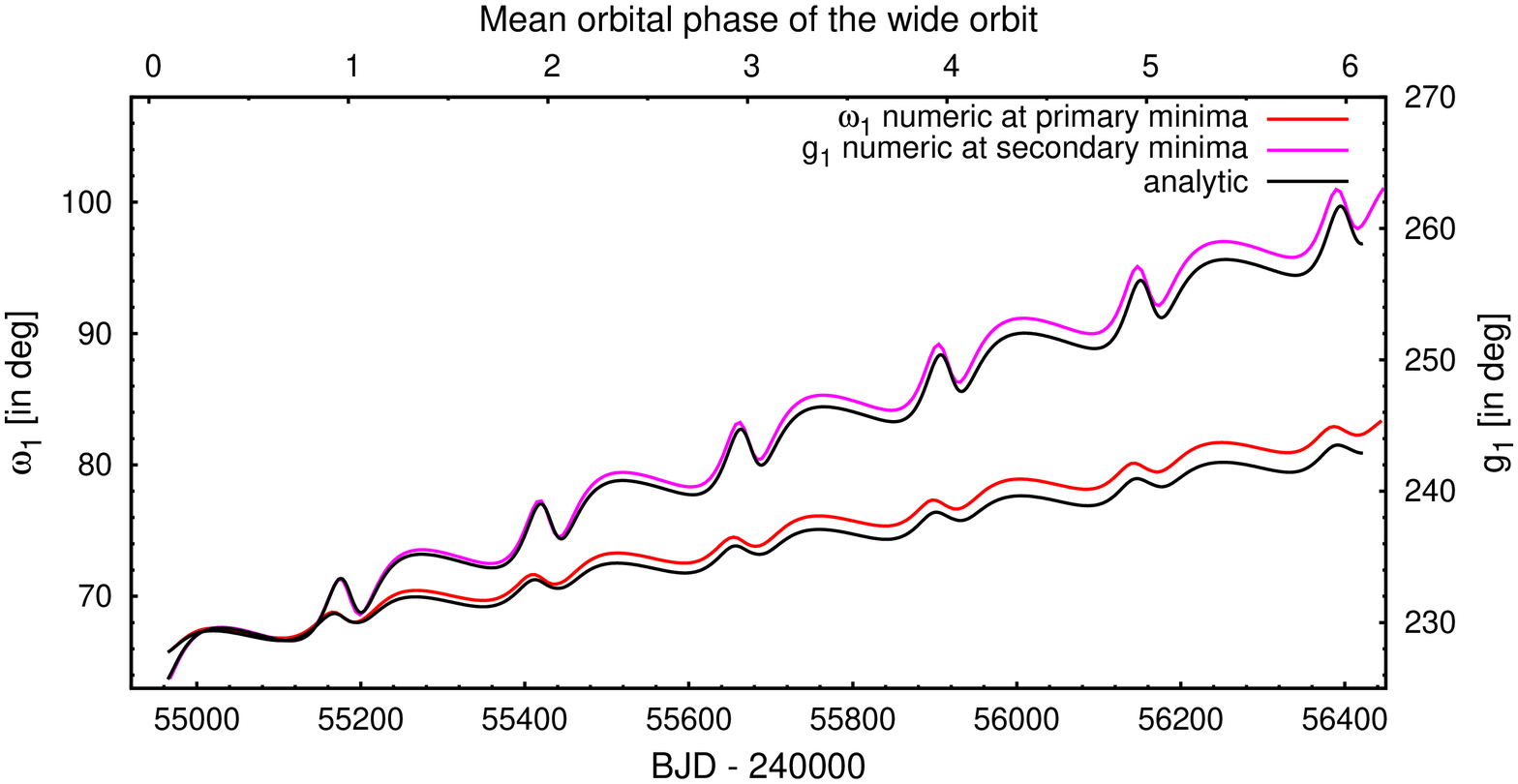}\includegraphics[width=86mm]{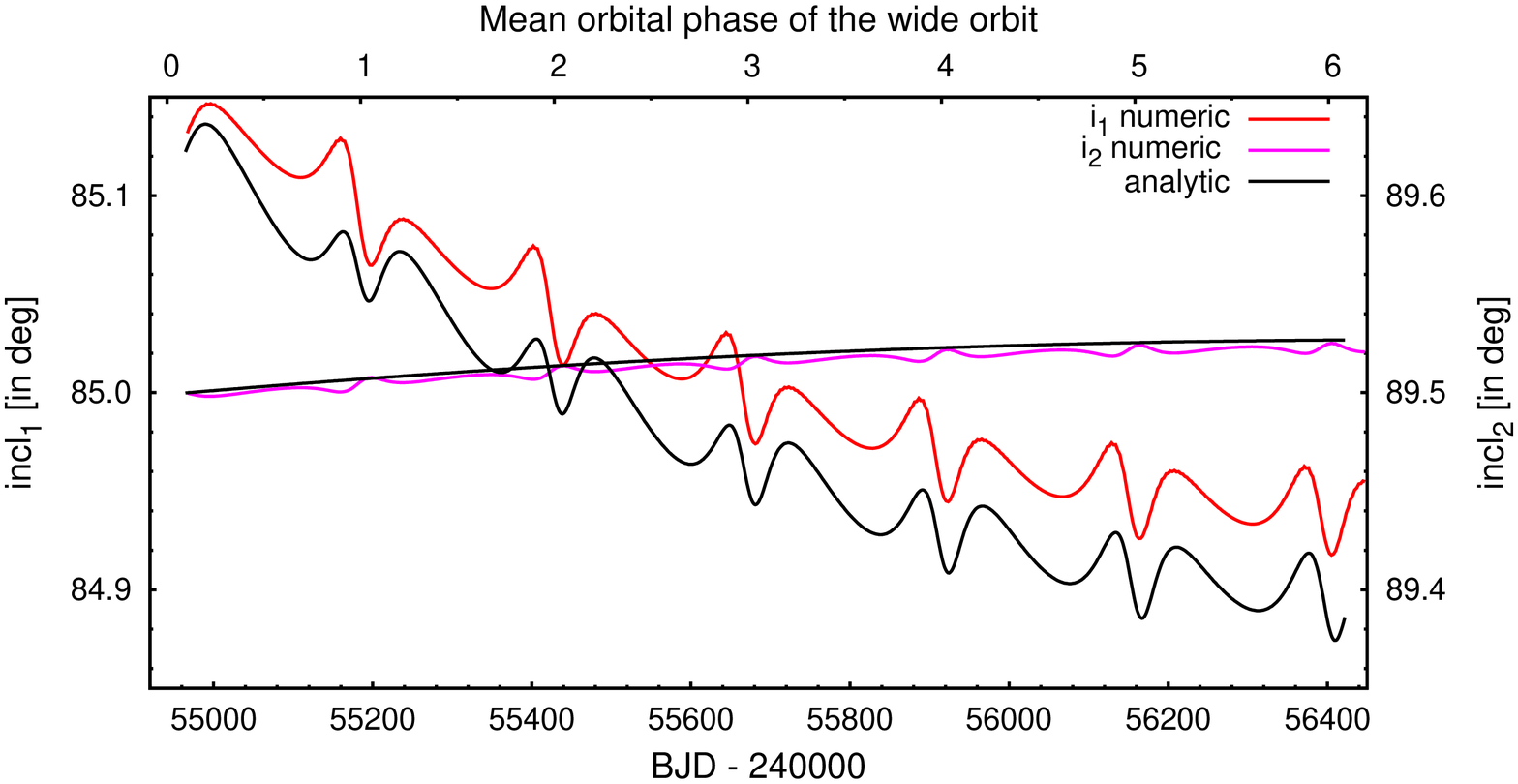}
\includegraphics[width=86mm]{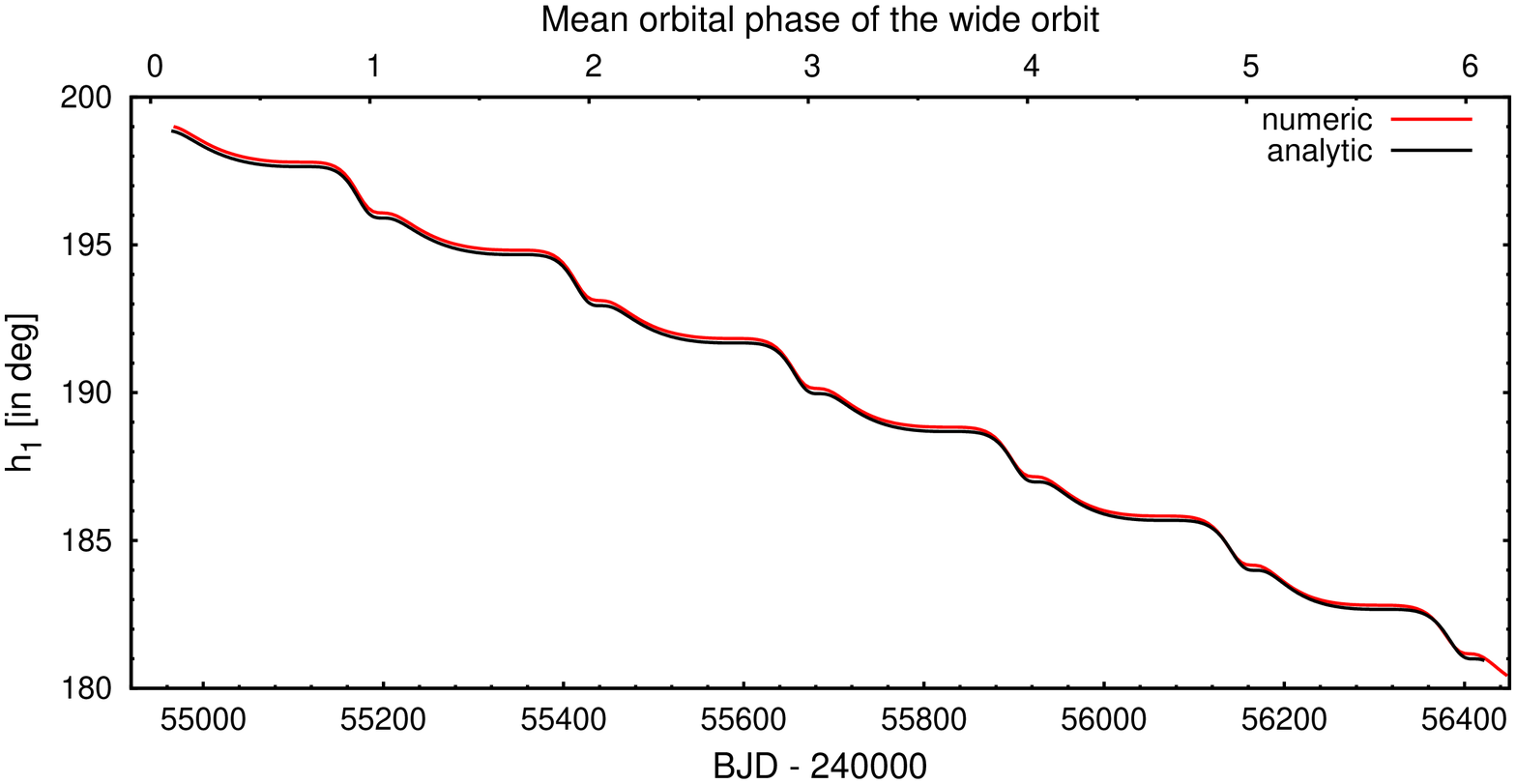}\includegraphics[width=86mm]{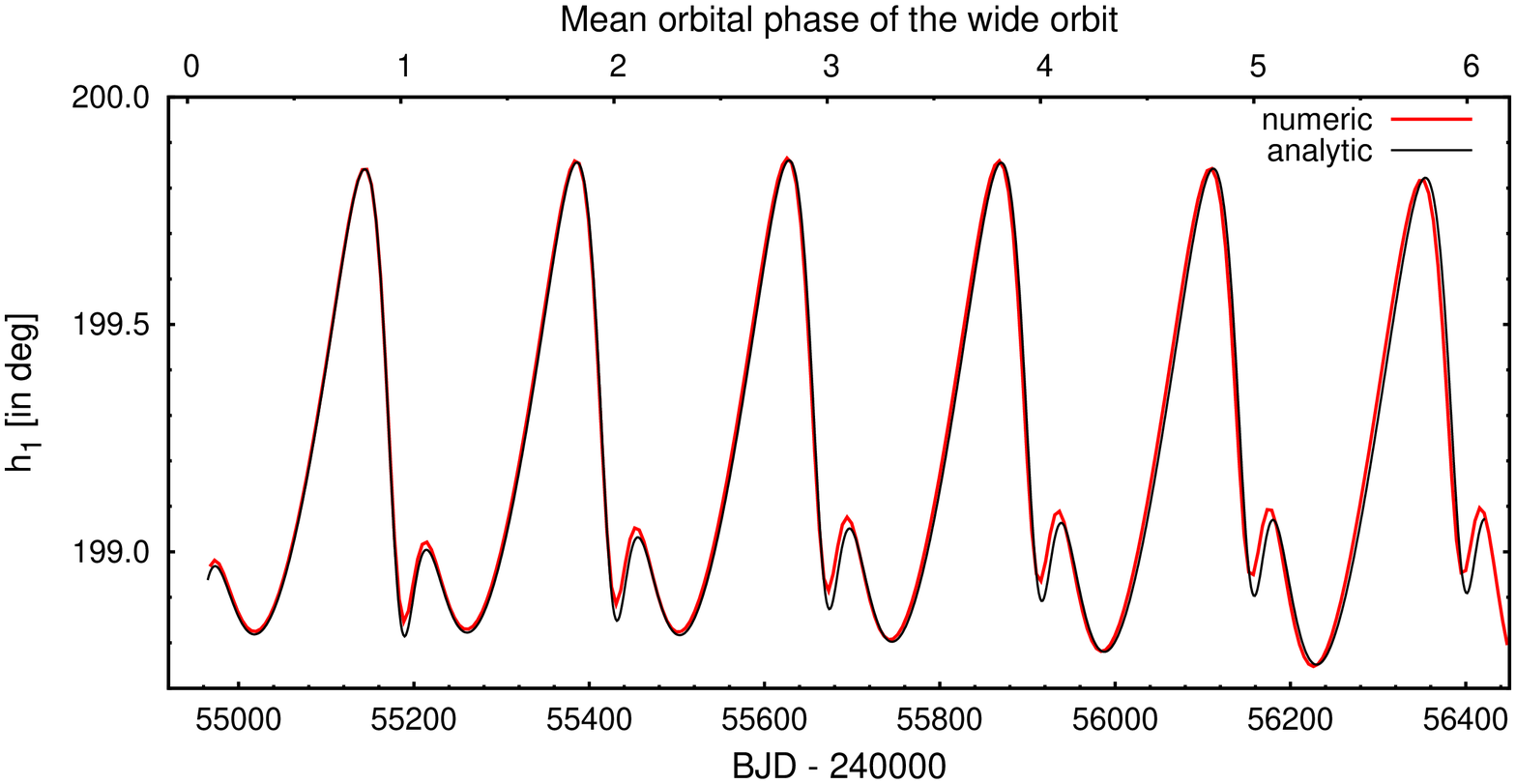}
\includegraphics[width=86mm]{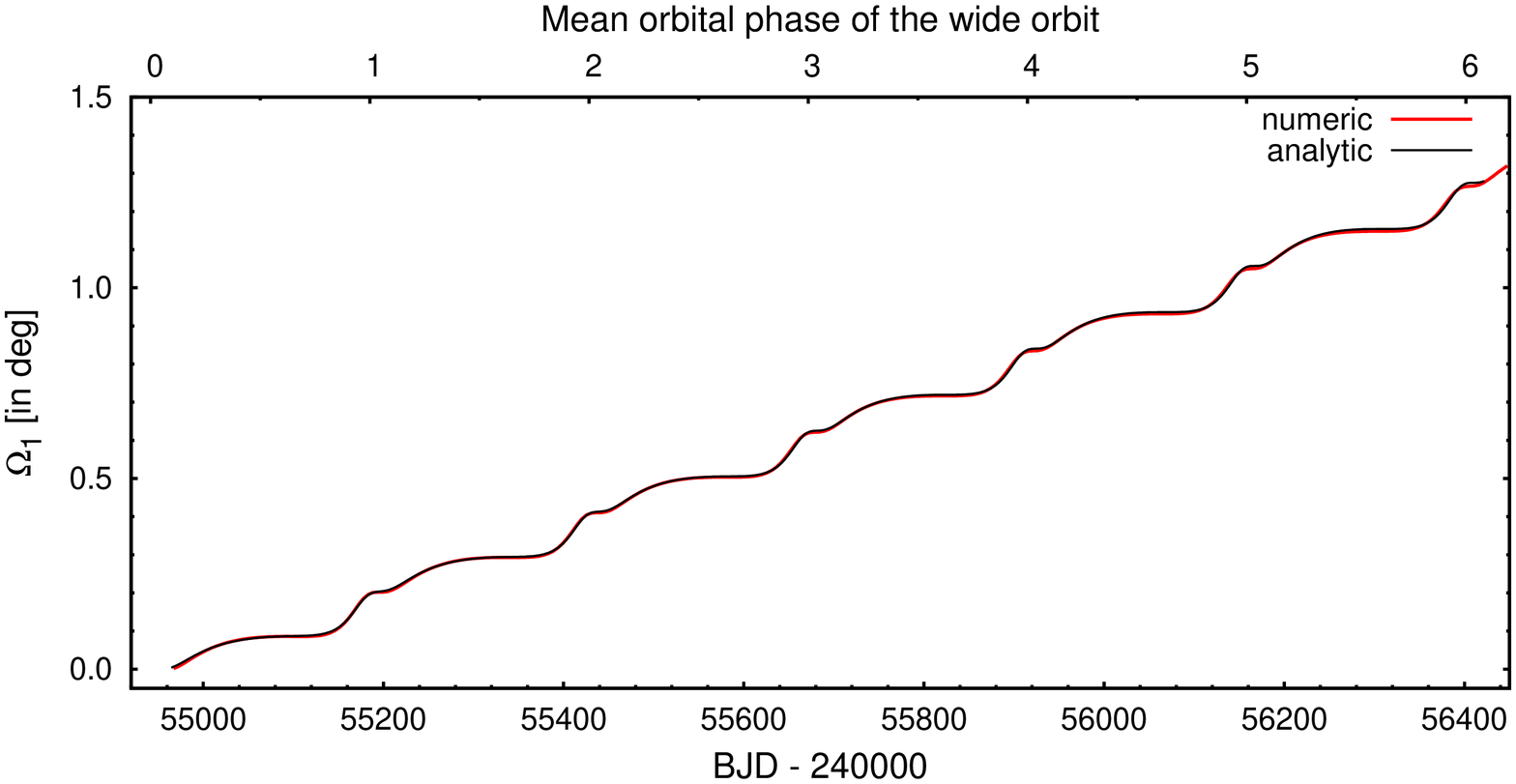}\includegraphics[width=86mm]{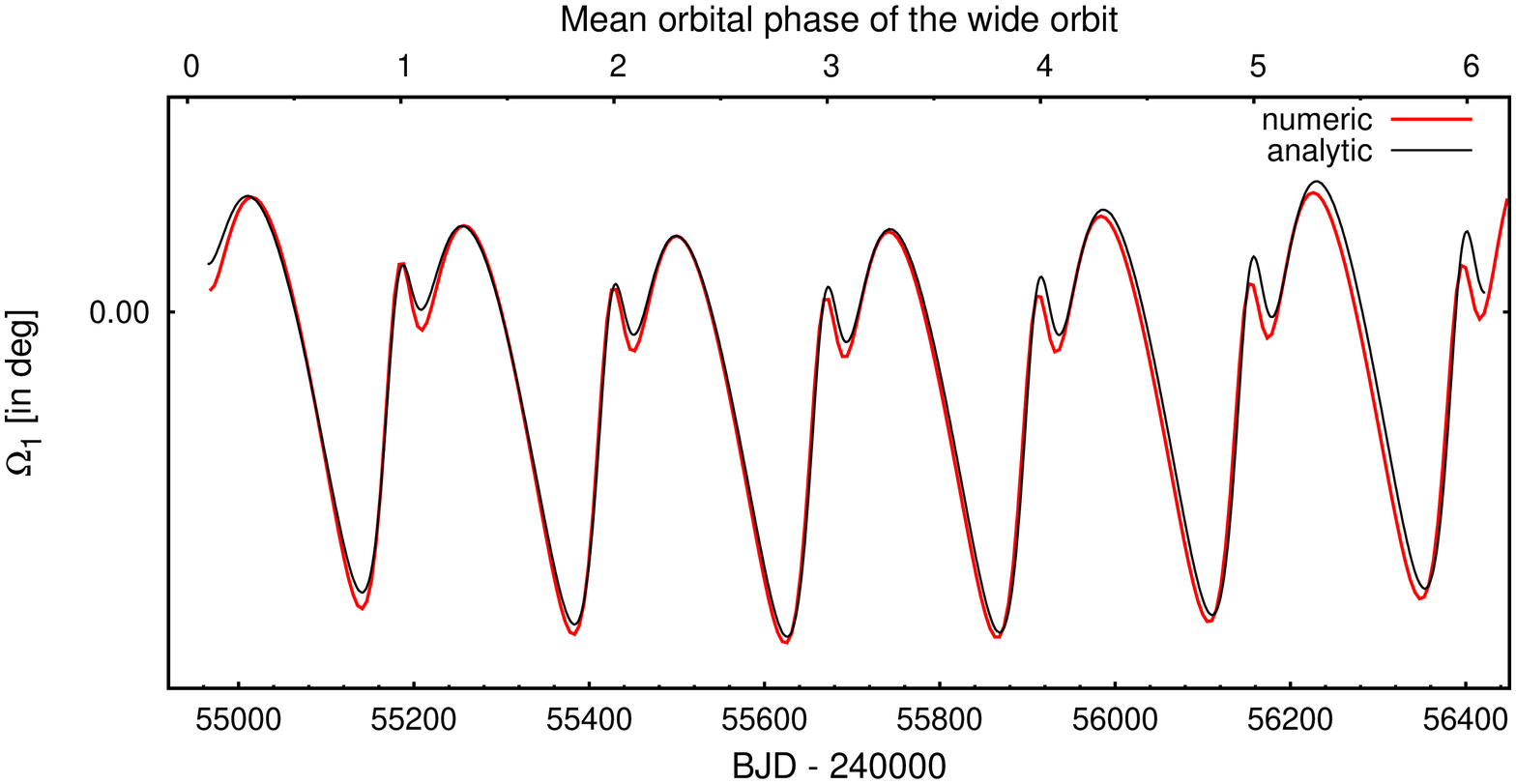}
 \caption{Continuation of Fig.~\ref{Fig:K7289157orbelems1}. The individual panels from left to right, and top to bottom are as follows: {\it first row:} the observable ($\omega_1$ -- red) and dynamical ($g_1$ -- magenta) arguments of periastron (where, in order to improve the appearance, the former were sampled at primary eclipses, and the latter at secondary eclipses, only); the observable inclinations ($i_1$ -- red and $i_2$ -- magenta); {\it second row:} the dynamical node ($h_1$); and the same, after subtracting its secular variation; {\it third row:} the observable node ($\Delta\Omega_1$); and the same, after subtracting its secular variation.}
 \label{Fig:K7289157orbelems2}
\end{figure*}

We now turn to the numerical realization of our model solution. In Fig.~\ref{Fig:K7289157numanalfits} we plot the observed ETV together with the fitted octupole (or second order) analytic solution.  For this particular fit, the third apsidal-motion model was used: `OS3'.\footnote{In naming the different solutions, the first letter, `O' or `N' refers to the observed vs.~numerically generated data series; the second letter, `F' or `S', denotes the first or second order (i.e., quadrupole or octupole) model solutions; while numbers `1'--`3' indicate the appropriate apsidal motion model that was applied. Where additional symbols were assigned to the name of a given solution, it is either self-explanatory or defined in the text.} 
% SR - why is the following necessary: ...and the numerically generated ETV was obtained from this solution. 
Some of the inner orbital elements calculated analytically during this process, as well as the same quantities obtained during the numerical integrations, are also plotted in Figs.~\ref{Fig:K7289157orbelems1},~\ref{Fig:K7289157orbelems2}. The numerically generated ETV curves show a somewhat shorter apsidal motion period which manifests itself in a bit faster convergence of the primary and secondary curves. Apart from this feature, it nicely matches the observed, and the analytical ETV as well. 

In the additional columns of Table~\ref{Tab:KIC7289157} we give analytic solutions, obtained with basically the same procedure, but using either different model approximations, or a different modification of the numerically generated data. Therefore, the Model \#N0F1 is a quadrupole (i.e., first-order model, using a constant dynamical apsidal advance rate, which was, however, unconstrained and was applied to the undistorted numerical curve (i.e., $\sigma=0$). Next, Model \#N0S3 was an octupole (second order) model, using the advanced apsidal motion modeling, and the same undistorted numerical data.  In the case of Model\#NsimS3 the same analytic model was applied, however, some simulated noise was added to the numerical data (a random scatter with $\sigma=0.001$ days, and also, 5\% of the data points were dropped out randomly).

In conclusion, the comparison of the different solutions reveals that our results for most of the parameters seem to be unique and robust, as was discussed in detail in Sect.~\ref{sec:parameterconstraints}.

\begin{figure*}
\includegraphics[width=84mm]{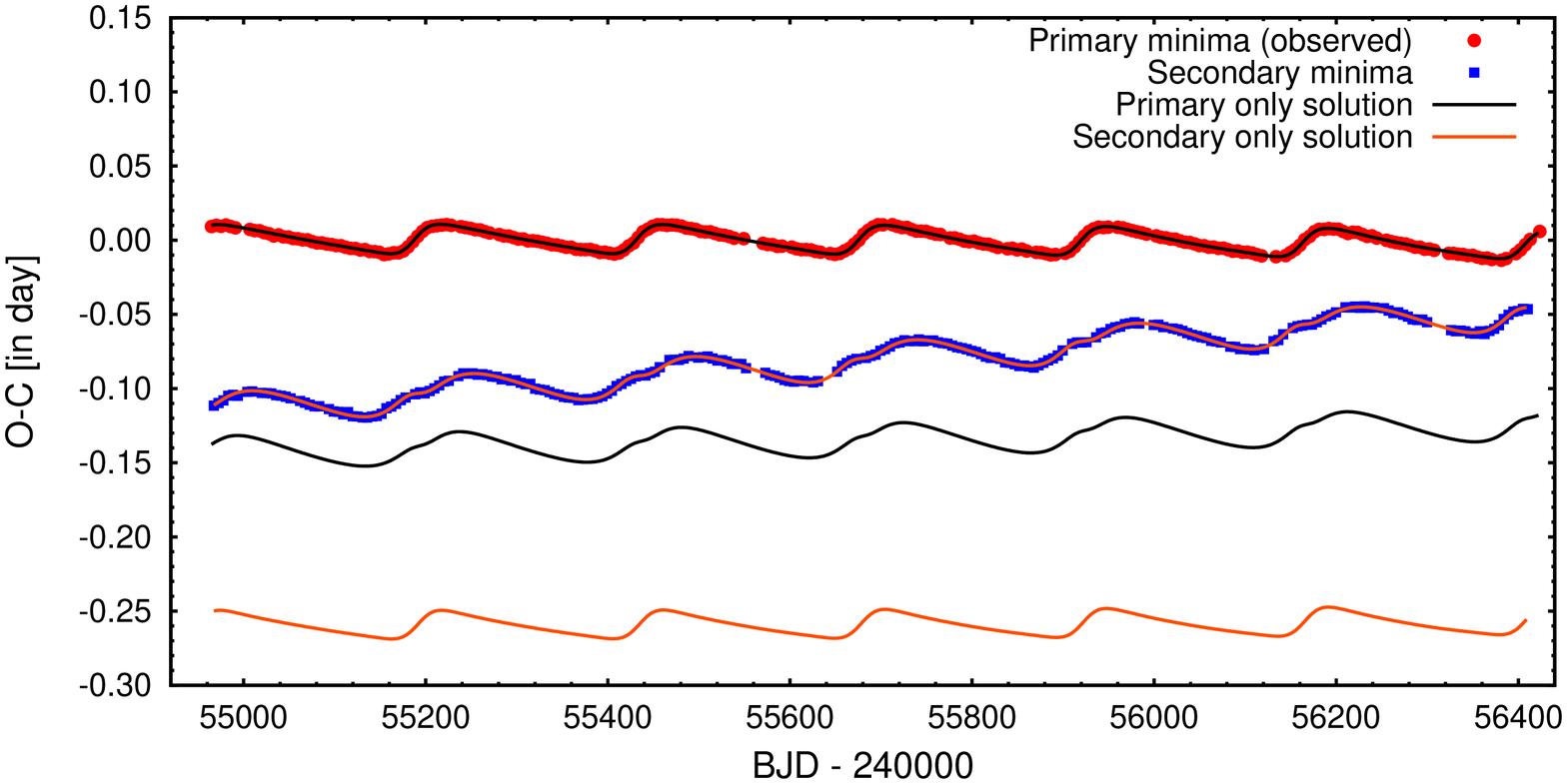}\includegraphics[width=84mm]{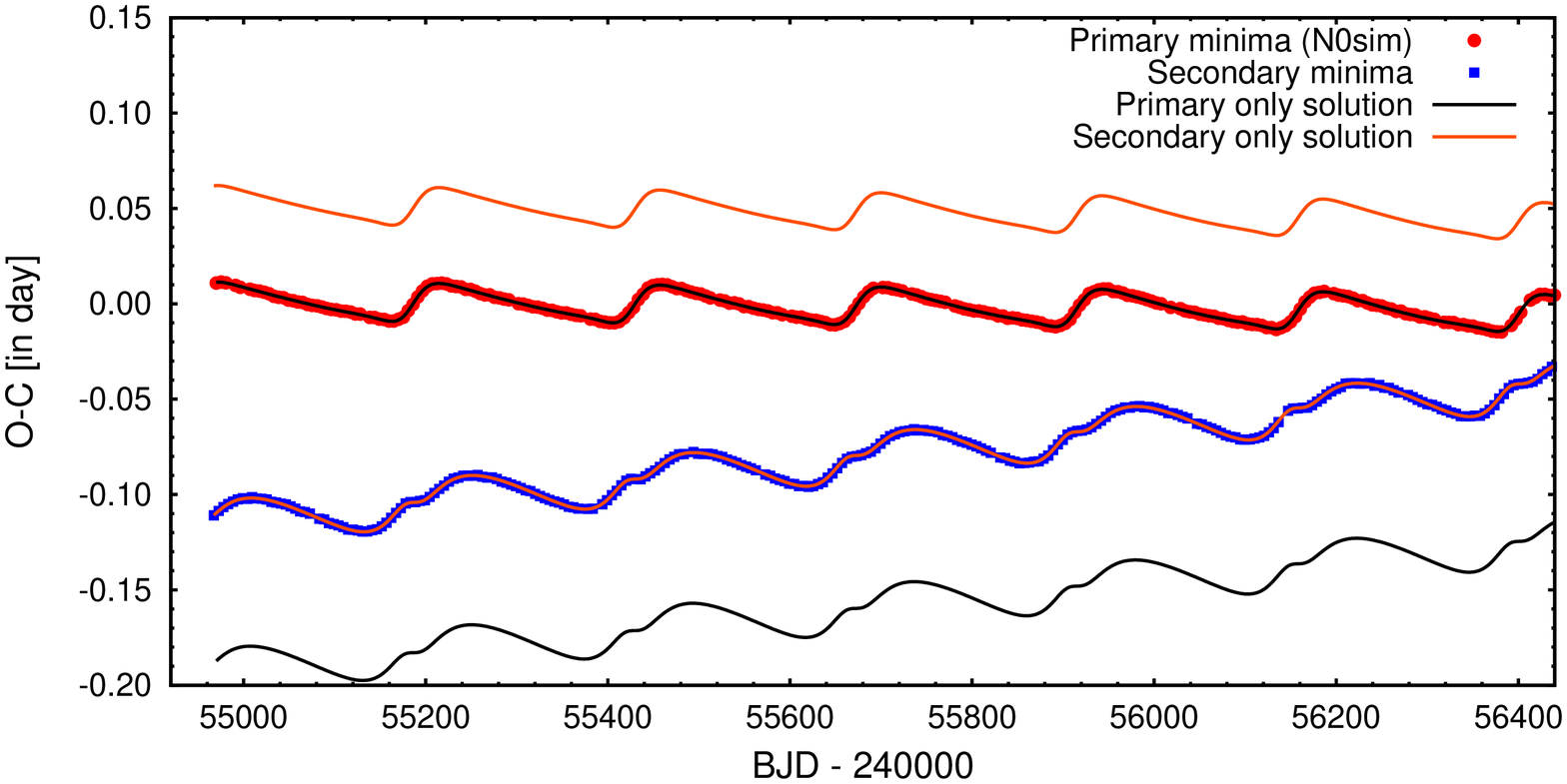}
 \caption{Separately fitted primary and secondary ETV solutions for the observed (left) and the numerically simulated (right) datasets. As one can see, when the primary ETV was used (OS3-pri, NsimS3-pri solutions), the solution fits the primary ETV curves well, but fails to fit the secondary's ETV curve, and vice versa. See text for further details.}
 \label{Fig:K7289157ETVprisecsep}
\end{figure*}

Finally, we performed an additional test to check what happens if only either the primary, or the secondary, ETV curves would be fitted. Therefore, similar adjustment processes were applied individually to both the primary and secondary eclipses of both the observed, and the numerically generated data with random added noise. (Note that, in this case, with the absence of any a priori information on apsidal motion, we could not constrain the initial value of $\omega_1$, and thus it was initialized with four, evenly spaced values.) The runs were carried out with all apsidal motion modes. The results of the apse mode 3 runs are listed in Table~\ref{Tab:KIC7289157b}, and also plotted in Fig.~\ref{Fig:K7289157ETVprisecsep}. As one can see, these ``results'' differ significantly from each other, and also from those which were previously obtained by simultaneously fitting both the primary and secondary ETV curves. 

\begin{table*}  
\begin{center}
\caption{Different model solutions for KIC 07289157, when primary and secondary minima were fitted separately.} 
\label{Tab:KIC7289157b}  
\begin{tabular}{lll|lll} 
\hline 
\multicolumn{2}{l}{Parameters} &OS3-pri&OS3-sec& NsimS3-pri& NsimS3-sec \\
\hline
$P_1$      & (d)       & 5.267728  & 5.267376  & 5.263624  & 5.263611  \\
$e_1$      &           & 0.0438    & 0.0902    & 0.1004    & 0.1016    \\
$\omega_1$ & ($\degr$) & 11.600    & 118.514   & 54.614    & 60.750    \\
$\tau_1$   & (MBJD)    & 54971.391 & 54967.577 & 54972.000 & 54972.155 \\
\hline
$P_2$      & (d)       & 243.303   & 243.111   & 242.495   & 242.424   \\
$a_2$      &(R$_\odot$)& 206.290   & 219.156   & 247.960   & 243.829   \\
$e_2$      &           & 0.329     & 0.304     & 0.321     & 0.324     \\
$\omega_2$ & ($\degr$) & 331.275   & 189.064   & 149.997   & 152.855   \\
$\tau_2$   & (MBJD)    & 54948.553 & 54946.823 & 54940.403 & 54941.209 \\
\hline
$\im$ & ($\degr$)          &  2.790 &  2.630   &  8.384 &  2.000 \\
$i_1$ & ($\degr$)          & 91.891 & 91.883   & 82.858 & 91.379 \\
$i_2$ & ($\degr$)          & 89.500 & 89.500   & 89.500 & 89.500 \\
$n_1$ & ($\degr$)          & 31.017 & 25.013   & 37.824 & 20.005 \\
$n_2$ & ($\degr$)          & 31.000 & 25.000   & 37.482 & 20.000 \\
$\Delta\Omega$ & ($\degr$) &$-1.437$&$-1.112$  &  5.130 & $-0.684$\\
\hline
$m_\mathrm{C}/m_\mathrm{ABC}$&& 0.582 & 0.388  & 0.371 & 0.361 \\
$m_\mathrm{B}/m_\mathrm{A}$ & & 0.430 & 0.460  & 0.430 & 0.470 \\
$m_\mathrm{A}$ & (M$_\odot$)  & 0.583 & 1.003  & 1.532 & 1.441 \\
$m_\mathrm{B}$ & (M$_\odot$)  & 0.251 & 0.461  & 0.659 & 0.677 \\
$m_\mathrm{C}$ & (M$_\odot$)  & 1.159 & 0.928  & 1.292 & 1.196 \\
\hline
$P_{\omega_1}$         & (y) &  59.48 & 92.57  & 94.01 & 96.36 \\
$P_{h}$                & (y) &  56.13 & 78.94  & 84.26 & 83.21  \\
\hline
$\chi^2$ &                   & 0.9154 & 1.5705& 0.3103 & 0.3095\\
\hline
\end{tabular} 
\end{center}  
\end{table*}

\subsection{The medium mutual inclination regime: KIC~08023317 and the retrograde KIC~07670617}

\begin{figure*}
\includegraphics[width=84mm]{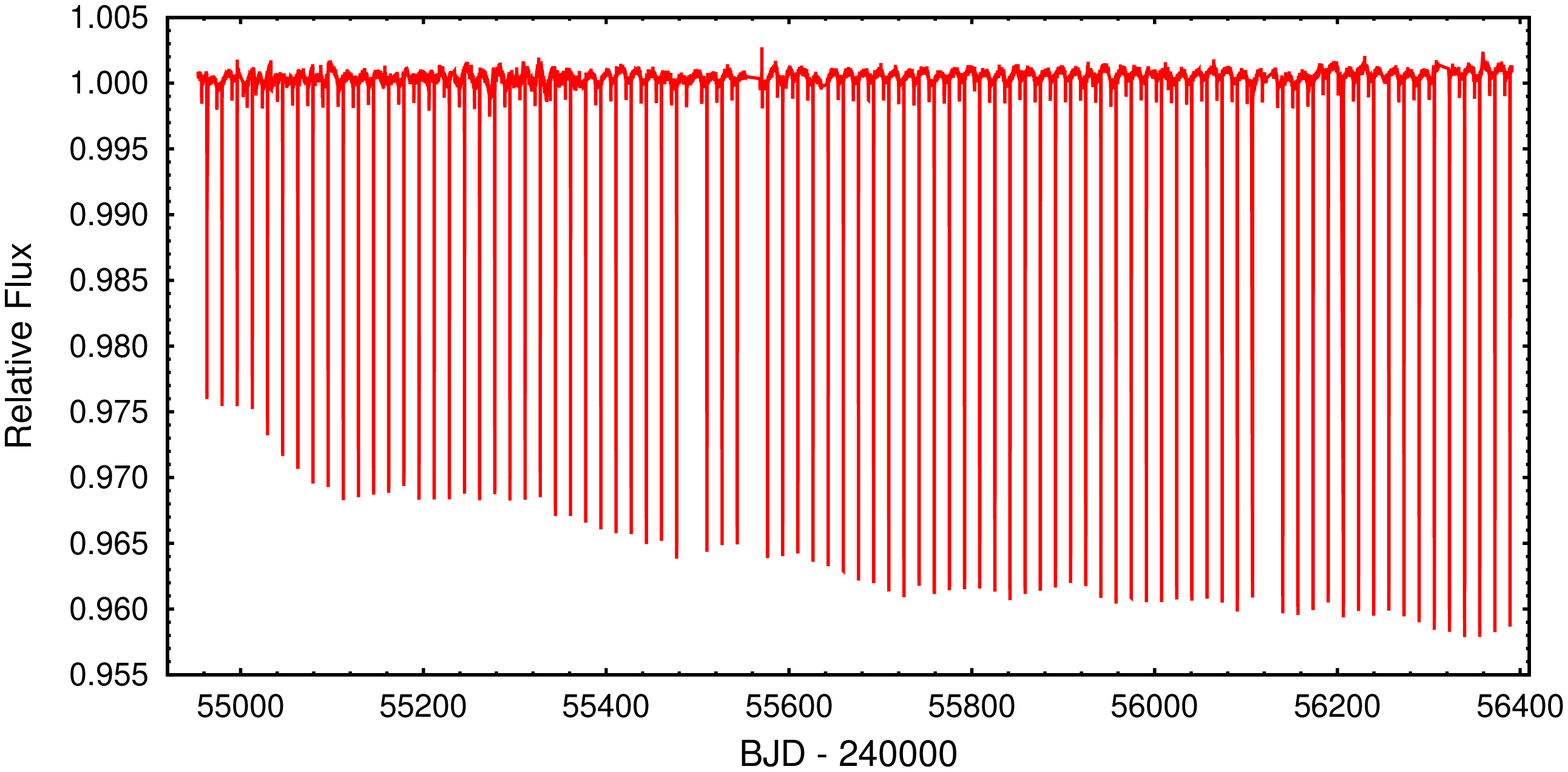}\includegraphics[width=84mm]{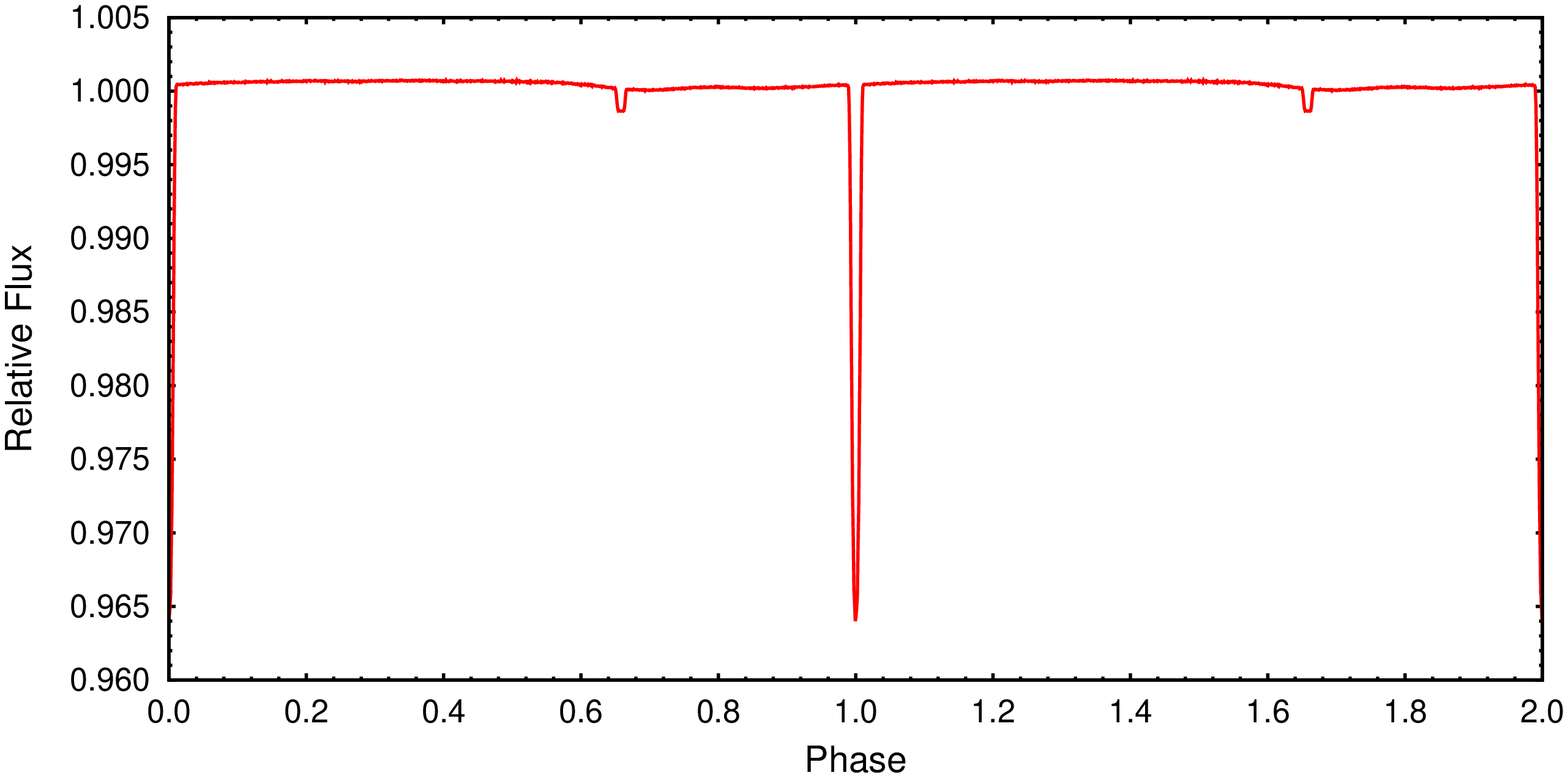}
 \caption{The Q0--Q17 long-cadence light curve of KIC~08023317 (left), and the phased light curve (right).}
 \label{Fig:K8023317lc}
\end{figure*}

KIC~08023317 exhibits low-amplitude, but total eclipses of the primary, and remarkably shallower and displaced secondary eclipses. During the 4-years of {\em Kepler} observations, the eclipse depth grew continuously (see Fig.\,\ref{Fig:K8023317lc}). Due to the shallowness and flat bottom of the secondary occultations, the times of secondary eclipses were determined with significantly lower accuracy. Therefore, for the case of fitting runs using equal, global ETV uncertainties, which are presented here, we set the value of $\sigma$ for the secondary ETV curve to ten times that of $\sigma$ for the primary curve. Our procedure was similar to that in the previous case, therefore, here we mainly concentrate only on the differences. We kept $i_1$ fixed at $89\degr$ due to the implications of total eclipses. Therefore, in this case $\im$, $\uvm$ and $i_1$ determined the spatial configuration of the triple. The total mass of the system was also initially constrained according to the parameters found in the KIC catalog. In the first three columns of Table\,\ref{Tab:KIC8023317} we present three different analytic fits which were obtained with different apsidal motion models, i.e., $\Delta\omega$ was considered to be constant and unconstrained (OS1); constant, but constrained (OS2); analytically computed from the first-order secular model (OS3), respectively. Note, in this latter model the program is also able to compute the secular variation of $e_1$ from the same analytical model, but after some preliminary checks, we disabled this option. For the first runs, the octupole models led to an unexpectedly high mass ratio ($q_1$) for the inner binary. However, we found, that this might be a consequence of the larger timing uncertainties for the secondary ETV curve. We made runs with different ratios of the relative global uncertainty for the secondary and primary ETV curves, and concluded that a smaller weight (i.e., larger uncertainty) for the secondary curve with respect to that of the primary resulted in a lower inner mass ratio. For equal uncertainties we found $q_1\approx4$. Therefore, for the final runs the mass ratio was arbitrarily fixed at $q_1=0.5$. 

\begin{figure*}
\includegraphics[width=86mm]{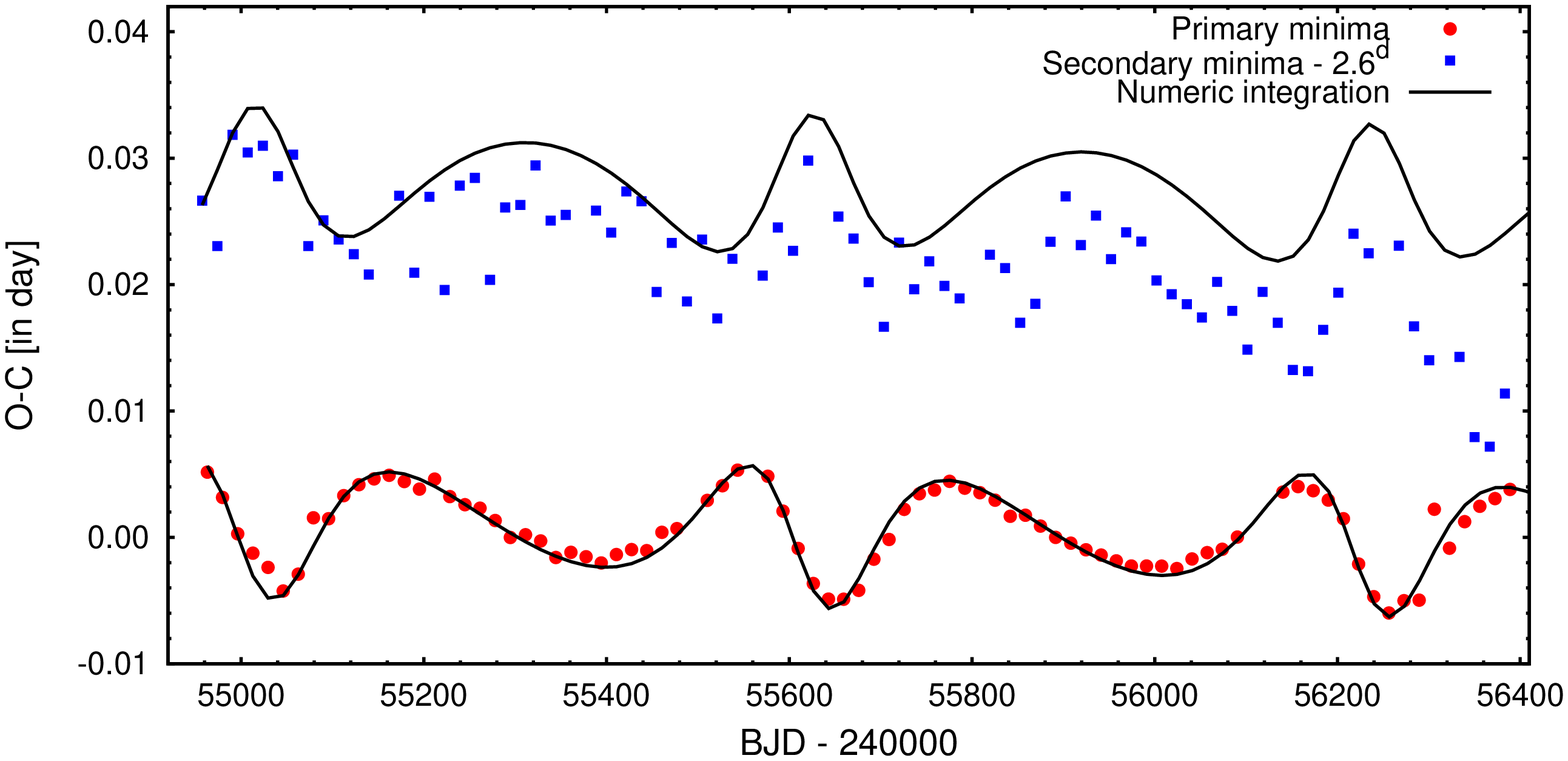}\includegraphics[width=86mm]{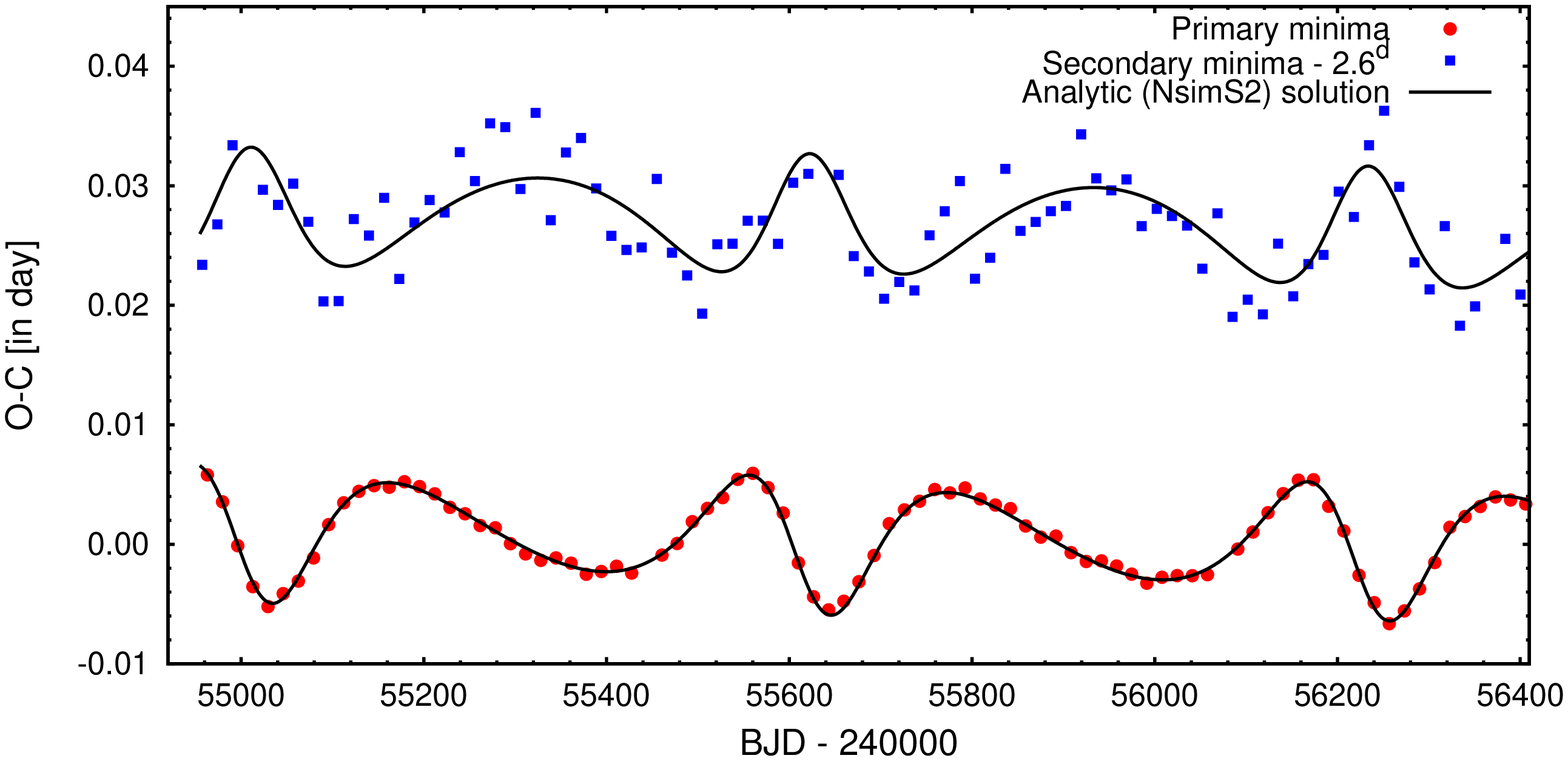}
 \caption{{\it Left panel: } The ETV curves of KIC 08023317 (filled red circles and blue boxes for primary and secondary eclipses, respectively) together with the numerically generated ETV curve for which the initial parameters were taken from  the analytical Model\#OS2 solution (see in Tab.\,\ref{Tab:KIC8023317}) (black lines). {\it Right panel: } The same, numerically generated curve with random timing noise added as the ``observed'' curve to be fitted (red circles and blue boxes), and the analytic ``NsimS2'' solution (black lines). }
 \label{Fig:K8023317numanalfits}
\end{figure*}

\begin{figure*}
\includegraphics[width=86mm]{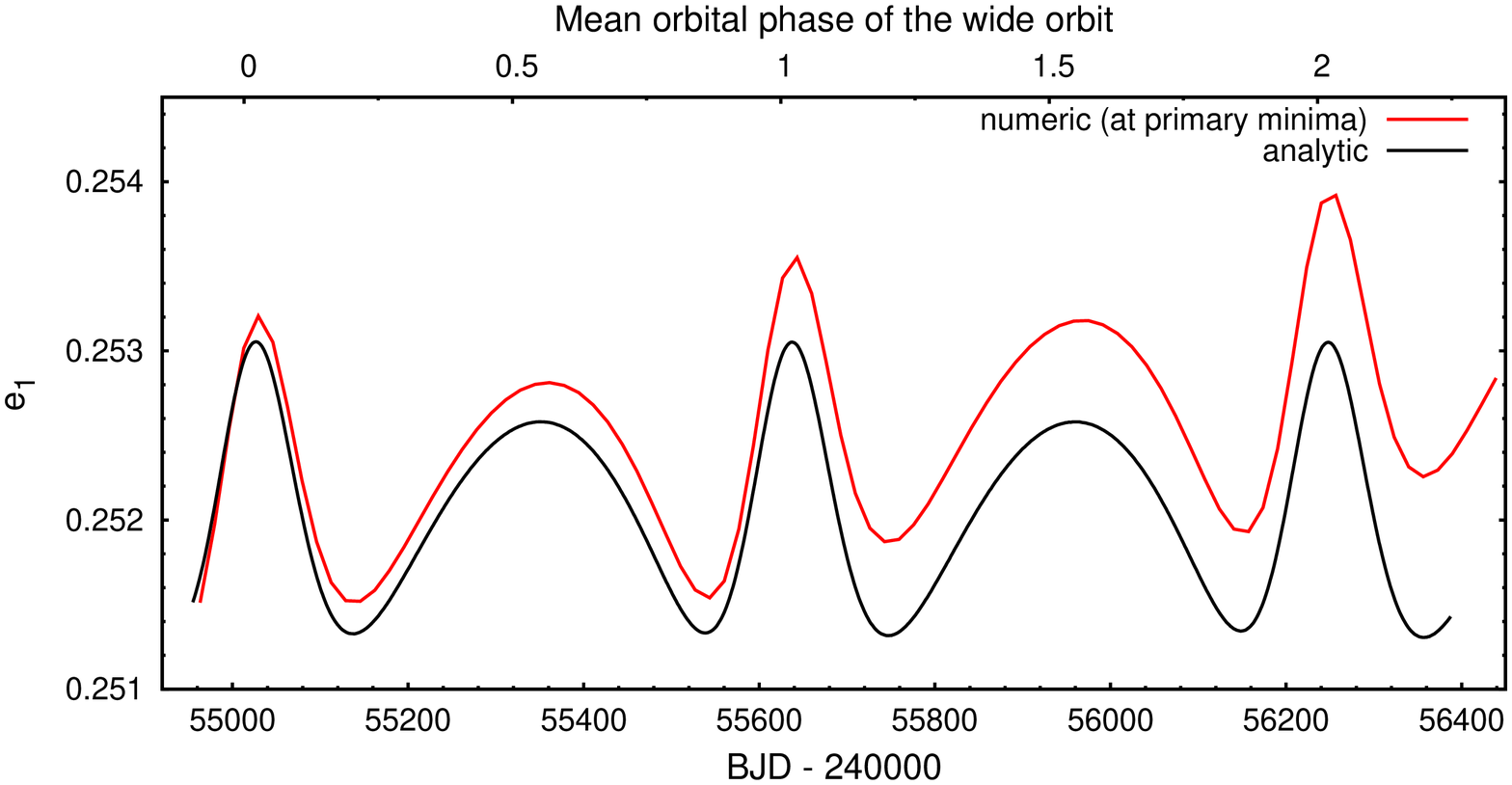}\includegraphics[width=86mm]{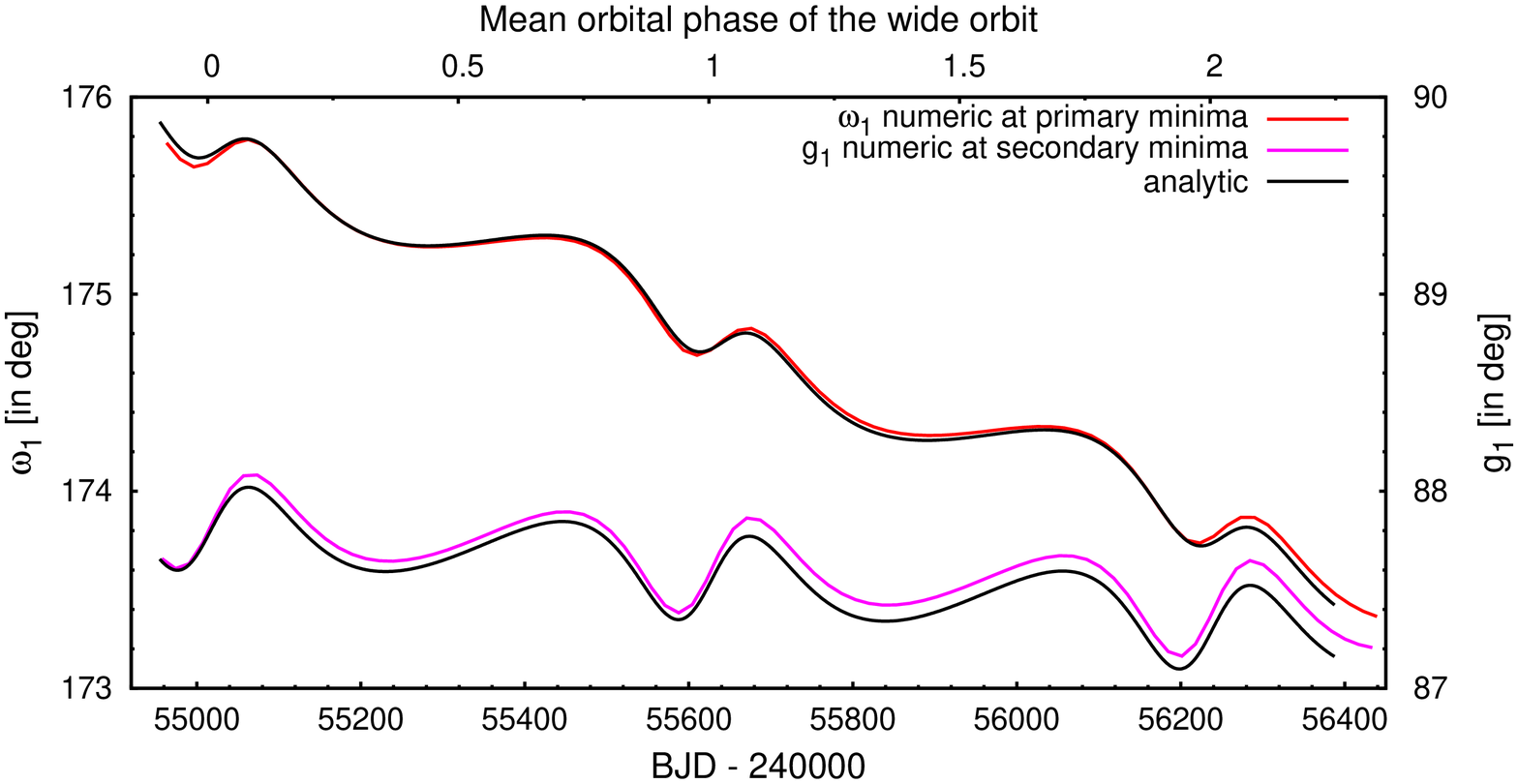}
 \caption{Analytically computed (black), and numerically integrated (red and magenta) orbital element variations for KIC 08023317 during the OS2 solution, and the corresponding numerical integration (NOS2) for KIC 08023317. Note the quick secular change in the eccentricity (left), and also the retrograde observable apsidal motion (right).}
 \label{Fig:K8023317excomg}
\end{figure*}

In a further analysis, the parameters obtained from solution OS2 were chosen as initial parameters for the numerical 3-body integration. In the left panel of Fig.\,\ref{Fig:K8023317numanalfits} the numerical output is plotted against the observed ETV curve. As one can see, while the individual numerical curves fit the corresponding observed curves quite well, the change in their relative displacement, which is a measure of the apsidal motion, varies much more slowly during the numerical integration. The reason can be readily seen in Fig.~\ref{Fig:K8023317excomg}. Although the linear, unconstrained apsidal motion model describes the observable retrograde(!) apsidal motion rate ($\Delta\omega_1$) very well, it misses correctly modeling the secular rapid eccentricity variation which would also be necessary for an accurate description of the net relative displacement variations between the primary -- secondary ETV curves. Fortunately, the additional fits to the numerically generated curves demonstrate clearly that the derived system parameters have only a minor sensitivity to this effect. The N0S2 column represents the model fit for the numerical ETV curve without added random errors, while the last column lists the solution for another numerical curve with simulated scatter with $\sigma_\mathrm{pri}=0.001$\,d and $\sigma_\mathrm{sec}=0.01$\,d. As one can see, apart from the mass parameters, the other quantities remain within a few percent uncertainty, independent of the applied apsidal motion model.

\begin{table*}  
\begin{center}
\caption{Different model solutions for KIC 08023317.} 
\label{Tab:KIC8023317}  
\begin{tabular}{lllll|ll} 
\hline 
\multicolumn{2}{l}{Parameters} &OS1 &    OS2    &    OS3    &   N0S2    &NsimS2 \\
\hline
$P_1$      & (d)       & 16.57770   & 16.57781  & 16.57903  & 16.58343  & 16.58343 \\
$e_1$      &           & 0.2519     & 0.2521    & 0.2558    & 0.2516    & 0.2515   \\
$\omega_1$ & ($\degr$) & 176.465    & 175.836   & 169.046   & 180.909   & 180.777  \\
$\tau_1$   & (MBJD)    & 54976.758  & 54976.732 & 54976.448 & 54976.951 & 54976.945 \\
\hline
$P_2$      & (d)       & 612.051    & 611.933   & 611.367   & 611.525   & 612.188 \\
$a_2$      &(R$_\odot$)& 347.543    & 351.101   & 363.151   & 357.239   & 322.756 \\
$e_2$      &           & 0.2472     & 0.2476    & 0.2504    & 0.2562    & 0.2649   \\
$\omega_2$ & ($\degr$) & 164.132    & 164.043   & 161.016   & 166.552   & 162.549 \\
$\tau_2$   & (MBJD)    & 55010.827  & 55011.182 & 55013.239 & 55010.385 & 55005.816\\
\hline
$\im$          & ($\degr$) & 51.962 &  52.109   &  52.258   &  51.015   & 51.326  \\
$i_1$          & ($\degr$) & 89.000 &  89.000   &  89.000   &  89.000   & 89.000  \\
$i_2$          & ($\degr$) & 88.303 &  87.788   &  84.197   &  90.324   & 89.007  \\
$n_1$          & ($\degr$) & 88.627 &  87.975   &  83.432   &  91.227   & 89.529  \\
$n_2$          & ($\degr$) & 89.942 &  89.545   &  86.762   &  91.549   & 90.486  \\
$\Delta\Omega$ & ($\degr$)&$-51.973$& $-52.118$ & $-52.151$ & $-51.000$ &$-51.334$ \\
\hline
$m_\mathrm{C}/m_\mathrm{ABC}$&&0.086&  0.086    &  0.083    &  0.095    &  0.093  \\
$m_\mathrm{B}/m_\mathrm{A}$ & &0.500&  0.500    &  0.500    &  0.570    &  0.580  \\
$m_\mathrm{A}$ & (M$_\odot$)  &0.917&  0.947    &  1.052    &  0.944    &  0.692  \\
$m_\mathrm{B}$ & (M$_\odot$)  &0.458&  0.473    &  0.526    &  0.538    &  0.401  \\
$m_\mathrm{C}$ & (M$_\odot$)  &0.130&  0.133    &  0.143    &  0.156    &  0.112  \\
\hline
$P_{\omega_1}$         & (y)&$-555.286$&$-609.64$&$-22480.12$&$-569.58$ & $-572.24$ \\
$P_{h}$                & (y) &699.42&  705.30   &  809.75   &  628.56   &  637.45 \\
\hline
$\chi^2$               &     &0.9494&  0.9498   &   0.9576  &  0.0016   &  0.3657 \\
\hline
\end{tabular} 
\end{center}  
\end{table*}

\begin{figure}
\includegraphics[width=84mm]{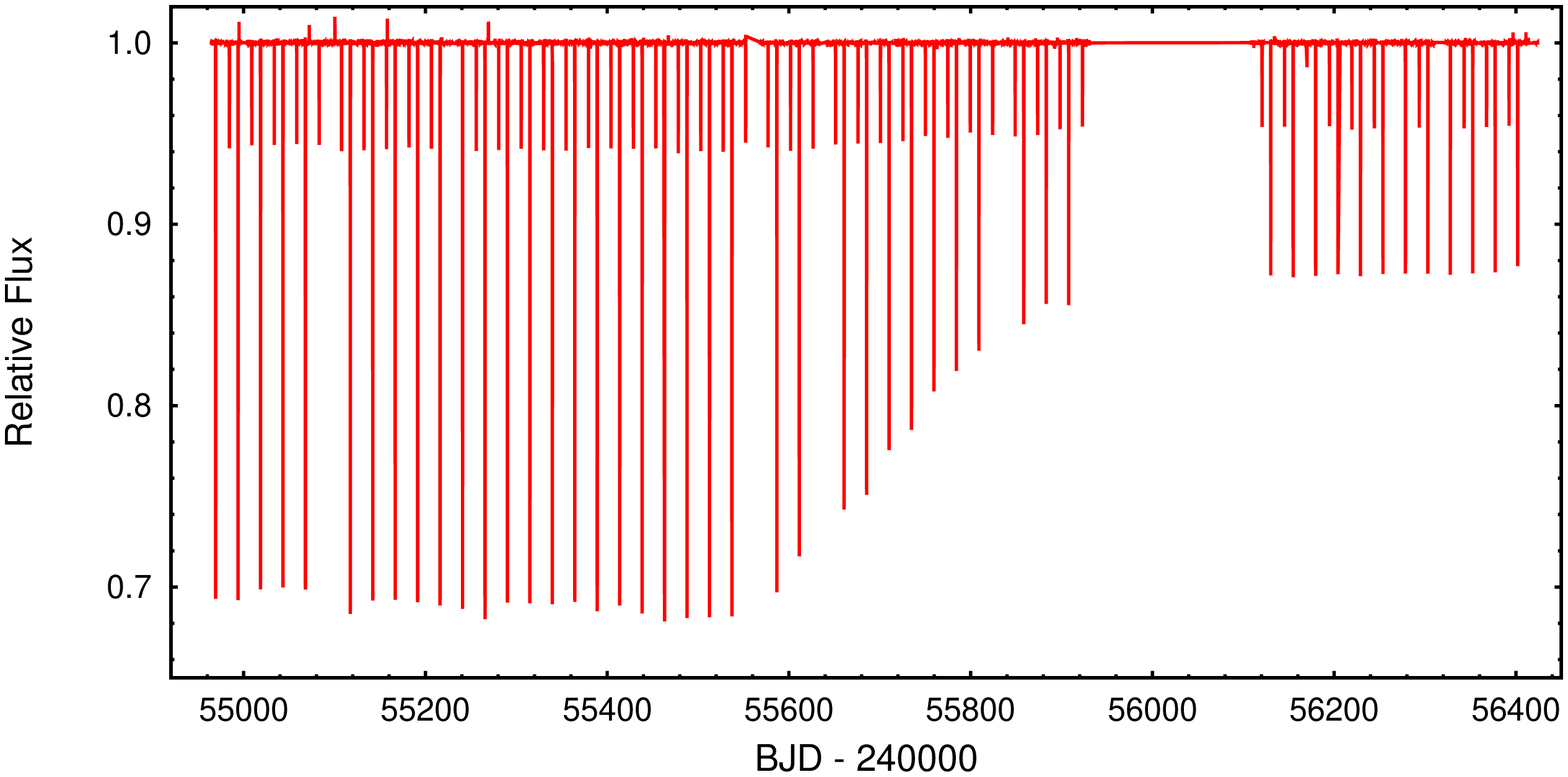}\includegraphics[width=84mm]{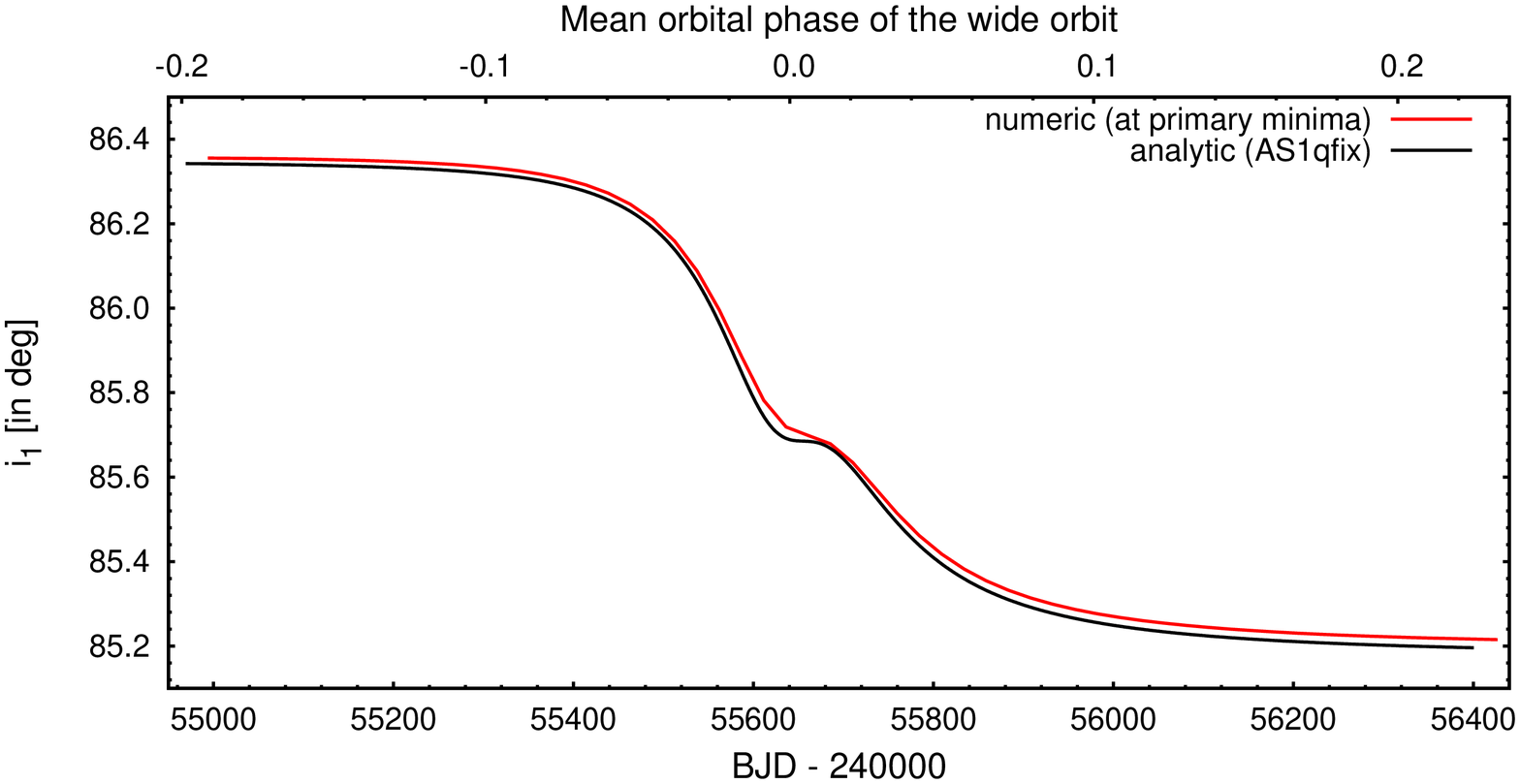}
 \caption{{\it Left panel: }The Q0--Q17 long-cadence light curve of KIC~07670617; {\it Right panel:} Inner inclination ($i_1$) variation of the eclipsing binary, according to the M\#OS1qfix solution (see Table~\ref{Tab:KIC7670617}), and the corresponding numerical integration. This figure explains well the sudden, jump-like eclipse depth variation of the system.}
 \label{Fig:K7670617lc}
\end{figure}

Another example of a system with a moderate mutual inclination angle is KIC~07670617. Its ETV is plotted in the fourth panel of Fig.~\ref{Fig:ETVall3a}. It reveals that the outer orbital period should be substantially longer than the length of the {\it Kepler}-observations. However, the marked variations around the periastron passage provides some hope for a satisfactory and reliable fit. The light curve reveals deep primary eclipses, and substantially shallower secondary ones. In the same interval, when the abrupt features occur in the ETVs, the depths of the primary eclipses are suddenly reduced to approximately half of their previous amplitude. Otherwise, the depths remain constant before and after this event. On the other hand, the secondary eclipse ETVs show only a relatively minor variation. Note, this fact suggests that the secondary eclipse should occur closer to periastron passage of the inner binary. The Q0--Q17 long cadence light curve is plotted in the left panel of Fig.~\ref{Fig:K7670617lc}. In the right panel of the same figure we also plot the variations in the observable inner binary inclination ($i_1$) according to our OS1 solution (see below).  This figure explains clearly the jump-like behavior seen in the eclipse depth variations.

\begin{table*}  
\begin{center}
\caption{Different model solutions for KIC 07670617.} 
\label{Tab:KIC7670617}  
\begin{tabular}{lllll|lll} 
\hline 
\multicolumn{2}{l}{Parameters} &OS1 & OS1q15    & OS3       & N01S1     & N02P2S1   & N02P2S3   \\
\hline
$P_1$      & (d)       & 24.70493   & 24.70497  & 24.70356  & 24.70666  & 24.70607  & 24.70555  \\
$e_1$      &           & 0.2465     & 0.2451    & 0.2318    & 0.2221    & 0.2563    & 0.2405    \\
$\omega_1$ & ($\degr$) & 135.852    & 136.191   & 140.341   & 144.606   & 134.466   & 138.630   \\
$\tau_1$   & (MBJD)    & 54961.528  & 54961.548 & 54961.813 & 54962.118 & 54961.482 & 54961.739 \\
\hline
$P_2$      & (d)       & 3377.992   & 3047.582  & 4250.000  & 3984.191  & 3378.620  & 3378.100  \\
$a_2$      &(R$_\odot$)& 1099.766   & 1093.065  & 1049.369  & 1091.749  & 1099.097  & 1048.099  \\
$e_2$      &           & 0.7105     & 0.6896    & 0.7565    & 0.7245    & 0.7020    & 0.6913    \\
$\omega_2$ & ($\degr$) & 85.463     & 85.200    & 82.295    & 80.294    & 86.483    & 87.300    \\
$\tau_2$   & (MBJD)    & 55640.396  & 55638.561 & 55643.246 & 55644.075 & 55646.098 & 55648.605 \\
\hline
$\im$          & ($\degr$) &147.174 & 147.895   & 143.605   & 146.158   & 148.640   &149.477    \\
$i_1$          & ($\degr$) & 86.000 &  86.000   &  86.000   &  88.000   &  88.000   & 89.000    \\
$i_2$          & ($\degr$) & 88.987 &  87.857   &  94.641   &  89.735   &  89.406   & 94.995    \\
$n_1$          & ($\degr$) & 81.889 &  79.516   &  87.603   &  86.538   &  85.570   & 96.449    \\
$n_2$          & ($\degr$) & 98.981 & 101.009   &  89.551   &  93.988   &  94.823   & 85.453    \\
$\Delta\Omega$ & ($\degr$)&$-147.537$&$-148.469$& 143.503   &$-146.227$ &$-148.743$ &$-149.563$ \\
\hline
$m_\mathrm{C}/m_\mathrm{ABC}$&&0.391&  0.366    &  0.411    &  0.431    &  0.346    &  0.340    \\
$m_\mathrm{B}/m_\mathrm{A}$ & &0.900& 15.000    &  0.900    &  0.900    &  0.900    &  0.900    \\
$m_\mathrm{A}$ & (M$_\odot$)  &0.502&  0.075    &  0.266    &  0.330    &  0.538    &  0.471    \\
$m_\mathrm{B}$ & (M$_\odot$)  &0.452&  1.122    &  0.240    &  0.297    &  0.484    &  0.424    \\
$m_\mathrm{C}$ & (M$_\odot$)  &0.612&  0.692    &  0.353    &  0.474    &  0.540    &  0.461    \\
\hline
$P_{\omega_1}$         & (y)&945.841&  953.708  & 7820.45   &  902.334  & 1458.947  & 2841.64   \\
$P_{h}$                & (y)&$-1693.24$&$-1455.80$&$-2157.89$&$-2026.56$&$-2020.12$ &$-2238.22$ \\
\hline
$\chi^2$               &     &1.6999&  0.9777   &  6.2518   &   0.8536  &  1.6840   &  5.2697   \\
\hline
\end{tabular} 
\end{center}  
\end{table*}

\begin{figure*}
\includegraphics[width=86mm]{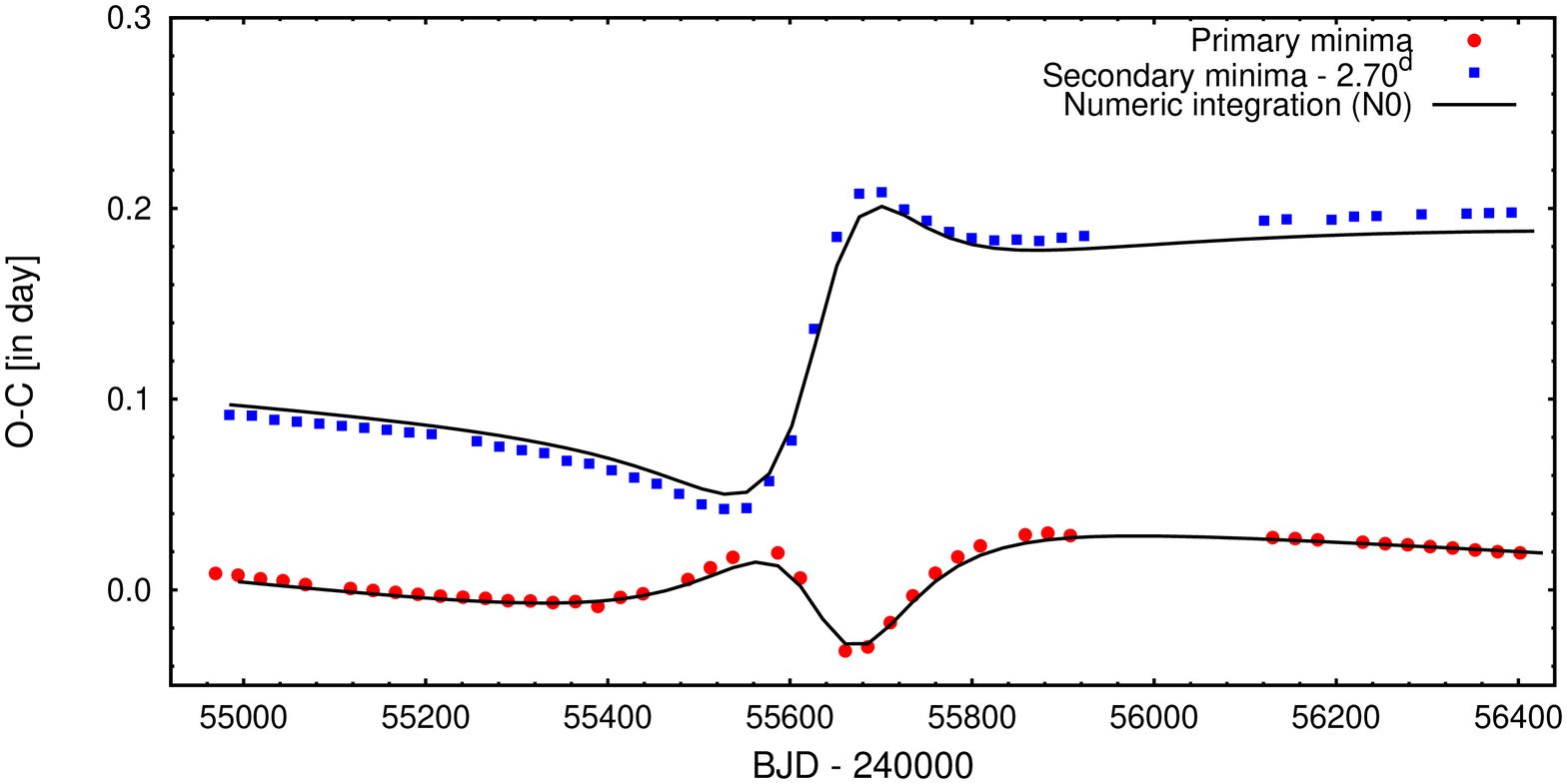}\includegraphics[width=86mm]{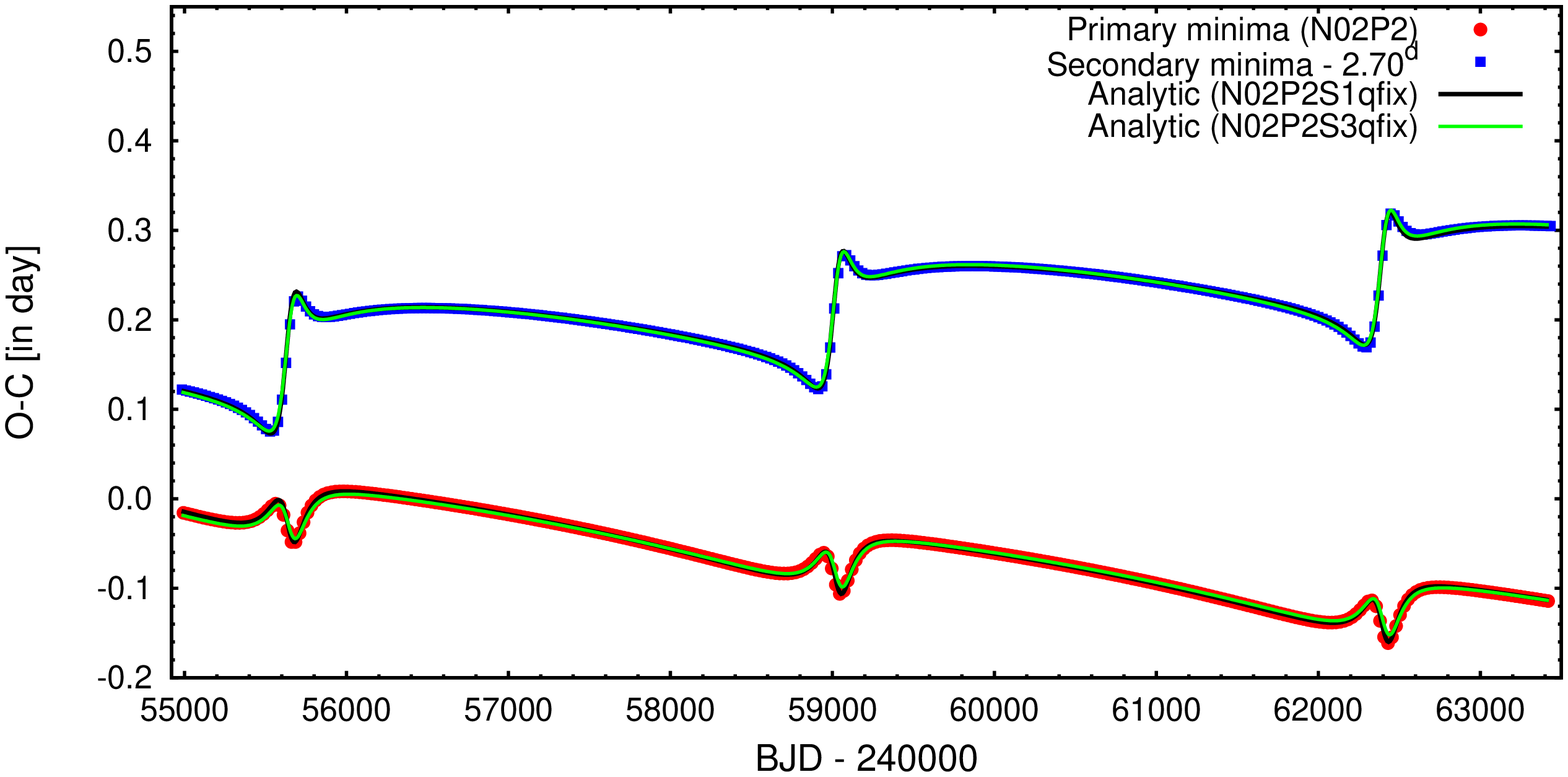}
 \caption{{\it Left panel: } The ETV curves of KIC 07670617 (filled red circles and blue boxes for primary and secondary eclipses, respectively) together with the numerically generated ETV curve for which the initial parameters were taken from  the analytical Model\#OS1 solution (see in Tab.\,\ref{Tab:KIC7670617}) (black lines). {\it Right panel: } The numerically generated $\gtrsim 8000$-day long ETV curve with two model solutions (black and green lines). }
 \label{Fig:K7670617numanalfits}
\end{figure*}

\begin{figure*}
\includegraphics[width=86mm]{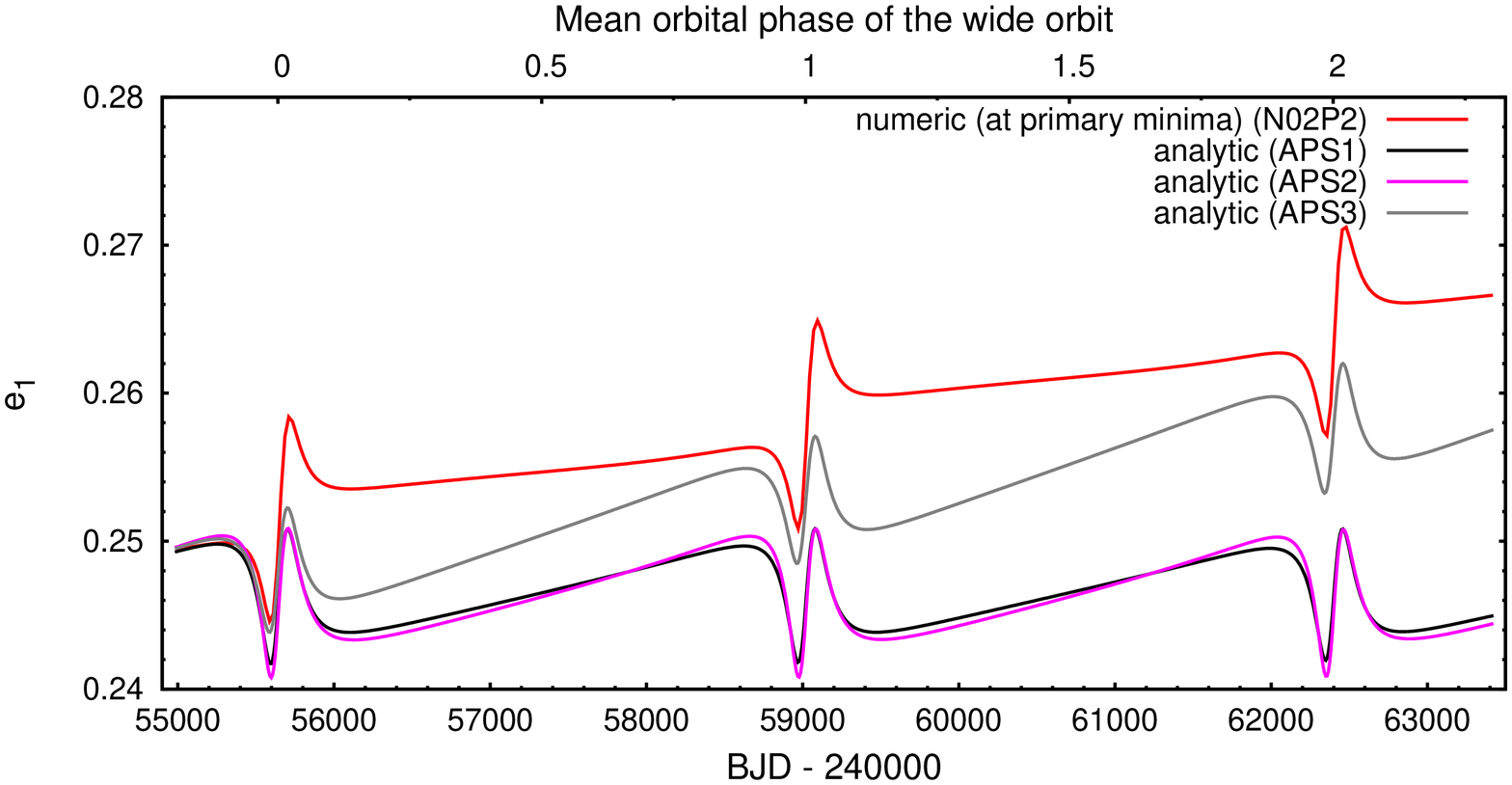}\includegraphics[width=86mm]{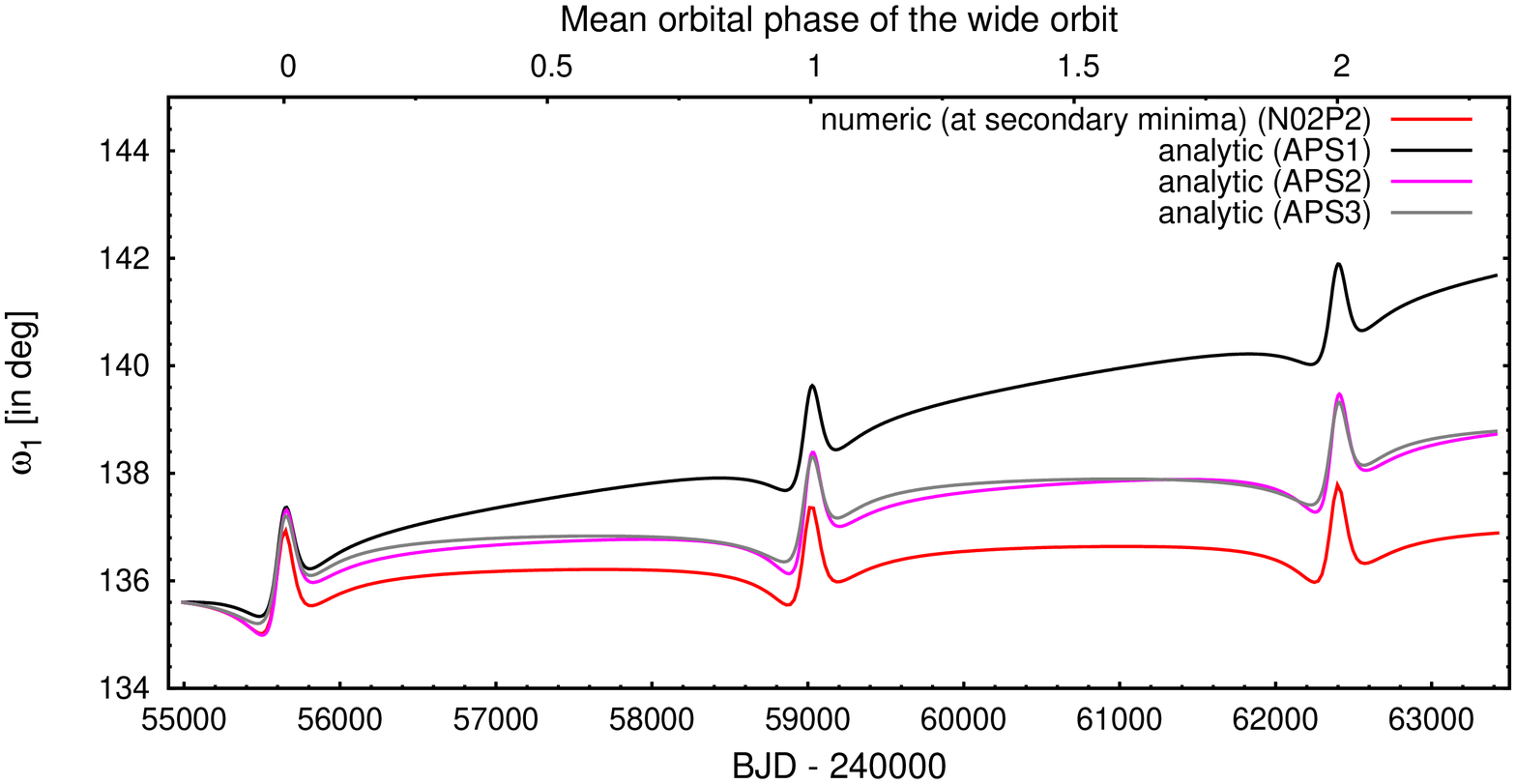}
 \caption{The secular (and long-term) inner eccentricity ($e_1$) and observable argument of periastron ($\omega_1$) variation of KIC~07670617 according to the OS1 solution. Red lines represent the results of the numerical integration, while black, magenta and grey stand for the analytically computed values according to the three different apsidal motion models. (See text for detals.)}
 \label{Fig:K7670617excom}
\end{figure*}

We present three analytic solutions for this system in the first three columns of Table~\ref{Tab:KIC7670617}. In the first two cases (OS1 and OS1q15) the apsidal advance rate was an unconstrained, but LM-adjusted constant, while in the third (OS3) it was constrained according to the first order analytic model. In this latter case the secular variation of the inner binary eccentricity was also calculated analytically. Since the inner binary mass ratio tended toward unrealistically high values in all runs (for which the OS1q15 solution is an example), we decided to fix it at $q_1=0.9$; however, the significant difference in depths between the primary and secondary eclipses suggests even smaller values.   For our follow-up test, the unconstrained apsidal motion solution was used. We generated numerical ETVs with this parameter set both for the same `short' (i.e., {\em Kepler}) time interval, and also for an $\sim8000$ day-long window, which covers more than two outer orbital periods (N0, and N02P2 datasets, respectively). Then, the entire analytic fitting process was reiterated for both the shorter and the longer datasets. The results are tabulated in the third to fifth columns of Table~\ref{Tab:KIC7670617}, and are shown in Fig.~\ref{Fig:K7670617numanalfits}. A comparison of the unconstrained apsidal motion solution for the short and the long numerical ETV (N01S1qfix, and N02P2S1qfix) columns reveals that all the orbital elements, including the mutual inclination, $\im$, remain within $\sim$10\% of the other value and, therefore, we may conclude that despite the short observational interval with respect to the orbital period, we can have confidence in our results. On the other hand, the first-order analytical apsidal motion model, unfortunately, seems to be insufficient for describing the secular variations of the orbital elements, even over such a short timescale. In Fig.~\ref{Fig:K7670617excom} the variations of $e_1$ and $\omega_1$ are plotted according to the numerical integration, and the three different apsidal motion models. These panels illustrate clearly that the displacement of the secondary eclipses with respect to the primary eclipses are affected not only by the apsidal advance rate, but by the secular variation of the inner eccentricity. The unconstrained apsidal motion model, therefore, gives a substantially faster apsidal advance rate, for counterbalancing the variation of the eccentricity. The other two apsidal motion models however, result in a more realistic apsidal advance rate. Despite this, in model AP2, due to the absence of eccentricity modeling, this more realistic apsidal advance rate produces only a very poor fit (not listed in Table~\ref{Tab:KIC7670617}), while in mode AP3 the analytically calculated eccentricity variation results in a somewhat better, but not so good fit.  Fortunately, this fact has only a minor influence on fitting the $P_2$ timescale terms.

In order to check how certain the retrograde solution is, we constructed numerically a prograde configuration with the same system parameters. It was done by changing $\omega_2\rightarrow180\degr-\omega_2$, $i_2\rightarrow180\degr-i_2$, $\Omega_2\rightarrow\Omega_2+180\degr$. Otherwise, the other initial parameters were set according to the OS1 solution. In the first three data columns of Table~\ref{Tab:KIC7670617b} different solutions are presented. Our findings are a bit contradictory. For the emulated prograde ETV dataset, with duration equal to the {\it Kepler} observations, we obtained only a weaker solution (i.e., higher $\chi^2$) than for the previously emulated retrograde curve and, furthermore, our fit resulted in a retrograde configuration (although in the same inclination regime). The situation, however, is not as bad as it might seem at first sight. When the inner mass ratio was allowed to take on large, unphysical values\footnote{In the amplitude of the octupole terms, the inner mass ratio appears in the form of $(1-q_1)/(1+q_1)$ which tends to $\mp1$ for extremely small or large mass ratios. Therefore, in our opinion, when an unphysical mass ratio occurs in a few of our solutions, it is an effect of some neglected higher-order terms of similar mathematical form which would give additional non-negligible contributions, e.g., in the case of extremely high eccentricities.}, the other solution parameters become much closer to the input values, and the mutual inclination was also changed to the prograde regime column (NproS1q15). As a comparison, during the analysis of the original, {\it Kepler} dataset, there was no case where the mutual inclination would have switched to the prograde domain. In the case of the longer simulated prograde dataset, the solution parameters naturally were closer to the initial values, but the solution both in $\chi^2$ and the parameter reconstruction, was found to be less robust than in the case of the simulated retrograde dataset. 

\begin{figure*}
\begin{center}
\includegraphics[width=86mm]{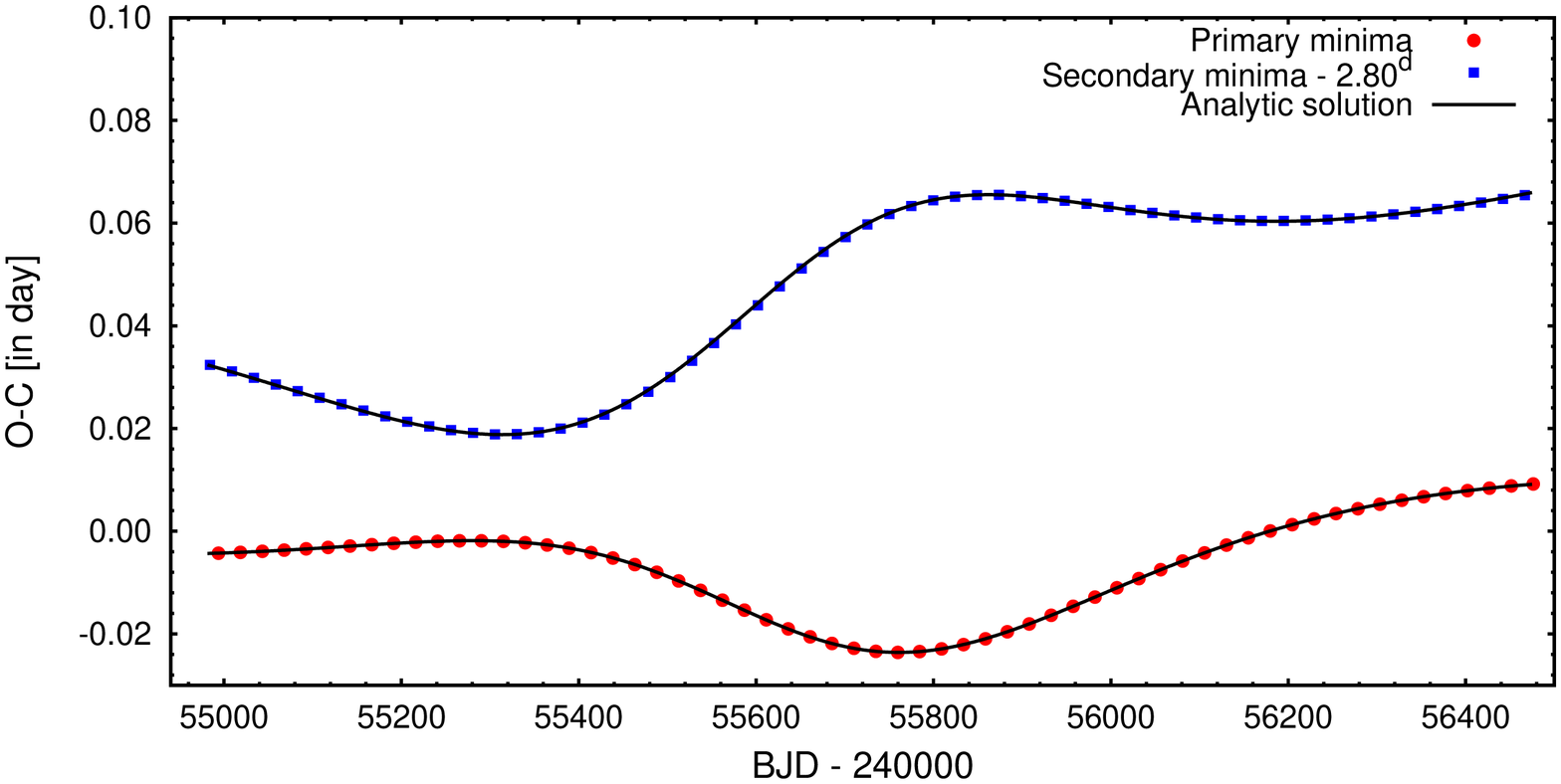}
 \caption{Moderately eccentric ($e_2=0.3$) simulated ETV curves for KIC~07670617, and analytical solutions for them. Other initial system parameters were taken from the OS1 solution of Table~\ref{Tab:KIC7670617}.}
 \label{Fig:K7670617ETVe23}
\end{center}
\end{figure*}

The origin of the less robust solutions, in our opinion, is to be found in the high outer eccentricity. In order to verify this, we made an additional test. In this case, returning to the retrograde OS1 solution, we modified the outer eccentricity to $e_2=0.3$, while all the other parameters were kept at the OS1 values. This run was used not only for testing the high outer eccentricity, but also to check whether the observation of only a single periastron passage in the smaller $e_2$ regime may result in sufficient information for parameter recovery. The analytic solution for the $\sim1450$ day-long $e_2=0.3$ ETV curve is presented in the last column of Table~\ref{Tab:KIC7670617b}, and also plotted in Fig.~\ref{Fig:K7670617ETVe23}. We feel that there is no need for additional comments on the robustness of this latter, smaller eccentricity solution of ``KIC~07670617''.

\begin{table*}  
\begin{center}
\caption{Prograde, and smaller outer eccentricity retrograde solutions for KIC 07670617.} 
\label{Tab:KIC7670617b}  
\begin{tabular}{lllll|l} 
\hline 
\multicolumn{2}{l}{Parameters}&NproS1&NproS1q15 &NproS12P2   & Ne23S1\\
\hline
$P_1$      & (d)       & 24.70798   & 24.70485  & 24.70589   & 24.70698 \\
$e_1$      &           & 0.2281     & 0.2818    & 0.2854     & 0.2237	\\
$\omega_1$ & ($\degr$) & 142.485    & 129.270   & 128.659    & 143.849  \\
$\tau_1$   & (MBJD)    & 54961.980  & 54961.160 & 54961.124  & 54962.056\\
\hline
$P_2$      & (d)       & 4000.795   & 4250.139  & 3268.371   & 3312.654 \\
$a_2$      &(R$_\odot$)&  963.460   & 1220.748  & 1181.821   & 979.366  \\
$e_2$      &           & 0.7441     & 0.7541    & 0.7126     & 0.3012	\\
$\omega_2$ & ($\degr$) & 53.213     & 92.121    & 90.872     & 79.529	\\
$\tau_2$   & (MBJD)    & 55643.097  & 55638.951 & 55638.259  & 55637.557\\
\hline
$\im$          & ($\degr$) &143.489 &  33.997   &  33.120    & 146.766  \\
$i_1$          & ($\degr$) & 86.000 &  86.000   &  86.000    & 86.000	\\
$i_2$          & ($\degr$) & 88.462 &  98.700   &  93.030    & 88.057	\\
$n_1$          & ($\degr$) & 81.972 &  67.983   &  78.220    & 80.285	\\
$n_2$          & ($\degr$) & 98.832 &  69.323   &  77.937    &100.320	\\
$\Delta\Omega$ & ($\degr$)&$-143.888$& 31.628   &  32.387    &$-147.282$\\
\hline
$m_\mathrm{C}/m_\mathrm{ABC}$&&0.519&  0.414    &  0.381     & 0.434	\\
$m_\mathrm{B}/m_\mathrm{A}$ & &0.900& 15.000    &  0.900     & 0.900	\\
$m_\mathrm{A}$ & (M$_\odot$)  &0.190&  0.049    &  0.676     & 0.342	\\
$m_\mathrm{B}$ & (M$_\odot$)  &0.171&  0.743    &  0.608     & 0.308	\\
$m_\mathrm{C}$ & (M$_\odot$)  &0.390&  0.561    &  0.792     & 0.499	\\
\hline
$P_{\omega_1}$         & (y)&796.945& 1414.539  & 1488.323   & 1943.051 \\
$P_{h}$                & (y)&$-1531.95$&1686.32 & 1271.16    &$-3505.63$\\
\hline
$\chi^2$               &     &1.7705&  0.6501   &  3.9799    & 0.0006	\\
\hline
\end{tabular} 
\end{center}  
\end{table*}

\subsection{The case of comparable LTTE and dynamical amplitudes: KIC~10268809}

Our solution process yielded the longest outer period ($P_2=7\,000$ days) for this triple. It is clear that a solution with 5 times longer period than the entire {\em Kepler} data set cannot to be taken too seriously. However, this solution yields an LTTE amplitude that is comparable to that of the dynamical quadrupole term (the ratio is $\sim0.64$; see Table~\ref{Tab:Massetal}).  Independent of the reliability of the solution, this makes it possible to probe whether we can obtain reliable individual masses in such cases. Therefore, we numerically integrated the equation of motion for the system according to our solution, and then emulated a $2P_2=14\,000$ day-long ETV. Then, this ETV curve was subjected to our parameter finding process. Furthermore, we carried out a second numerical integration, where the only modification to the initial parameters was the reduction of the outer eccentricity from $e_2=0.74$ to $e_2=0.30$. This was necessary because, as we illustrated above in the case of KIC 10268809, for such a high outer eccentricity our model gives a somewhat weaker solution. 

\begin{table*}  
\begin{center}
\caption{Solutions for a 14\,000 day-long numerically emulated ETV-s of KIC 102687809.} 
\label{Tab:KIC10268809}  
\begin{tabular}{l|llll|ll} 
\hline 
\multicolumn{2}{l}{Parameters}&OS1  &   N2P2S1  & N2P2S3     & Ne232P2S1 & Ne232P2S3 \\
\hline
$P_1$      & (d)       & 24.70934   & 24.70917  & 24.70886   & 24.70887  & 24.70879  \\
$e_1$      &           & 0.3205     & 0.3126    & 0.2624     & 0.3402	 & 0.2889    \\
$\omega_1$ & ($\degr$) & 141.505    & 144.627   & 168.045    & 137.854   & 152.426   \\
$\tau_1$   & (MBJD)    & 54965.474  & 54965.658 & 54967.067  & 54965.288 & 54966.143 \\
\hline
$P_2$      & (d)       & 7000.000   & 6999.169  & 7000.950   & 7003.080  & 7000.080  \\
$a_\mathrm{AB}\sin i_2$&(R$_\odot$)&1000.000&960.156&1058.717& 972.894   & 1012.024  \\
$a_2$      &(R$_\odot$)& 2125.679   & 1838.509  & 1712.516   & 2043.912  & 2225.400  \\
$e_2$      &           & 0.7381     & 0.7403    & 0.7328     & 0.3029	 & 0.3334    \\
$\omega_2$ & ($\degr$) & 291.841    & 290.327   & 303.448    & 287.244	 & 293.996   \\
$\tau_2$   & (MBJD)    & 56147.399  & 56139.445 & 56133.369  & 56116.054 & 56078.512 \\
\hline
$\im$          & ($\degr$) & 24.300 &  28.104   &  28.694    & 32.259    & 20.305    \\
$i_1$          & ($\degr$) & 84.000 &  84.000   &  84.000    & 84.000	 & 84.000    \\
$i_2$          & ($\degr$) & 94.819 &  92.650   &  88.310    & 98.427	 & 81.344    \\
$n_1$          & ($\degr$) & 64.020 &  72.812   &  97.484    & 63.729	 & 81.255    \\
$n_2$          & ($\degr$) & 63.791 &  72.013   &  99.433    & 64.361	 & 83.872    \\
$\Delta\Omega$ & ($\degr$) & 21.792 &  26.777   &$-28.441$   & 28.937    & $-20.300$ \\
\hline
$m_\mathrm{C}/m_\mathrm{ABC}$&&0.472&  0.523    &  0.618     & 0.481	 & 0.460     \\
$m_\mathrm{B}/m_\mathrm{A}$ & &0.700&  0.700    &  0.700     & 0.700	 & 0.700     \\
$m_\mathrm{A}$ & (M$_\odot$)  &0.818&  0.478    &  0.309     & 0.713	 & 0.960     \\
$m_\mathrm{B}$ & (M$_\odot$)  &0.572&  0.335    &  0.216     & 0.500	 & 0.672     \\
$m_\mathrm{C}$ & (M$_\odot$)  &1.243&  0.891    &  0.851     & 1.125	 & 1.390     \\
\hline
$P_{\omega_1}$ (unconstrained)&(y)& 1904& 5333  &    $-$     & 20733     & $-$      \\
$P_{\omega_1}$ (constrained)  &(y)&14808&22007  &   19315    &$-207924$  & 23413    \\
$P_{h}$                & (y)  & 3311&  3199     &    3155    & 9801      & 9928     \\
\hline
$\chi^2$               &     &0.8870&  1.2434   &  3.3243    & 0.1443    & 0.9575   \\
\hline
\end{tabular} 
\end{center}  
\end{table*}

\begin{figure*}
\includegraphics[width=86mm]{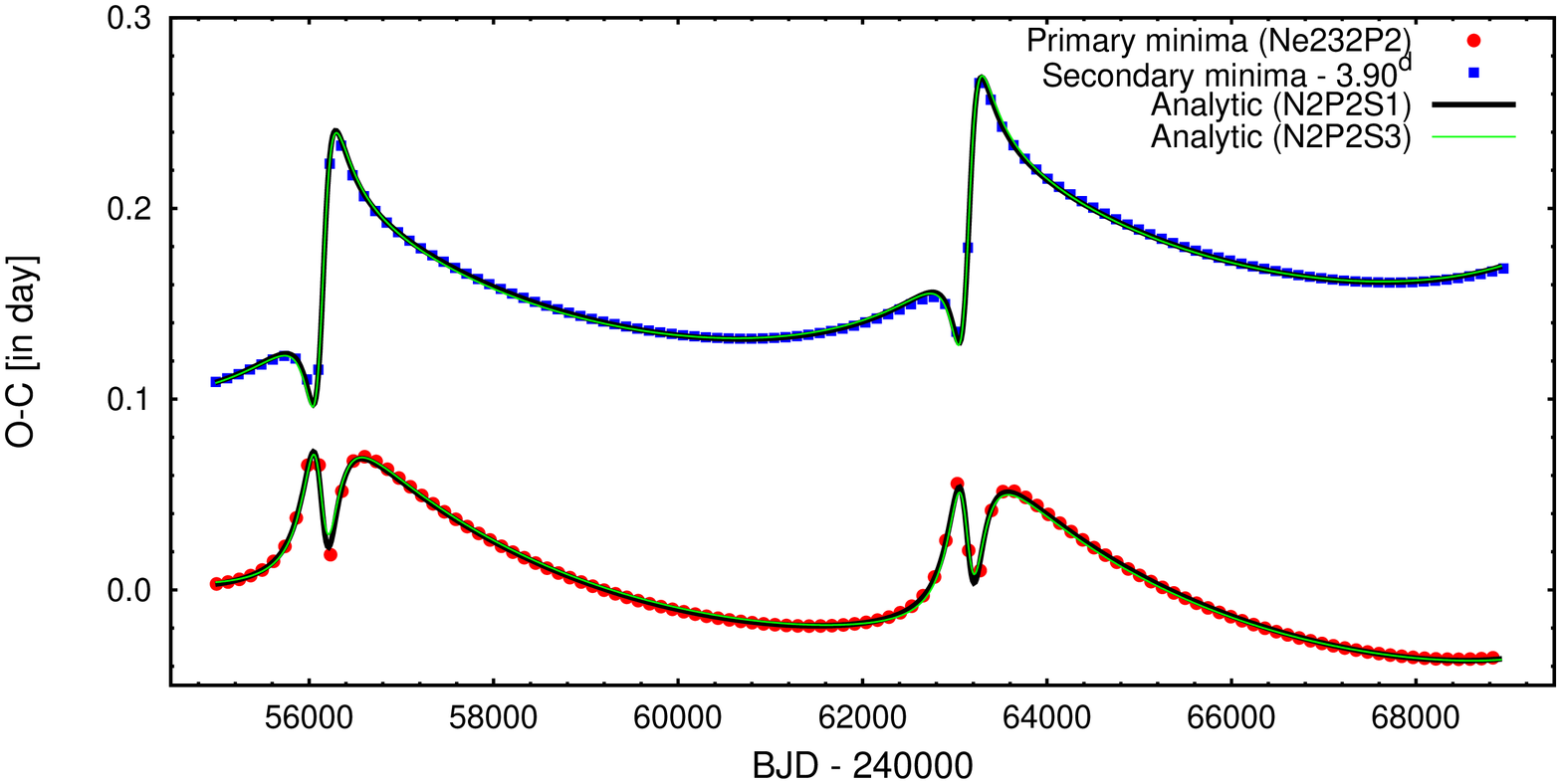}\includegraphics[width=86mm]{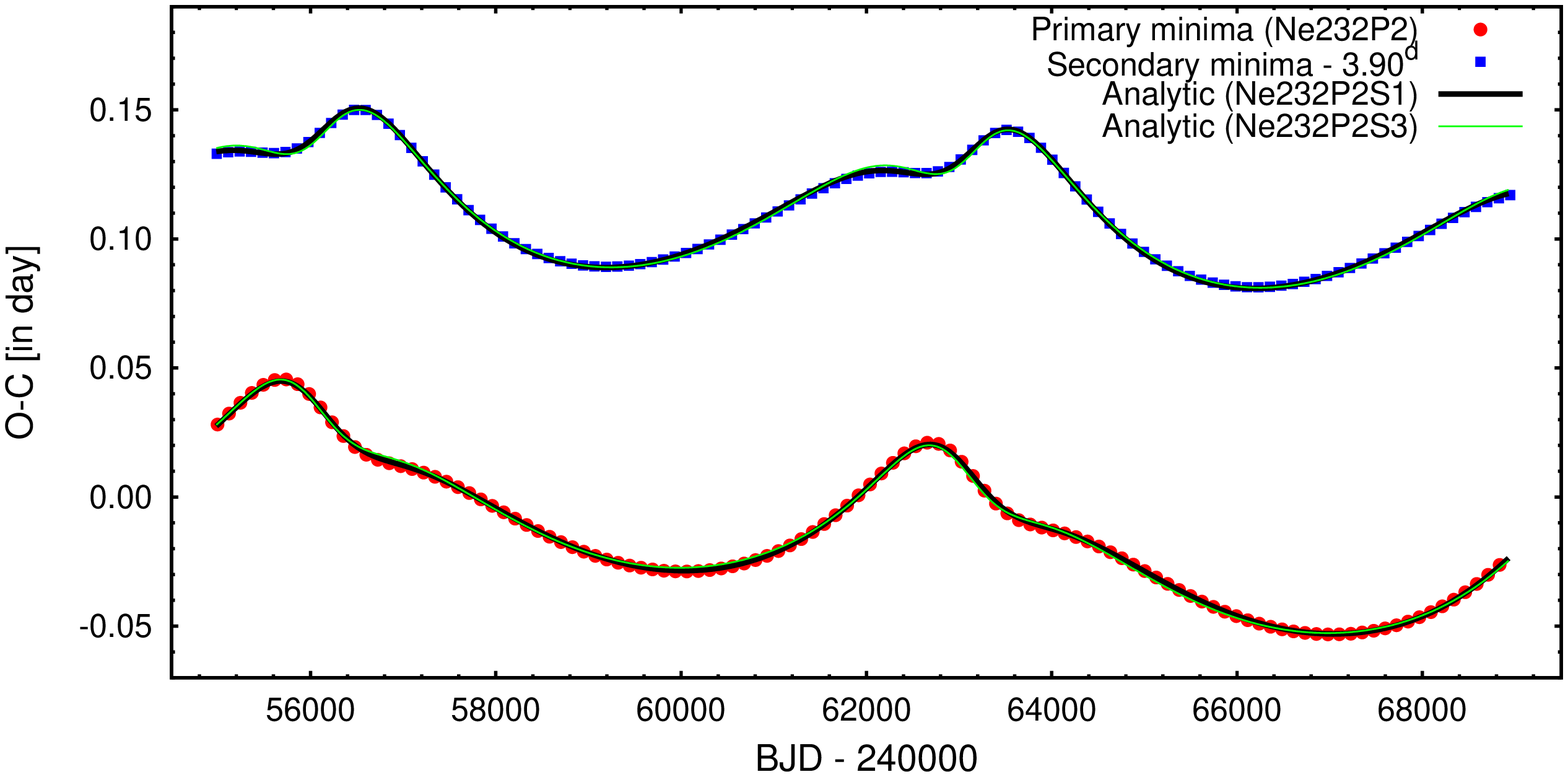}
 \caption{Left panel: Numerically emulated $2P_2=14\,000$ day-long ETV curves (red and blue symbols) with the initial parameters of our $OS1$ solution for KIC~10268809, as well as the corresponding analytic fits which are tabulated in Table~\ref{Tab:KIC10268809}. Right panel: The same, but the outer eccentricity was reduced to $e_2=0.3$. Note, for the sake of clarity, only every fifth eclipse point is displayed.}
 \label{Fig:K10268809numanalfits}
\end{figure*}

In Table~\ref{Tab:KIC10268809} we list our solutions obtained with apsidal motion modes AP1 and AP3 both for the ``original'' and the reduced outer eccentricity numerically emulated ETV. For an easier comparison with the initial parameters we also give the relevant parameters of the $OS1$ solution of the observed ETV. The numerically generated ETVs and the fits are also plotted in Fig.~\ref{Fig:K10268809numanalfits}. A column-by-column comparison reveals again that our solutions reproduce the initial values for most parameters to within a few percent uncertainty.  Furthermore, as was expected, we obtained a better solution for the lower eccentricity case. Now, we concentrate on the individual masses. The two relevant parameters for these are $a_\mathrm{AB}\sin{i_2}$, which is strongly related to the LTTE amplitude, and the outer mass ratio $m_\mathrm{C}/m_\mathrm{ABC}$. As one can see for the $e_2\sim0.74$ case, the latter was obtained to within $4-6\%$, while the discrepancy in the $e_3=0.3$ case remains $1-3\%$. In contrast, the dynamical amplitudes, and apsidal motion period, which are related to the outer mass ratio, were not well reproduced in the high eccentricity case, but were found to be within $\sim 3\%$ of their actual value in the moderate eccentricity case. Thus, even in this latter case, the resultant discrepancy in the masses remained above $10\%$. 

In conclusion, we can state that for comparable magnitudes of the dynamical and LTTE terms, we can deduce individual masses from our fitting process; however, only with moderate accuracy.

\clearpage

\section{Tables of Determined Times of Minima for all the 26 Systems}
\label{app:ToMtables}

{In this Appendix we tabulate the individual $O-C$ times for each of the primary and secondary eclipses for all 26 compact hierarchical triple systems that we considered in this study.  The uncertainty in each individual $O-C$ determination is also listed.}

%\clearpage

\begin{table*}
\caption{Times of minima of KIC 04940201}
 \label{Tab:KIC 04940201_ToM}
\begin{tabular}{@{}lrllrllrl}
\hline
BJD & Cycle  & std. dev. & BJD & Cycle  & std. dev. & BJD & Cycle  & std. dev. \\ 
$-2\,400\,000$ & no. &   \multicolumn{1}{c}{$(d)$} & $-2\,400\,000$ & no. &   \multicolumn{1}{c}{$(d)$} & $-2\,400\,000$ & no. &   \multicolumn{1}{c}{$(d)$} \\ 
\hline
54967.284492 &    0.0 & 0.000084 & 55359.628854 &   44.5 & 0.000161 & 55923.876985 &  108.5 & 0.000157 \\ 
54971.702667 &    0.5 & 0.000164 & 55364.027915 &   45.0 & 0.000080 & 55928.280615 &  109.0 & 0.000060 \\ 
54976.103341 &    1.0 & 0.000084 & 55368.444204 &   45.5 & 0.000160 & 55932.696764 &  109.5 & 0.000157 \\ 
54980.519311 &    1.5 & 0.000184 & 55465.415898 &   56.5 & 0.000140 & 55937.099270 &  110.0 & 0.000075 \\ 
54984.919328 &    2.0 & 0.000073 & 55469.816917 &   57.0 & 0.000062 & 55941.515436 &  110.5 & 0.000173 \\ 
54989.333952 &    2.5 & 0.000162 & 55474.230689 &   57.5 & 0.000162 & 55945.916169 &  111.0 & 0.000076 \\
\hline
\end{tabular}
\end{table*}

\addtocounter{table}{-1}
\begin{table*}
\caption{Times of minima of KIC 04940201 (continued)}
%\label{Tab:KIC_04940201_ToM}
\begin{tabular}{@{}lrllrllrl}
\hline
BJD & Cycle  & std. dev. & BJD & Cycle  & std. dev. & BJD & Cycle  & std. dev. \\ 
$-2\,400\,000$ & no. &   \multicolumn{1}{c}{$(d)$} & $-2\,400\,000$ & no. &   \multicolumn{1}{c}{$(d)$} & $-2\,400\,000$ & no. &   \multicolumn{1}{c}{$(d)$} \\ 
\hline
55214.141172 &   28.0 & 0.000086 & 55699.061007 &   83.0 & 0.000069 & 56276.541128 &  148.5 & 0.000149 \\ 
55218.559018 &   28.5 & 0.000157 & 55703.476861 &   83.5 & 0.000156 & 56280.941097 &  149.0 & 0.000062 \\ 
55222.961795 &   29.0 & 0.000085 & 55707.877434 &   84.0 & 0.000071 & 56285.356786 &  149.5 & 0.000143 \\ 
55227.376697 &   29.5 & 0.000154 & 55712.292798 &   84.5 & 0.000176 & 56289.758976 &  150.0 & 0.000065 \\ 
55236.197744 &   30.5 & 0.000171 & 55716.692618 &   85.0 & 0.000070 & 56294.176238 &  150.5 & 0.000136 \\ 
55240.598897 &   31.0 & 0.000074 & 55721.105992 &   85.5 & 0.000153 & 56298.577361 &  151.0 & 0.000061 \\
\hline
\end{tabular}
\end{table*}

\begin{table*}
\caption{Times of minima of KIC 05255552}
 \label{Tab:KIC 05255552_ToM}
\begin{tabular}{@{}lrllrllrl}
\hline
BJD & Cycle  & std. dev. & BJD & Cycle  & std. dev. & BJD & Cycle  & std. dev. \\ 
$-2\,400\,000$ & no. &   \multicolumn{1}{c}{$(d)$} & $-2\,400\,000$ & no. &   \multicolumn{1}{c}{$(d)$} & $-2\,400\,000$ & no. &   \multicolumn{1}{c}{$(d)$} \\ 
\hline
54970.588895 &    0.0 & 0.000208 & 55443.010892 &   14.5 & 0.000151 & 55911.806709 &   29.0 & 0.000238 \\ 
54988.785838 &    0.5 & 0.000146 & 55457.395193 &   15.0 & 0.000227 & 55930.412697 &   29.5 & 0.000172 \\ 
55003.039896 &    1.0 & 0.000221 & 55475.468042 &   15.5 & 0.000172 & 55944.275563 &   30.0 & 0.000216 \\ 
55021.256127 &    1.5 & 0.000160 & 55489.829350 &   16.0 & 0.000241 & 55962.854231 &   30.5 & 0.000167 \\ 
55035.504910 &    2.0 & 0.000231 & 55507.932891 &   16.5 & 0.000168 & 55976.742813 &   31.0 & 0.000205 \\ 
55053.705977 &    2.5 & 0.000162 & 55522.259406 &   17.0 & 0.000251 & 56009.207289 &   32.0 & 0.000187 \\
\hline
\end{tabular}
\end{table*}

\begin{table*}
\caption{Times of minima of KIC 05653126}
 \label{Tab:KIC 05653126_ToM}
\begin{tabular}{@{}lrllrllrl}
\hline
BJD & Cycle  & std. dev. & BJD & Cycle  & std. dev. & BJD & Cycle  & std. dev. \\ 
$-2\,400\,000$ & no. &   \multicolumn{1}{c}{$(d)$} & $-2\,400\,000$ & no. &   \multicolumn{1}{c}{$(d)$} & $-2\,400\,000$ & no. &   \multicolumn{1}{c}{$(d)$} \\ 
\hline
54985.877118 &    0.0 & 0.001065 & 55693.781088 &   18.5 & 0.007507 & 56063.792943 &   28.0 & 0.000618 \\ 
55024.375156 &    1.0 & 0.001010 & 55717.229991 &   19.0 & 0.000695 & 56102.265065 &   29.0 & 0.001850 \\ 
55062.864411 &    2.0 & 0.001085 & 55732.241048 &   19.5 & 0.007570 & 56116.966241 &   29.5 & 0.004294 \\ 
55101.343642 &    3.0 & 0.001037 & 55755.739417 &   20.0 & 0.000641 & 56140.730795 &   30.0 & 0.000569 \\ 
55139.815064 &    4.0 & 0.000998 & 55794.256058 &   21.0 & 0.000641 & 56155.479420 &   30.5 & 0.004062 \\ 
55178.278681 &    5.0 & 0.001038 & 55809.140030 &   21.5 & 0.007592 & 56179.188630 &   31.0 & 0.000661 \\
\hline
\end{tabular}
\end{table*}

\begin{table*}
\caption{Times of minima of KIC 06545018}
 \label{Tab:KIC 06545018_ToM}
\begin{tabular}{@{}lrllrllrl}
\hline
BJD & Cycle  & std. dev. & BJD & Cycle  & std. dev. & BJD & Cycle  & std. dev. \\ 
$-2\,400\,000$ & no. &   \multicolumn{1}{c}{$(d)$} & $-2\,400\,000$ & no. &   \multicolumn{1}{c}{$(d)$} & $-2\,400\,000$ & no. &   \multicolumn{1}{c}{$(d)$} \\ 
\hline
54965.834001 &    0.0 & 0.000033 & 55438.837987 &  118.5 & 0.000044 & 55923.785390 &  240.0 & 0.000032 \\ 
54967.838089 &    0.5 & 0.000044 & 55440.826435 &  119.0 & 0.000032 & 55925.787266 &  240.5 & 0.000042 \\ 
54969.827218 &    1.0 & 0.000037 & 55442.829147 &  119.5 & 0.000041 & 55927.775882 &  241.0 & 0.000030 \\ 
54971.831876 &    1.5 & 0.000047 & 55444.817114 &  120.0 & 0.000031 & 55929.777895 &  241.5 & 0.000043 \\ 
54973.821372 &    2.0 & 0.000035 & 55446.819891 &  120.5 & 0.000040 & 55933.768208 &  242.5 & 0.000040 \\ 
54975.826148 &    2.5 & 0.000046 & 55448.807591 &  121.0 & 0.000032 & 55935.757118 &  243.0 & 0.000033 \\
\hline
\end{tabular}
\end{table*}

\addtocounter{table}{-1}
\begin{table*}
\caption{Times of minima of KIC 06545018 (continued)}
% \label{Tab:KIC 06545018_ToM}
\begin{tabular}{@{}lrllrllrl}
\hline
55041.669246 &   19.0 & 0.000033 & 55510.675735 &  136.5 & 0.000040 & 56007.607978 &  261.0 & 0.000035 \\ 
55043.672628 &   19.5 & 0.000045 & 55512.665316 &  137.0 & 0.000034 & 56009.609536 &  261.5 & 0.000044 \\ 
55045.661140 &   20.0 & 0.000032 & 55514.669442 &  137.5 & 0.000039 & 56011.598277 &  262.0 & 0.000032 \\ 
55047.664721 &   20.5 & 0.000041 & 55516.659669 &  138.0 & 0.000030 & 56013.599854 &  262.5 & 0.000041 \\ 
55049.653308 &   21.0 & 0.000032 & 55518.663625 &  138.5 & 0.000038 & 56017.590185 &  263.5 & 0.000043 \\ 
55051.656882 &   21.5 & 0.000041 & 55520.653982 &  139.0 & 0.000032 & 56019.579209 &  264.0 & 0.000036 \\ 
55053.645467 &   22.0 & 0.000034 & 55522.657441 &  139.5 & 0.000041 & 56021.580699 &  264.5 & 0.000049 \\ 
55055.649492 &   22.5 & 0.000046 & 55524.647008 &  140.0 & 0.000030 & 56023.569849 &  265.0 & 0.000036 \\
\hline
\end{tabular}
\end{table*}

\addtocounter{table}{-1}
\begin{table*}
\caption{Times of minima of KIC 06545018 (continued)}
% \label{Tab:KIC 06545018_ToM}
\begin{tabular}{@{}lrllrllrl}
\hline
55169.407164 &   51.0 & 0.000034 & 55658.358547 &  173.5 & 0.000042 & 56139.322264 &  294.0 & 0.000033 \\ 
55171.410303 &   51.5 & 0.000041 & 55660.347025 &  174.0 & 0.000033 & 56141.323906 &  294.5 & 0.000044 \\ 
55173.397974 &   52.0 & 0.000033 & 55662.349121 &  174.5 & 0.000042 & 56143.314520 &  295.0 & 0.000031 \\ 
55175.400890 &   52.5 & 0.000044 & 55664.337788 &  175.0 & 0.000033 & 56145.316804 &  295.5 & 0.000041 \\ 
55177.388368 &   53.0 & 0.000032 & 55666.339787 &  175.5 & 0.000043 & 56147.307774 &  296.0 & 0.000032 \\ 
55179.391514 &   53.5 & 0.000042 & 55668.328778 &  176.0 & 0.000034 & 56149.310422 &  296.5 & 0.000043 \\ 
55181.378803 &   54.0 & 0.000034 & 55670.330795 &  176.5 & 0.000043 & 56151.302356 &  297.0 & 0.000035 \\ 
55187.372196 &   55.5 & 0.000043 & 55672.320038 &  177.0 & 0.000034 & 56153.305202 &  297.5 & 0.000044 \\
\hline
\end{tabular}
\end{table*}

\addtocounter{table}{-1}
\begin{table*}
\caption{Times of minima of KIC 06545018 (continued)}
% \label{Tab:KIC 06545018_ToM}
\begin{tabular}{@{}lrllrllrl}
\hline
55305.105983 &   85.0 & 0.000033 & 55788.078670 &  206.0 & 0.000032 & 56275.036861 &  328.0 & 0.000033 \\ 
55307.108603 &   85.5 & 0.000042 & 55790.082255 &  206.5 & 0.000043 & 56277.038043 &  328.5 & 0.000041 \\ 
55311.099731 &   86.5 & 0.000043 & 55792.072896 &  207.0 & 0.000035 & 56279.027331 &  329.0 & 0.000032 \\ 
55313.088675 &   87.0 & 0.000035 & 55794.076000 &  207.5 & 0.000047 & 56281.028385 &  329.5 & 0.000042 \\ 
55315.091338 &   87.5 & 0.000046 & 55796.066198 &  208.0 & 0.000034 & 56283.017665 &  330.0 & 0.000030 \\ 
55317.080487 &   88.0 & 0.000033 & 55798.068644 &  208.5 & 0.000042 & 56285.018699 &  330.5 & 0.000038 \\ 
55319.083365 &   88.5 & 0.000042 & 55800.057987 &  209.0 & 0.000032 & 56287.008144 &  331.0 & 0.000030 \\ 
55321.072467 &   89.0 & 0.000033 & 55802.060154 &  209.5 & 0.000041 & 56289.009163 &  331.5 & 0.000038 \\
\hline
\end{tabular}
\end{table*}

\begin{table*}
\caption{Times of minima of KIC 07289157}
 \label{Tab:KIC 07289157_ToM}
\begin{tabular}{@{}lrllrllrl}
\hline
BJD & Cycle  & std. dev. & BJD & Cycle  & std. dev. & BJD & Cycle  & std. dev. \\ 
$-2\,400\,000$ & no. &   \multicolumn{1}{c}{$(d)$} & $-2\,400\,000$ & no. &   \multicolumn{1}{c}{$(d)$} & $-2\,400\,000$ & no. &   \multicolumn{1}{c}{$(d)$} \\ 
\hline
54964.709593 &   -1.0 & 0.000132 & 55428.144031 &   87.0 & 0.000159 & 55925.753742 &  181.5 & 0.001335 \\ 
54967.222224 &   -0.5 & 0.001008 & 55430.687213 &   87.5 & 0.001266 & 55928.461344 &  182.0 & 0.000210 \\ 
54969.976404 &    0.0 & 0.000132 & 55433.413934 &   88.0 & 0.000160 & 55931.021829 &  182.5 & 0.001435 \\ 
54972.490479 &    0.5 & 0.001008 & 55435.953952 &   88.5 & 0.001144 & 55933.730208 &  183.0 & 0.000192 \\ 
54975.242597 &    1.0 & 0.000156 & 55438.683856 &   89.0 & 0.000164 & 55936.290238 &  183.5 & 0.001422 \\ 
54977.758571 &    1.5 & 0.001081 & 55441.220543 &   89.5 & 0.001174 & 55938.997573 &  184.0 & 0.000189 \\
\hline
\end{tabular}
\end{table*}

\addtocounter{table}{-1}
\begin{table*}
\caption{Times of minima of KIC 07289157 (continued)}
%\label{Tab:KIC 07289157_ToM}
\begin{tabular}{@{}lrllrllrl}
\hline
BJD & Cycle  & std. dev. & BJD & Cycle  & std. dev. & BJD & Cycle  & std. dev. \\ 
$-2\,400\,000$ & no. &   \multicolumn{1}{c}{$(d)$} & $-2\,400\,000$ & no. &   \multicolumn{1}{c}{$(d)$} & $-2\,400\,000$ & no. &   \multicolumn{1}{c}{$(d)$} \\ 
\hline
55117.421214 &   28.0 & 0.000147 & 55593.943518 &  118.5 & 0.001314 & 56081.175755 &  211.0 & 0.000210 \\ 
55119.941377 &   28.5 & 0.001110 & 55599.210320 &  119.5 & 0.001117 & 56083.744099 &  211.5 & 0.001508 \\ 
55122.687092 &   29.0 & 0.000165 & 55601.932582 &  120.0 & 0.000202 & 56086.441420 &  212.0 & 0.000233 \\ 
55125.207851 &   29.5 & 0.001165 & 55604.476498 &  120.5 & 0.001096 & 56089.009599 &  212.5 & 0.001313 \\ 
55127.953040 &   30.0 & 0.000143 & 55607.198489 &  121.0 & 0.000155 & 56091.707447 &  213.0 & 0.000218 \\
55130.473121 &   30.5 & 0.001258 & 55609.742848 &  121.5 & 0.001193 & 56094.275057 &  213.5 & 0.001362 \\
\hline
\end{tabular}
\end{table*}

\addtocounter{table}{-1}
\begin{table*}
\caption{Times of minima of KIC 07289157 (continued)}
%\label{Tab:KIC 07289157_ToM}
\begin{tabular}{@{}lrllrllrl}
\hline
BJD & Cycle  & std. dev. & BJD & Cycle  & std. dev. & BJD & Cycle  & std. dev. \\ 
$-2\,400\,000$ & no. &   \multicolumn{1}{c}{$(d)$} & $-2\,400\,000$ & no. &   \multicolumn{1}{c}{$(d)$} & $-2\,400\,000$ & no. &   \multicolumn{1}{c}{$(d)$} \\ 
\hline
55270.157968 &   57.0 & 0.000153 & 55757.229420 &  149.5 & 0.001322 & 56239.177426 &  241.0 & 0.000245 \\ 
55272.694898 &   57.5 & 0.001008 & 55759.934617 &  150.0 & 0.000210 & 56241.762604 &  241.5 & 0.001525 \\ 
55277.960568 &   58.5 & 0.001430 & 55762.494823 &  150.5 & 0.001242 & 56244.443008 &  242.0 & 0.000240 \\ 
55280.689576 &   59.0 & 0.000151 & 55765.200290 &  151.0 & 0.000189 & 56252.294868 &  243.5 & 0.001578 \\ 
55283.226144 &   59.5 & 0.001126 & 55767.761026 &  151.5 & 0.001416 & 56254.974160 &  244.0 & 0.000261 \\ 
55285.955305 &   60.0 & 0.000157 & 55773.026248 &  152.5 & 0.001414 & 56257.560129 &  244.5 & 0.001458 \\
\hline
\end{tabular}
\end{table*}

\begin{table*}
\caption{Times of minima of KIC 07812175}
 \label{Tab:KIC 07812175_ToM}
\begin{tabular}{@{}lrllrllrl}
\hline
BJD & Cycle  & std. dev. & BJD & Cycle  & std. dev. & BJD & Cycle  & std. dev. \\ 
$-2\,400\,000$ & no. &   \multicolumn{1}{c}{$(d)$} & $-2\,400\,000$ & no. &   \multicolumn{1}{c}{$(d)$} & $-2\,400\,000$ & no. &   \multicolumn{1}{c}{$(d)$} \\ 
\hline
55749.941579 &   42.0 & 0.000097 & 55963.494896 &   54.0 & 0.000072 & 56184.357181 &   66.5 & 0.000469 \\ 
55757.348342 &   42.5 & 0.000486 & 55970.865112 &   54.5 & 0.000405 & 56194.834144 &   67.0 & 0.000085 \\ 
55767.737873 &   43.0 & 0.000076 & 55981.281063 &   55.0 & 0.000070 & 56202.153194 &   67.5 & 0.000487 \\ 
55775.131187 &   43.5 & 0.000471 & 55988.662911 &   55.5 & 0.000603 & 56212.624796 &   68.0 & 0.000193 \\ 
55785.535592 &   44.0 & 0.000074 & 55999.073257 &   56.0 & 0.000099 & 56219.951403 &   68.5 & 0.001116 \\ 
55792.921032 &   44.5 & 0.000459 & 56006.456895 &   56.5 & 0.000321 & 56230.412406 &   69.0 & 0.000237 \\
\hline
\end{tabular}
\end{table*}

\begin{table*}
\caption{Times of minima of KIC 08023317}
 \label{Tab:KIC 08023317_ToM}
\begin{tabular}{@{}lrllrllrl}
\hline
BJD & Cycle  & std. dev. & BJD & Cycle  & std. dev. & BJD & Cycle  & std. dev. \\ 
$-2\,400\,000$ & no. &   \multicolumn{1}{c}{$(d)$} & $-2\,400\,000$ & no. &   \multicolumn{1}{c}{$(d)$} & $-2\,400\,000$ & no. &   \multicolumn{1}{c}{$(d)$} \\ 
\hline
54957.481039 &   -1.5 & 0.004864 & 55438.275599 &   27.5 & 0.005723 & 55919.065074 &   56.5 & 0.008751 \\ 
54963.158283 &   -1.0 & 0.000657 & 55443.943368 &   28.0 & 0.000363 & 55924.734286 &   57.0 & 0.000329 \\ 
54979.735488 &    0.0 & 0.000483 & 55454.848437 &   28.5 & 0.008557 & 55941.312919 &   58.0 & 0.000346 \\ 
54990.646340 &    0.5 & 0.005723 & 55460.523121 &   29.0 & 0.000362 & 55952.225768 &   58.5 & 0.011009 \\ 
54996.312016 &    1.0 & 0.000476 & 55471.431640 &   29.5 & 0.008218 & 55957.891019 &   59.0 & 0.000350 \\ 
55007.226717 &    1.5 & 0.006334 & 55477.103292 &   30.0 & 0.000343 & 55968.802452 &   59.5 & 0.007464 \\
\hline
\end{tabular}
\end{table*}

\begin{table*}
\caption{Times of minima of KIC 08210721}
 \label{Tab:KIC 08210721_ToM}
\begin{tabular}{@{}lrllrllrl}
\hline
BJD & Cycle  & std. dev. & BJD & Cycle  & std. dev. & BJD & Cycle  & std. dev. \\ 
$-2\,400\,000$ & no. &   \multicolumn{1}{c}{$(d)$} & $-2\,400\,000$ & no. &   \multicolumn{1}{c}{$(d)$} & $-2\,400\,000$ & no. &   \multicolumn{1}{c}{$(d)$} \\ 
\hline
54971.154947 &    0.0 & 0.000086 & 55437.837056 &   20.5 & 0.002469 & 55923.412250 &   42.0 & 0.000083 \\ 
54984.387903 &    0.5 & 0.003046 & 55447.317490 &   21.0 & 0.000089 & 55946.082371 &   43.0 & 0.000096 \\ 
54993.828520 &    1.0 & 0.000078 & 55470.003141 &   22.0 & 0.000091 & 55968.754474 &   44.0 & 0.000091 \\ 
55007.043799 &    1.5 & 0.001685 & 55483.178695 &   22.5 & 0.001799 & 55981.960914 &   44.5 & 0.002498 \\ 
55029.709993 &    2.5 & 0.002147 & 55492.679349 &   23.0 & 0.000096 & 55991.423291 &   45.0 & 0.000098 \\ 
55039.174277 &    3.0 & 0.000083 & 55505.858853 &   23.5 & 0.003050 & 56004.628906 &   45.5 & 0.003112 \\
\hline
\end{tabular}
\end{table*}

\begin{table*}
\caption{Times of minima of KIC 08938628}
 \label{Tab:KIC 08938628_ToM}
\begin{tabular}{@{}lrllrllrl}
\hline
BJD & Cycle  & std. dev. & BJD & Cycle  & std. dev. & BJD & Cycle  & std. dev. \\ 
$-2\,400\,000$ & no. &   \multicolumn{1}{c}{$(d)$} & $-2\,400\,000$ & no. &   \multicolumn{1}{c}{$(d)$} & $-2\,400\,000$ & no. &   \multicolumn{1}{c}{$(d)$} \\ 
\hline
54966.608306 &    0.0 & 0.000100 & 55505.273267 &   78.5 & 0.000185 & 55965.039724 &  145.5 & 0.000280 \\ 
54970.027946 &    0.5 & 0.000179 & 55508.715768 &   79.0 & 0.000134 & 55968.480865 &  146.0 & 0.000157 \\ 
54973.470272 &    1.0 & 0.000100 & 55512.135163 &   79.5 & 0.000204 & 55971.900319 &  146.5 & 0.000249 \\ 
54976.890029 &    1.5 & 0.000159 & 55515.577670 &   80.0 & 0.000125 & 55975.342936 &  147.0 & 0.000193 \\ 
54980.332415 &    2.0 & 0.000109 & 55518.996970 &   80.5 & 0.000250 & 55978.762464 &  147.5 & 0.000236 \\ 
54983.751969 &    2.5 & 0.000146 & 55522.439519 &   81.0 & 0.000119 & 55982.206415 &  148.0 & 0.000344 \\
\hline
\end{tabular}
\end{table*}

\addtocounter{table}{-1}
\begin{table*}
\caption{Times of minima of KIC 08938628 (continued)}
% \label{Tab:KIC 08938628_ToM}
\begin{tabular}{@{}lrllrllrl}
\hline
BJD & Cycle  & std. dev. & BJD & Cycle  & std. dev. & BJD & Cycle  & std. dev. \\ 
$-2\,400\,000$ & no. &   \multicolumn{1}{c}{$(d)$} & $-2\,400\,000$ & no. &   \multicolumn{1}{c}{$(d)$} & $-2\,400\,000$ & no. &   \multicolumn{1}{c}{$(d)$} \\ 
\hline
55289.131287 &   47.0 & 0.000113 & 55742.037234 &  113.0 & 0.000145 & 56184.636009 &  177.5 & 0.000262 \\ 
55292.550782 &   47.5 & 0.000211 & 55745.456530 &  113.5 & 0.000202 & 56188.078567 &  178.0 & 0.000235 \\ 
55295.993644 &   48.0 & 0.000110 & 55748.899197 &  114.0 & 0.000182 & 56191.497681 &  178.5 & 0.000285 \\ 
55299.413321 &   48.5 & 0.000184 & 55752.318798 &  114.5 & 0.000199 & 56194.940222 &  179.0 & 0.000217 \\ 
55302.856021 &   49.0 & 0.000121 & 55755.761219 &  115.0 & 0.000160 & 56198.359538 &  179.5 & 0.000322 \\ 
55306.275487 &   49.5 & 0.000172 & 55759.180366 &  115.5 & 0.000230 & 56201.801955 &  180.0 & 0.000197 \\
\hline
\end{tabular}
\end{table*}

\begin{table*}
\caption{Times of minima of KIC 09714358}
 \label{Tab:KIC 09714358_ToM}
\begin{tabular}{@{}lrllrllrl}
\hline
BJD & Cycle  & std. dev. & BJD & Cycle  & std. dev. & BJD & Cycle  & std. dev. \\ 
$-2\,400\,000$ & no. &   \multicolumn{1}{c}{$(d)$} & $-2\,400\,000$ & no. &   \multicolumn{1}{c}{$(d)$} & $-2\,400\,000$ & no. &   \multicolumn{1}{c}{$(d)$} \\ 
\hline
54967.385151 &    0.0 & 0.000092 & 55440.012945 &   73.0 & 0.000102 & 55838.242361 &  134.5 & 0.001075 \\ 
54970.671400 &    0.5 & 0.001074 & 55443.306248 &   73.5 & 0.001174 & 55841.420653 &  135.0 & 0.000111 \\ 
54973.866577 &    1.0 & 0.000096 & 55446.483141 &   74.0 & 0.000107 & 55844.713830 &  135.5 & 0.001315 \\ 
54977.158186 &    1.5 & 0.001277 & 55449.778819 &   74.5 & 0.001205 & 55847.891311 &  136.0 & 0.000099 \\ 
54980.350770 &    2.0 & 0.000099 & 55452.953299 &   75.0 & 0.000108 & 55851.187210 &  136.5 & 0.001112 \\ 
54983.641005 &    2.5 & 0.001099 & 55456.251168 &   75.5 & 0.001372 & 55854.361177 &  137.0 & 0.000127 \\
\hline
\end{tabular}
\end{table*}

\addtocounter{table}{-1}
\begin{table*}
\caption{Times of minima of KIC 09714358 (continued)}
% \label{Tab:KIC 09714358_ToM}
\begin{tabular}{@{}lrllrllrl}
\hline
BJD & Cycle  & std. dev. & BJD & Cycle  & std. dev. & BJD & Cycle  & std. dev. \\ 
$-2\,400\,000$ & no. &   \multicolumn{1}{c}{$(d)$} & $-2\,400\,000$ & no. &   \multicolumn{1}{c}{$(d)$} & $-2\,400\,000$ & no. &   \multicolumn{1}{c}{$(d)$} \\ 
\hline
55168.080840 &   31.0 & 0.000098 & 55660.128327 &  107.0 & 0.000095 & 56220.211105 &  193.5 & 0.001266 \\ 
55171.374607 &   31.5 & 0.001375 & 55663.428309 &  107.5 & 0.001416 & 56223.394090 &  194.0 & 0.000104 \\ 
55174.557467 &   32.0 & 0.000098 & 55666.598157 &  108.0 & 0.000088 & 56226.695468 &  194.5 & 0.001266 \\ 
55177.847857 &   32.5 & 0.001090 & 55669.899388 &  108.5 & 0.001105 & 56229.872579 &  195.0 & 0.000100 \\ 
55181.039506 &   33.0 & 0.000101 & 55673.069933 &  109.0 & 0.000096 & 56233.170360 &  195.5 & 0.001349 \\ 
55278.143935 &   48.0 & 0.000096 & 55676.371038 &  109.5 & 0.001338 & 56236.348310 &  196.0 & 0.000103 \\
\hline
\end{tabular}
\end{table*}
\clearpage

\begin{table*}
\caption{Times of minima of KIC 05771589}
 \label{Tab:KIC 05771589_ToM}
\begin{tabular}{@{}lrllrllrl}
\hline
BJD & Cycle  & std. dev. & BJD & Cycle  & std. dev. & BJD & Cycle  & std. dev. \\ 
$-2\,400\,000$ & no. &   \multicolumn{1}{c}{$(d)$} & $-2\,400\,000$ & no. &   \multicolumn{1}{c}{$(d)$} & $-2\,400\,000$ & no. &   \multicolumn{1}{c}{$(d)$} \\ 
\hline
54956.780276 &   -0.5 & 0.003637 & 55413.136525 &   42.0 & 0.004471 & 55933.937183 &   90.5 & 0.005525 \\ 
54962.077912 &    0.0 & 0.001693 & 55418.470867 &   42.5 & 0.005591 & 55939.371842 &   91.0 & 0.002458 \\ 
54967.515367 &    0.5 & 0.003108 & 55423.901897 &   43.0 & 0.003676 & 55944.666491 &   91.5 & 0.009295 \\ 
54972.862093 &    1.0 & 0.001443 & 55429.251396 &   43.5 & 0.005180 & 55955.383948 &   92.5 & 0.004924 \\ 
54978.298611 &    1.5 & 0.003318 & 55434.705218 &   44.0 & 0.002785 & 55960.827866 &   93.0 & 0.002881 \\ 
54983.651538 &    2.0 & 0.001564 & 55439.998277 &   44.5 & 0.007918 & 55966.110482 &   93.5 & 0.004330 \\
\hline
\end{tabular}
\end{table*}

\begin{table*}
\caption{Times of minima of KIC 06964043}
 \label{Tab:KIC 06964043_ToM}
\begin{tabular}{@{}lrllrllrl}
\hline
BJD & Cycle  & std. dev. & BJD & Cycle  & std. dev. & BJD & Cycle  & std. dev. \\ 
$-2\,400\,000$ & no. &   \multicolumn{1}{c}{$(d)$} & $-2\,400\,000$ & no. &   \multicolumn{1}{c}{$(d)$} & $-2\,400\,000$ & no. &   \multicolumn{1}{c}{$(d)$} \\ 
\hline
55190.173588 &    0.0 & 0.000305 & 55603.078799 &   38.5 & 0.000188 & 56010.587115 &   76.5 & 0.000195 \\ 
55195.460530 &    0.5 & 0.000237 & 55608.431983 &   39.0 & 0.000145 & 56015.844256 &   77.0 & 0.000242 \\ 
55200.886326 &    1.0 & 0.000305 & 55613.819266 &   39.5 & 0.000168 & 56021.303286 &   77.5 & 0.000192 \\ 
55206.172452 &    1.5 & 0.000336 & 55619.165008 &   40.0 & 0.000175 & 56026.563995 &   78.0 & 0.000210 \\ 
55211.601150 &    2.0 & 0.000314 & 55624.552763 &   40.5 & 0.000177 & 56032.020670 &   78.5 & 0.000341 \\ 
55222.317438 &    3.0 & 0.000308 & 55629.889603 &   41.0 & 0.000153 & 56037.290799 &   79.0 & 0.000223 \\
\hline
\end{tabular}
\end{table*}

\addtocounter{table}{-1}
\begin{table*}
\caption{Times of minima of KIC 06964043 (continued)}
% \label{Tab:KIC 06964043_ToM}
\begin{tabular}{@{}lrllrllrl}
\hline
BJD & Cycle  & std. dev. & BJD & Cycle  & std. dev. & BJD & Cycle  & std. dev. \\ 
$-2\,400\,000$ & no. &   \multicolumn{1}{c}{$(d)$} & $-2\,400\,000$ & no. &   \multicolumn{1}{c}{$(d)$} & $-2\,400\,000$ & no. &   \multicolumn{1}{c}{$(d)$} \\ 
\hline
55549.311845 &   33.5 & 0.000202 & 55967.726462 &   72.5 & 0.000163 & 56386.110902 &  111.5 & 0.000365 \\ 
55570.764524 &   35.5 & 0.000171 & 55972.979333 &   73.0 & 0.000187 & 56396.827476 &  112.5 & 0.000451 \\ 
55576.159983 &   36.0 & 0.000181 & 55978.439783 &   73.5 & 0.000204 & 56402.001462 &  113.0 & 0.000402 \\ 
55581.552540 &   36.5 & 0.000182 & 55983.693744 &   74.0 & 0.000196 & 56407.547455 &  113.5 & 0.000406 \\ 
55586.935079 &   37.0 & 0.000170 & 55989.154588 &   74.5 & 0.000220 & 56412.719525 &  114.0 & 0.000371 \\ 
55592.366837 &   37.5 & 0.000168 & 55999.871173 &   75.5 & 0.000196 & 56423.431712 &  115.0 & 0.000571 \\ 
55597.676834 &   38.0 & 0.000162 & 56005.126585 &   76.0 & 0.000247 &&& \\ 
\hline
\end{tabular}
\end{table*}

\begin{table*}
\caption{Times of minima of KIC 07668648}
 \label{Tab:KIC 07668648_ToM}
\begin{tabular}{@{}lrllrllrl}
\hline
BJD & Cycle  & std. dev. & BJD & Cycle  & std. dev. & BJD & Cycle  & std. dev. \\ 
$-2\,400\,000$ & no. &   \multicolumn{1}{c}{$(d)$} & $-2\,400\,000$ & no. &   \multicolumn{1}{c}{$(d)$} & $-2\,400\,000$ & no. &   \multicolumn{1}{c}{$(d)$} \\ 
\hline
54977.209230 &    0.5 & 0.000254 & 55450.183461 &   17.5 & 0.000252 & 55936.919306 &   35.0 & 0.000130 \\ 
54991.150587 &    1.0 & 0.000311 & 55464.008112 &   18.0 & 0.000262 & 55964.780375 &   36.0 & 0.000120 \\ 
55005.016478 &    1.5 & 0.000331 & 55477.970686 &   18.5 & 0.000263 & 55979.044185 &   36.5 & 0.000100 \\ 
55018.947371 &    2.0 & 0.000353 & 55491.814436 &   19.0 & 0.000187 & 55992.636571 &   37.0 & 0.000088 \\ 
55032.788866 &    2.5 & 0.000223 & 55505.868120 &   19.5 & 0.000161 & 56006.909113 &   37.5 & 0.000134 \\ 
55046.745774 &    3.0 & 0.000256 & 55519.691026 &   20.0 & 0.000246 & 56020.426099 &   38.0 & 0.000101 \\
\hline
\end{tabular}
\end{table*}

\begin{table*}
\caption{Times of minima of KIC 07955301}
 \label{Tab:KIC 07955301_ToM}
\begin{tabular}{@{}lrllrllrl}
\hline
BJD & Cycle  & std. dev. & BJD & Cycle  & std. dev. & BJD & Cycle  & std. dev. \\ 
$-2\,400\,000$ & no. &   \multicolumn{1}{c}{$(d)$} & $-2\,400\,000$ & no. &   \multicolumn{1}{c}{$(d)$} & $-2\,400\,000$ & no. &   \multicolumn{1}{c}{$(d)$} \\ 
\hline
54960.452670 &   -0.5 & 0.002977 & 55435.608663 &   30.5 & 0.000881 & 55918.228094 &   62.0 & 0.000568 \\ 
54967.990412 &    0.0 & 0.003681 & 55443.016479 &   31.0 & 0.000780 & 55926.194364 &   62.5 & 0.000647 \\ 
54975.771287 &    0.5 & 0.002811 & 55450.918502 &   31.5 & 0.000876 & 55933.569311 &   63.0 & 0.000610 \\ 
54983.291100 &    1.0 & 0.003409 & 55458.331839 &   32.0 & 0.000891 & 55941.542393 &   63.5 & 0.000717 \\ 
54991.084261 &    1.5 & 0.002998 & 55466.231895 &   32.5 & 0.000777 & 55948.896869 &   64.0 & 0.000645 \\ 
55006.396341 &    2.5 & 0.002751 & 55473.656481 &   33.0 & 0.000733 & 55956.866918 &   64.5 & 0.000703 \\
\hline
\end{tabular}
\end{table*}

\begin{table*}
\caption{Times of minima of KIC 04769799}
 \label{Tab:KIC 04769799_ToM}
\begin{tabular}{@{}lrllrllrl}
\hline
BJD & Cycle  & std. dev. & BJD & Cycle  & std. dev. & BJD & Cycle  & std. dev. \\ 
$-2\,400\,000$ & no. &   \multicolumn{1}{c}{$(d)$} & $-2\,400\,000$ & no. &   \multicolumn{1}{c}{$(d)$} & $-2\,400\,000$ & no. &   \multicolumn{1}{c}{$(d)$} \\ 
\hline
54956.352796 &   -0.5 & 0.029526 & 55329.116635 &   16.5 & 0.009746 & 55823.755992 &   39.0 & 0.000629 \\
54968.515191 &    0.0 & 0.000560 & 55341.308516 &   17.0 & 0.000523 & 55845.684186 &   40.0 & 0.000616 \\
54990.441825 &    1.0 & 0.000541 & 55351.038969 &   17.5 & 0.015473 & 55867.612976 &   41.0 & 0.000604 \\
55012.368264 &    2.0 & 0.000549 & 55363.239476 &   18.0 & 0.000553 & 55889.539465 &   42.0 & 0.000601 \\
55022.112405 &    2.5 & 0.013832 & 55372.971413 &   18.5 & 0.011921 & 55911.467004 &   43.0 & 0.000605 \\
55034.295863 &    3.0 & 0.000558 & 55385.168431 &   19.0 & 0.000544 & 55933.394965 &   44.0 & 0.000635 \\
\hline
\end{tabular}
\end{table*}

\begin{table*}
\caption{Times of minima of KIC 05003117}
 \label{Tab:KIC 05003117_ToM}
\begin{tabular}{@{}lrllrllrl}
\hline
BJD & Cycle  & std. dev. & BJD & Cycle  & std. dev. & BJD & Cycle  & std. dev. \\ 
$-2\,400\,000$ & no. &   \multicolumn{1}{c}{$(d)$} & $-2\,400\,000$ & no. &   \multicolumn{1}{c}{$(d)$} & $-2\,400\,000$ & no. &   \multicolumn{1}{c}{$(d)$} \\ 
\hline
54965.117553 &   -0.5 & 0.022936 & 55491.646335 &   13.5 & 0.004898 & 55963.938558 &   26.0 & 0.000328 \\ 
54986.058450 &    0.0 & 0.000542 & 55512.663403 &   14.0 & 0.000462 & 55980.572670 &   26.5 & 0.008580 \\ 
55023.674982 &    1.0 & 0.000581 & 55529.259663 &   14.5 & 0.006331 & 56001.544298 &   27.0 & 0.000359 \\ 
55040.349903 &    1.5 & 0.007223 & 55550.271711 &   15.0 & 0.000484 & 56018.169318 &   27.5 & 0.013006 \\ 
55061.292921 &    2.0 & 0.000595 & 55587.879247 &   16.0 & 0.000432 & 56039.147356 &   28.0 & 0.000281 \\ 
55098.911800 &    3.0 & 0.000584 & 55604.471413 &   16.5 & 0.010641 & 56055.772815 &   28.5 & 0.006041 \\
\hline
\end{tabular}
\end{table*}

\begin{table*}
\caption{Times of minima of KIC 05731312}
 \label{Tab:KIC 05731312_ToM}
\begin{tabular}{@{}lrllrllrl}
\hline
BJD & Cycle  & std. dev. & BJD & Cycle  & std. dev. & BJD & Cycle  & std. dev. \\ 
$-2\,400\,000$ & no. &   \multicolumn{1}{c}{$(d)$} & $-2\,400\,000$ & no. &   \multicolumn{1}{c}{$(d)$} & $-2\,400\,000$ & no. &   \multicolumn{1}{c}{$(d)$} \\ 
\hline
54966.175075 &   -0.5 & 0.000258 & 55389.253239 &   53.0 & 0.000019 & 55816.439608 &  106.5 & 0.000391 \\ 
54968.092761 &    0.0 & 0.000019 & 55395.282538 &   53.5 & 0.000266 & 55818.355067 &  107.0 & 0.000024 \\ 
54974.122050 &    0.5 & 0.000236 & 55397.199631 &   54.0 & 0.000019 & 55824.385890 &  107.5 & 0.000359 \\ 
54976.039127 &    1.0 & 0.000019 & 55403.228913 &   54.5 & 0.000234 & 55826.301094 &  108.0 & 0.000022 \\ 
54982.068903 &    1.5 & 0.000226 & 55405.146081 &   55.0 & 0.000019 & 55832.333141 &  108.5 & 0.000357 \\ 
54983.985463 &    2.0 & 0.000019 & 55411.175125 &   55.5 & 0.000228 & 55834.247096 &  109.0 & 0.000032 \\
\hline
\end{tabular}
\end{table*}

\addtocounter{table}{-1}
\begin{table*}
\caption{Times of minima of KIC 05731312 (continued)}
% \label{Tab:KIC 05731312_ToM}
\begin{tabular}{@{}lrllrllrl}
\hline
BJD & Cycle  & std. dev. & BJD & Cycle  & std. dev. & BJD & Cycle  & std. dev. \\ 
$-2\,400\,000$ & no. &   \multicolumn{1}{c}{$(d)$} & $-2\,400\,000$ & no. &   \multicolumn{1}{c}{$(d)$} & $-2\,400\,000$ & no. &   \multicolumn{1}{c}{$(d)$} \\ 
\hline
55214.431837 &   31.0 & 0.000026 & 55649.564014 &   85.5 & 0.000369 & 56253.496328 &  161.5 & 0.000470 \\ 
55220.462171 &   31.5 & 0.000278 & 55651.483076 &   86.0 & 0.000020 & 56255.406420 &  162.0 & 0.000023 \\ 
55222.378315 &   32.0 & 0.000021 & 55657.509993 &   86.5 & 0.000247 & 56261.443196 &  162.5 & 0.000471 \\ 
55228.408542 &   32.5 & 0.000239 & 55659.429124 &   87.0 & 0.000021 & 56263.352962 &  163.0 & 0.000023 \\ 
55236.355244 &   33.5 & 0.000240 & 55665.456850 &   87.5 & 0.000314 & 56269.389212 &  163.5 & 0.000475 \\ 
55238.271157 &   34.0 & 0.000023 & 55667.375141 &   88.0 & 0.000021 & 56271.299378 &  164.0 & 0.000024 \\
\hline
\end{tabular}
\end{table*}

\begin{table*}
\caption{Times of minima of KIC 07670617}
 \label{Tab:KIC 07670617_ToM}
\begin{tabular}{@{}lrllrllrl}
\hline
BJD & Cycle  & std. dev. & BJD & Cycle  & std. dev. & BJD & Cycle  & std. dev. \\ 
$-2\,400\,000$ & no. &   \multicolumn{1}{c}{$(d)$} & $-2\,400\,000$ & no. &   \multicolumn{1}{c}{$(d)$} & $-2\,400\,000$ & no. &   \multicolumn{1}{c}{$(d)$} \\ 
\hline
54969.147865 &    0.0 & 0.000541 & 55379.507551 &   16.5 & 0.001365 & 55799.579499 &   33.5 & 0.001800 \\ 
54984.282574 &    0.5 & 0.001598 & 55389.084292 &   17.0 & 0.000619 & 55809.069767 &   34.0 & 0.000895 \\ 
54993.850120 &    1.0 & 0.000557 & 55404.207213 &   17.5 & 0.001583 & 55824.281472 &   34.5 & 0.001743 \\ 
55008.985311 &    1.5 & 0.001413 & 55413.792149 &   18.0 & 0.000528 & 55848.984978 &   35.5 & 0.001760 \\ 
55018.551378 &    2.0 & 0.000442 & 55428.906547 &   18.5 & 0.001525 & 55858.481864 &   36.0 & 0.000975 \\ 
55033.686304 &    2.5 & 0.001642 & 55438.497230 &   19.0 & 0.000514 & 55873.687560 &   36.5 & 0.001691 \\
\hline
\end{tabular}
\end{table*}

\begin{table*}
\caption{Times of minima of KIC 08143170}
 \label{Tab:KIC 08143170_ToM}
\begin{tabular}{@{}lrllrllrl}
\hline
BJD & Cycle  & std. dev. & BJD & Cycle  & std. dev. & BJD & Cycle  & std. dev. \\ 
$-2\,400\,000$ & no. &   \multicolumn{1}{c}{$(d)$} & $-2\,400\,000$ & no. &   \multicolumn{1}{c}{$(d)$} & $-2\,400\,000$ & no. &   \multicolumn{1}{c}{$(d)$} \\ 
\hline
54954.736584 &   -0.5 & 0.000814 & 55472.871297 &   17.5 & 0.000852 & 55948.841491 &   34.0 & 0.000368 \\ 
54970.117733 &    0.0 & 0.000347 & 55488.265941 &   18.0 & 0.000382 & 55962.227443 &   34.5 & 0.000902 \\ 
54983.521447 &    0.5 & 0.000900 & 55501.656937 &   18.5 & 0.000852 & 55977.627461 &   35.0 & 0.000389 \\ 
55012.305733 &    1.5 & 0.000920 & 55517.052226 &   19.0 & 0.000398 & 55991.015318 &   35.5 & 0.001040 \\ 
55027.690786 &    2.0 & 0.000393 & 55530.442539 &   19.5 & 0.000907 & 56006.411096 &   36.0 & 0.000429 \\ 
55041.091233 &    2.5 & 0.000849 & 55545.838044 &   20.0 & 0.000416 & 56019.801248 &   36.5 & 0.000949 \\
\hline
\end{tabular}
\end{table*}

\begin{table*}
\caption{Times of minima of KIC 09715925}
 \label{Tab:KIC 09715925_ToM}
\begin{tabular}{@{}lrllrllrl}
\hline
BJD & Cycle  & std. dev. & BJD & Cycle  & std. dev. & BJD & Cycle  & std. dev. \\ 
$-2\,400\,000$ & no. &   \multicolumn{1}{c}{$(d)$} & $-2\,400\,000$ & no. &   \multicolumn{1}{c}{$(d)$} & $-2\,400\,000$ & no. &   \multicolumn{1}{c}{$(d)$} \\ 
\hline
55373.471096 &   59.5 & 0.003969 & 55623.441514 &   99.0 & 0.000306 & 55859.208163 &  136.5 & 0.003308 \\ 
55377.421033 &   60.0 & 0.000359 & 55625.806474 &   99.5 & 0.009576 & 55869.466085 &  138.0 & 0.000296 \\ 
55379.786092 &   60.5 & 0.003585 & 55629.749909 &  100.0 & 0.000316 & 55871.827502 &  138.5 & 0.005181 \\ 
55383.729648 &   61.0 & 0.000247 & 55632.112640 &  100.5 & 0.003235 & 55875.774257 &  139.0 & 0.000302 \\ 
55386.092949 &   61.5 & 0.002409 & 55642.365520 &  102.0 & 0.000321 & 55878.135102 &  139.5 & 0.004363 \\ 
55390.036982 &   62.0 & 0.000329 & 55644.722929 &  102.5 & 0.003856 & 55882.082842 &  140.0 & 0.000211 \\
\hline
\end{tabular}
\end{table*}

\addtocounter{table}{-1}
\begin{table*}
\caption{Times of minima of KIC 09715925 (continued)}
% \label{Tab:KIC 09715925_ToM}
\begin{tabular}{@{}lrllrllrl}
\hline
BJD & Cycle  & std. dev. & BJD & Cycle  & std. dev. & BJD & Cycle  & std. dev. \\ 
$-2\,400\,000$ & no. &   \multicolumn{1}{c}{$(d)$} & $-2\,400\,000$ & no. &   \multicolumn{1}{c}{$(d)$} & $-2\,400\,000$ & no. &   \multicolumn{1}{c}{$(d)$} \\ 
\hline
55585.592150 &   93.0 & 0.000321 & 55825.305596 &  131.0 & 0.000305 & 56172.262480 &  186.0 & 0.000192 \\ 
55587.953183 &   93.5 & 0.002425 & 55827.670968 &  131.5 & 0.001954 & 56174.625267 &  186.5 & 0.002963 \\ 
55591.900440 &   94.0 & 0.000305 & 55831.614533 &  132.0 & 0.000217 & 56178.569711 &  187.0 & 0.000245 \\ 
55598.209137 &   95.0 & 0.000293 & 55837.923892 &  133.0 & 0.000319 & 56180.938148 &  187.5 & 0.003715 \\ 
55600.572543 &   95.5 & 0.003230 & 55840.286257 &  133.5 & 0.003141 & 56184.878500 &  188.0 & 0.000276 \\ 
55604.516466 &   96.0 & 0.000325 & 55844.232887 &  134.0 & 0.000326 & 56191.186336 &  189.0 & 0.000281 \\ 
55606.880699 &   96.5 & 0.002296 & 55846.595972 &  134.5 & 0.003098 & 56193.547240 &  189.5 & 0.002422 \\ 
55610.826000 &   97.0 & 0.000392 & 55850.540711 &  135.0 & 0.000312 & 56197.495444 &  190.0 & 0.000271 \\ 
55613.185160 &   97.5 & 0.002205 & 55852.900128 &  135.5 & 0.004199 & 56199.861644 &  190.5 & 0.003518 \\ 
55617.132915 &   98.0 & 0.000296 & 55856.849224 &  136.0 & 0.000217 & 56203.803666 &  191.0 & 0.000264 \\ 
55619.494326 &   98.5 & 0.003300 &&&&& \\ 
\hline
\end{tabular}
\end{table*}

\begin{table*}
\caption{Times of minima of KIC 09963009}
 \label{Tab:KIC 09963009_ToM}
\begin{tabular}{@{}lrllrllrl}
\hline
BJD & Cycle  & std. dev. & BJD & Cycle  & std. dev. & BJD & Cycle  & std. dev. \\ 
$-2\,400\,000$ & no. &   \multicolumn{1}{c}{$(d)$} & $-2\,400\,000$ & no. &   \multicolumn{1}{c}{$(d)$} & $-2\,400\,000$ & no. &   \multicolumn{1}{c}{$(d)$} \\ 
\hline
54967.176456 &   -0.5 & 0.000597 & 55327.817624 &    8.5 & 0.000661 & 55827.478806 &   21.0 & 0.000186 \\ 
54986.013827 &    0.0 & 0.000177 & 55346.645727 &    9.0 & 0.000176 & 55947.687791 &   24.0 & 0.000168 \\ 
55007.246116 &    0.5 & 0.000642 & 55367.888912 &    9.5 & 0.000648 & 55968.969648 &   24.5 & 0.000771 \\ 
55026.084742 &    1.0 & 0.000188 & 55386.714735 &   10.0 & 0.000198 & 56009.038907 &   25.5 & 0.000813 \\ 
55047.317754 &    1.5 & 0.000656 & 55407.959910 &   10.5 & 0.000710 & 56027.827026 &   26.0 & 0.000171 \\ 
55066.155794 &    2.0 & 0.000186 & 55426.783904 &   11.0 & 0.000199 & 56049.110172 &   26.5 & 0.000841 \\
\hline
\end{tabular}
\end{table*}

\begin{table*}
\caption{Times of minima of KIC 10268809}
 \label{Tab:KIC 10268809_ToM}
\begin{tabular}{@{}lrllrllrl}
\hline
BJD & Cycle  & std. dev. & BJD & Cycle  & std. dev. & BJD & Cycle  & std. dev. \\ 
$-2\,400\,000$ & no. &   \multicolumn{1}{c}{$(d)$} & $-2\,400\,000$ & no. &   \multicolumn{1}{c}{$(d)$} & $-2\,400\,000$ & no. &   \multicolumn{1}{c}{$(d)$} \\ 
\hline
54972.004648 &    0.0 & 0.000518 & 55441.468207 &   19.0 & 0.000539 & 55951.964162 &   39.5 & 0.000530 \\ 
54988.320182 &    0.5 & 0.000484 & 55457.793089 &   19.5 & 0.000452 & 55960.390706 &   40.0 & 0.000761 \\ 
54996.713272 &    1.0 & 0.000557 & 55466.176973 &   20.0 & 0.000572 & 55976.669382 &   40.5 & 0.000591 \\ 
55013.028636 &    1.5 & 0.000526 & 55482.501443 &   20.5 & 0.000470 & 55985.103504 &   41.0 & 0.000771 \\ 
55021.421595 &    2.0 & 0.000422 & 55490.886734 &   21.0 & 0.000504 & 56001.374294 &   41.5 & 0.000553 \\ 
55037.738227 &    2.5 & 0.000467 & 55507.210243 &   21.5 & 0.000498 & 56009.817414 &   42.0 & 0.000741 \\
\hline
\end{tabular}
\end{table*}

\begin{table*}
\caption{Times of minima of KIC 10319590}
 \label{Tab:KIC 10319590_ToM}
\begin{tabular}{@{}lrllrllrl}
\hline
BJD & Cycle  & std. dev. & BJD & Cycle  & std. dev. & BJD & Cycle  & std. dev. \\ 
$-2\,400\,000$ & no. &   \multicolumn{1}{c}{$(d)$} & $-2\,400\,000$ & no. &   \multicolumn{1}{c}{$(d)$} & $-2\,400\,000$ & no. &   \multicolumn{1}{c}{$(d)$} \\ 
\hline
54965.709958 &    0.0 & 0.000075 & 55104.429240 &    6.5 & 0.000254 & 55253.675734 &   13.5 & 0.000317 \\ 
54976.508639 &    0.5 & 0.000230 & 55114.971599 &    7.0 & 0.000074 & 55264.234574 &   14.0 & 0.000107 \\ 
54987.043911 &    1.0 & 0.000073 & 55125.742347 &    7.5 & 0.000254 & 55275.008860 &   14.5 & 0.000319 \\ 
54997.835895 &    1.5 & 0.000314 & 55136.282104 &    8.0 & 0.000077 & 55285.572163 &   15.0 & 0.000109 \\ 
55008.376422 &    2.0 & 0.000074 & 55147.056224 &    8.5 & 0.000268 & 55296.333398 &   15.5 & 0.000289 \\ 
55019.158496 &    2.5 & 0.000292 & 55157.592709 &    9.0 & 0.000087 & 55306.895914 &   16.0 & 0.000106 \\ 
55029.704721 &    3.0 & 0.000072 & 55168.371034 &    9.5 & 0.000276 & 55317.645614 &   16.5 & 0.000354 \\ 
55040.481870 &    3.5 & 0.000322 & 55178.906945 &   10.0 & 0.000095 & 55328.202717 &   17.0 & 0.000161 \\ 
55051.029560 &    4.0 & 0.000057 & 55189.689480 &   10.5 & 0.000259 & 55338.949932 &   17.5 & 0.000531 \\ 
55061.798825 &    4.5 & 0.000239 & 55200.226280 &   11.0 & 0.000113 & 55349.504280 &   18.0 & 0.000227 \\ 
55072.345750 &    5.0 & 0.000070 & 55211.012012 &   11.5 & 0.000317 & 55360.265468 &   18.5 & 0.001346 \\ 
55083.115168 &    5.5 & 0.000266 & 55221.554686 &   12.0 & 0.000112 & 55370.813909 &   19.0 & 0.000559 \\ 
55093.659956 &    6.0 & 0.000072 & 55242.891584 &   13.0 & 0.000119 &&& \\ 
\hline
\end{tabular}
\end{table*}

\clearpage

\begin{table*}
\caption{Times of minima of KIC 10979716}
 \label{Tab:KIC 10979716_ToM}
\begin{tabular}{@{}lrllrllrl}
\hline
BJD & Cycle  & std. dev. & BJD & Cycle  & std. dev. & BJD & Cycle  & std. dev. \\ 
$-2\,400\,000$ & no. &   \multicolumn{1}{c}{$(d)$} & $-2\,400\,000$ & no. &   \multicolumn{1}{c}{$(d)$} & $-2\,400\,000$ & no. &   \multicolumn{1}{c}{$(d)$} \\ 
\hline
54967.091539 &    0.0 & 0.000163 & 55383.750208 &   39.0 & 0.000174 & 55816.626831 &   79.5 & 0.000239 \\ 
54972.573426 &    0.5 & 0.000244 & 55389.240914 &   39.5 & 0.000253 & 55821.815682 &   80.0 & 0.000147 \\ 
54977.776258 &    1.0 & 0.000178 & 55394.435276 &   40.0 & 0.000169 & 55827.311352 &   80.5 & 0.000227 \\ 
54983.257761 &    1.5 & 0.000248 & 55405.118123 &   41.0 & 0.000157 & 55832.497204 &   81.0 & 0.000202 \\ 
54988.459713 &    2.0 & 0.000169 & 55410.608964 &   41.5 & 0.000249 & 55837.995807 &   81.5 & 0.000266 \\ 
54993.941291 &    2.5 & 0.000272 & 55415.801531 &   42.0 & 0.000150 & 55843.183902 &   82.0 & 0.000182 \\
\hline
\end{tabular}
\end{table*}

\addtocounter{table}{-1}
\begin{table*}
\caption{Times of minima of KIC 10979716 (continued)}
% \label{Tab:KIC 10979716_ToM}
\begin{tabular}{@{}lrllrllrl}
\hline
BJD & Cycle  & std. dev. & BJD & Cycle  & std. dev. & BJD & Cycle  & std. dev. \\ 
$-2\,400\,000$ & no. &   \multicolumn{1}{c}{$(d)$} & $-2\,400\,000$ & no. &   \multicolumn{1}{c}{$(d)$} & $-2\,400\,000$ & no. &   \multicolumn{1}{c}{$(d)$} \\ 
\hline
55298.283197 &   31.0 & 0.000157 & 55741.839381 &   72.5 & 0.000238 & 56345.318952 &  129.0 & 0.000159 \\ 
55303.770310 &   31.5 & 0.000221 & 55747.025914 &   73.0 & 0.000151 & 56350.817171 &  129.5 & 0.000249 \\ 
55314.454283 &   32.5 & 0.000292 & 55752.523753 &   73.5 & 0.000232 & 56356.002181 &  130.0 & 0.000159 \\ 
55319.650804 &   33.0 & 0.000163 & 55757.710861 &   74.0 & 0.000149 & 56361.501217 &  130.5 & 0.000259 \\ 
55325.137671 &   33.5 & 0.000248 & 55763.207156 &   74.5 & 0.000243 & 56366.686229 &  131.0 & 0.000245 \\ 
55330.334121 &   34.0 & 0.000163 & 55768.394537 &   75.0 & 0.000154 & 56372.185402 &  131.5 & 0.000267 \\ 
55335.821342 &   34.5 & 0.000234 & 55773.891675 &   75.5 & 0.000286 & 56377.369271 &  132.0 & 0.000174 \\ 
55341.017216 &   35.0 & 0.000160 & 55779.078655 &   76.0 & 0.000164 & 56382.869043 &  132.5 & 0.000215 \\ 
55346.505189 &   35.5 & 0.000253 & 55784.575851 &   76.5 & 0.000200 & 56388.053365 &  133.0 & 0.000170 \\ 
55351.700769 &   36.0 & 0.000156 & 55789.762905 &   77.0 & 0.000156 & 56393.552667 &  133.5 & 0.000196 \\ 
55357.189814 &   36.5 & 0.000267 & 55795.260399 &   77.5 & 0.000288 & 56398.736451 &  134.0 & 0.000151 \\ 
55362.384155 &   37.0 & 0.000150 & 55800.447482 &   78.0 & 0.000155 & 56404.236231 &  134.5 & 0.000263 \\ 
55367.873338 &   37.5 & 0.000200 & 55805.942897 &   78.5 & 0.000238 & 56409.419438 &  135.0 & 0.000204 \\ 
55373.067647 &   38.0 & 0.000150 & 55811.131821 &   79.0 & 0.000145 & 56420.102998 &  136.0 & 0.000154 \\ 
55378.556462 &   38.5 & 0.000280 &&&&& \\ 
\hline
\end{tabular}
\end{table*}

\begin{table*}
\caption{Times of minima of KIC 11519226}
 \label{Tab:KIC 11519226_ToM}
\begin{tabular}{@{}lrllrllrl}
\hline
BJD & Cycle  & std. dev. & BJD & Cycle  & std. dev. & BJD & Cycle  & std. dev. \\ 
$-2\,400\,000$ & no. &   \multicolumn{1}{c}{$(d)$} & $-2\,400\,000$ & no. &   \multicolumn{1}{c}{$(d)$} & $-2\,400\,000$ & no. &   \multicolumn{1}{c}{$(d)$} \\ 
\hline
54959.289801 &   -0.5 & 0.000457 & 55446.870324 &   21.5 & 0.000414 & 55934.375959 &   43.5 & 0.000474 \\ 
54972.973573 &    0.0 & 0.000541 & 55460.548055 &   22.0 & 0.000383 & 55948.104251 &   44.0 & 0.000421 \\ 
54981.444703 &    0.5 & 0.000419 & 55469.029171 &   22.5 & 0.000366 & 55956.538133 &   44.5 & 0.000435 \\ 
54995.147925 &    1.0 & 0.000453 & 55482.708646 &   23.0 & 0.000371 & 55970.267421 &   45.0 & 0.000405 \\ 
55003.600088 &    1.5 & 0.000458 & 55491.188956 &   23.5 & 0.000388 & 55978.698399 &   45.5 & 0.000355 \\ 
55017.325868 &    2.0 & 0.000429 & 55504.869658 &   24.0 & 0.000396 & 55992.428223 &   46.0 & 0.000613 \\
\hline
\end{tabular}
\end{table*}

\begin{table*}
\caption{Times of minima of KIC 12356914}
 \label{Tab:KIC 12356914_ToM}
\begin{tabular}{@{}lrllrllrl}
\hline
BJD & Cycle  & std. dev. & BJD & Cycle  & std. dev. & BJD & Cycle  & std. dev. \\ 
$-2\,400\,000$ & no. &   \multicolumn{1}{c}{$(d)$} & $-2\,400\,000$ & no. &   \multicolumn{1}{c}{$(d)$} & $-2\,400\,000$ & no. &   \multicolumn{1}{c}{$(d)$} \\ 
\hline
54965.165041 &   -0.5 & 0.001429 & 55386.125528 &   15.0 & 0.000149 & 55795.740758 &   30.0 & 0.000151 \\ 
54976.503539 &    0.0 & 0.000134 & 55402.097749 &   15.5 & 0.001560 & 55811.716472 &   30.5 & 0.001532 \\ 
54992.471254 &    0.5 & 0.001380 & 55413.432583 &   16.0 & 0.000154 & 55823.058749 &   31.0 & 0.000150 \\ 
55003.812342 &    1.0 & 0.000146 & 55429.405573 &   16.5 & 0.001534 & 55839.016170 &   31.5 & 0.001706 \\ 
55019.780553 &    1.5 & 0.001607 & 55440.739496 &   17.0 & 0.000149 & 55850.376768 &   32.0 & 0.000162 \\ 
55031.121111 &    2.0 & 0.000147 & 55456.715283 &   17.5 & 0.001473 & 55866.322654 &   32.5 & 0.001656 \\
\hline
\end{tabular}
\end{table*}

\end{document}